\documentclass[12pt,preprint]{aastex}
\usepackage{import}
\usepackage{graphicx}
\pdfoutput=1
\usepackage[dvipsnames,svgnames]{xcolor} 
\usepackage{natbib} 
\usepackage{hyperref}
\usepackage{amsmath}
\usepackage{multirow}
\usepackage{accents}
\hypersetup{    
  colorlinks      = true,
  linkcolor       = {black},
  linkbordercolor = {white},
  citecolor       = {black},
  citebordercolor = {white},
  urlcolor        = {blue},
  urlbordercolor  = {white},
}
\usepackage[figure,figure*]{hypcap}
\usepackage{soul} 
\usepackage{amsmath}
\usepackage{xspace}
\usepackage{lineno}
\usepackage{verbatim}

\usepackage{aas_macros}
\newcommand{\simlt}{\lower.5ex\hbox{$\; \buildrel < \over \sim \;$}}
\newcommand{\simgt}{\lower.5ex\hbox{$\; \buildrel > \over \sim \;$}}

\def\pvm#1{
}

\begin{document}
\begin{titlepage}
\pagestyle{empty}


\title{The Dark Energy Camera}
 
\author{
B.~Flaugher\altaffilmark{1},
H.~T.~Diehl\altaffilmark{1,*},
K.~Honscheid\altaffilmark{2,3},
T.~M.~C.~Abbott\altaffilmark{4},
O.~Alvarez\altaffilmark{1},
R.~Angstadt\altaffilmark{1},
J.~T.~Annis\altaffilmark{1},
M.~Antonik\altaffilmark{5},
O.~Ballester\altaffilmark{6},
L.~Beaufore\altaffilmark{3}
% ...
\\ \vspace{0.2cm} (The DES Collaboration) \\
}
 
\altaffiltext{*}{diehl@fnal.gov}
\altaffiltext{1}{Fermi National Accelerator Laboratory, P.~O.~Box 500, Batavia, IL 60510, USA}
\altaffiltext{2}{Center for Cosmology and Astro-Particle Physics, The Ohio State University, Columbus, OH 43210, USA}
\altaffiltext{3}{Department of Physics, The Ohio State University, Columbus, OH 43210, USA}
\altaffiltext{4}{Cerro Tololo Inter-American Observatory, National Optical Astronomy Observatory, Casilla 603, La Serena, Chile}
\altaffiltext{5}{Department of Physics \& Astronomy, University College London, Gower Street, London, WC1E 6BT, UK}
\altaffiltext{6}{Institut de F\'{\i}sica d'Altes Energies, Universitat Aut\`onoma de Barcelona, E-08193 Bellaterra, Barcelona, Spain}
% ...

\clearpage
 
\begin{abstract}
The Dark Energy Camera is a new imager with a 2.2-degree diameter field of view mounted at the prime focus of the Victor M. Blanco 
4-meter telescope on Cerro Tololo near La Serena, Chile.  The camera was designed and constructed by the Dark Energy Survey Collaboration, and meets or exceeds the stringent requirements designed for the wide-field and supernova surveys for which the collaboration uses it.   The camera consists of a five element optical corrector, seven filters, a shutter with a 60 cm aperture, and a CCD focal plane of 250-$\mu$m thick fully-depleted CCDs cooled inside a vacuum Dewar.  The 570 Mpixel focal plane comprises  62 2k$\times$4k CCDs for imaging and 12 2k$\times$2k  CCDs for guiding and focus.  The CCDs have  $15 \mu {\rm m} \times 15 \mu {\rm m}$ pixels with a plate scale of 0.263\arcsec per pixel. A hexapod system provides state-of-the-art focus and alignment capability. The camera is read out in 20 seconds with 6-9 electrons readout noise.   This paper provides a technical description of the camera's engineering, construction, installation,  and current status.
\end{abstract}

\maketitle

\end{titlepage}
\cleardoublepage
\pagenumbering{arabic}          

\section{Introduction}\label{s1:Intro}

\begin{figure}[h]
\begin{center}
\includegraphics[scale=0.4]{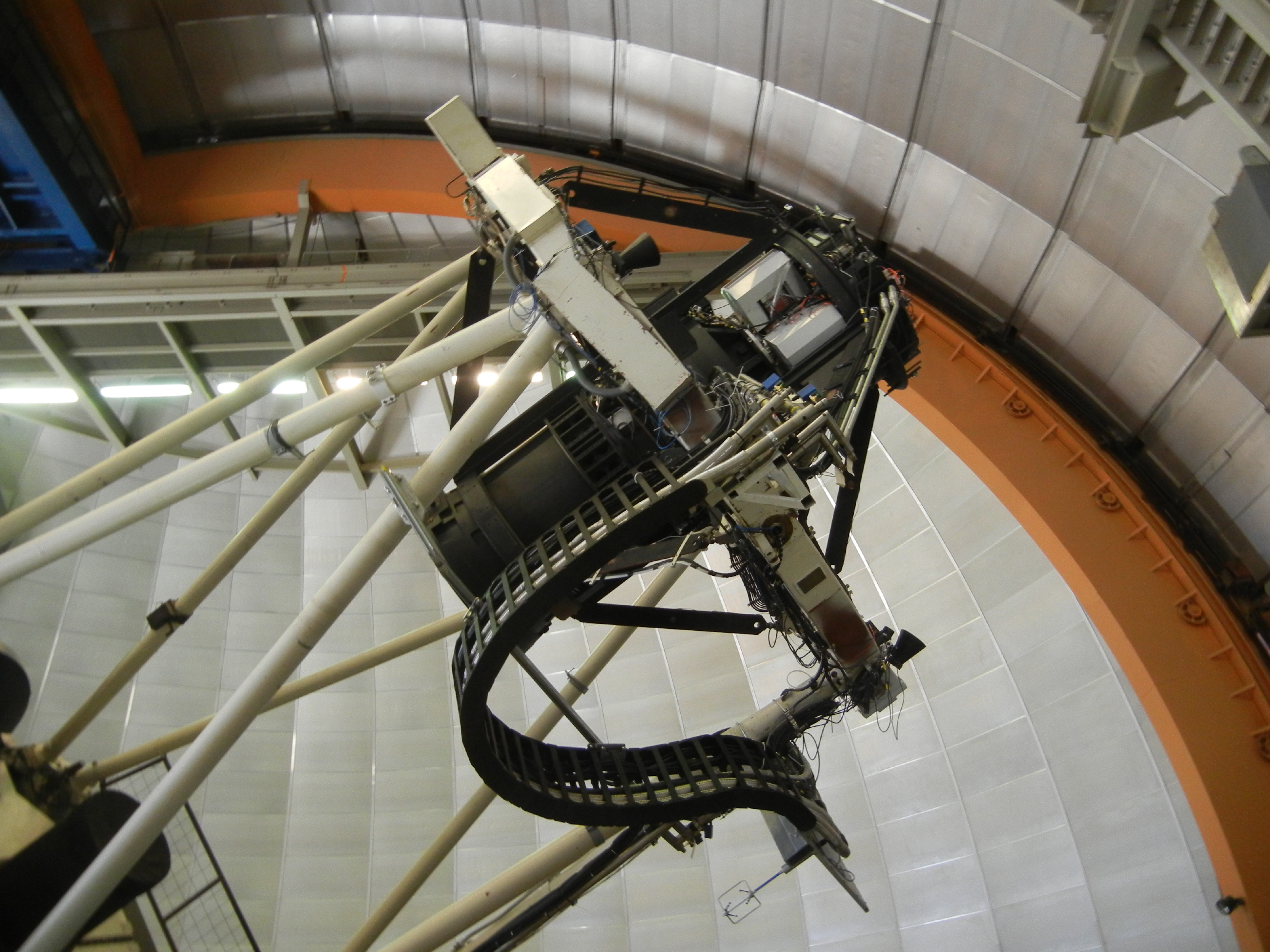}
\caption{The Dark Energy Camera is mounted at the Prime Focus of the Blanco 4m telescope at CTIO.  The primary mirror is just out of the photo, low and to the left. The camera assembly, including the support cage, is approximately 3.6 meters  long and is secured to the inner telescope ring. The camera, not including the support cage and counterweights, weighs approximately 4350 kgs. }
\label{fig:camerasideview} 
\end{center}
\end{figure}

The Dark Energy Camera, DECam, is a 570 Mpixel, 2.2-degree field-of-view camera currently installed and operating as a survey and community instrument on the 4-meter Victor M.  Blanco telescope at the Cerro Tololo Inter American Observatory (CTIO). See Fig.~\ref{fig:camerasideview}.  DECam was designed and constructed by the Dark Energy Survey (DES) collaboration with the primary goal of studying the nature of dark energy using four complementary probes:  galaxy clusters, weak lensing, Type Ia supernovae and baryon acoustic oscillations. In exchange for the camera, the DES collaboration was allocated 105 nights per year of telescope time over the next 5 years to perform a deep and wide photometric survey of the southern Galactic cap.  This Dark Energy Survey consists of a wide field survey of 5000 sq. deg. and a 30 sq. deg. area for detection of supernovae.   The survey field was designed to include complete overlap with the SZ cluster survey area covered by the South Pole Telescope~\citep{spt2009} (SPT) to provide additional constraints on the clusters measured by both surveys. It also overlaps  with part of SDSS~\citep{sdssdr9} stripe 82 to provide tight constraints on the survey photometric calibration.  DES will obtain photometric redshifts out to redshift of $\sim 1.2$ for over 300 million galaxies, 100,000 galaxy clusters and about 3000 type Ia SNe.  DES represents an increase in volume over SDSS by roughly a factor of 7.  In the parlance of the Dark Energy Task Force~\citep{detf}, DES is a Stage III project and will improve the Dark Energy Task Force figure of merit, the area of the ellipse formed by the  un-excluded limits of $w$ and $w^{\prime}$, by a factor of 3-5 over stage 2 projects.  

\begin{figure}[h]
\begin{center}
\includegraphics[scale=0.7]{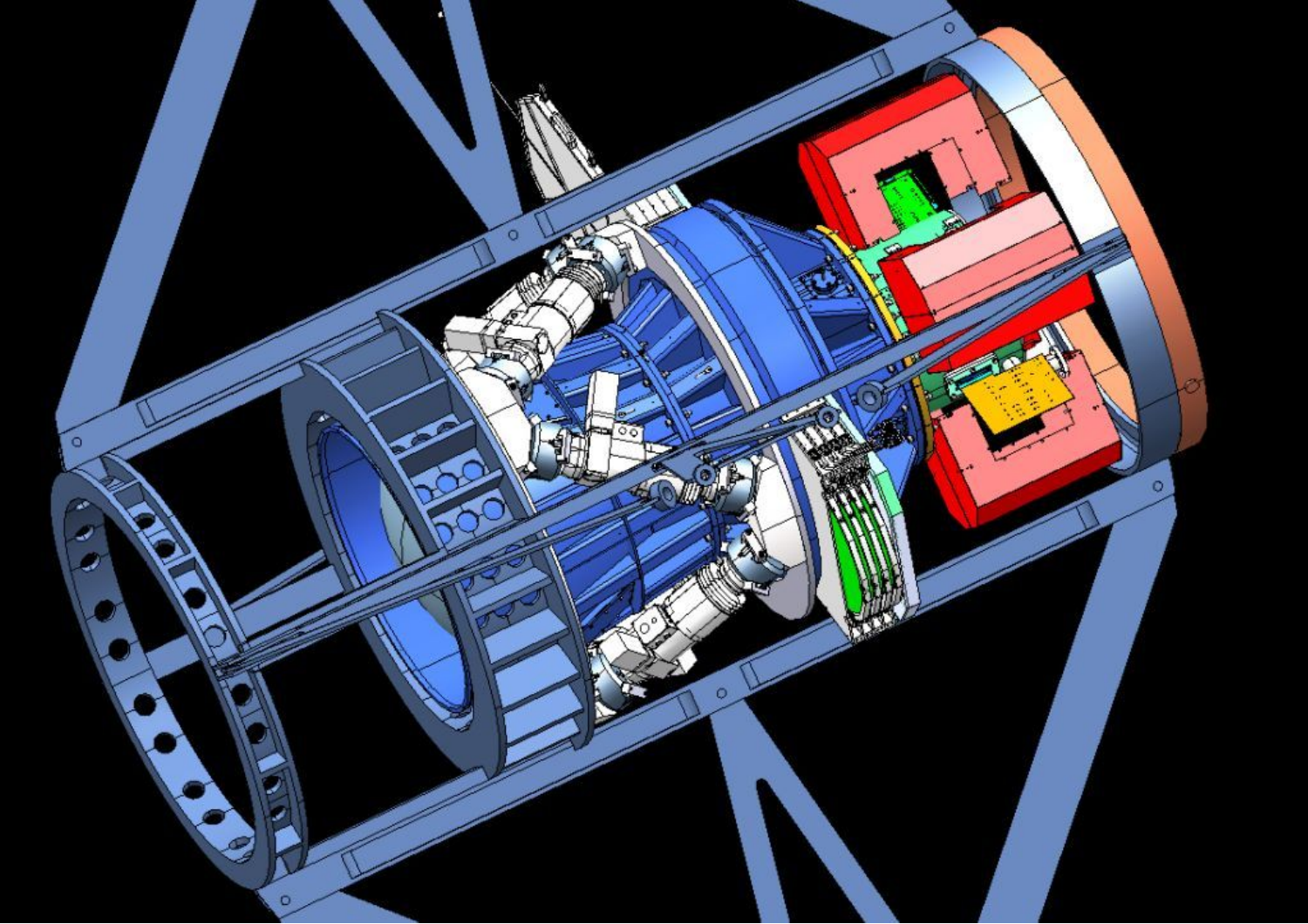}
\caption{The Dark Energy Camera and Prime Focus Cage. The primary mirror (not shown) is to the left side of the camera. Listing major components starting from the right side of the diagram, the imager vessel is green. The electronics crates are red. The optical elements are supported by the barrel (blue).  The filter changer (grey) has the sides removed so that the filters (green) can be seen in the out position. The arms of the hexapod are white. The crown of the 1st corrector element  (C1) can be seen at the left side of the barrel. The camera is attached to the cage at the heavy-duty ``hexapod ring". The cage is attached to the telescope by the four ``fin" structures, which are also shown.}
\label{fig:cage} 
\end{center}
\end{figure}

The high level requirements on the DECam design were driven by the need to survey a 5000 sq. deg. area in a total of 525 nights, with excellent image quality, high sensitivity in the near infrared, and low readout noise.   To meet these requirements the new camera has a 3 sq. deg. field of view, a new 5 lens optical corrector, and 250-micron thick fully-depleted red-sensitive CCDs.  Photometric redshifts are obtained using 5 filters (g, r, i, z, and Y-band) that span the wavelength range from 400-1065 nm.   The focal plane includes 62 of the $2\rm{k} \times 4 \rm{k}$ CCDs that are used for imaging, and 12 smaller format $2\rm{k} \times 2 \rm{k}$ CCDs for guiding and focus/alignment. The five lens optical corrector is supported in a steel barrel and mounted to the prime focus cage with a hexapod that provides focus, lateral positioning, and tip/tilt capabilities.  Figure \ref{fig:cage} shows a schematic of DECam in the new prime focus cage. 

The shutter and filter changer are located between the 3rd and 4th lenses.  The CCDs are cooled with a closed loop liquid nitrogen system and housed in a vacuum vessel mounted to the corrector barrel.  The fifth lens of the corrector also serves as the window of the vessel.  The CCD electronics are housed in thermally controlled crates mounted to the CCD vessel.  The prime focus cage was redesigned to provide greater stiffness and other features specific to DECam while maintaining the ability to support operations with a secondary mirror providing a Cassegrain focus, as an alternative to DECam.  

Funding for the DECam construction was provided primarily by the Department of Energy, with significant contributions from international partners and US universities.  The DES Collaboration formed in 2004 and began R\&D as well as the review and approval processes of the various agencies and funding sources.  In 2007 DECam received funding from STFC (UK) and in 2008 the project received approval from DOE to initiate construction.  In 2012 DECam was completed and installed on the Blanco telescope, with first light in September 2012. 

This paper describes the DECam design and construction, testing and performance in the lab and the installation at the prime focus of the Blanco 4m telescope. Some of the experiences gained during the first year of operations are also
included. The DECam on-sky performance will be covered in detail in a forthcoming publication. The sections below will follow the path of the light through DECam.  Section 2 describes the optical corrector including the lenses, filters, and support structure. Section 3 describes the filter changer, shutter, and active optics system.  Section 4 provides details about the focal plane detectors, which are charge-coupled devices (CCDs). Section 5 describes the readout electronics. Section 6 describes the camera structure and infrastructure, including the prime focus cage.  Section 7 describes the system controls, and the observer and telescope interfaces.  Section 8 covers systems external to the camera such as the calibration system, and the auxiliary systems.  Section 9 discusses the integration and installation, including the initial camera performance.   In each section we provide the present status of the systems and note where there have been improvements or other changes since the original construction.

\clearpage
\section{Optical Corrector}\label{s3:Optics}

The DES science goals require a large area survey with accurate photometry of faint sources in g-, r-, i-, and z-bands as well as accurate shape measurements~\citep{bernjar}, particularly in r-band and i-band.   To meet these goals the DECam optical system was designed~\citep{SPIEkent2006, SPIEdoel2008} to have a wide field of view, high throughput over wavelength range 400-1000 nm and good image quality~\citep{SPIEantonik2009} over the entire field of view.  In addition, the design also had to satisfy tight budget constraints, and accommodate an aggressive construction schedule requiring that fabrication risks be minimized. 

\begin{figure}[h]
\begin{center}
\includegraphics[scale=0.6]{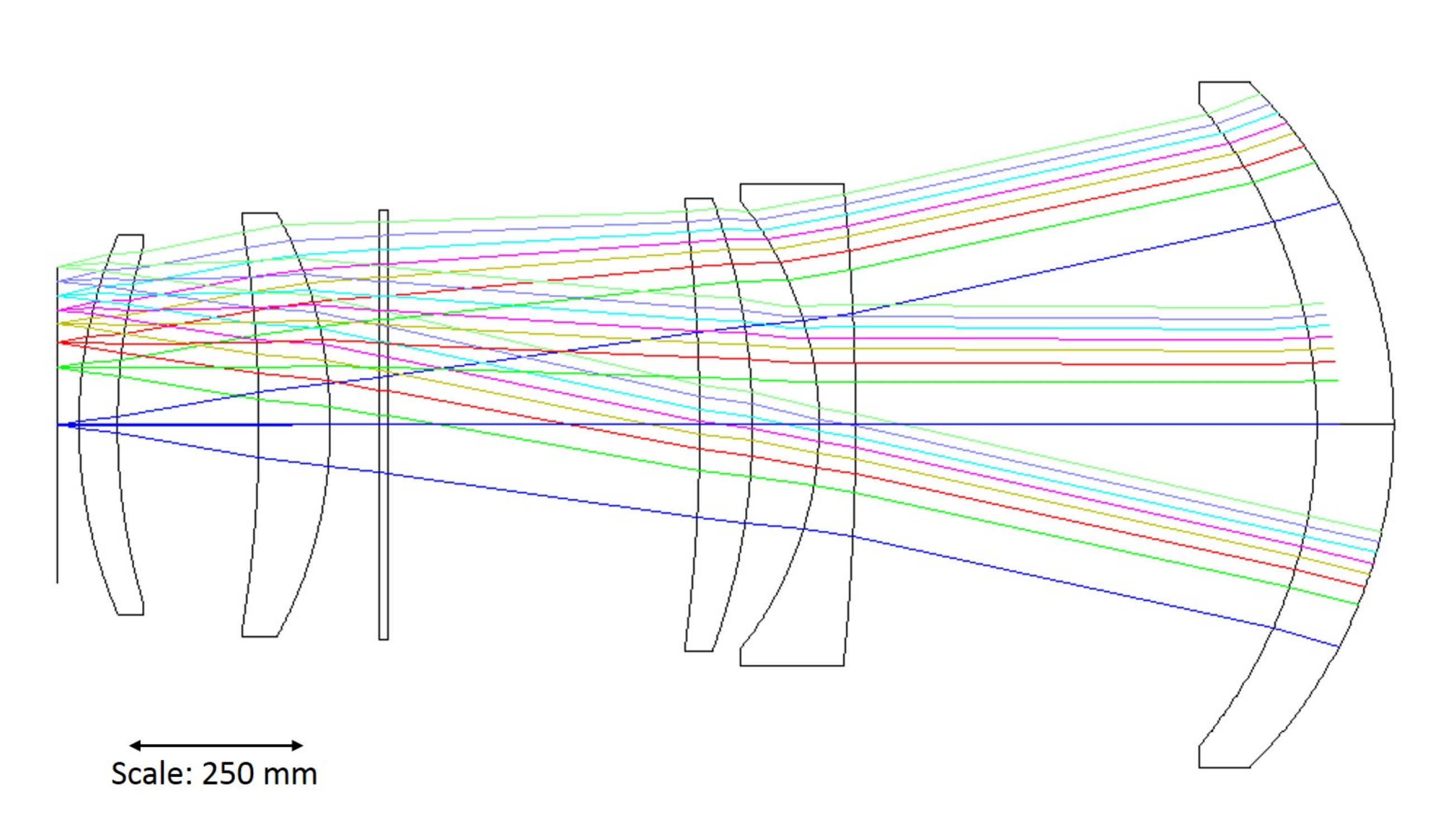}
\caption{The baseline optical design for DECam. The elements, from right to left, are C1, C2, C3, a plano-plano filter (1 of 4 positions is shown), C4, C5 (Dewar window), and the focal plane array.  The primary mirror is approximately 8.9 m from the vertex of C1. The total length of the camera from C1 to the focal plane is approximately 1.9 m. The full range of incidence angles on the filters is $0 - 14 ^{\circ}$.}
\label{fig:decamzemax} 
\end{center}
\end{figure}

Figure~\ref{fig:decamzemax} shows the overall optical design.  Fused silica  was chosen as the material for all 5 lenses to provide good performance over the full wavelength range required for DES while also providing good performance in the u-band. Only one of the four filter positions is shown in the figure, but all were used in the ghosting analysis and the lens optimization. The design is nearly achromatic for $\lambda > 500$ nm.   The smallest lens, C5, is curved into the vacuum vessel providing field flattening and minimizing ghosting, and it functions as the vacuum vessel window. This section will describe the design and as-built results for the lenses and coatings, the cells that provide the interface between the lenses and the barrel, the filters and the barrel as well as the overall assembly and alignment. 

\subsection{Optical Specifications and Performance Goals}

\subsubsection{Field of View and Pixel Scale}
The field of view of the camera was specified at 2.2 degrees in diameter based on the desired survey area, the available observing time, and the specification for the image qualty.  It may have been possible to specify a corrector with a larger field of view and good image quality by using a larger C1 and/or more aspheric lenses, but that would have entailed higher costs and greater manufacturing risks.  

The focal ratio at prime focus of the Blanco 4-meter telescope is f/2.7 (52 $\mu$m/arcsec). The optical designs that we explored fell naturally into the range f/2.9--f/3.0 (56-57$\mu$m/arcsec), slightly slower than the primary mirror. This pixel scale was designed to be well-matched to the expected image quality; 2 pixels corresponds to 0.52\arcsec FWHM, which is roughly the convolution of the best quartile of seeing at CTIO ($\sim 0.4$\arcsec FWHM) convolved with the as--built performance goal of the optics alone ($\sim 0.33\arcsec$ FWHM, see below). Note that the typical best quartile of realized image quality of the previous prime focus camera (MOSAIC II) at the Blanco over a 0.6$^{\circ}$ diameter field in the r-band filter was 0.89\arcsec with a median of 0.99\arcsec~\citep{BCS}.

\subsubsection{Image Quality and Wavelength Range}
The goal for the {\it as}-{\it built} contribution (including lens sag, alignment errors, etc ... ) to the FWHM for DECam optics was 0.33\arcsec \  FWHM, or 18 $\mu$m. The goal for the {\it as}-{\it designed} image quality is 0.27\arcsec \ FWHM, or 15$\mu$m. For Gaussian images, this corresponds to an RMS radius of ${\rm R_{RMS}}\sim 9 \mu$m. Note that ${\rm R_{RMS}} = 2^{1/2}\sigma   = (2^{1/2}/2.35)$ FWHM = 0.60 FWHM for a 2D Gaussian.  The 80\% encircled energy radius is R80=0.76 FWHM, so that the ${\rm R_{RMS}} = 0.78$ R80.

The optical prescription for DECam was optimized for the wavelength range  400 nm to 1000 nm with four filters with nominal wavelength ranges: 
g-band (400\textendash 550 nm), r-band (560\textendash 710 nm),  i-band (700\textendash 850 nm), and z-band (830\textendash 1000 nm).  After the optical design was finalized, DES added a Y-band (950\textendash 1065 nm) filter to the DECam system primarily for the identification of high-redshift quasars. This addition did not impact the optical design. Weak gravitational  lensing measurements will be made primarily in r-, i-, and z-bands. The image quality has therefore been optimized to favor these bands to the extent that it is possible without violating the requirements in g-band. The best image quality for weak lensing is delivered by optimizing the camera for the smallest RMS image sizes.  Accordingly, the images were quantified in terms of the average RMS image size uniformly weighted over the full field.  

Fused Silica, which also has high transmission into the u-band, was the preferred material for the lenses because of the excellent homogeneity and because it has a negligible residual radioactivity (making it particularly suitable for the Dewar vacuum window).  While it is not of primary interest to DES, the u-band is of interest to the astronomy user community and CTIO contributed a u-band filter.  As the image quality optimization included a compromise between the blue and red image quality, adequate imaging is also achieved while using the u-band filter.  Although u-band images are noticeably worse than in the g-band, they do still have $\rm{R_{RMS}} < 6\mu {\rm m}$  (0.17\arcsec FWHM) over the full field of view of the previous corrector on the Blanco (0.6$^{\circ}$  diameter), and so represent an improvement in image quality (0.25--0.5\arcsec) over the previous corrector.  In March 2014 a VR-band (500 to 760 nm) filter was purchased by CTIO and added to DECam. This filter is primarily of interest to those searching for or studying objects in our solar system. 

\subsubsection{Ghosting and Surface Coatings}
Accurate photometry requires accurate flat fielding. A predictable complication for prime focus cameras in this regard is the pupil ghosting. Without coating, the reflective loss at a single lens surface is $R=((n_s-n_0)/(n_s+n_0))^2$, where $n_s$ and $n_0$ are the indices of refraction for the lens material and for air. Because fused silica has an index of refraction of $\sim 1.46$ at optical wavelengths, the reflective losses for each lens in the DECam corrector would be $\sim 7\%$.   An example of such ghosting and its mitigation using surface coatings  is described~\citep{SPIEjacoby1998} for the Mayall prime focus camera.  To mitigate against light loss we required that the reflectance be less than 1.5\% in the wavelength range 340 to 1080 nm and less than 1.2\% in the wavelength range 480 to 690 nm. The non-uniformity was required to be less than 0.7\%. To mitigate the data-reduction problem introduced by ghosting, we required that the gradient in the pupil ghost intensity must be smaller than 3\% across the long dimension of one CCD (61mm, or 0.3 degrees). 

To reduce the ghosting, all lens surfaces apart from those of C1 were coated by the polishing vendor. The coatings were chosen to minimize pupil and stellar images ghosting, and to maximize throughput. Also required was that the coatings must be mechanically robust and not degrade in the environmental conditions each lens will encounter over the expected lifetime of the instrument.  C1 was not coated because of a combination of the difficulty of identifying a vendor who would guarantee sufficient uniformity of the coating for a lens of that size and depth of curvature and the risk incurred from additional shipping of the lens.

\subsubsection{Lens Dimensions and the Use of Aspherical Surfaces}
To minimize figure errors and improve mechanical robustness (therefore reducing fabrication cost and risk), a minimum aspect ratio of 1:10 (axial thickness: lens diameter) was chosen.  Thicker elements were allowed when driven by the image quality.  C2 was near the critical thickness limit for a blank fabricated by standard methods. Indeed, C1 proved to be even thicker and was slumped after casting to achieve the necessary  curvature. 

The design includes two aspheric surfaces. The figure and placement of these elements was constrained based on feedback from several vendors. The most important characteristic regarding fabrication was not the total deviation from the best-fitted sphere, but rather the slope in this quantity with radius. We limited this slope to 1mm/50mm, which the vendors indicated was in the range that would be straightforward to fabricate.  Different vendors indicated preferences for testing concave and convex aspheres, but all vendors suggested that both are readily fabricated and tested in elements similar to those discussed here.  Note that here we used the phrase ``best-fitted spherical deviation" in the practical sense of the millimeters of material that one would need to remove from the glass after figuring the surface to the best fitting spherical approximation. The surfaces were finished and tested at THALES SESO~\citep{SPIEfappani2011}.

\subsection{Characteristics of the DECam Optical Design}
\begin{figure}[ht]
\begin{center}
\includegraphics[scale=0.60]{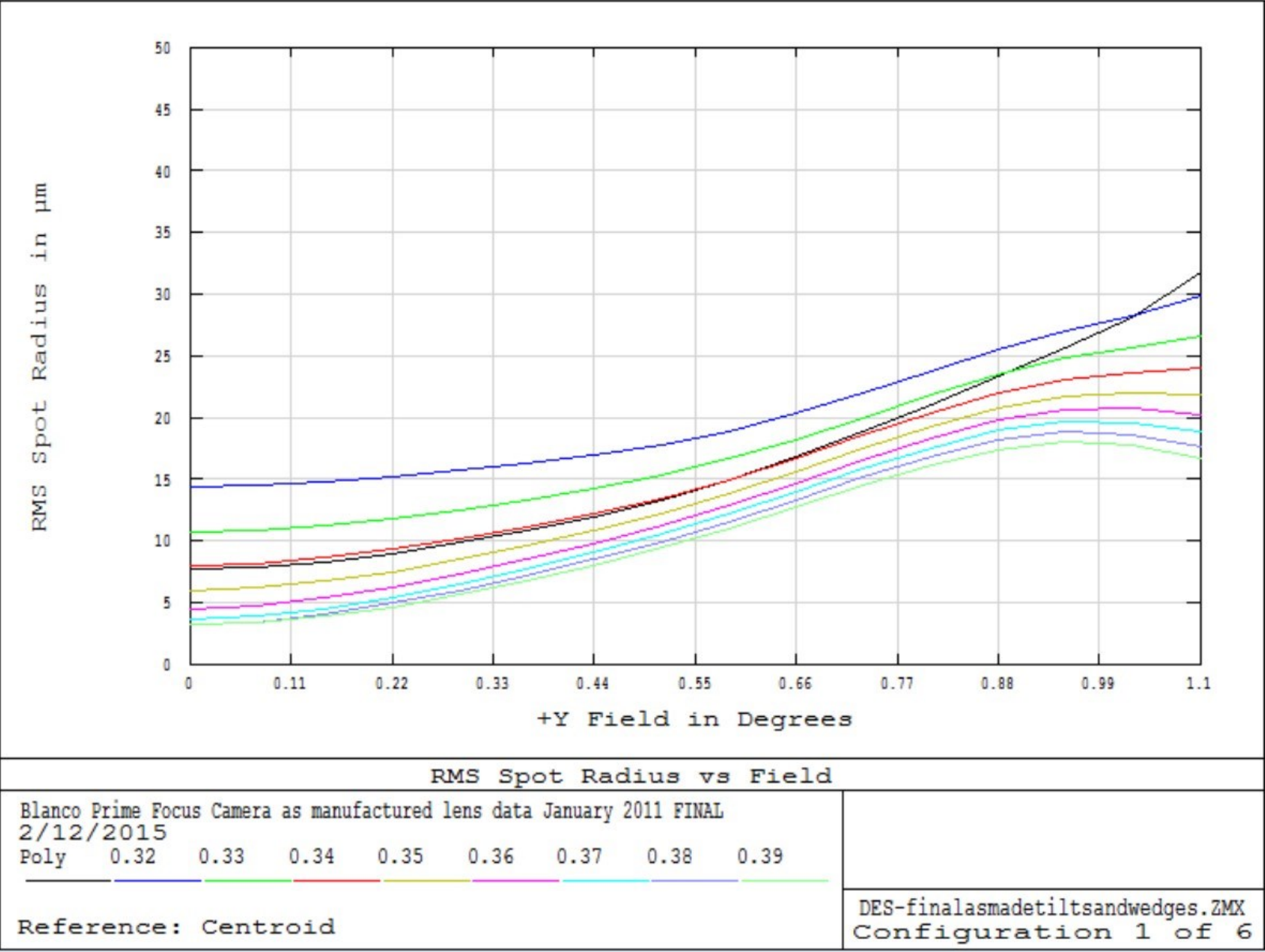}
\includegraphics[scale=0.60]{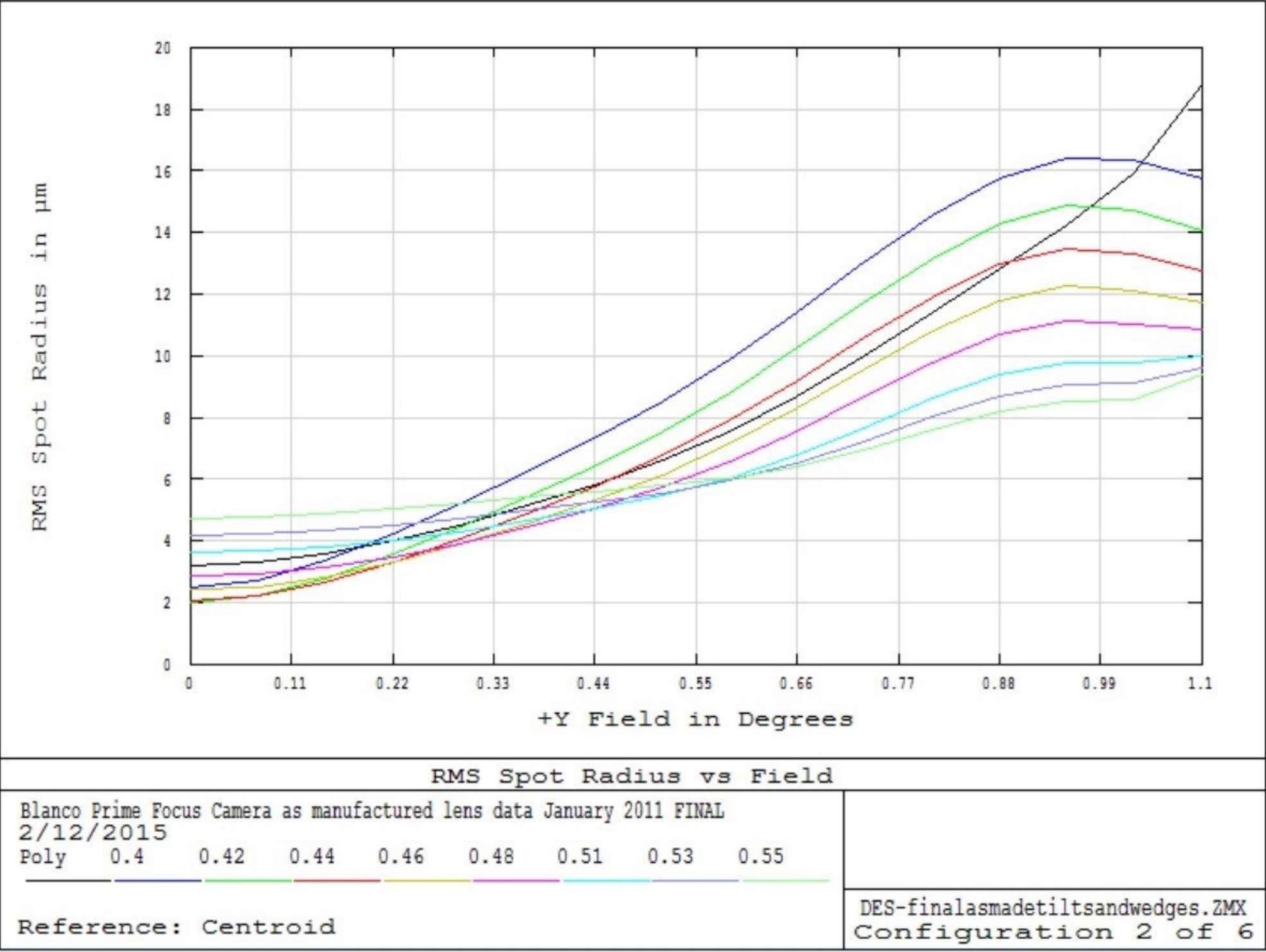}
\caption{The rms image radius as a function of field position. The plots show the image radius for the  u- (upper) and g-band (lower).  The full y-scale on each plot depends on the filter. For u-band (g-band) that is $50\mu$m ($20\mu$m).   The different colors show the wavelengths at which the images were traced within the bandpass; these are listed in the lower left panel of each plot.  }
\label{fig:opticalresult-ug} 
\end{center}
\end{figure}

\begin{figure}[ht]
\begin{center}
\includegraphics[scale=0.60]{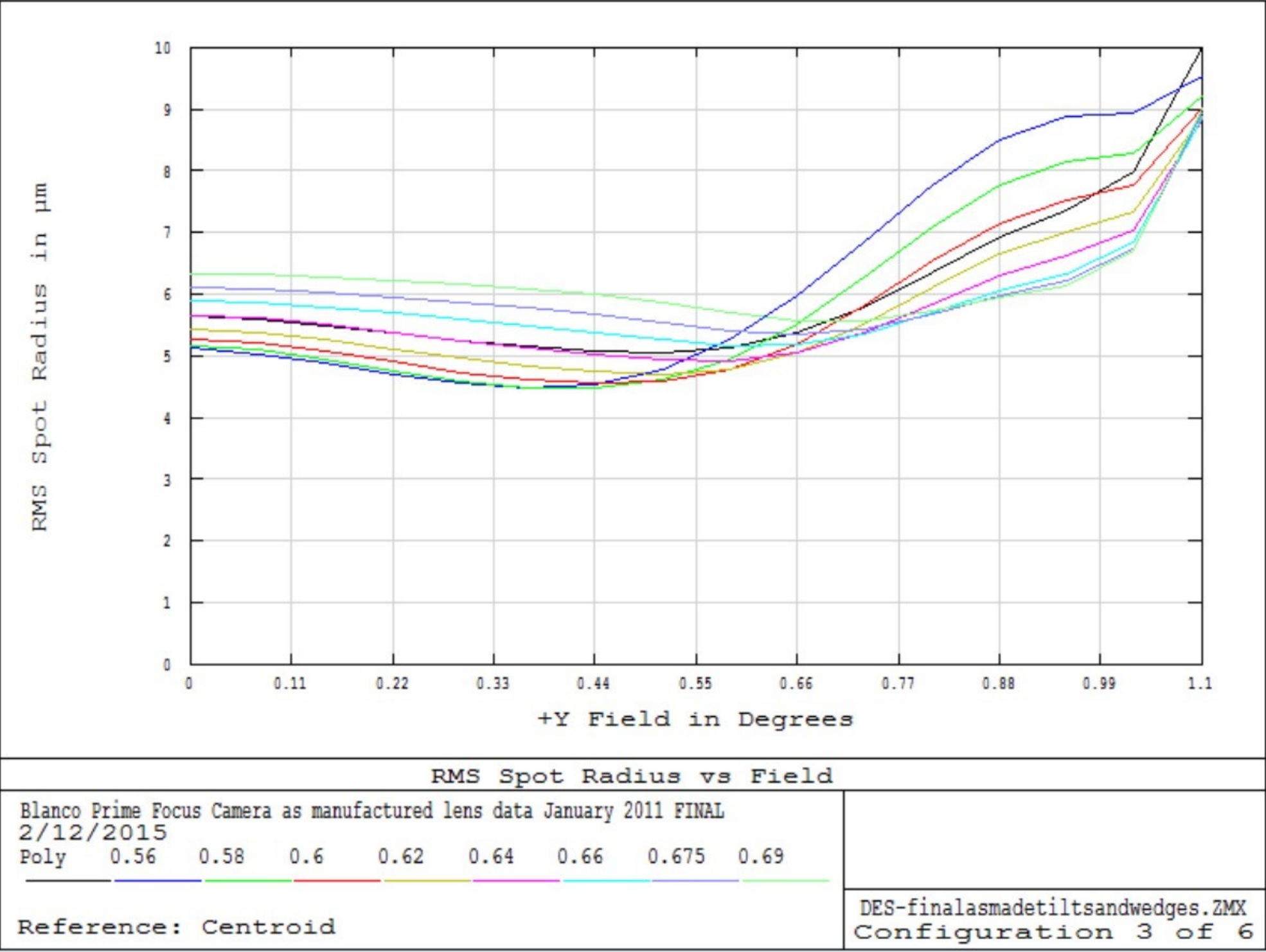}
\includegraphics[scale=0.60]{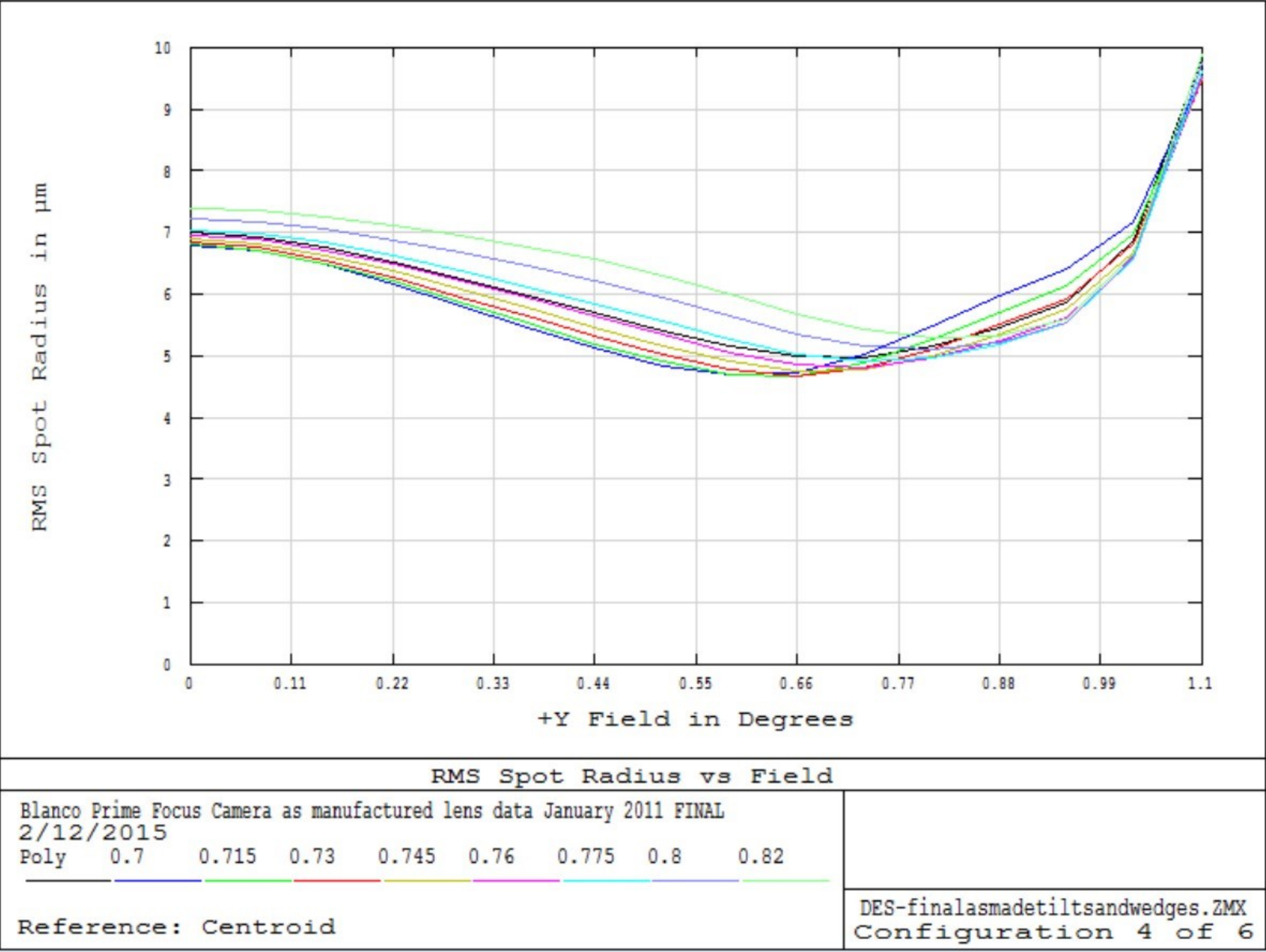}
\caption{The rms image radius as a function of field position. The plots show the image radius for the  r- (upper) and i-band (lower).  The full y-scale on each plot is $10\mu$m. The different colors show the wavelengths at which the images were traced within the bandpass; these are listed in the lower left panel of each plot.  }
\label{fig:opticalresult-ri} 
\end{center}
\end{figure}

\begin{figure}[ht]
\begin{center}
\includegraphics[scale=0.60]{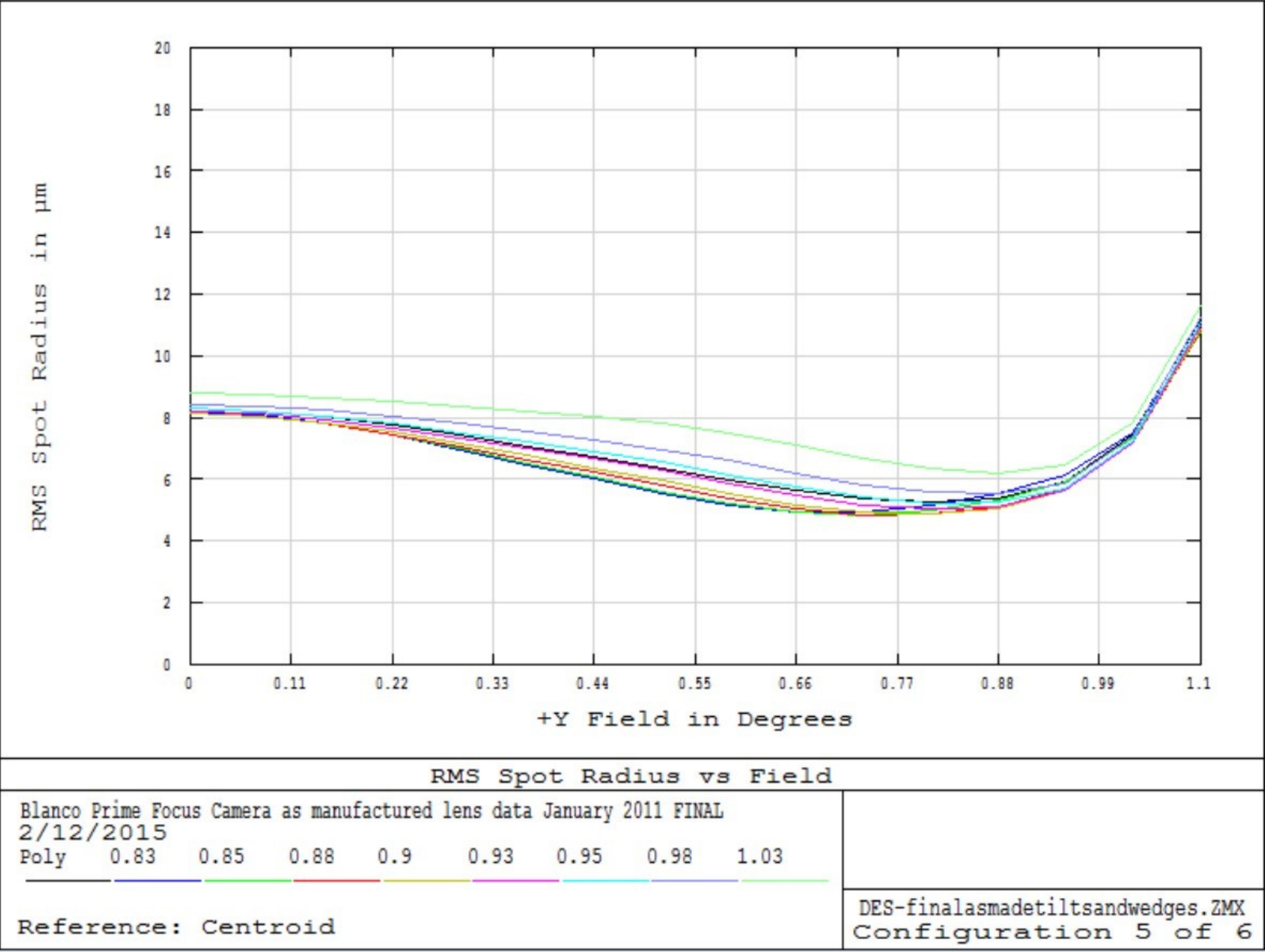}
\includegraphics[scale=0.60]{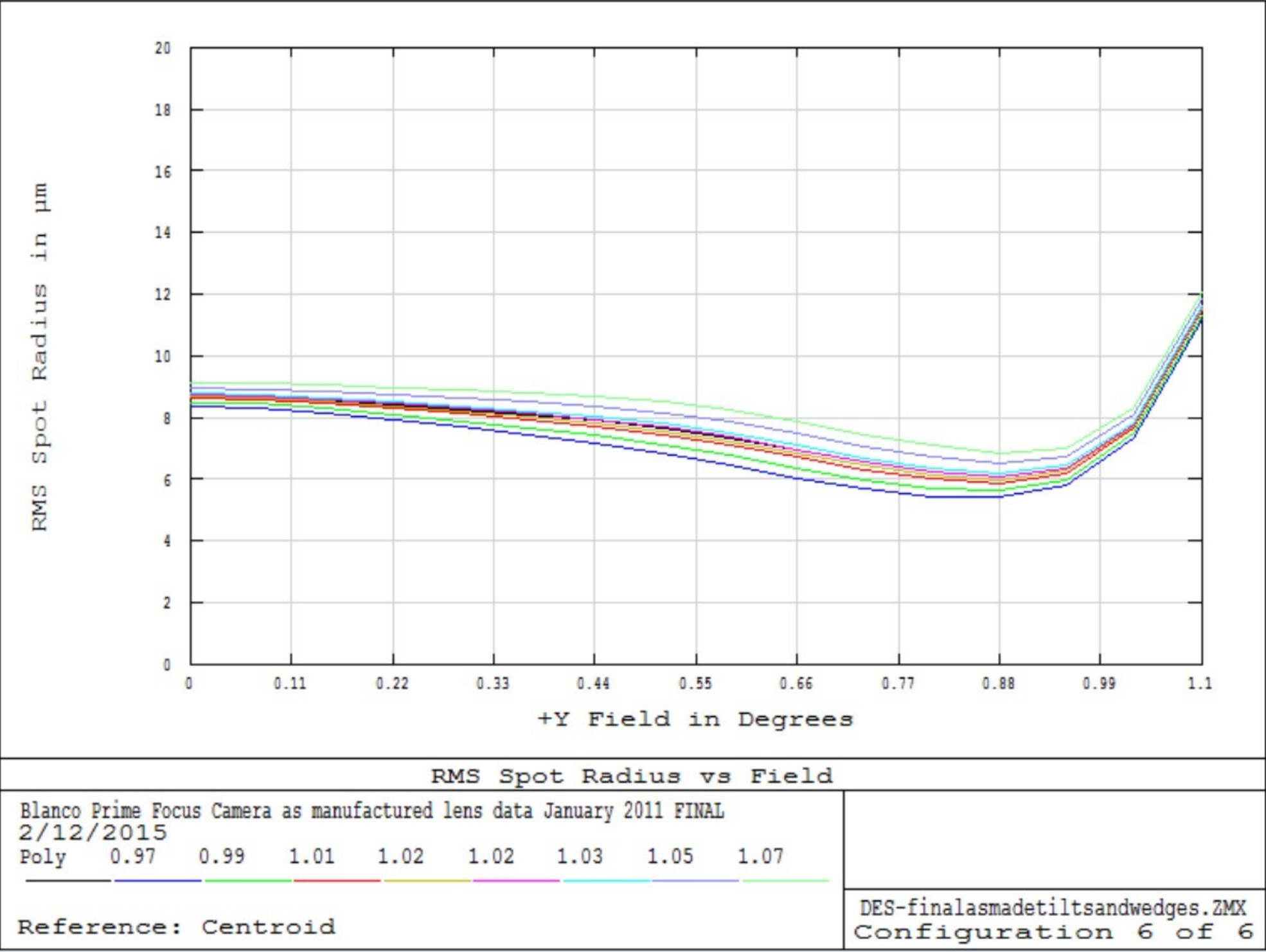}
\caption{The rms image radius as a function of field position. The plots show the image radius for the  z- (upper) and Y-band (lower).  The full y-scale on each plot is $20\mu$m.   The different colors show the wavelengths at which the images were traced within the bandpass; these are listed in the lower left panel of each plot.  }
\label{fig:opticalresult-zy} 
\end{center}
\end{figure}

The baseline optical design for DECam is shown in  Fig.~\ref{fig:decamzemax}.  The optical prescription for the camera is given in Table~\ref{tab-ZEMAX}. The effective focal ratio is f/3.0. The pixel scale is 56.88 $\mu$m per arcsecond at the center of the focal plane and 57.12 $\mu$m per arcsecond at the edge.  This scale was not constrained during the optimization.  The prescriptions for the aspheric surfaces are listed in Table~\ref{tab-asphere}.  Figs.~\ref{fig:opticalresult-ug} to \ref{fig:opticalresult-zy} show the RMS image radius as a function of field position and wavelength for the u, g, r, i, z, and Y-band filters.  This design results in lenses with the mechanical characteristics listed in Table~\ref{tab-lensparameters}.

\begin{table}
\begin{center}
\caption{\label{tab-ZEMAX} The optical prescription for DECam from the primary mirror to the focal plane. M1 refers to the primary mirror. The sign convention is that of ZEMAX, whereby distance along the optical axis in the direction of ray propagation is positive from the object to the primary mirror and negative after reflection off the mirror. The optically clear radii of the lenses are about 15 mm smaller than the lens radii (shown in the table) to avoid edge effects from polishing.}
\begin{tabular}{|c|c|c|c|c|c|c|} \hline
Element & Radius of & Thickness & Material & Radius    & Conic Const. \\ 
        & Curvature (mm) & (mm)      &          & (mm)      &              \\ \hline \hline
M1      & -21311.600&           & Cer-Vit  & 1905   & -1.0976 \\
        &           & -8875.037 &          &        &         \\ \hline
C1      & -685.980  & -110.540  & Fused Silica
                                           & 490    & 0       \\
        & -711.870  & -658.094  &          & 460    & 0       \\ \hline
C2      & -3385.600 & -51.136   & Fused Silica 
                                           & 345    & 0  \\
        & -506.944  & -94.607   &          & 320    & 0 \\ \hline
C3      & -943.600  & -75.590   & Fused Silica  
                                           & 326    & 0 \\
        & -2416.850 & -325.107  &          & 313    & 0 \\ \hline
Filter  & planar & -13.000      & Fused Silica 
                                           & 307    & 0 \\
1 of 4 positions
        & planar    & -191.490  &          & 307    & 0 \\ \hline
C4      & -662.430  
                    & -101.461  & Fused Silica
                                           & 302    & 0 \\ 
        & -1797.280 & -202.125  &          & 292    & 0 \\ \hline
C5 (Dewar Window)
        & 899.815   & -53.105   & Fused Silica 
                                           & 256    & 0 \\
        & 685.010   & -29.900   &          & 271    & 0 \\ \hline
Focal Plane
        &           & 0.000     &          & 225.8  & 0 \\ \hline
\end{tabular}
\end{center}
\end{table}

\begin{table}
\begin{center}
\caption{\label{tab-asphere} The prescription for the aspheric surfaces of elements C2 and C4. The values are the constants multiplying the indicated terms in the usual definition for an even asphere.  The sign convention is that of ZEMAX.}
\begin{tabular}{|c|c|c|c|} \hline 
Surface      & R$^4$         & R$^6$    & R$^8$     \\ \hline \hline
C2 surface 1(convex) 
             & $1.579e^{-10}$& $1.043e^{-16}$
                                        & $-1.351e^{-22}$ \\ \hline
C4 surface 2(concave)
             & $-1.798e^{-10}$&$-1.126e^{-15}$
                                        & $-7.907e^{-21}$ \\ \hline 
\end{tabular}
\end{center}
\end{table}

\begin{table}
\begin{center}
\caption{\label{tab-lensparameters} Dimensions and weight of the five DECam lenses, as-built. }
\begin{tabular}{|c|c|c|c|c|} \hline 
     &Center         &Dome to          & Diameter of     & Approx. \\ 
Lens &Thickness (mm) &Flat (mm)     & Surface 1 (mm)  & Weight (kg)  \\ \hline \hline
C1   & 110.54        & 278.02        & 980.74          & 172.7 \\ \hline         
C2   & 51.136        & 164.091      & 690.12          & 87.2 \\ \hline
C3   & 75.59         & 96.80           & 652.547         & 42.1 \\ \hline
C4   & 101.461       & 125.99       & 604.99          & 49.6 \\ \hline
C5   & 53.105        & 88.808        & 501.9            & 24.3 \\ \hline
\end{tabular}
\end{center}
\end{table}



\subsection{Filters} 
The filters are interference filters on 13mm thick plano-plano fused silica substrates.  The housing holds a total of eight filters.  Two filters are located opposite to each other at four positions (one is shown in Fig.\ref{fig:cage}).  Filters are changed by moving perpendicular to the optical beam. See Section~\ref{ss:filterchanger} for more information about the filter changer mechanism. 

\begin{figure}[h]
\begin{center}
\includegraphics[scale=0.6]{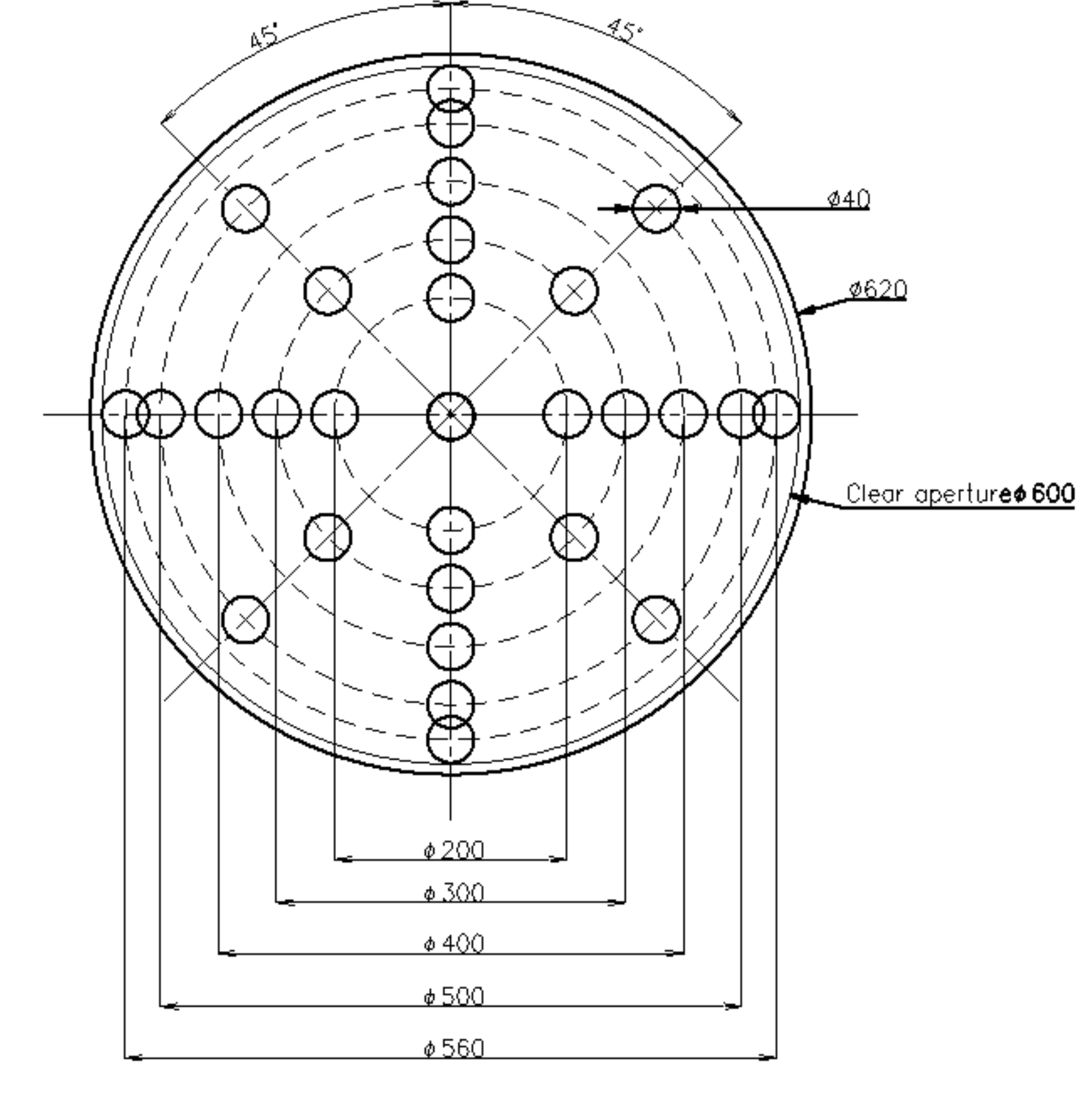}
\caption{Measurement positions for filter uniformity determination.}
\label{fig:filtspots} 
\end{center}
\end{figure}

\begin{figure}[h]
\begin{center}
\includegraphics[scale=0.9]{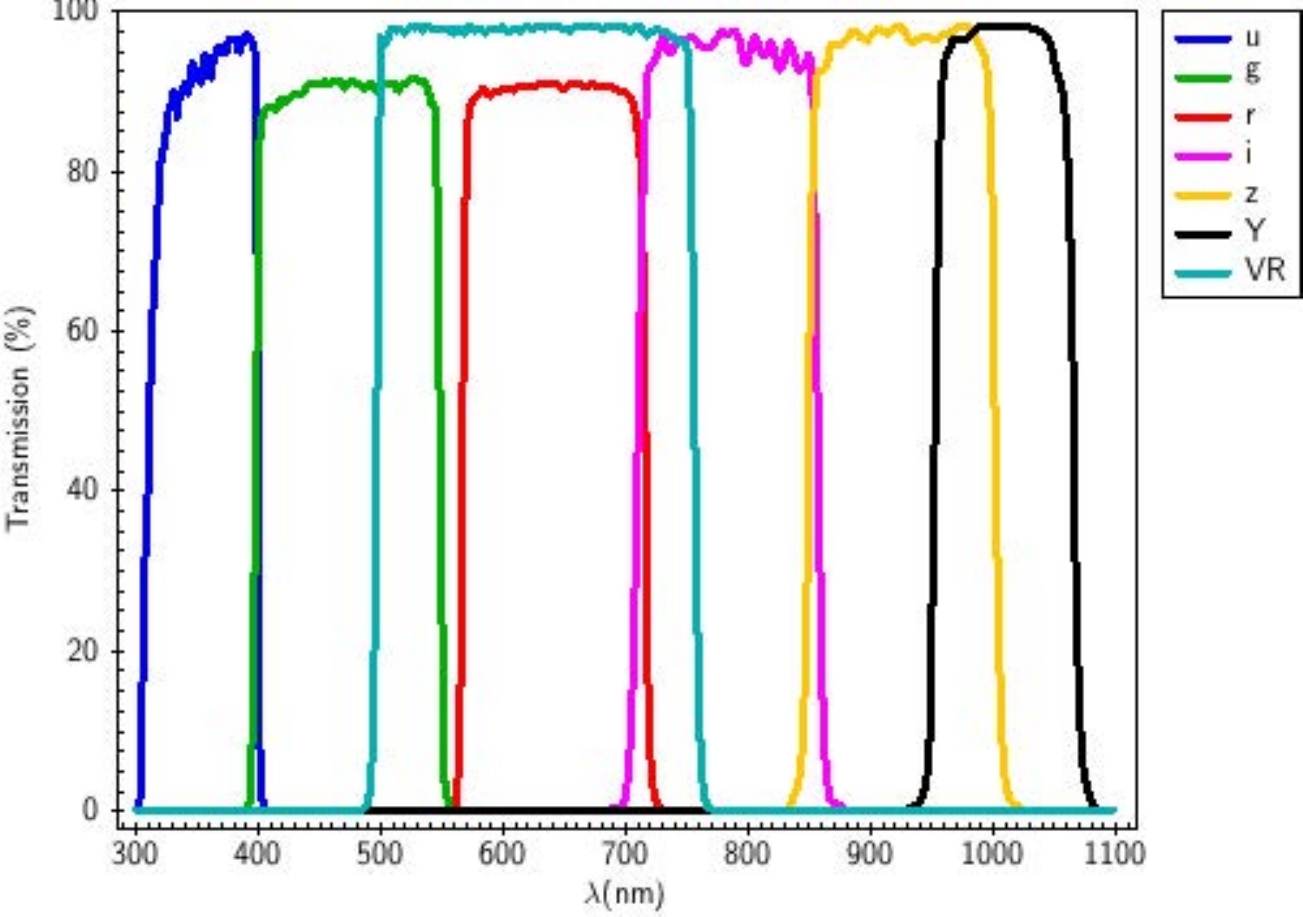}
\caption{The DES and DECam filter set delivered transmissions.}
\label{fig:DECamFilterBandpasses} 
\end{center}
\end{figure}

The DES and DECam filters presented a significant fabrication challenge.  With a diameter of 620mm and tight uniformity requirements, no vendors had demonstrated capability prior to production of the DECam filters.  A detailed evaluation of the available vendors and their proposed cost and schedules led us to select ASAHI Spectra to fabricate the DECam filter set.   The interference filter coatings were applied using a magnetron sputtering technique similar to that used to coat large telescope mirrors.  Transmission of the filters turned out to be excellent, exceeding the DECam requirement of $> 85\%$ by a substantial amount. The absolute transmission and uniformity of the filter was measured using a 70mm diameter beam in 29 positions on the filter.  Figure~\ref{fig:filtspots} shows the locations of the measurement positions. The most difficult part was to achieve the uniformity over the filter in transmission, and on the turn-ons and cut-offs of the band passes.   Table~\ref{tab-filtgen} shows the general characteristics of the DECam filters and Figure~\ref{fig:DECamFilterBandpasses} shows the delivered bandpasses of the DECam filters (u,g,r,i,z,Y, and VR)\footnote{\url{http://www.ctio.noao.edu/noao/content/dark-energy-camera-decam} has a table of filter transmission versus wavelength}.  The DES requirement of excellent photometry drove tight constraints on the filter uniformity.    Table~\ref{tab-filtunitran} shows the specifications for the uniformity and slopes of turn-on and cut-off transitions for the DECam filters. The delivered filters met the specifications in almost all cases. When specifications were missed it was only by a small amount. The first two filters produced did not meet the  uniformity specification as follows. The wavelength of the r-band filter cut-off  has a  radial dependance. The inner $r<0.3 \; {\rm R_{max}}$ has a cut-off  wavelength 25 nm greater than the outer $r>0.3 \; {\rm R_{max}}$.  The i-band filter turn-on has a radial dependance of  $\sim 50$ nm width over the full radius of the filter with the outer radii turning-on at the longer wavelength~\citep{jenetal}.  Evaluation of the violations of the specifications showed that the impact on the DES science will be negligible.

The DES and DECam filter specifications required that in the wavelength range $310< \lambda <1100$ nm the average transmission of out-of-band light must be less than 0.01\% with less than 0.1\% absolute transmission at any wavelength.  All of the DES flters met this requirement.   

\begin{table}\begin{center}
\caption{\label{tab-filtgen} Characteristics of the filters available in DECam. These are {\it as-built} area-weighted averages for each filter. The turn-on and cut-off wavelengths are for 50\% absolute transmission.  The values for the FWHM are indistinguishable from the difference between the 50\% turn-on and cut-off wavelengths.   These filters are nominally 13 mm thick and have a diameter of 620 mm. Their mass is about $9.95$ kilograms each. The u-band and VR-band filters were purchased by CTIO. }
\begin{tabular}{|c|c|c|c|c|c|} \hline
                 & Central               & Blue  Turn-on       & Red  Cut-off      &                   & Peak Absolute       \\
Filter         & $\lambda$ (nm)  & $\lambda$ (nm)   & $\lambda$ (nm) & FWHM (nm)& Transmission (\%)  \\ \hline \hline
DECam u   & 355                     & 312                     & 400                    & 88              & 96-97        \\ \hline
DES g        & 473                     & 398                     & 548                    & 150            & 91-92 \\ \hline 
DES r        & 642                      & 568                     &716                     & 148            &  90-91 \\ \hline 
DES i         & 784                      & 710                   & 857                      & 147            &  96-97 \\ \hline 
DES z        & 926                       & 850                  & 1002                    & 152            & 97-98 \\ \hline 
DES Y        & 1009                  & 953                   & 1065                     & 112            & 98-99 \\ \hline 
DECam VR  & 626                   & 497                    & 756                     & 259            & 98-99    \\ \hline
\end{tabular}
\end{center}
\end{table}

\begin{table} \begin{center}
\caption{\label{tab-filtunitran} Uniformity specifications for DES and DECam filters (difference between area-weighted transmission curve and any 70 mm diameter spot on filter).  Transition specifications for the DES and DECam filters. These are the specifications for the wavelengths spanned by the best-fit line in the filter transitions between 10\% and 90\% for the turn-on edge and 90\% and 10\% for the cut-off edge. The filters met the transition specifications.}
\begin{tabular}{|c|c|c|c|c|c|} \hline
                 & Uniformity                                        & Uniformity                                      &Uniformity      &\multicolumn{2}{c|}{Transition}    \\
Filter         & $\lambda({\rm Blue \; turn-on}$)    & $\lambda({\rm  Red \; cut-off}$)   &Allowable       &\multicolumn{2}{c|}{$\Delta \lambda (10\% \; {\rm to} \; 90\%)$}    \\ 
                 & (nm)                                                 &(nm)                                                &Gradient (\%)&Blue (nm)     &Red (nm)   \\ \hline \hline
DECam u   &   None                                              & $\pm 3$                                         &$\pm 5$          & None       &$<10$                  \\ \hline
DES g        & $\pm 2$                                           & $\pm 2$                                         &$\pm 5$          & $<4$        & $<5$         \\ \hline 
DES r         & $\pm 3$                                           & $\pm 3$                                        &$\pm 7$          &  $<5$       & $<7$         \\ \hline 
DES i          & $\pm 3$                                           & $\pm 4$                                        &$\pm 5$          &  $<7$        & $<9$        \\ \hline 
DES z         & $\pm 4$                                           & $\pm 5$                                        &$\pm 9$         & $<9$         & $<10$       \\ \hline 
DES Y        & $\pm 5$                                           & None                                             &$\pm 9$           &$<10$      & None           \\ \hline 
DECam VR & $\pm 3$                                          &$\pm 3$                                          & $\pm 5$          &  $<10$     &$<10$          \\ \hline
\end{tabular} \end{center}
\end{table}

\subsection{Barrel}
\begin{figure}[h]
\begin{center}
\includegraphics[scale=0.9]{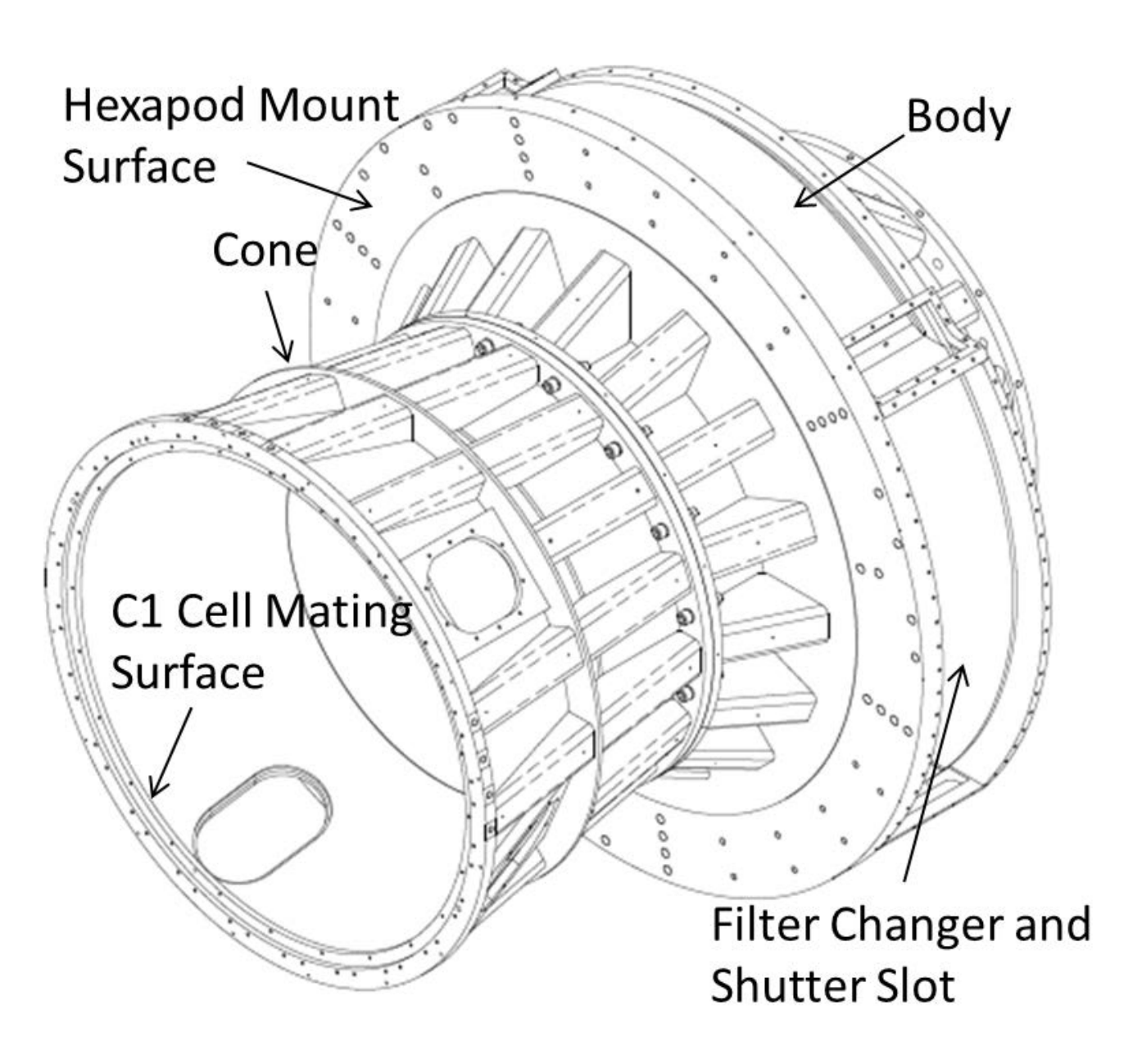}
\caption{The barrel body and cone. The ring of bolts marks the location of the body-cone joining surface. Of course, the C1 end of the barrel is oriented towards the primary mirror.  The body weldment weighs approximately 1395 kgs. The cone weldment weighs approximately 260 kgs. The length of the assembly is 1.442 meters. The outer diameter of the cone is 1.078 meters.   }
\label{fig:bodyandcone} 
\end{center}
\end{figure}

\begin{figure}[h]
\begin{center}
\includegraphics[scale=0.9]{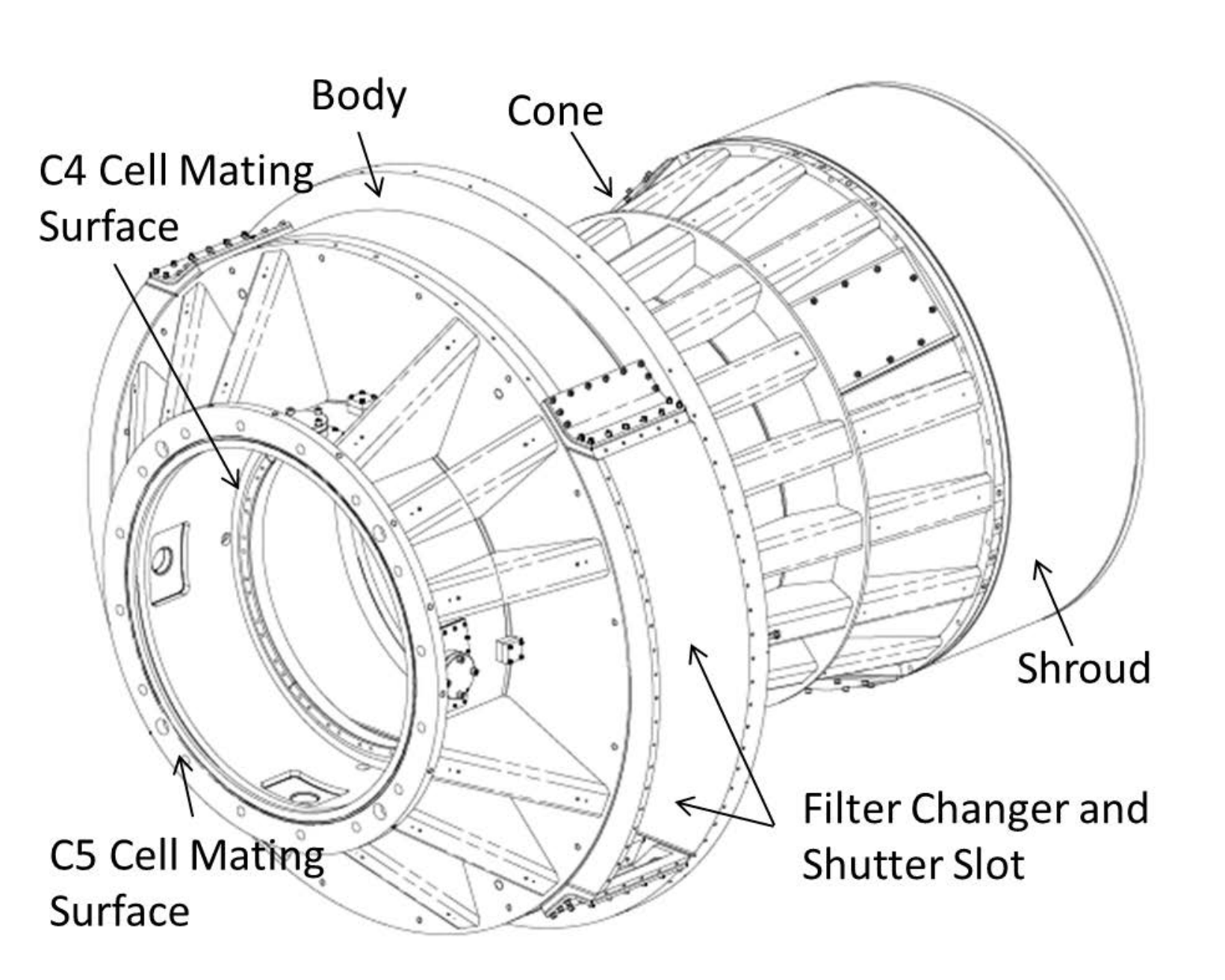}
\caption{The barrel assembly. The C5 end of the barrel is on the end away from the primary mirror. The filter-changer and shutter fit into the slot with some clearance between them and the C4 lens. Some of the small covers are also shown. The shroud weighs approximately 45 kgs. All barrel components were coated, inside and out, with anti-reflective black paint.}
\label{fig:barrelassy} 
\end{center}
\end{figure}

The barrel comprises two steel structures: a larger ``body" and a smaller ``cone". The body and cone together provide a very stiff support for the lenses.  The upper end of the body supports the DECam Dewar, where the C5 cell is bolted to it (maintaining electrical isolation of the barrel from the Dewar). It also supports the C4 cell and C2/C3 cell assembly. A slot through the body provides a mounting surface for the filter-changer and shutter. A large steel ring provides the mounting surface to which the hexapod is bolted. The other side of the hexapod is bolted to the cage. The cone is bolted to the body. It supports the C1 cell as well as the thin steel ``shroud", which surrounds the optical path and provides a lightweight protective shield.  Figure~\ref{fig:bodyandcone} shows an isometric view of the body and cone assembly. Figure~\ref{fig:barrelassy} also shows an isometric view, but looking from the opposite direction. This drawing also shows the shroud, as well as covers over some of the small access ports. 

The barrel components are weldments with precisely machined flanges provided for the cell mating surfaces.  After manufacture the barrel elements and lens cells (sans lenses) were measured using a long-reach coordinate measuring machine (CMM). The flange positions were within $\pm 7.5 \mu {\rm m}$ of the design positions and were very flat.  Using the measured dimensions, the cells were oriented to their optimal position for centering their respective lenses and then drilled and pinned so that their positions could be reproduced with the lenses in them. The body and cone were then aligned and keyed. These parts were shipped to UCL for lens installation and assembly.

\subsection{C1 to C4 Lens Cells}
The DECam lenses were mounted into their respective lens cells and then those assemblies were mounted into the main body (barrel) of the camera. The lenses in the camera had to be mounted and held to a high precision and had to maintain this position over a wide temperature range (-5 to 27$^{\circ}$C) and differing gravity vector. Both axial and radial supports of the lenses are required. Table~\ref{tab-lenscells} shows the decenter and alignment tolerances of the DECam lenses. 

The thermal stability of the lens positions is a key design element, and was ensured by the following design choices. The 
high ($\sim 12$ ppm/$^{\circ}$K) coefficient of thermal expansion (CTE) of the barrel steel compared to that of the fused silica lenses ($\sim 0.57$ ppm/$^{\circ}$K) was solved by using nickel/iron alloy cells that capture the lenses using radial and axial silicon rubber (RTV560) pads. This solution is similar to a design for wide-field correctors for the MMT~\citep{fabricant1998}. The CTE of  the Ni/Fe alloy with 38\% Ni is ($\sim 3.0$ ppm/$^{\circ}$K) matches well to that of the fused silica.  RTV was chosen as the material for the support pads because it is sufficiently tough but not very hard.  Because RTV560 has a high CTE (200-300 ppm/$^{\circ}$K), a proper selection of the pad thickness could compensate for the different thermal expansion of the fused silica compared to the cells. This alloy was a better match than INVAR 36, which has a CTE of ($\sim 1$ ppm/$^{\circ}$K), because with the latter the pads would be too thin. Lastly, pads made from RTV560 can be manufactured to high dimensional accuracy.   The cells, with lenses installed, are then coupled to the barrel flanges using thin steel rings with spacers that adjust the position, and with thin flexures that compensate for the large differential thermal expansion between those assemblies.  In the final assembly of the cells, thin annular rings were inserted on the lens cell to provide stray light baffling of the lenses.

\begin{table}
\caption{\label{tab-lenscells} The de-center and alignment tolerances of the DECam lenses. The lens pair separation tolerances (e.g. C1 to C2) are all $50 \mu {\rm m}$.  }
\begin{center}
\begin{tabular}{|c|c|c|} \hline
          & Decenter                & Tilt         \\
Lens      & Tolerance $(\pm \mu{\rm m})$
                                    & Tolerance \arcsec
                                                   \\ \hline \hline
C1        & 100                     &  10          \\ \hline
C2        &  50                     &  17          \\ \hline 
C3        & 100                     &  20          \\ \hline 
Filter    & 500                     & 200          \\ \hline 
C4        & 100                     &  20          \\ \hline 
C5        & 200                     &  40          \\ \hline 
\end{tabular}
\end{center}
\end{table}

\begin{figure}[h]
\begin{center}
\includegraphics[scale=1.0]{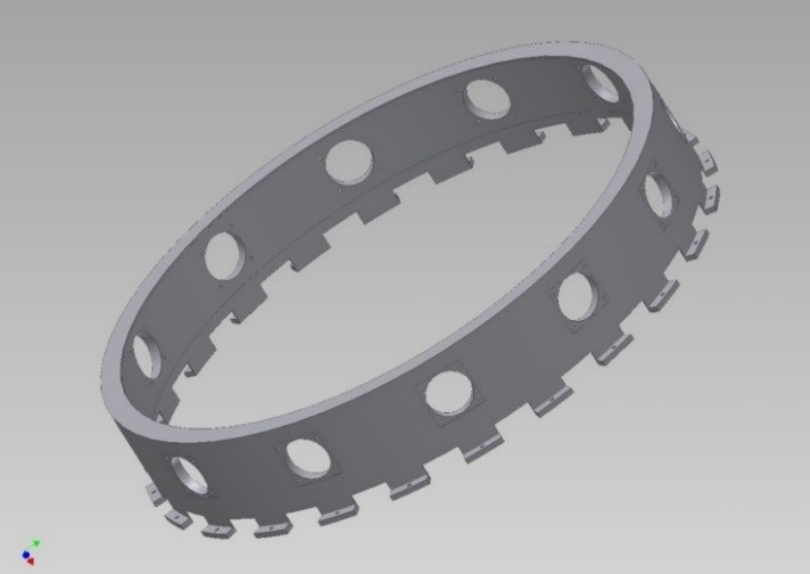}
\caption{The lens cell for optical element C4.  There are 12 holes round holes for the pad inserts.  The are 24 flexures that are bolted to the barrel through the thin steel rings and spacers. The designs of the other cells are similar to this one.  }
\label{fig:c4cellnew} 
\end{center}
\end{figure}

\begin{figure}[h]
\begin{center}
\includegraphics[scale=0.6]{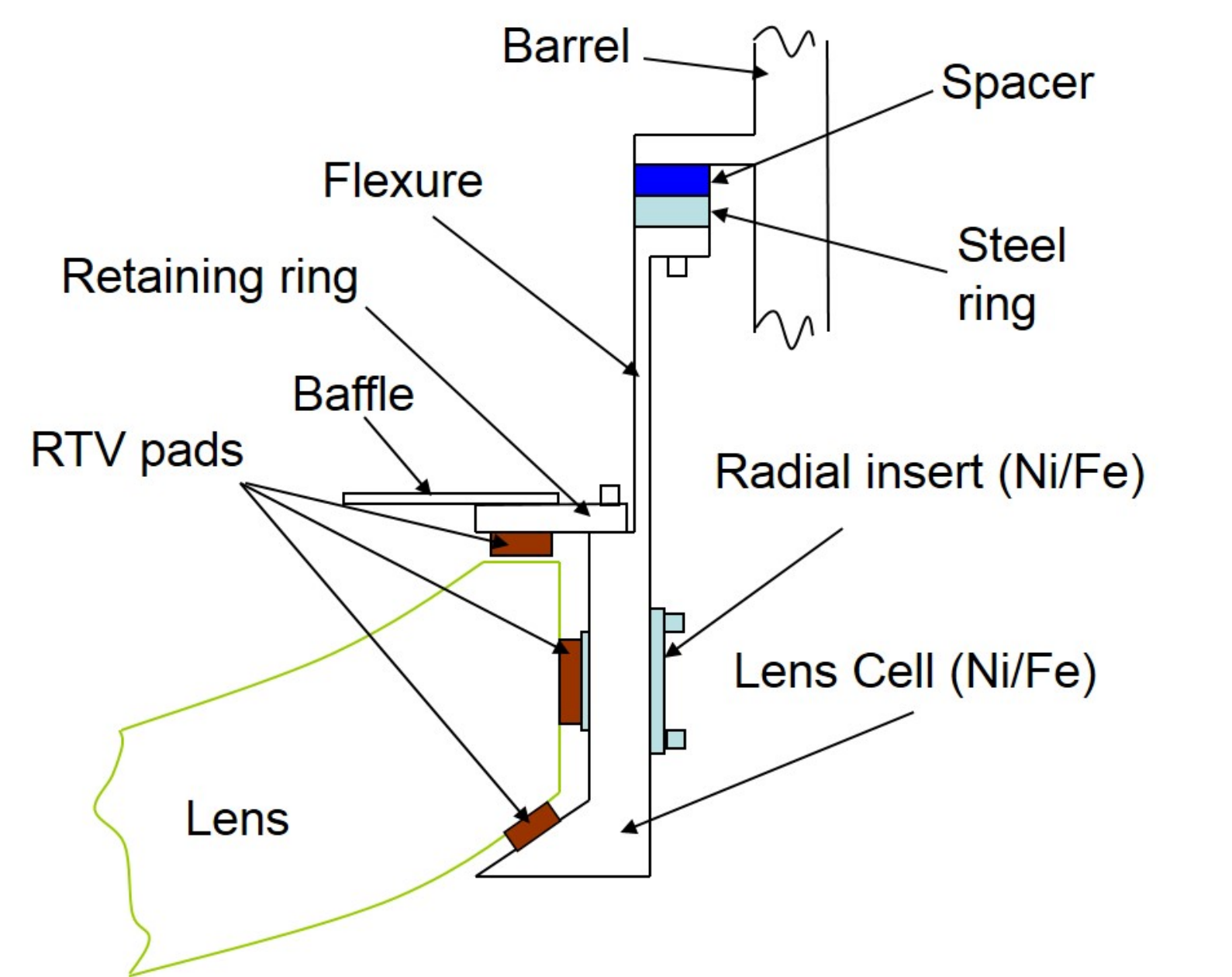}
\caption{Design of the C1 to C4 lens cell assemblies. This general schematic shows the major components, how they secure the lens, and how the assembly is attached to the barrel. }
\label{fig:cellschematic} 
\end{center}
\end{figure}

The C1 cell, which has to support the largest optical element, is coupled to the lens by 24 radial and axial RTV pads. The C2, C3, and C4 cells use 12 radial and axial pads to support each lens. The C2 and C3 cells are joined into an assembly that is mounted into the barrel. The C5 cell also serves as the Dewar vacuum window, so it is described in more detail in the next subsection. Figure~\ref{fig:c4cellnew} shows a drawing of the C4 cell.  Figure~\ref{fig:cellschematic} shows a schematic labeling the parts of the C1 to C4  lens cells. 

\subsection{C5 Cell and Interface Flange}
\label{ss:c5cell}
\begin{figure}[h]
\begin{center}
\includegraphics[scale=0.9]{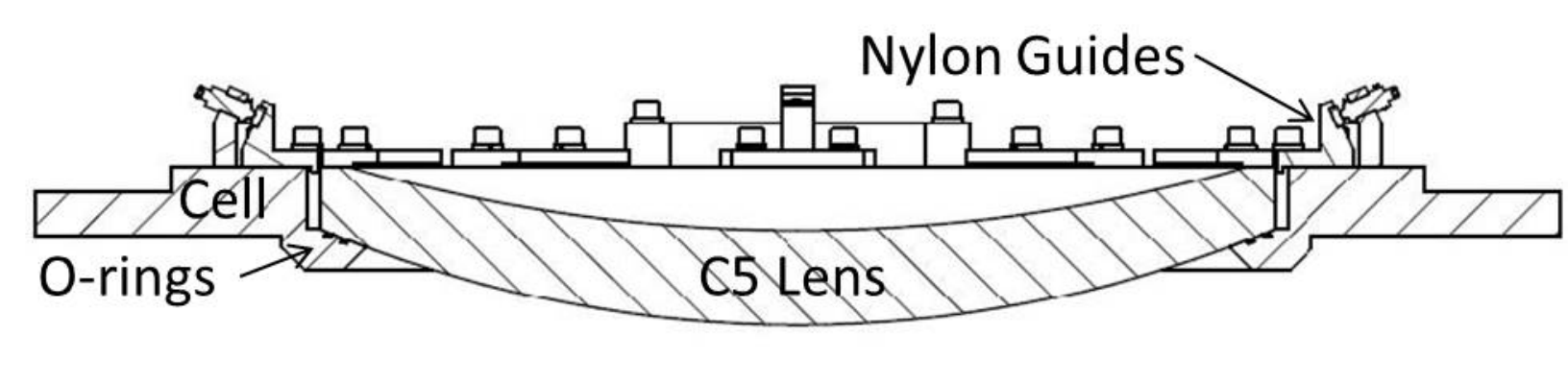}
\caption{Schematic of the C5 lens and cell. The lens is constrained by the nylon guides when the vacuum is not applied. The adjustment screws mounted on the cell positions the guides and the lens. }
\label{fig:c5cell} 
\end{center}
\end{figure}

The C5 cell holds the curved lens that is also used as the vacuum window to the instrument Dewar. The cell is manufactured from stainless steel and has a sloped surface to match the curvature of the lens where the lens makes contact with the cell. RTV pads are not used. Instead two o--rings are held in grooves in the cell, making the vacuum seal with the lens. There is also a $50\mu {\rm m}$ thick and 5 mm wide Mylar ring, mounted on the cell at the edge of the optical surface of the lens, preventing the glass from contacting the metal cell when the vacuum is applied. The lens is held in place by nylon restraints when the imager is not under vacuum.  Four radial restraints are used to center the lens in the cell, and to keep the lens on-center when the vacuum is cycled. The eight axial and radial combined restraints provide enough force to keep the lens in contact with the o--ring seal at atmospheric pressure.  After the cell is assembled, it is then aligned to the interface flange. The interface flange is the interface for the barrel, the C5 cell, and the imager vacuum vessel. Figure~\ref{fig:c5cell}  illustrates the components in the cell. There is an epoxy-fiberglass (g-10) spacer that electrically insulates the C5 cell from the barrel.  

The C5 lens was designed with the gravity, vacuum and thermal operational conditions in mind. The maximum stress calculated in the lens under 1 atmosphere loading, and at operating temperature, is 2.8 MPa. The fused silica tensile strength is 54 MPa which is a factor of 19 greater than the calculated stress in the lens. The maximum deflection in the lens is calculated as 30 microns. The lens is thermally coupled to the focal plane by thermal radiation. The thermal load is significant and cools the outside of the C5 lens to below freezing at the center of the lens. A dry gas purge of about 180 standard cubic feet per hour (85 liters per minute)  is blown into the space between the C4 and C5 lenses to keep the lens warm to prevent condensation from forming on the lens.  That dry gas vents out of the barrel at C1. 

An interface flange was used to set up an alignment coordinate system between the imager vessel, the C5 cell, and the corrector body. At Fermilab, a coordinate system was set up on the imager vessel using dowel pin holes in the front face of the imager vessel mounting flange. The pins in the vessel flange were used as a coordinate system to align the focal plate and CCDs with respect to it. Then the interface flange with the C5 cell was separately aligned and pinned to the corrector body. The interface flange stays with the imager vessel.  The barrel and imager vessel are later joined by re-pinning the interface flange to the barrel. The repeatability tolerance in the pinned joints is $\pm 12.5$ microns.

\subsection{Alignment and Assembly}
\begin{figure}[h]
\begin{center}
\includegraphics[scale=0.9]{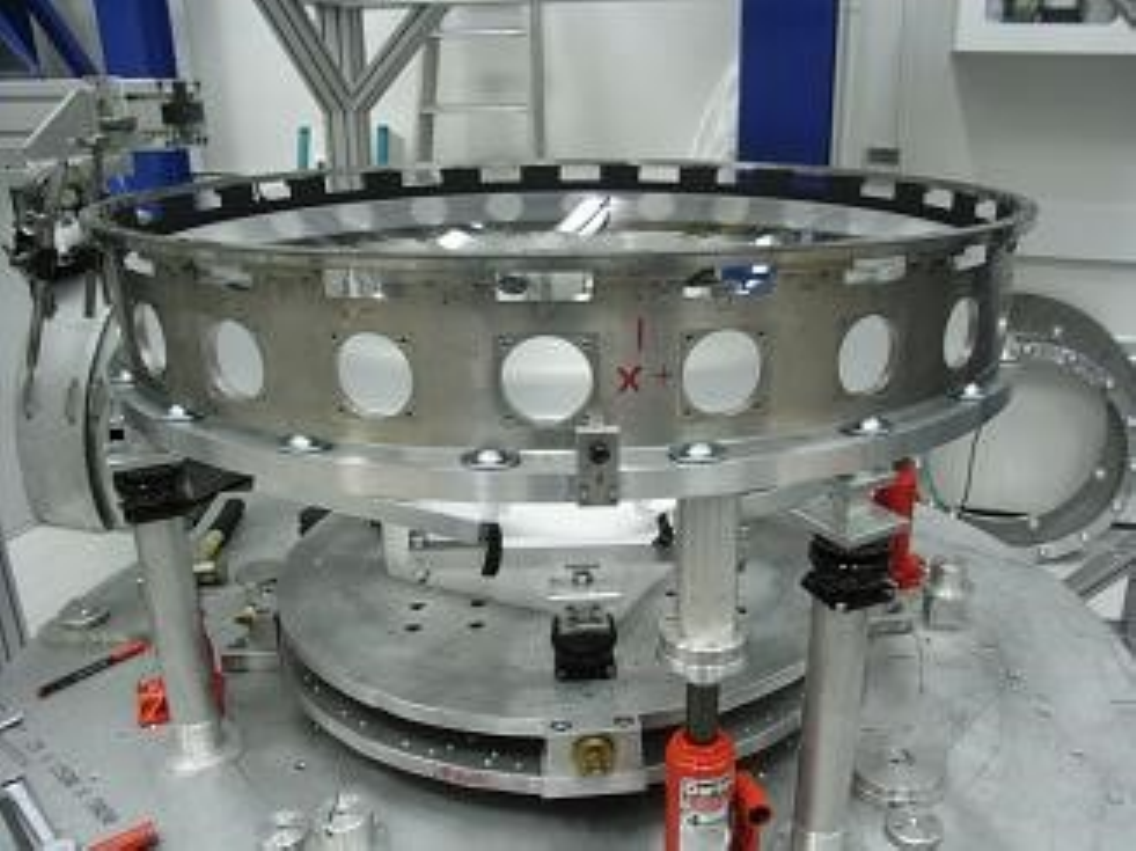}
\caption{ The C1 lens is partially installed in the C1 cell. The lens is supported from underneath on tip and tilt stages that are mounted on a  rotary table. The cell is supported on jacks and its own tip/tilt system. This cell already has the thin steel ring mounted on its flexures.   }
\label{fig:c1rotary} 
\end{center}
\end{figure}

The optical assembly and alignment was done at University College London (UCL). First each lens was installed into its cell. The lens was supported on a rotary table using a set of tip-tilt stages with plastic pads that conformed to the curvature of the lens. The rotary table was used to check that components were centered and level. The cell, supported on a separate x-y stage and tip-tilt system, was then aligned with the lens. The lens was then clocked into the optimal rotation, and the cell was carefully jacked into position so that the full weight of the lens rested on the axial RTV pads. The alignment of the lens could then be checked relative to fiducial surfaces  on the cell. If the lens was within tolerances, the radial pads, on their cell inserts, were now glued into place.  Finally the lens was constrained by a safety retaining ring with RTV pads. In order that the lens  not be over constrained, the RTV pads on the retaining ring do not touch the glass but are held $\sim 50 \mu {\rm m}$ from the lens surface.  Figure~\ref{fig:c1rotary} shows the C1 lens and cell  being mated together on the rotary table.

The C2 and C3 lenses were installed separately into their respective cells, and then the C2 and C3 cell/lens combinations were mated together to form the assembly.  The assembly was inserted into the body with the body held vertically. The distance from the end of C3 to the C5 flange was measured. The assembly was then removed and spacers were inserted to set the C3 to C5 gap to the nominal spacing. The C1 lens and cell were inserted into the body using a similar procedure. The deflections of the cells under a $30\deg$ angle tilt were measured and found to be small and nominal.  The C4 lens/cell assembly was inserted with the body oriented so that the assembly could be lowered from above. After it was inserted, the distance from C4 to C3 was measured. Spacers were used to set the C4 lens at the correct separation from C3.

The body and cone were then assembled, with spacers used to set the distance from C1 to C2. Keys were used to preserve the assembly positions.  Laser alignment tests were performed at each stage of the assembly of lenses into their respective barrel components and during the mating of the body and cone. The laser system provided a measurement of the tip/tilt and decenters of the optical surfaces by comparing the centroids the various reflected and through-going spots.  Additional detail about all these assembly and alignment procedures is available~\citep{SPIEdoel2012}.

\subsection{Prime Focus Cage}
\begin{figure}[h]
\begin{center}
\includegraphics[scale=0.7]{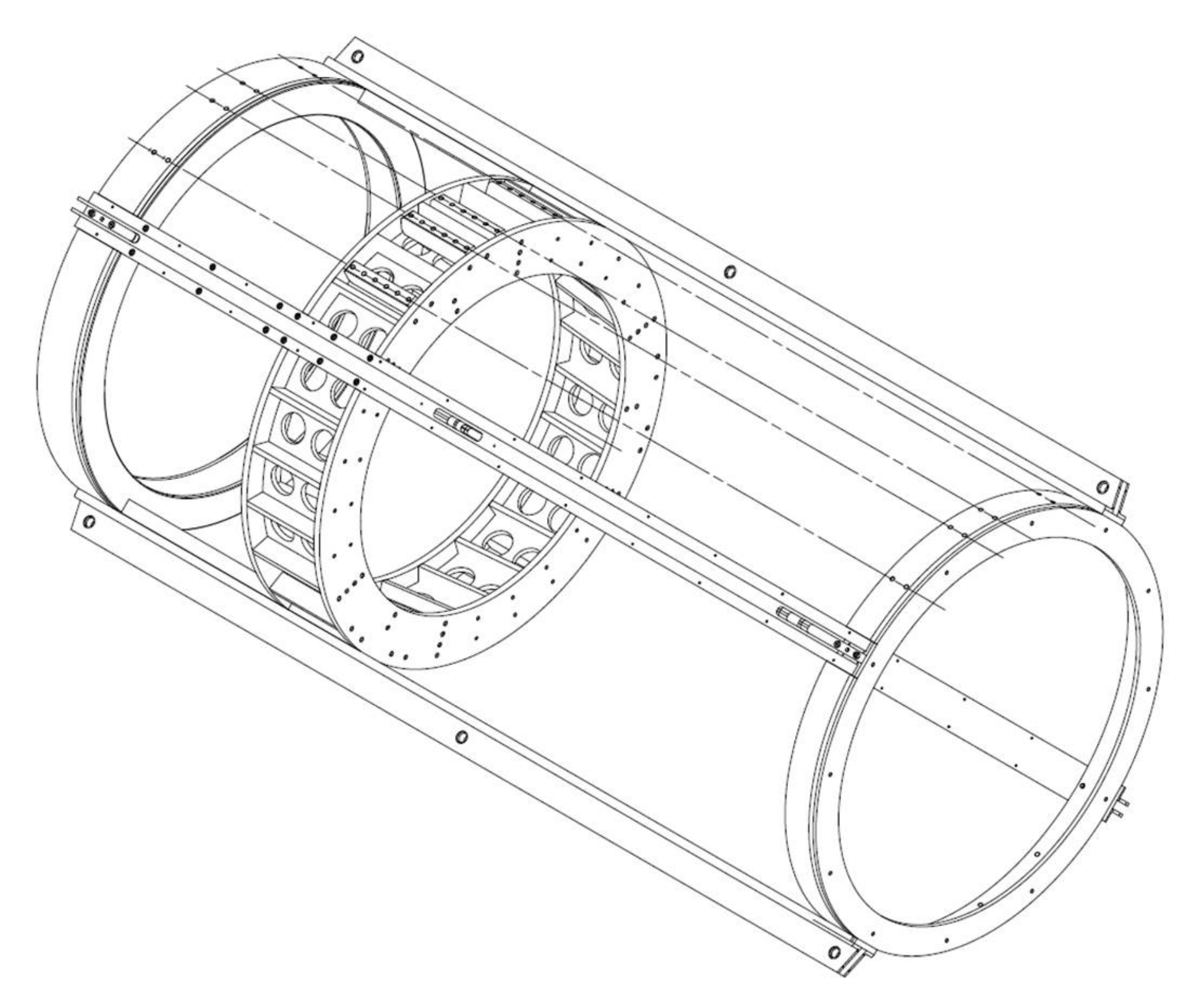}
\caption{Isometric view of the DECam cage.  The end of the cage in the upper left points in the direction of the primary mirror. The hexapod is bolted to the near side of the heavy ring towards the middle of the cage. Slots in the cage rails accommodate the pins that connect the cage to the fins. The weight of the cage components shown is 2239 kgs.}
\label{fig:isometriccage} 
\end{center}
\end{figure}

The whole camera is connected to the prime focus cage, or just ``the cage", by the hexapod, as is described in Section~\ref{ss:hexapod}. The cage, in turn, is connected to the upper rings of the telescope by the four-plane``fins" structure that supported the previous instruments. Every part interior to the fins was replaced by DECam.  Due to the asymmetric loading required by the DECam design, it was necessary to redesign and fabricate a new, sturdier cage. The new cage, shown in Fig.~\ref{fig:isometriccage}, is a four-beam steel structure with reinforced ring weldments where the base of the hexapod makes contact. In addition to thicker rails, the inner diameter of the cage end rings was increased slightly to allow more access for imager installation and maintenance.   The attachment joint between the cage and the fins was modified to be electrically isolating by capturing the 1.5 inch-wide steel attachment pins within G10 sleeves and washers. 

A light baffle is attached to the cage in front of the corrector.  It consists of a series of 8 concentric annuli that decrease in diameter as the distance from the front (mirror side) of the cage increases.  The size of the annuli was set to match the clear aperture of the corrector.  A flexible black material is used between the baffle and the front of the corrector to allow relative motion of the corrector (for focus and alignment) with respect to the baffle. 

Some community observers may want to use instruments at the Blanco Cassegrain focus. To this end,  a secondary mirror may be mounted on  the front end of the Prime Focus cage.  When DECam is used, a new annular counterweight of the same mass as the secondary mirror assembly is mounted on the front of the cage in front of the DECam optics. It   also acts as a light baffle.  Removable counterweights on back end of the cage are used to balance the cage around its connections to the telescope rings.  A cage cap provides thermal and physical protection at the back end of the cage and cage covers seal the area around the imager.  The components are painted with Aeroglaze \textregistered \  Z-306, an anti-reflective black polyurethane coating. 

\clearpage
\section{Focal Plane Detectors}\label{s4:CCDs}

\begin{figure}[h]
\begin{center}
\includegraphics[scale=1.0]{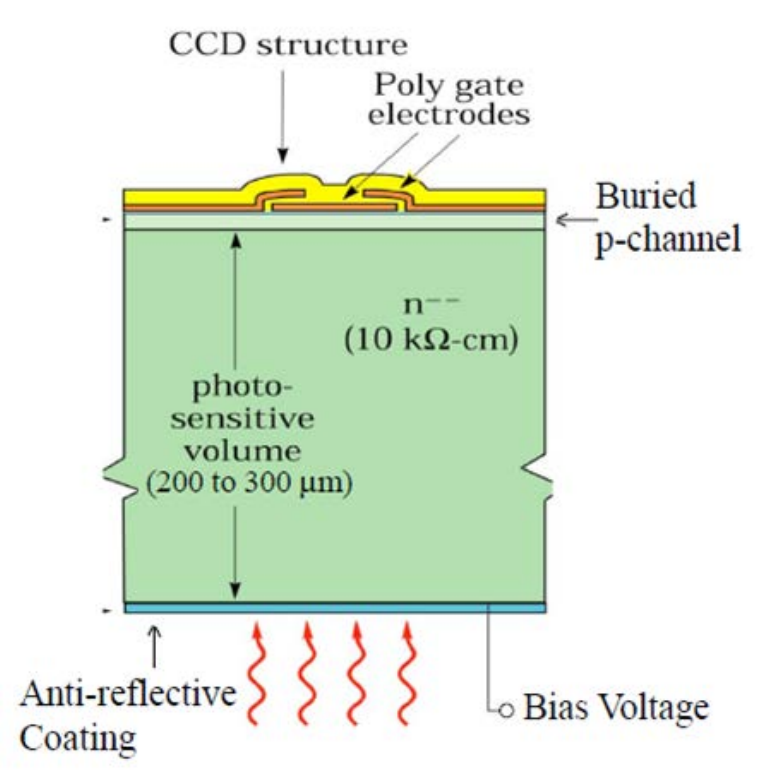}
\caption{DECam uses the LBNL--designed, thick, fully--depleted CCDs. The DECam CCDs are $250\mu {\rm m}$ thick. Photons (red wavy lines) are shown impinging from the underneath (back--side illuminated). The positive bias voltage sweeps positively--charged holes to the poly gate electrodes. The CCDs are operated in 2-phase collection mode.}
\label{fig:thickccd} 
\end{center}
\end{figure}

The DES technical requirements demanded CCDs with low dark-current, low noise, and high quantum efficiency (QE). In particular DES requires a QE$>65\%$ in the z-band.   Table~\ref{tab-ccdspecs} lists the complete set of technical requirements. Lawrence Berkeley National Laboratory (LBNL) developed fully-depleted red-sensitive back-illuminated CCDs~\citep{LBLccd2003, LBLccd2007} that met the requirements. These CCDs are p-channel, fabricated on a high-resistivity n-type substrate. See Fig.~\ref{fig:thickccd}. The CCDs are $250\mu {\rm m}$ thick with 15~$\mu$m pixels and are fully-depleted by a 40V substrate voltage. The positively-charged holes are collected  in the depletion region in buried channels located a few $\mu$m under the gate electrodes. An anti-reflective coating formed from indium-tin oxide and SiO2 is applied to the back side. Each CCD has two serial registers and corresponding output amplifiers that can be readout simultaneously.   

LBNL supplied the CCDs, diced from 6" wafers, to Fermilab.  Each wafer had four $2048\times 4096$ CCDs, one $2048\times 2048$ CCD, and eight 
small test CCDs.   

\begin{table}
\begin{center}
\caption{\label{tab-ccdspecs} The technical requirements for the DECam CCDs and the corresponding characteristics of the LBNL fully-depeleted red-sensitive devices. The requirement for QE stability with time sets a technical requirement on the temperature stability of the focal plane of $0.5^{\circ}$K over the same time period.   While the cosmetic requirements for DECam CCDs were that  no individual CCD shall has more than 2.5\% bad pixels, an additional criterion was applied to the average of the focal plane. The whole focal plane was required to have  no more than 0.5\% bad pixels.The requirement for the flatness of the CCDs came from astrometric science requirements. Within a given 1 sq-cm surface the RMS deviation from the mean is  $< 3 \mu {\rm m}$. Next, adjacent 1 sq-cm regions  have mean elevations that are within $10 \mu {\rm m}$ of each other.  }
\begin{tabular}{|l|c|c|} \hline
                  & DECam                     & LBNL CCD            \\ 
                  & Requirements          & Performance       \\ \hline \hline
Pixel array & $2048\times 4096$ & $2048\times 4096$ \\ \hline            
Pixel size  & $15 \mu$m                & $15 \mu$m          \\ \hline
Readout Channels
                  & 2                             & 2                 \\ \hline
QE(g,r,i,z) & $60\%$, $75\%$, $60\%$, $65\%$ 
                                &$70\%$, $90\%$, $90\%$, $75\%$ \\ \hline
QE Instability
            & $< 0.3\%$ in 12--18 hrs
                                & Stable (see caption)          \\ \hline                                  
QE Uniformity & $< 5\%$ over 18 hrs
                                & Adequate          \\ \hline
Full Well   & $> 130,000$ e$^-$ & $> 170,000$ e$^-$\\ \hline
Dark Current& $< 25 {\rm e}^-$/hr/pixel
                                & Achieved at T $< 180$ $^{\circ}$K                \\ \hline
Persistence & No residual image & Erase mechanism  \\ \hline
Amplifier Crosstalk   
            & $< 0.001\%$       & $< 0.0001\%$     \\ \hline 
Read Noise  & $< 15 {\rm e}^-$ at 250 kpixel/s
                                &  $< 7 {\rm e}^-$ at 250 kpixel/s \\ \hline
Charge Transfer
Inefficiency& $< 10^{-5}$       & $< 10^{-6}$      \\ \hline
Charge Diffusion
            & $\sigma<7.5 \mu$m  & $\sigma = $5--6 $\mu$m       \\ \hline
Cosmetic  Defects  & See caption      & Adequate                  \\ \hline
Non-linearity   & $< 1\%$             & $< 1\%$          \\ \hline
Package Flatness
            & See caption       & Adequate         \\ \hline
\end{tabular}
\end{center}
\end{table}

\subsection{Packaging and Testing $2048 \times 4096$ Imaging CCDs}
\begin{figure}[h]
\begin{center}
\includegraphics[scale=0.5]{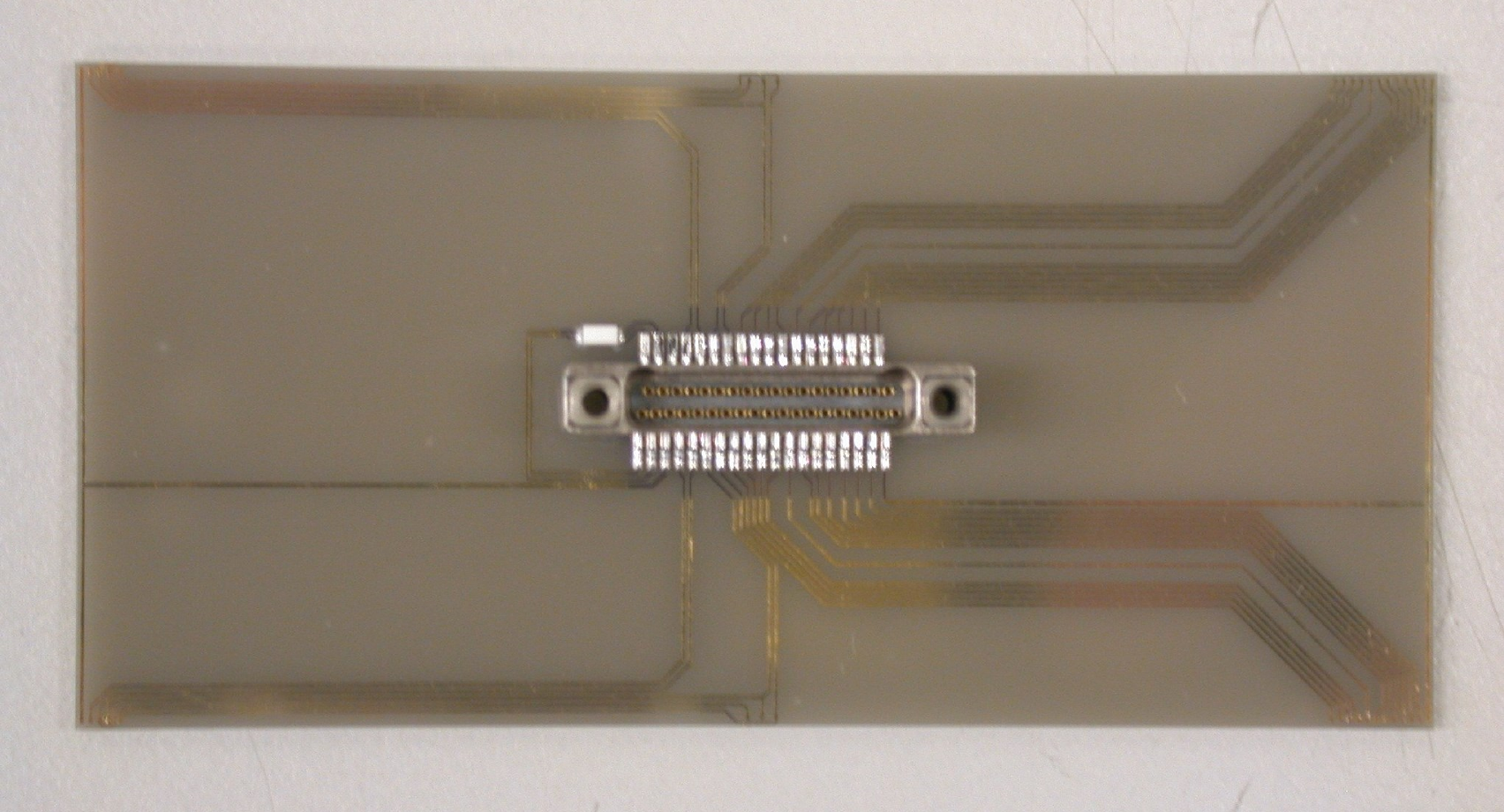}
\caption{The aluminum nitride card with Airborne 37-pin connector (NK-2B2-037-225-TH00-010) soldered to it. The overall dimensions of the AlN card are 61.72  $\times$ 29.97 $\times$ 1.0 mm. The gold traces can be seen. The CCD is glued circuit-side up to the underside of the AlN card and then wirebonded. }
\label{fig:alncard} 
\end{center}
\end{figure}

\begin{figure}[h]
\begin{center}
\includegraphics[scale=1.0]{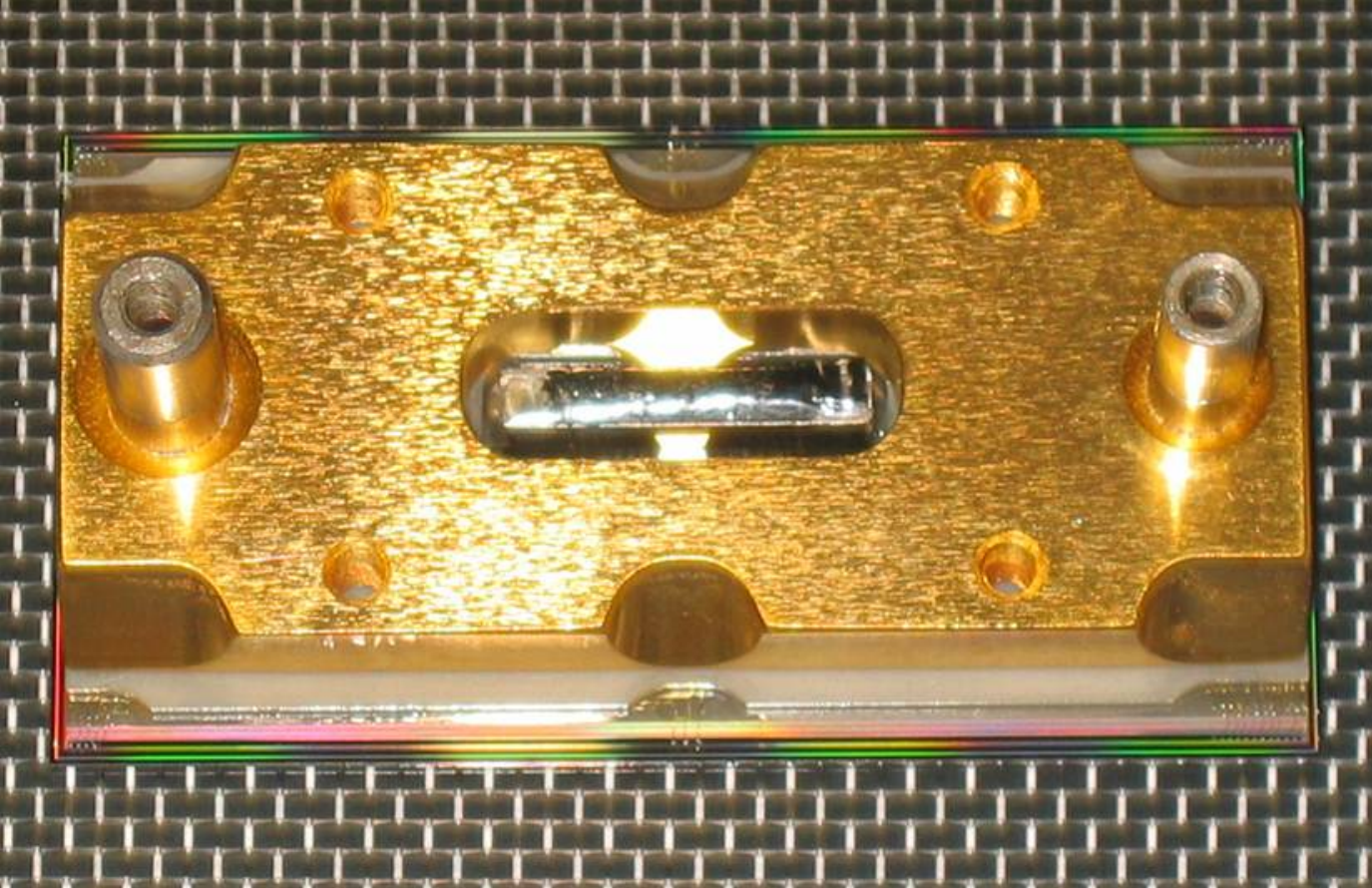}
\caption{A DECam CCD is shown with the illuminated back-side facing down. The CCD is at the bottom of the stack. Wirebonds in six places can be seen connecting the CCD to the aluminum nitride circuit. The gold-plated foot is at the top of the stack. The connector is shown in the center of the assembly. It has a shorting-plug inserted in it for electro-static protection. The two alignment pins are shown. Note that the one shown on the left has a larger diameter. That enforces orientation of the CCD on the focal plane. Also note the 4 small, threaded holes in the foot. These are for handling the CCD, particularly when the device is inserted into the focal plane.}
\label{fig:ccdback} 
\end{center}
\end{figure}

\begin{figure}[h]
\begin{center}
\includegraphics[scale=0.9]{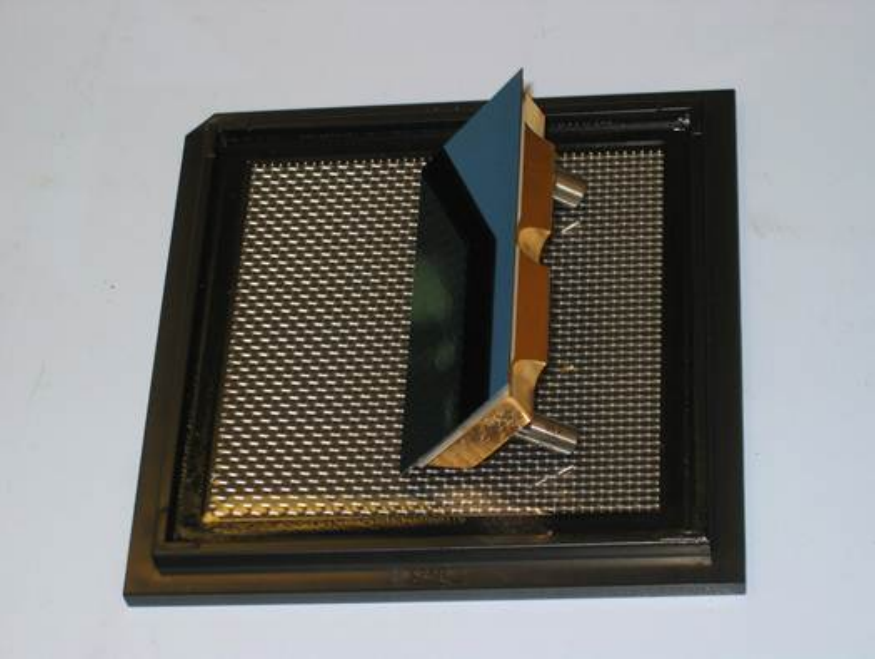}
\caption{A DECam CCD viewed from the side. The back-side, with the anti-reflective coating on it, is facing to the left. }
\label{fig:ccdside}
\end{center}
\end{figure}

The diced CCDs were packaged~\citep{SPIEgd2006} at Fermilab.  The assembly was performed in a series of steps. Alignment and positioning were accomplished using precision tooling.  First, an aluminum nitride (AlN) circuit board was assembled that had a 37-pin micro-connector soldered to the circuitry and an AlN spacer, with a rectangular hole through it, glued to the circuit side so that the connector protruded through the hole.  Figure~\ref{fig:alncard} shows the AlN board with the connector soldered to it.  The front surface of the CCD was glued to the AlN assembly while the CCD was held tightly against a vacuum jig with a very flat surface. Next, aluminum wirebonds were used to connect the CCD to the AlN circuit.  After that the CCD plus AlN assembly was held flat and glued to a gold-plated Invar pedestal or ``foot" that had two alignment/mounting pins pressed into it. In all cases the glue used in assembly was Epotek 301-2, which was found suitable for this work because of its low viscosity and good cryogenic properties~\citep{ccdglue}. A CCD flatness measurement, which involved a time-consuming surface scan of the CCDs at operating temperature, was performed on a small sample of the devices. That test established the capability of the packaging process~\citep{SPIEgd2006} to meet the required flatness constraints.  Fig.~\ref{fig:ccdback} shows the CCD assembly and all of these components. Fig.~\ref{fig:ccdside} shows the CCD assembly as viewed from the side and front.

\begin{figure}[h]
\begin{center}
\includegraphics[scale=0.6]{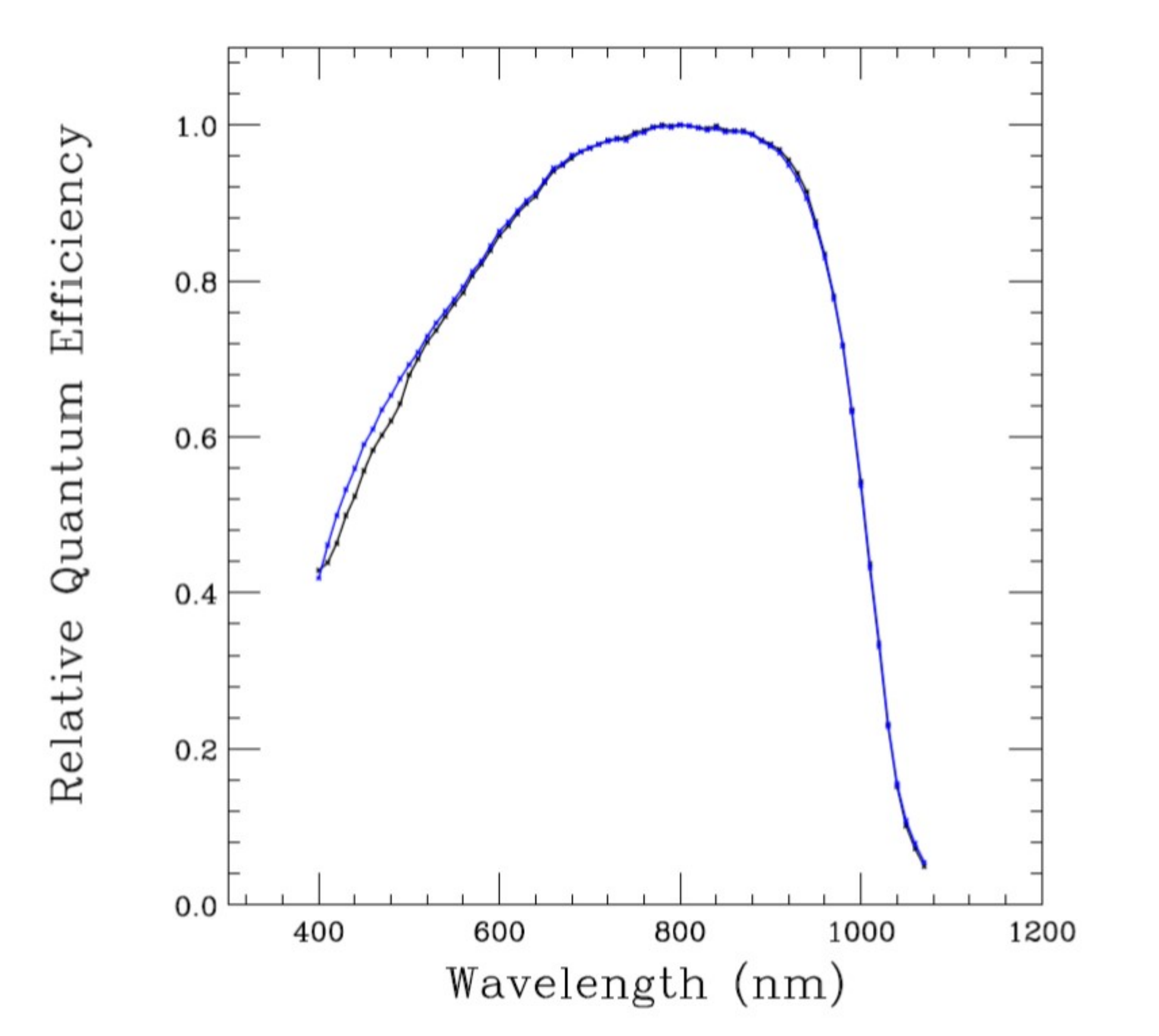}
\caption{This is the quantum efficiency versus wavelength for two typical CCDs in the DECam focal plane. The QE shown are both relative to what was measured at the peak (800 nm). }
\label{fig:qe2ccds}
\end{center}
\end{figure}

CCD testing~\citep{SPIEhtd2008} was performed at Fermilab using pre-production versions of the electronics. After each detector was packaged it was installed in a CCD test Dewar where all of the technical requirements were verified at the nominal operating temperature of -100$^{\circ}$C. The two-stage test process~\citep{SPIEdk2010} takes about three days to complete. The first stage tests the basic functionality of the CCD, determines the QE versus $\lambda$ (see Fig.~\ref{fig:qe2ccds}), and counts the number of hot/dead pixels with a response that is more than $20\%$ different from the average. Such pixels are considered ``defective" and the CCD should not have more than $2.5 \%$ defective pixels. If the CCD didn't pass these criteria, it was removed from the sample of possible science-grade CCDs, so it ws not studied further. The second stage tests required more significant manual setup and included determination of the charge diffusion, charge-transfer efficiency, and dark current and QE as a function of temperature during the warm-up. All of the CCD modules were inspected for thickness and flatness using an optical microscope at room temperature. Additional details of the testing hardware, procedures, analyses, and results are available~\citep{SPIEje2010,SPIEderylo2010}.  

CCDs that passed all of the technical requirements were denoted as ``science grade".  In total, CCD production and testing resulted in 124 science grade $2048\times 4096$ pixel CCDs~\citep{htd-tipp,bebek2012}, for a yield of 25\%. 

\subsection{$2048 \times 2048$ CCDs for Alignment and Guiding}
We chose to use $2048 \times 2048$ CCDs for alignment and guiding applications. These made efficient use of the partially-vignetted areas of the focal plane around the edges of the imaging CCDs. There are eight CCDs used for focus and four used for guiding. Aside from the size the detectors are identical to the larger CCDs.   They are assembled in a pedestal package with a design similar to the larger devices. One key difference is that the overall thickness of the package is $1500 \mu {\rm m}$ less than that of imaging CCDs. The height of the focus CCDs are adjusted so that they are $1500 \mu {\rm m}$ above or below the focal plane. The elevations of the surface of the CCDs are set by AlN  shims, $1500$ and $3000 \mu {\rm m}$ thick, as necessary for guiding ($1500 \mu {\rm m}$ shim) or focus (no shim or $3000 \mu {\rm m}$ shim) roles. 

\subsection{Selection of CCDs for the Focal Plane}
\begin{figure}[h]
\begin{center}
\includegraphics[scale=0.9]{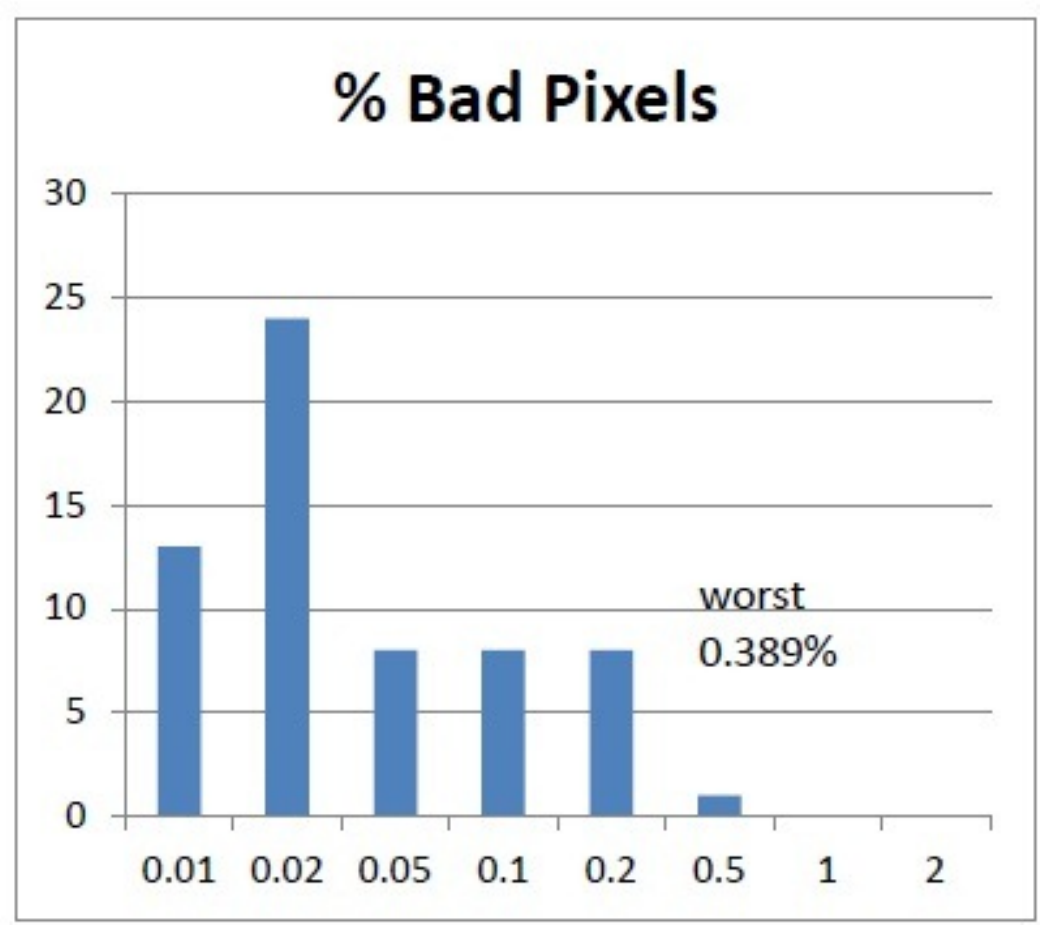}
\caption{The distribution of the percentage of defective pixels for the 62 CCDs installed onto the focal plane.}
\label{fig:cosmetics}
\end{center}
\end{figure}

\begin{figure}[h]
\begin{center}
\includegraphics[scale=0.7]{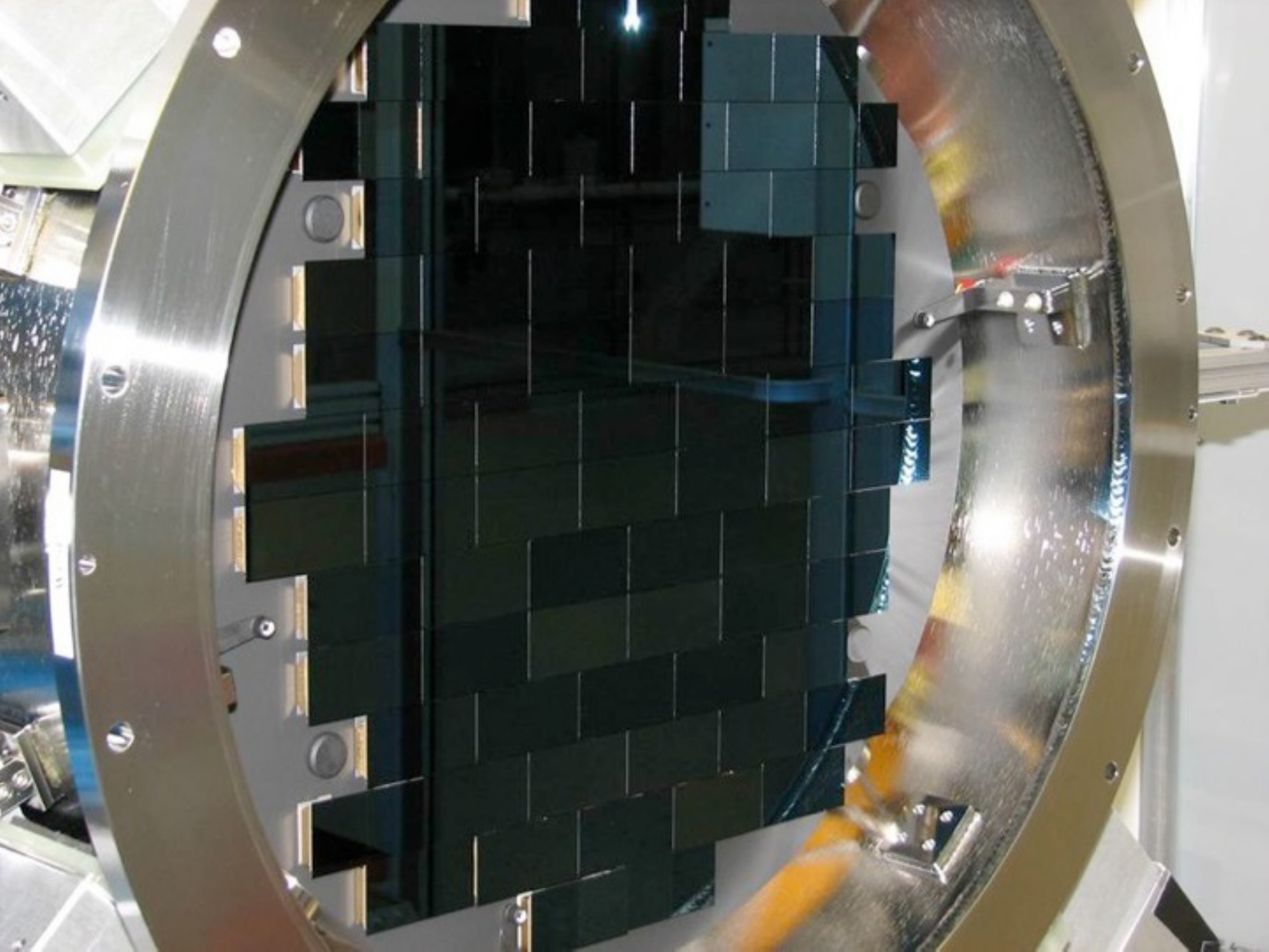}
\caption{The DECam Focal Plane showing the 62 $2 {\rm k} \times 4 {\rm k}$ and 12 $2 {\rm k} \times 2 {\rm k}$ CCDs. The four button-like temperature sensors are also shown. Three photodiodes (see Section~\ref{subsec:slowcontrols}) are mounted on the inside of the camera vessel and stick out over the edges of the focal plane support plate. From this angle and with this lighting, the focal plane support plate doesn't appear as black as the Aeroglaze \textregistered \  Z306 paint actually makes it.  }
\label{fig:CCDsAndMore}
\end{center}
\end{figure}

\begin{figure}[h]
\begin{center}
\includegraphics[scale=0.5]{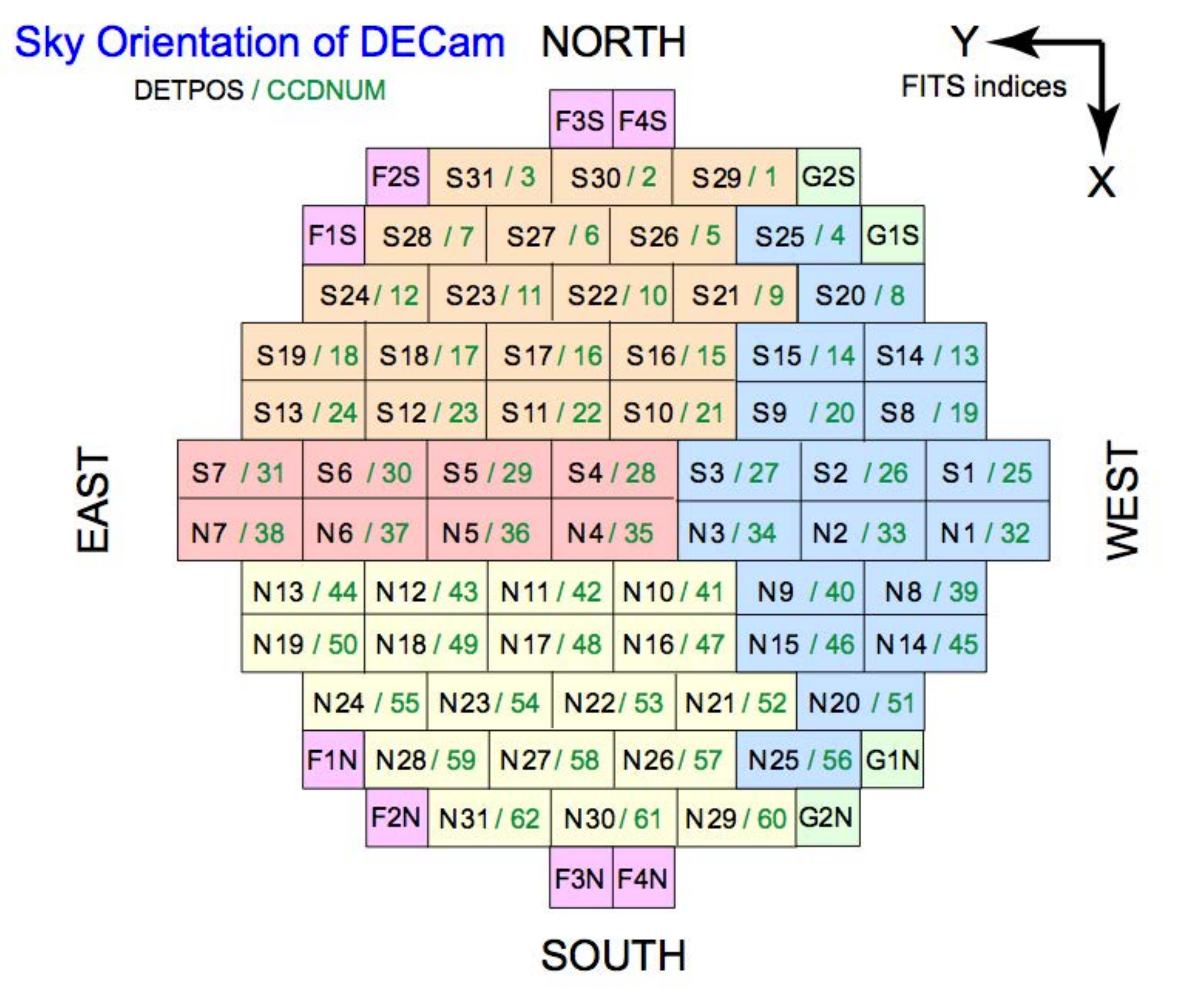}
\caption{The DECam Focal Plane showing the 62 $2 {\rm k} \times 4 {\rm k}$, 8 $2 {\rm k} \times 2 {\rm k}$ CCDs (labeled ``F") for the adaptive optics system, and 4 $2 {\rm k} \times 2 {\rm k}$ (labeled ``G") for guiding.  The orientation of the sky is indicated.  The label (e.g. S30) indicates a position on the focal plane. The label (e.g. 2) indicates  the number of the CCD as is in the multi-extension FITS header.  When the focal plane is viewed with the real-time display at the telescope and also 
with default SAOImage DS9 settings, the direction labeled ``North" is displayed to the left and ``East" at the top. The background colors of the CCDs indicate the electronics backplane that reads them out  (see Section~\ref{subsec:readoutcrate}). }
\label{fig:decamorientation}
\end{center}
\end{figure}

We populated the surface of the focal plane with $2048\times 4096$ CCDs for imaging in all places where more than one-half of the device was unvignetted. Having done that, it happened that none of the science chips was vignetted.  The 62 CCDs used were chosen from the 124 (science grade) devices that passed all of the post-production tests. The selection criteria were, in order, an especially high full well (FW $>$ 180,000 e$^-$), high QE, and lastly a low fraction of defective pixels ($< 0.4\%$ bad pixels). Table~\ref{tab-ccdspecs} listed the requirements on cosmetic defects. Recall that the whole focal plane was required to have no more than $0.5\%$ bad pixels.  Figure~\ref{fig:cosmetics} shows the distribution of the percentage of defective pixels for the 62 CCDs selected for the focal plane. The worst CCD had $0.389\%$ bad pixels. Over the 62 CCDs on the focal plane just $0.049\%$ are considered bad pixels, more than $10\times$ better than the requirement.   The CCDs are operated with the same clock voltages and sequences as was used when they were tested. A photograph of the DECam focal plane is shown in Fig.~\ref{fig:CCDsAndMore}. A schemetic drawing that indicates the orientation of the focal plane on the sky and in  SAOImage DS9~\footnote{http://ds9.si.edu/site/Home.html} displays is shown in Fig.~\ref{fig:decamorientation}.

Similar criteria were used for the guide and focus CCDs. Some of theses CCDs are partially vignetted. Spare CCDs were chosen, as well, for delivery with the camera.

\clearpage
\section{DECam Imager Dewar}\label{s2:DECam}

The DECam CCD imager~\citep{SPIEcease2008, SPIEderylo2010} is a 24--inch diameter cylindrical stainless steel vessel. The imager vessel houses the focal plane support plate, the CCDs and their electronic connections,  a liquid nitrogen heat exchanger and focal plane thermal control connections, sensors, and heaters.  Because the CCDs are operated at $-100 \; ^{\circ}$C, the imager vessel is by necessity a vacuum Dewar.  This section describes the Dewar and its contents: the focal plate assembly and CCD support, the liquid nitrogen circulation system, the Dewar vacuum system, and the instrument ``slow controls" system.

\begin{figure}[h]
\begin{center}
\includegraphics[scale=0.6]{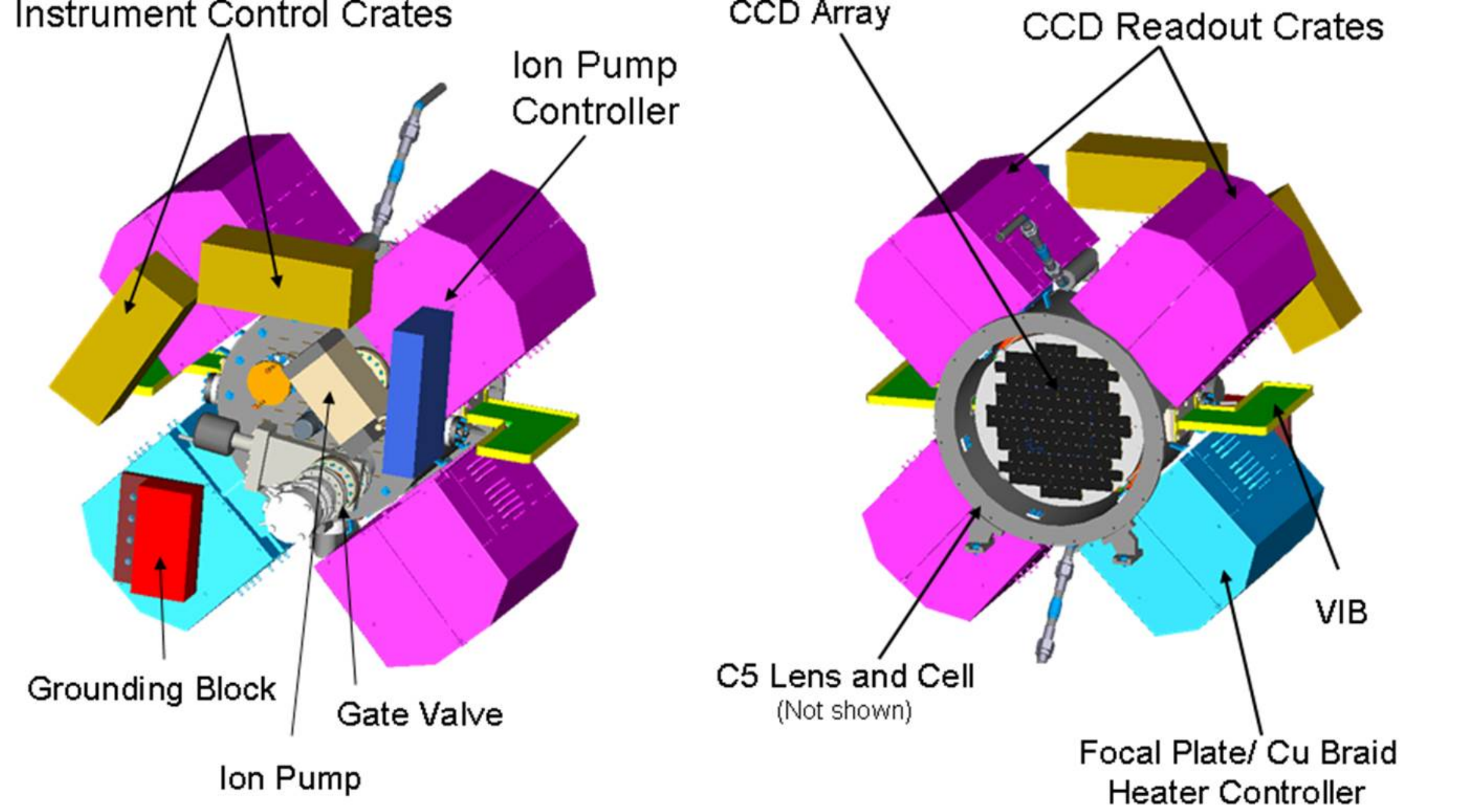}
\caption{Isometric view of the fully instrumented imager, rear (left) and front (right) views. The imager vessel rear flange surface shows the gate valve between the turbo pump and the Dewar, the ion pump, the vacuum gauge (gray), and a positive pressure relief mechanism (orange cap). The front surface shows the focal plane, including the CCD array.}
\label{fig:dewar-frontandrear} 
\end{center}
\end{figure}

The front of the vessel consists of the interface flange that serves as the cell for the final (C5) lens of the optical corrector (described in Section~\ref{ss:c5cell}).   The back of the imager vessel is a stainless steel flange. The walls and back of the imager vessel provide a mechanical mounting structure for the three CCD readout crates, one temperature control crate, and the vacuum systems.  Ports in the sides and in the back cover provide access for the LN2 to the heat exchanger, the vacuum pumps, the electronic and control signals, and the pressure relief (safety) valve.   Two Vacuum Interface Boards (VIBs) route all the CCD electrical signals through the wall of the vacuum vessel. The temperature sensors and heater control signals exit the vacuum vessel through two separate fittings near the VIB.    To minimize signal path lengths to the three CCD readout crates (described in Section~\ref{s5:Electronics}), approximately 2/3 of the CCDs are read out on one side of the imager and the other 1/3 on the opposite side.  Figure~\ref{fig:dewar-frontandrear} shows a schematic of the imager vessel as viewed from the front and rear.

\subsection{Focal Plate Assembly}
\begin{figure}[h]
\begin{center}
\includegraphics[height = 2. in]{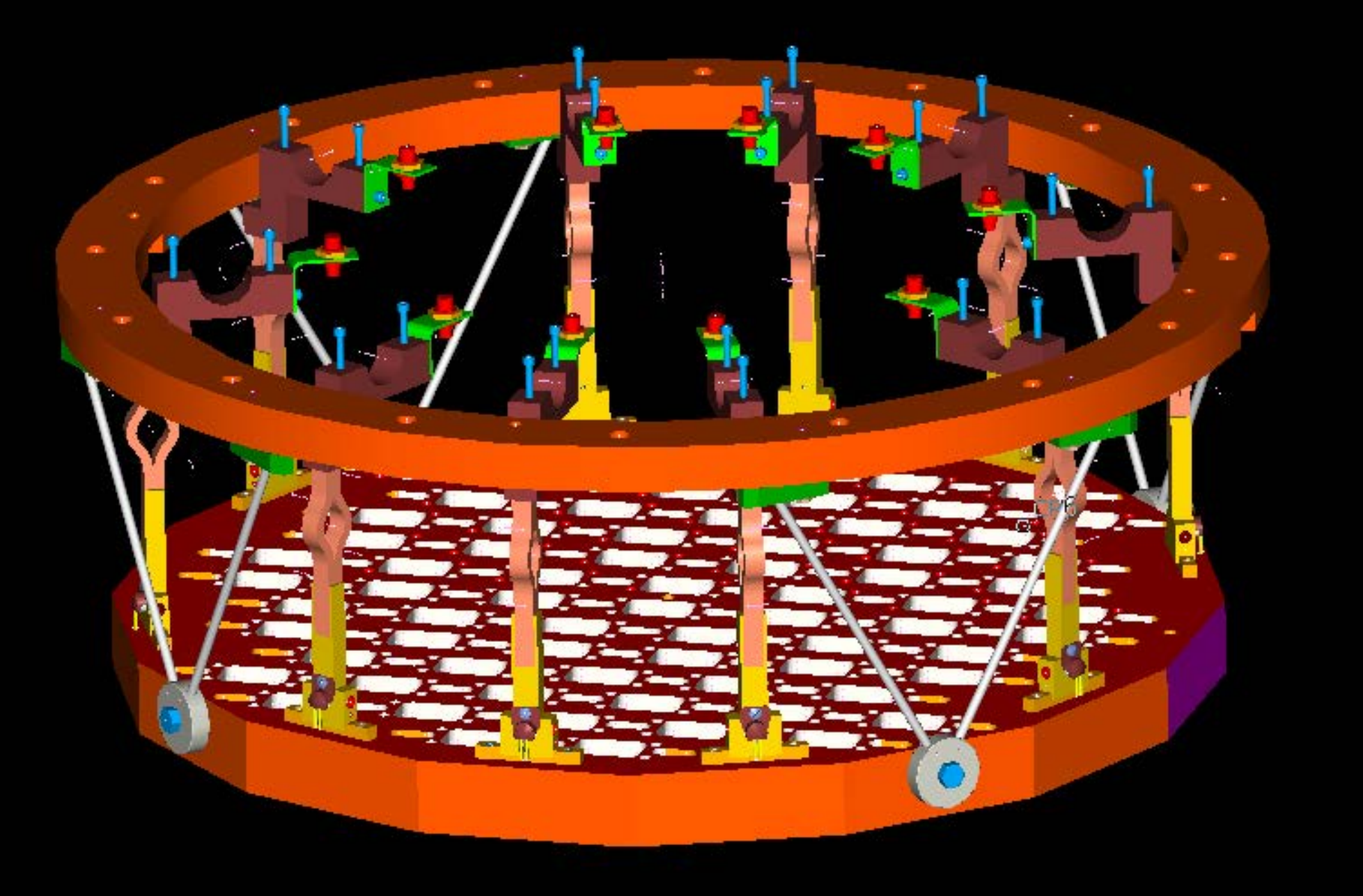}
\includegraphics[height = 2. in]{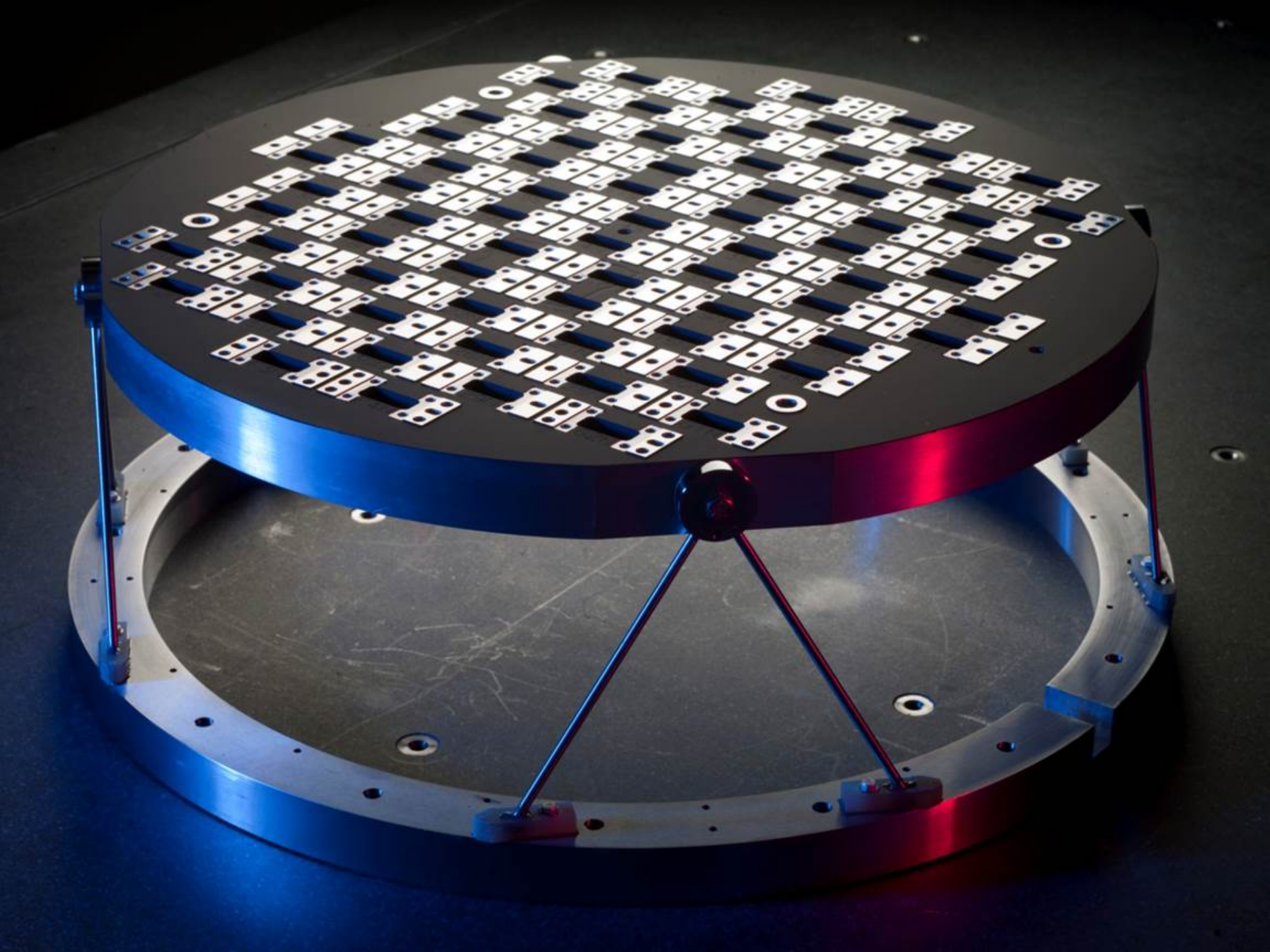}
\caption{The focal plate assembly. The schematic on the left shows the underside of the focal plate assembly including the ring that is bolted to the inside of the imager, the 4 bipods (gray), and the 10 copper braids. The photograph on the right shows the top of the focal plate assembly including the FPSP with raised CCD contact surfaces (metal), the bipods, and the support ring. The surface of the FPSP that is not in contact with the CCDs has been painted with anti-reflective black paint to reduce light scattered from the surface. }
\label{fig:focalplateassy} 
\end{center}
\end{figure}

The focal plate assembly includes the focal plane support plate (FPSP) on which all of the CCDs are mounted, the bipod supports for the FPSP, and the copper braids for cooling and thermal control.  The assembly interfaces with the heat exchanger, the cooling system and the VIB, and is installed by attaching to the internal mounting ring inside the imager vessel. The C5 cell, the rear flange on the imager, and the internal heat exchanger must be removed to gain access to the CCDs. The VIB stays in place. There is a  segmented alignment ring between the focal plate assembly and the mounting ring inside the imager vessel. The alignment ring segments are individually machined to set the dimension between the CCD array and the C5 optical window as well as the parallelism between the two. Fig~\ref{fig:focalplateassy} is an illustration of the focal plate assembly.

The FPSP is supported using four bipod assemblies. The bipod material is Ti-6Al-4V, a titanium alloy, which is a low thermal-expansion, low thermal-conductivity, high-strength metal. Four bipods are used instead of three due to the symmetry of the CCD array and the VIB. The bipods are all supported off of the bipod support ring to make handling the assembly easier. Though we used an electrically-insulating material between the bipods and the bipod support ring, during initial tests we connected the focal plane support plate to the imager vessel interior to improve the noise performance.

\begin{figure}[h]
\begin{center}
\includegraphics[height = 2. in]{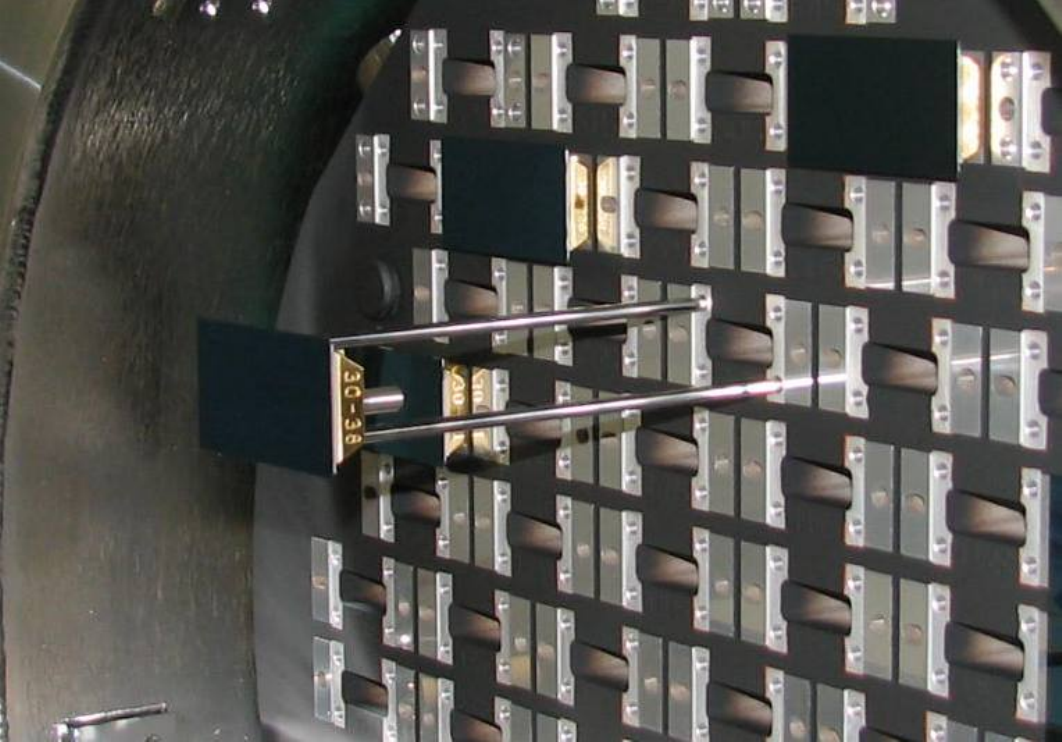} $\:$
\includegraphics[height = 2. in]{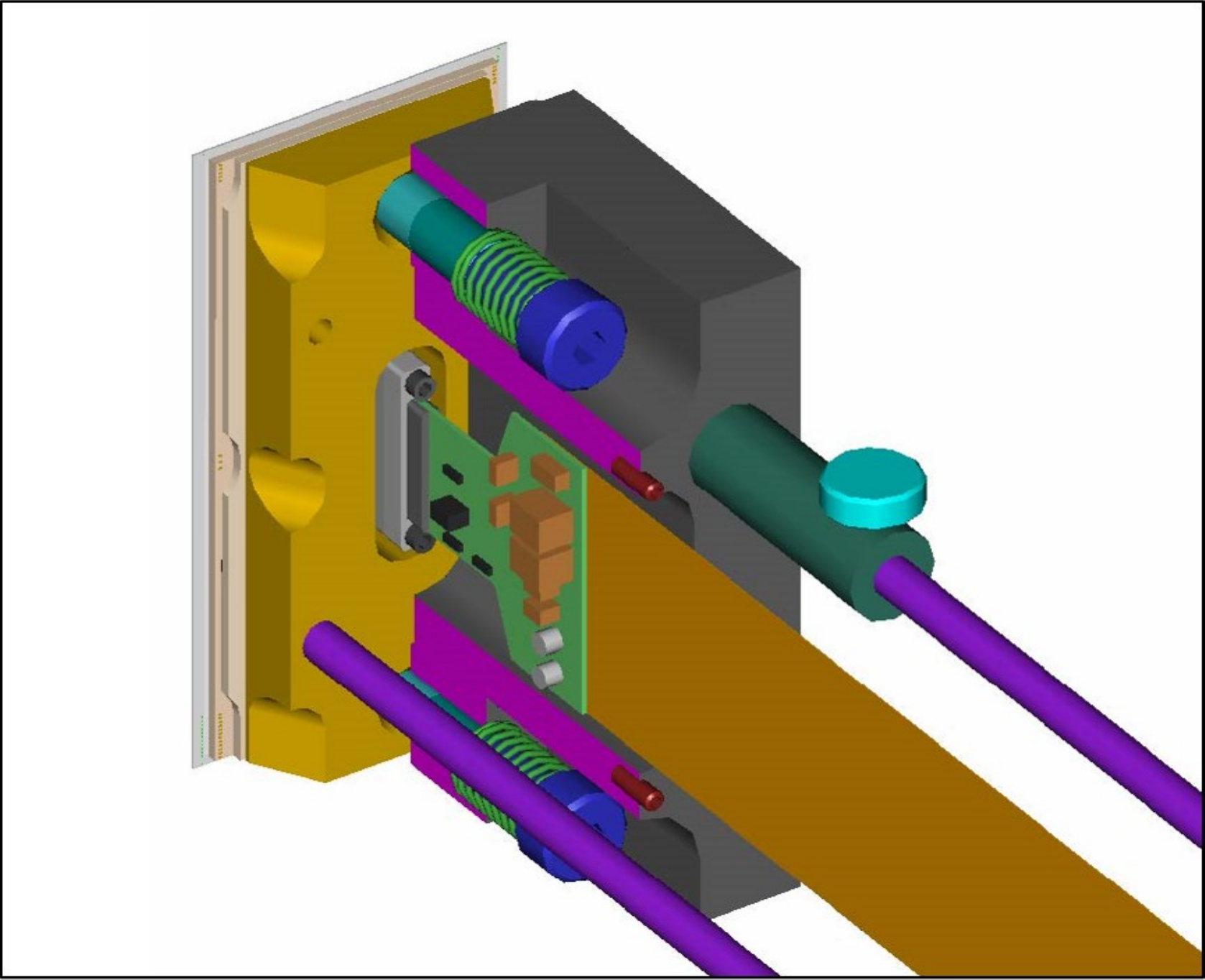}
\caption{The CCD  is held (left) on two steel installation rods that screw into threaded holes on the CCD foot. The rods are inserted into two holes on the FPSP and pulled so that the CCD contacts the (unpainted) raised aluminum surfaces on the FPSP. As  the alignment pins on the CCD mate with the holes in the FPSP, the CCD is precisely positioned so that it doesn't contact its neighbors.  The graphic on the right shows the CCD foot (gold) pulled against the front of the FPSP (gray). The two installation rods are shown (purple) as well as a safety clamp that fits around the one of the rods so the the CCD cannot slip out of place. The two spring-loaded screws that secure the CCD to the FPSP are shown (dark blue). Finally a Kapton flex-cable (see Section~\ref{subsec:kapton}) is shown plugged into the connector on the back of the CCD assembly.   }
\label{fig:holdccds} 
\end{center}
\end{figure}

The FPSP supports the entire CCD array and provides a cold surface for controlling the temperature of the CCDs. The plate is constructed of aluminum MIC-6 tooling plate. There are raised pads on the front side that contact the undersides of the CCDs. These pads form a flat surface.   Cast aluminum MIC-6 was chosen because it has good dimensional stability after machining. Aluminum also has a high thermal conductivity, thus minimizing temperature gradients across the focal plate.   Finite element analysis performed on the focal plate design showed a temperature gradient of $1.5^{\circ}$ C with a flatness, after cooling to operating temperature, within 10 microns.  The CCD assemblies (shown in Figs.~\ref{fig:ccdback} and \ref{fig:ccdside}) are mounted to the aluminum focal plane support plate with one mounting pin through a hole and the other mounting pin through a slot. A spring-loaded fastener screws into the ends of each mounting pin from the back of the focal plane support plate, securing the CCD but allowing the contact surfaces to slip. This combination allows for the different thermal expansion coefficients between the Invar CCD packages and the aluminum focal plate.  After the CCDs are cooled we found the CCD focal surface flat within $\pm 25 \mu {\rm m}$~\citep{SPIEderylo2010,SPIEjhao2010}.  Figure~\ref{fig:holdccds} shows a CCD as it is being mounted on the front of the focal plane support plate as well as how it is secured from the back of the focal plane support plate.

\subsection{Focal Plane Cooling System}
The DECam CCDs must be held at a stable operating temperature of -100 $^{\circ}$C. Warmer temperatures result in increased dark current. Colder temperatures reduce the QE in the near-infrared wavelengths.  The primary requirements for the focal plane cooling system  include that it be able to maintain the mean CCD operating temperature, that the temperature uniformity be $< 10 ^{\circ}$C across the focal plane, that the temperature stability be $\pm 0.25 ^{\circ}$C over a 12-hour period, and that warm up and cool-down times be $<12$ hours.   The heat load from the CCD electronics and the front window of the imager vessel were estimated to be slightly more than 110 W. When the telescope is at zenith the instrument is 12 meters above the pump/cryocooler station.  

For DECam, the combined requirements of high heat load, temperature stability, low vibration, operation in any orientation, high liquid nitrogen cost and limited  available space led to the design of a pumped, closed loop, circulating nitrogen system with a heat exchanger inside the imager Dewar~\citep{SPIEcease2010, SPIEcease2012}.

\begin{figure}[h]
\begin{center}
\includegraphics[height=3in]{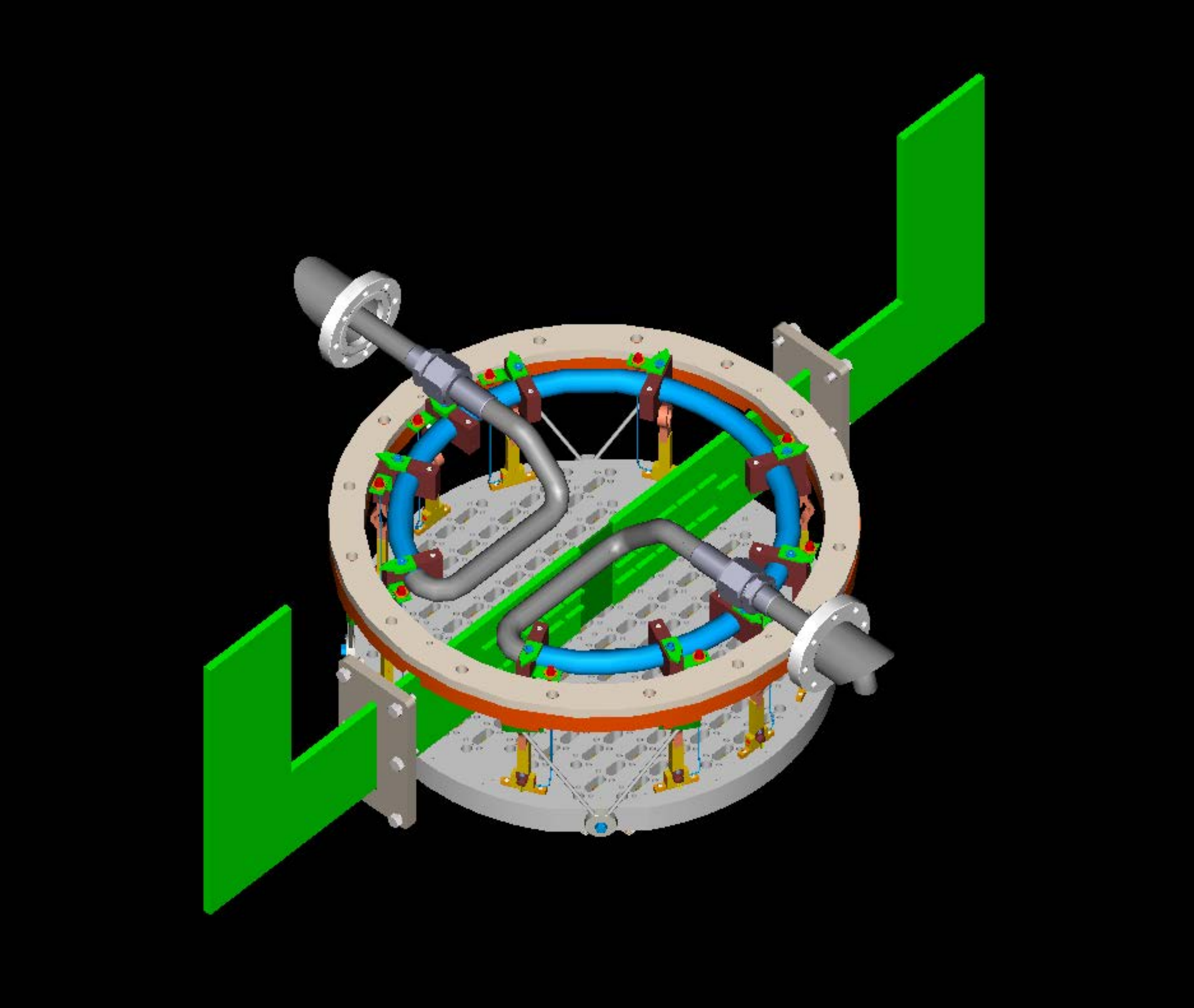}
\includegraphics[height=3in]{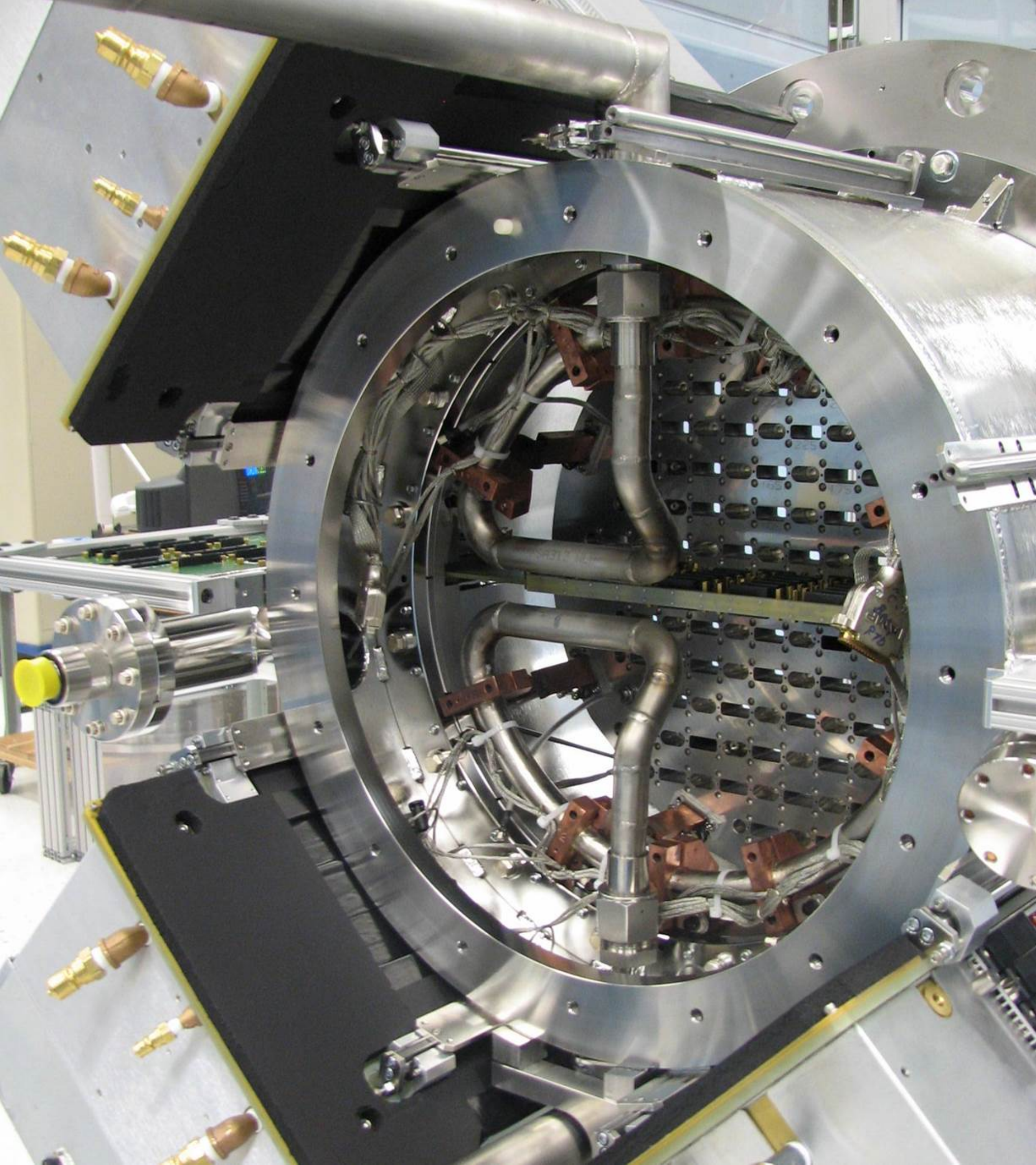}
\caption{Isometric view (left) of the focal plate assembly showing the heat exchange coil (blue and gray) and copper braids that connect the focal plane support plate to the heat exchange coil. These components are identifiable in the photograph (right), which also illustrates the location of the focal plane assembly within the imager Dewar.}
\label{fig:ccdheatexchanger} 
\end{center}
\end{figure}

The heat exchanger, shown in Fig.~\ref{fig:ccdheatexchanger}, is used to remove heat from the focal plate. The focal plate operating temperature is $-100^{\circ}$C. In order to cool the focal plane support plate, the heat exchanger uses liquid nitrogen at a temperature of  $-173^{\circ}$C and as the refrigerant. The heat exchanger is a simple 1-inch stainless steel tube that makes a single loop around the inside of the imager vessel.  Ten copper braid assemblies are mounted around the circumference of the backside of the focal plate for thermal transfer.  To gain access to the CCDs, the heat exchanger can be removed from the imager vessel by disconnecting the copper braids and the metal gasket (VCR) fittings on the tubing. The assembly is then removed through the back of the vessel. 

Each copper braid assembly is a thermal strap between the heat exchanger and the focal plate. Each assembly has a resistive heater and resistive temperature detector (RTD) element  mounted in the lug on the focal plate end for thermal control. The copper braid thicknesses were trimmed to adjust the overall cooling capacity.  A $55^{\circ}$ bi-metallic thermal cutout switch is mounted to the braid and is used in series with each heater for protection against overheating.  An aluminum mounting block is used to attach the copper braid assembly to the focal plate.    A layer of 50 $\mu$m-thick adhesive between the copper lug and the aluminum mounting block provides electrical isolation of the braid from the focal plane.   The wires for the heater and RTD are routed up the side of the copper braid assembly and terminated in a 7 pin connector at the top of the braid. Making the copper braid assembly modular allows for easy replacement of the braid assembly in the event a heater or RTD is damaged.

The thermal connection between the copper lugs and the heat exchanger tube proved to be somewhat challenging.  Custom fitting of the lug-tube joints and Indium foil were used instead of grease to improve the connections because grease had the potential to migrate on to the CCDs.  

The LN2 used to cool the CCDs is stored in a 200 L tank supported on the roof of the old console room in the dome, close by the telescope. The nitrogen is in 2-phase state and is pumped by a submerged pump in the tank up to the camera Dewar and returned in vacuum-jacketed hoses. Flexible hose is used at the polar axis wrap located near the back of the telescope, at the wrap around declination axis on the west side of the telescope, and at the back of the camera. The vacuum-jacketed line segments that cross from the outer ring of the telescope to the camera are only 1.5 inches in diameter and are partially shadowed by the fins, thus minimizing the obscuration of the primary mirror.     There is a total of about 160 feet of hard pipe and about 75 feet of flexible pipe in each of the two lines, supply and return, for a total of about 470 feet of vacuum-jacketed piping.    Two 300W helium cryo-coolers that penetrate the top of the 200 L tank are used to re-condense the nitrogen gas, making a closed LN2 system. There is 70W to 80W of cooling headroom provided by the cryo-coolers. Two resistive heaters in the LN2 tank provide a heat load so that we maintain 2-phase nitrogen. 

The cryogenic system requires routine maintenance.  The submerged pump that forces circulation of the LN2 is being refurbished and replaced on a roughly 7 month cycle because of wear to the bearings, requiring a warm-up of the camera.  We are working to extend the lifetime of the pump to 12 months or more. The vacuum-jacketed line segments are evacuated with each pump replacement.  We have also found that from time-to-time the 70W of cooling headroom has been helpful due to temporary extra heat loads on the system.  If the heat load exceeds the cooling capacity of the cryocoolers, N2 is vented. In that case, the 200 L tank can be manually topped-up at a convenient time.  During operations over the past two years we found the cool-down time is about 4 hours. The camera warms up to +8 $\deg$ C within 24 hours without using the focal plane heaters to speed the warm-up.  

\subsection{Dewar Vacuum System}
The DECam imager vacuum system consists of a roughing and turbo pump system primarily for initial pump-down, cryo-pumping by the cold surfaces (such as the heat exchanger internal to the vessel), and an ion pump for maintenance of the vacuum after good vacuum ($10^{-6}$ Torr) has been established. A full range vacuum gauge is used to monitor the pressure of the imager vessel.  All of the vacuum components are mounted to the rear flange of the imager vessel.  Inside the vessel the flange ports have baffles to prevent debris from falling into the vacuum components and to eliminate any light leaks.  Fig.~\ref{fig:dewar-frontandrear} also shows the rear flange of the imager vessel with the vacuum components.  The turbo pump is attached to the gate valve and a flexible line runs from the output of the turbo pump to a roughing pump mounted in the Cassegrain cage. Vibration isolating mounts eliminate mechanical coupling between the roughing pump and the telescope. There is no molecular sieve (zeolite or activated charcoal) within the Dewar. 

Gas loads on the vacuum system are caused primarily from outgassing and permeation through seals. Major components that caused outgassing were the surfaces inside the imager vessel including the  CCD Kapton cables and the VIBs, which consists of two G--10 multilayer boards that penetrate the vessel walls (see Section~\ref{s5:Electronics}).  All of the large seals use O--rings.  Copper gaskets are used for flanges such as instrumentation feed-throughs that are rarely opened.  The total expected initial gas load from outgassing  (now finished) was $7.5\times 10^{-4}$  Torr--L/s and was dominated by water vapor outgassing from the Kapton cables. The total expected gas load from permeation is $5 \times 10^{-6}$ Torr--L/sec coming through the O--ring type seals.  

Initially the turbo/roughing pump system was used to bring the pressure in the vessel from atmospheric to $2\times 10^{-4}$ Torr  prior to cooling the CCDs.  The combination of cryopumping plus either the ion pump or the turbo ensured a good vacuum.   At present, the pressure within the Dewar is $\sim 1\times 10^{-6}$ Torr when the focal plane is at room temperature  and  $\sim 1.5 \times 10^{-7}$ Torr when the focal plane is at operating temperature and we are running the ion pump.

\subsection{DECam Instrument Slow Controls System} \label{subsec:slowcontrols}
The instrument ``slow" controls system (ICS) controls and monitors critical systems described in this section as well as the crate monitor board within each front-end crate (see Section~\ref{subsec:readoutcrate}). Control loops and monitor functions are programmed in LabVIEW  and use a mixture of National Instrument Compact RIO and FieldPoint programmable automation controllers. On-camera hardware is located behind the Dewar near the back of the cage. The controls for the LN2 system are located nearby the LN2 tank.  These systems may generate alarms when abnormal conditions are detected. Appropriate alert levels are generated, depending on the severity of the condition, ranging from an email alert (low priority) to auto-dialed phone call tree (high priority). The slow controls are also programmed to implement protective actions that protect the equipment when abnormal conditions persist.  For instance, the focal plane has three photodiodes mounted  in unvignetted locations outside of the positions of some CCDs (see Fig.~\ref{fig:CCDsAndMore}). The photodiodes will detect if there is a light level that is dangerous to the CCDs and the slow controls will take a protective action.  These slow controls systems interface to the data acquisition system (SISPI) to archive alarm messages and telemetry information in the DECam database (see Section~\ref{subsubsec:telemetry}).

\clearpage
\section{Front-End Electronics}\label{s5:Electronics}

The design and development of the DECam electronics~\citep{SPIEcardiel2008, SPIEcampa2008, SPIEcastilla2010, SPIEshaw2010,SPIEshaw2012} was a joint effort among multiple institutions in the U.S. and Spain. The main challenge was to read out the focal plane at a rate of 250k pixels/second with less than 15 e$^-$ RMS of readout noise. This was complicated by the limited amount of space available at the top of the prime focus cage, which required that the readout electronics be very compact. 

The design is based on the NOAO Monsoon CCD controller architecture~\citep{Monsoon2004}. The electronics control the following sequence of events that occurs during observations. The CCD is flushed of any residual charge using a reset procedure.  The shutter opens and the CCD accumulates charge in the potential wells of the pixels.  The four CCDs that supply telescope guiding information may be read out while the other CCDs are integrating. After the shutter closes, the Clock \& Bias Boards, which are situated in the readout crates, control a sequence that reads out the CCDs in parallel, first shifting each half-row of the CCD's array of pixels onto one of the two serial registers, then shifting the serial registers, one pixel at a time onto the amplifier nodes, which provide the video-output signal. This is digitized in the 12-channel Acquisition Cards that reside in the readout crates.  By using the correlated-double-sampling (CDS) technique, the noise baseline is removed from the charge integration.  The digitized signal (16 bits) is stored in the Master Control Board (MCB) until after all of the pixels have been digitized.  At that time the digital information is sent to data collection computers (see Section~\ref{s7:sispi}) over optical link.  The Master Control Boards are synchronized so that each is performing the same step of the procedure at the same time, essential for keeping the readout noise small. After the CCDs have been readout we perform an erase and clear sequence that removes any remaining residual charge due to image persistence.

In this section we describe the electronics and supporting infrastructure. We trace the electronics from the connections to the CCDs, to the vacuum-interface boards that penetrate the Dewar vessel wall, to the readout crates with the clock boards, video cards, and master readout controls.  We describe key components of the infrastructure that provides services to the electronics.

\subsection{Electronics Inside the Vacuum Dewar: CCD to Vacuum Interface Board} \label{subsec:kapton}

\begin{figure}[h]
\begin{center}
\includegraphics[scale=0.4]{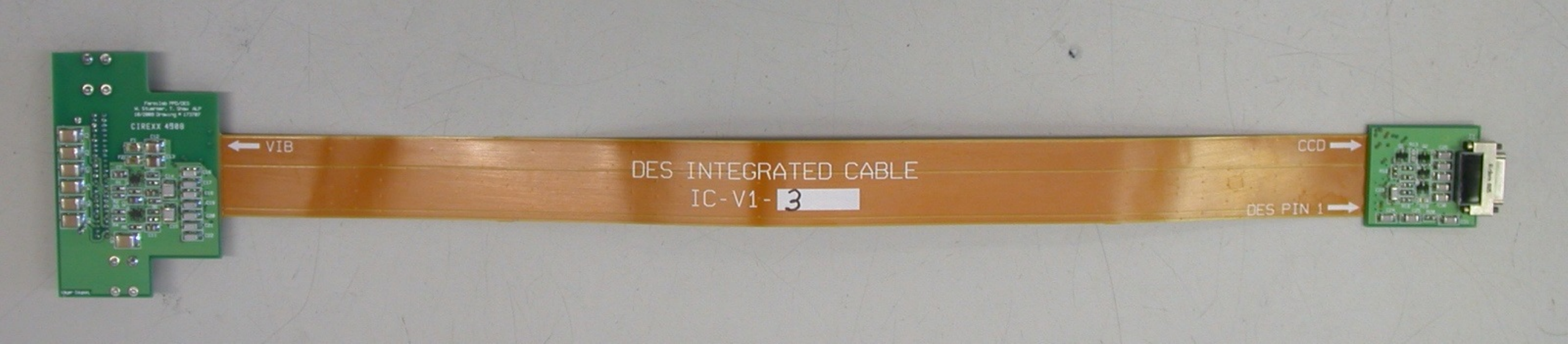}
\caption{One of the Kapton cables that connects each CCD to the vacuum interface board. The JFET source follower is on the right. The video preamplifier is on the left . The connecter for the preamplifier side is on the other side of the card.}
\label{fig:kaptoncable} 
\end{center}
\end{figure}

\begin{figure}[h]
\begin{center}
\includegraphics[scale=0.65]{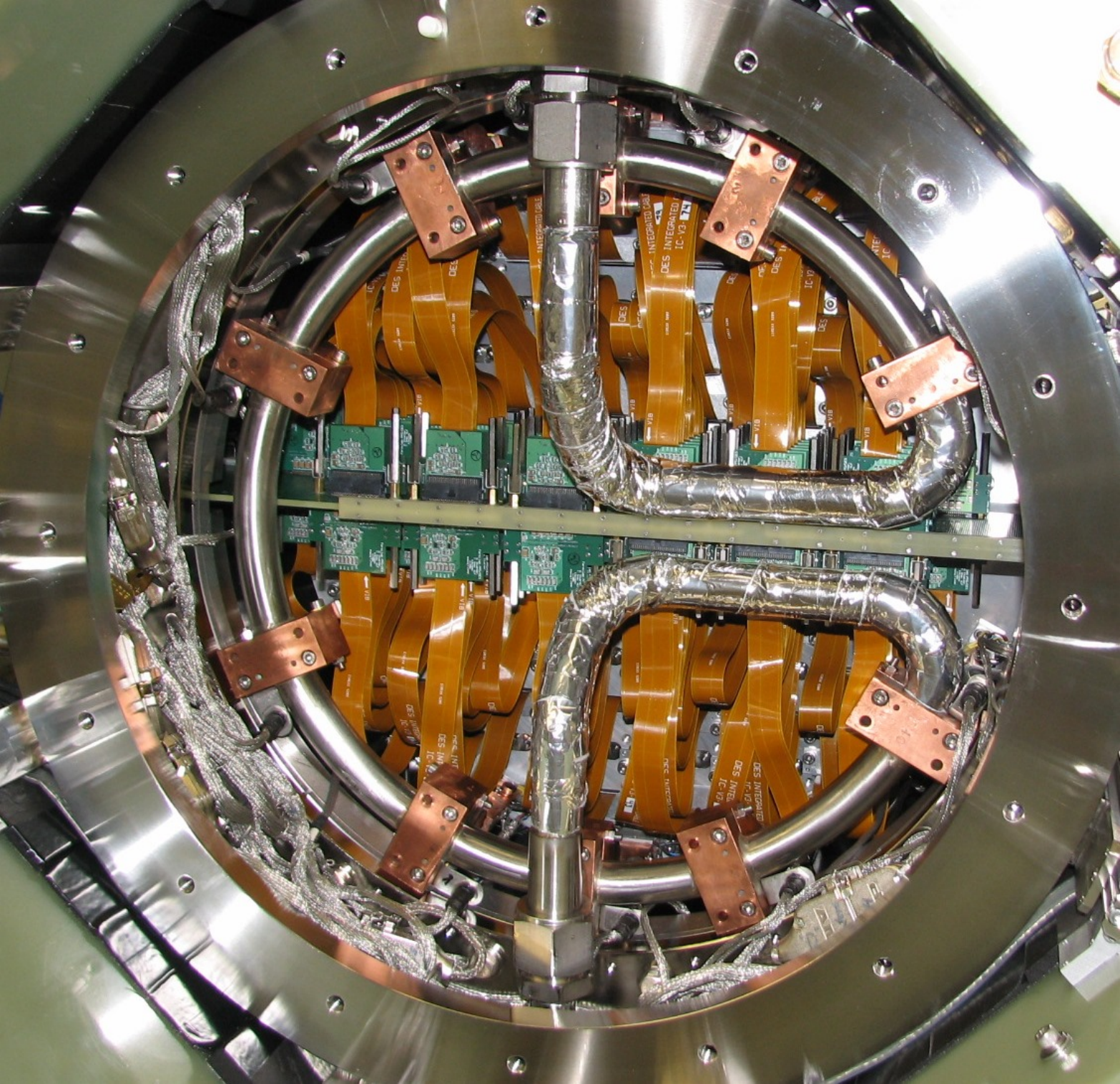}
\caption{DECam Dewar interior with the back cover removed. Some of the 74 Kapton flex cables with preamplifiers are seen plugged into the VIB. The ``pretzel-shaped" tube with the ten copper cooling-braids connected to it is for the LN2. These cooling braids are also coupled to the focal plane support plate.}
\label{fig:StuffedImagerBackside} 
\end{center}
\end{figure}

\begin{figure}[h]
\begin{center}
\includegraphics[scale=0.65]{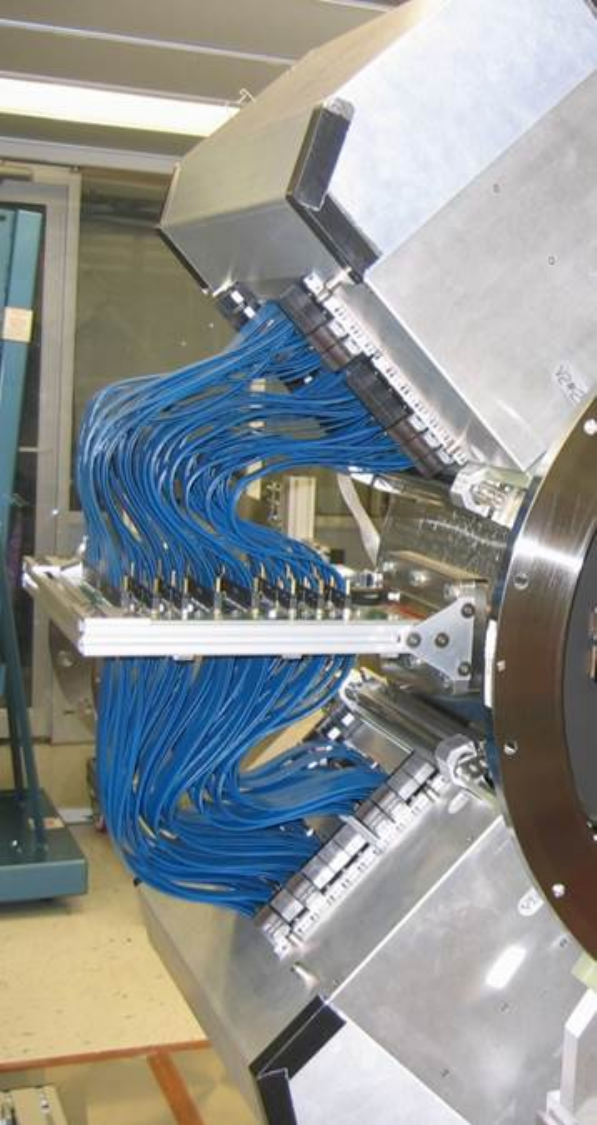}
\caption{The multi-conductor cables that connect one VIB to two of the DECam electronics crates. These cables are normally covered by an aluminum cover that provides shielding against electrical noise and that blocks light that could pass by or through the VIB.}
\label{fig:VIBCables} 
\end{center}
\end{figure}

An 8-layer board connected to a flexible cable is plugged directly into the connector on the back of the CCD. It carries the clock and bias levels to the CCD and the video output signals out. The video signals are transmitted using a dual JFET source follower circuit that reduces the large driver impedance of the CCD video output amplifier. The flexible cable is roughly 10 inches long and has 3 layers. The outer two layers provide shielding and the inner layer carries the clock and bias levels to the CCD and the video outputs to a small ``preamplifier" card. Figure~\ref{fig:kaptoncable} shows the flex cable with the cards on either end. The preamplifier cards on each flexible cable are plugged into either of two Vacuum Interface Boards (VIBs) and drive the video signal to the readout crates. Figure~\ref{fig:StuffedImagerBackside} shows the view of the imager Dewar with the back cover removed. It shows the Kapton flex cables as those are plugged into the VIB. It also shows the LN2 cooling system. The VIBs are mounted into vacuum flanges such that one section of each board is inside the vacuum within the imager vessel while the other side is on the outside of the vessel. Care is taken to form a continuous copper shield against external electrical noise and to block light that might otherwise make its way from the edge of the VIB to the inside of the vessel. The VIBs are connected to the DECam electronics crates using multi-conductor coaxial cables for the video signal  and bias voltages, and multi-conductor twinaxial cable for the clock signals, as shown in Figure~\ref{fig:VIBCables}. 

\subsection{DECam Readout Crate Electronics} \label{subsec:readoutcrate}
\begin{figure}[h]
\begin{center}
\includegraphics[scale=0.1]{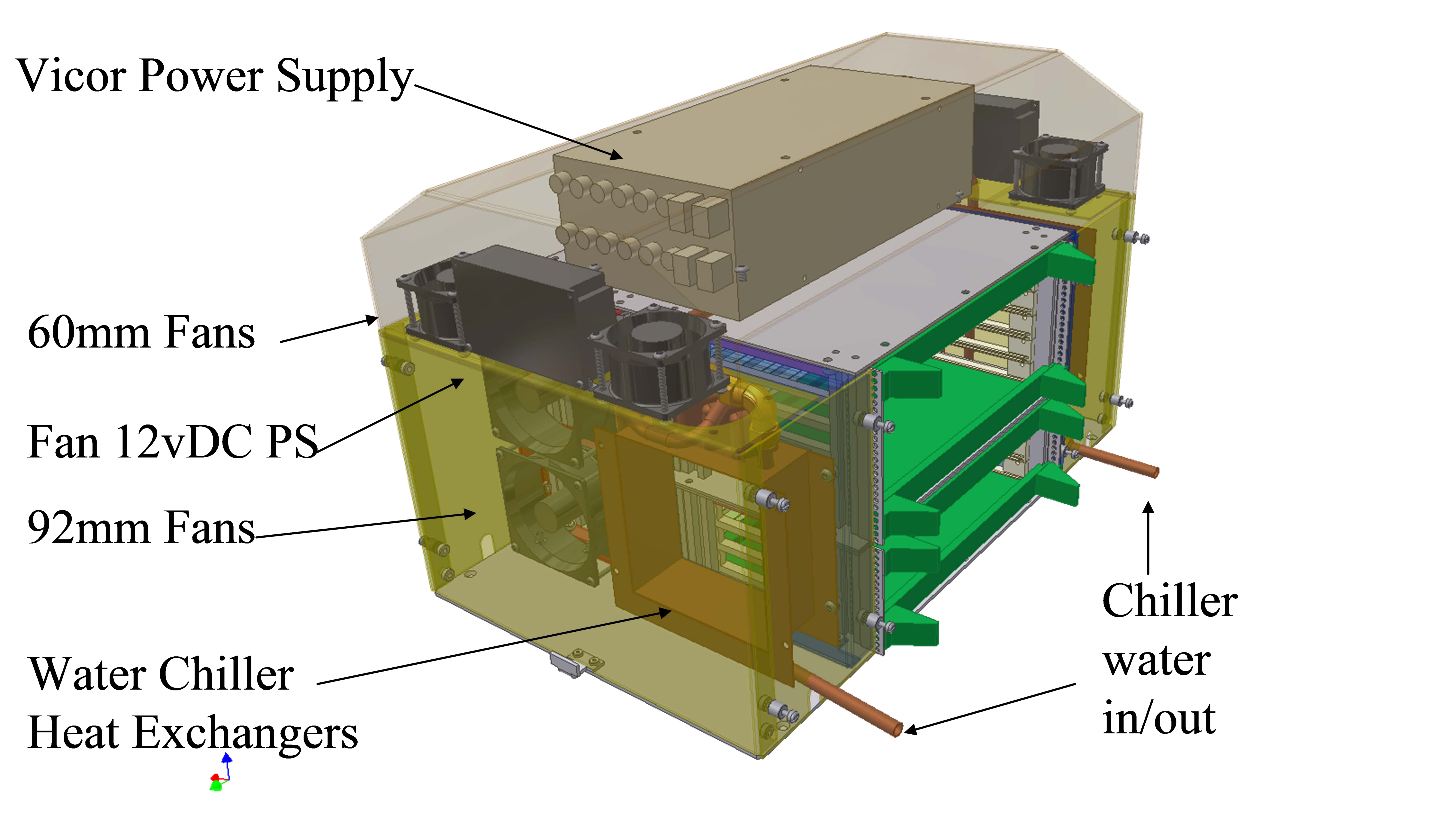}
\caption{DECam Readout Crate. The front-side of the crate is on the right.  The Master Control Board, Clock and Bias Boards, and Video Boards plug into the front side. The cables between the VIBs and the crate plug into Transition Cards, which are on the back side of the crate. The location of the crate infrastructure is highlighted in the figure. An insulating shield covers the crates after they are installed on the camera. }
\label{fig:DECamCrate} 
\end{center}
\end{figure}

DECam has three readout crates. The unit DECam crate, shown in Figure~\ref{fig:DECamCrate}, has dual 6-slot and 4-slot backplanes for a total of 10 slots of main (front-side) Monsoon modules, and ten slots for 120mm transition cards on the back side.  At both ends of the crate there are air plenums to re-circulate the air through the Monsoon modules and transition cards.  There is also an air plenum for the power supplies.  There is a water-cooled heat exchanger at each end of the Monsoon modules, and two fans at each end of the transition cards.  Two more fans at each end of the power supply plenum force some of the cooled air through the power supplies; the rest blows through the transition cards. Separate DC supplies are used to power the fans, which must be powerful enough to overcome the pressure drops in the heat exchangers.  The DC supplies also power an independent internal crate monitor board that  communicates real-time (slow) controls and monitoring information to an interface computer and a telemetry database. 

\begin{figure}[h]
\begin{center}
\includegraphics[scale=0.4]{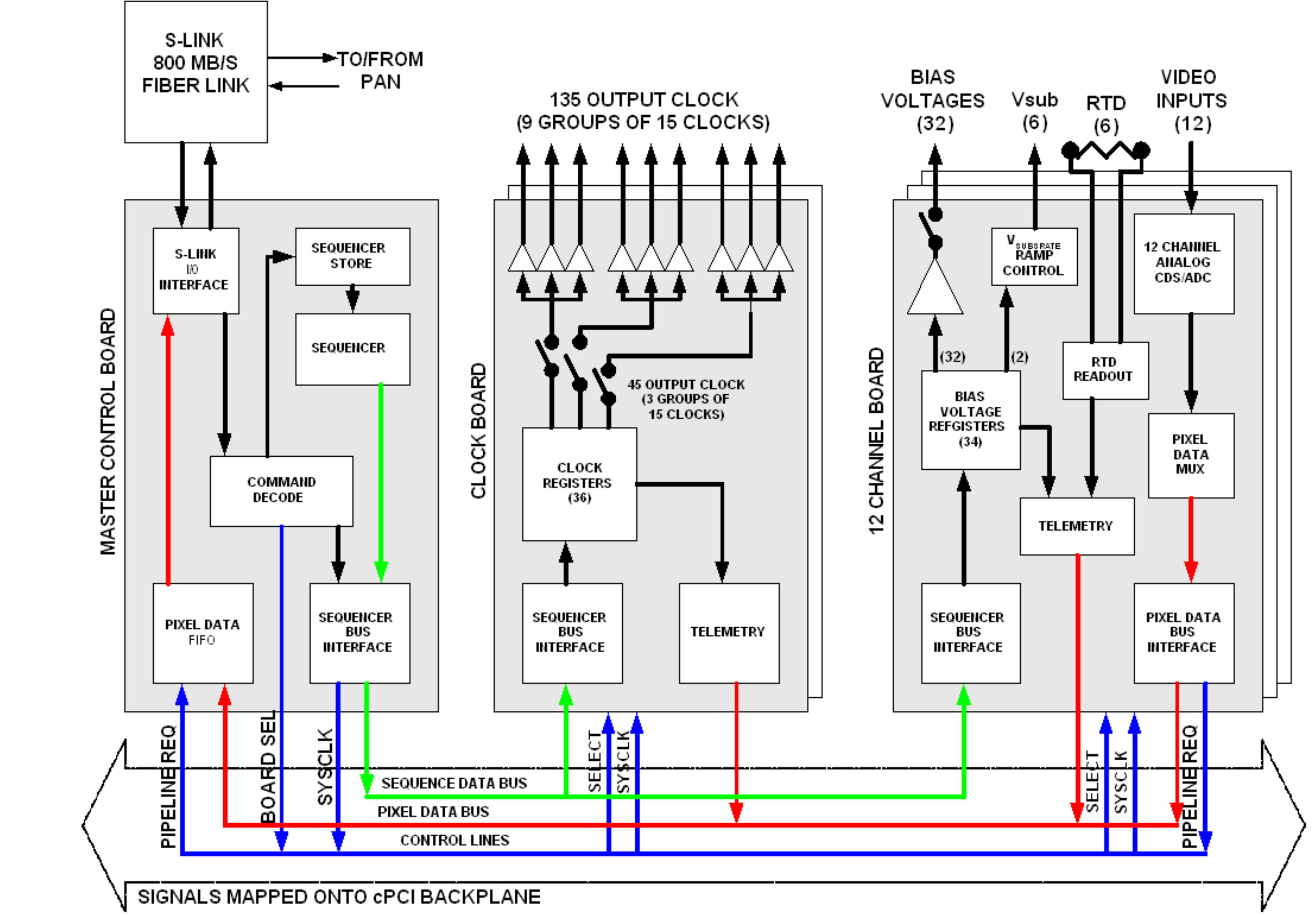}
\caption{Block diagram of the DECam readout crate showing the 3 different cards and the I/O interfaces.}
\label{fig:Electronics-Block-Diagram} 
\end{center}
\end{figure}

There are three kinds of cards on the front side of the readout crate: one Master Control Board, Clock and Bias Boards, and Acquisition Cards.  All DECam main (front side) modules have the format of a 6U 160mm cPCI card.  All transition, or rear, modules have a 6U 120mm format.  Just as with the original Monsoon modules, a proprietary cPCI backplane is used; however most of the pin functions have been reassigned for the DECam design. Figure~\ref{fig:Electronics-Block-Diagram} shows a block diagram of a crate with a 6-slot backplane that allows for the readout of up to 18 CCDs.  This implementation makes use of a Master Control Board, two Clock Boards and three 12-Channel Acquisition Boards.  DECam also makes use of a 4-slot backplane which contains a Master Control Board, one Clock Board and two 12-Channel Acquisition Boards which can read out up to nine CCDs.  DECam uses three 6-slot backplanes and one 4-slot backplane to readout the 62 2kx4k imaging CCDs. It uses one 4-slot backplane to readout the 8 focus CCDs and another to readout the 4 guide CCDs.  

While most read and write operations are controlled by the MCB, the backplane protocol allows for multiple peripheral boards to arbitrate prioritized high-speed block transfers of pixel data through the MCB to the Pixel Access Node (PAN) computer.  The DECam backplane is synchronous to the rising edge of a 40MHz clock generated by the MCB.  Each peripheral board slot receives a dedicated, independently-controlled buffered copy of the MCB system clock.

\subsubsection{Master Control Boards}
\begin{figure}[h]
\begin{center}
\includegraphics[width=3in]{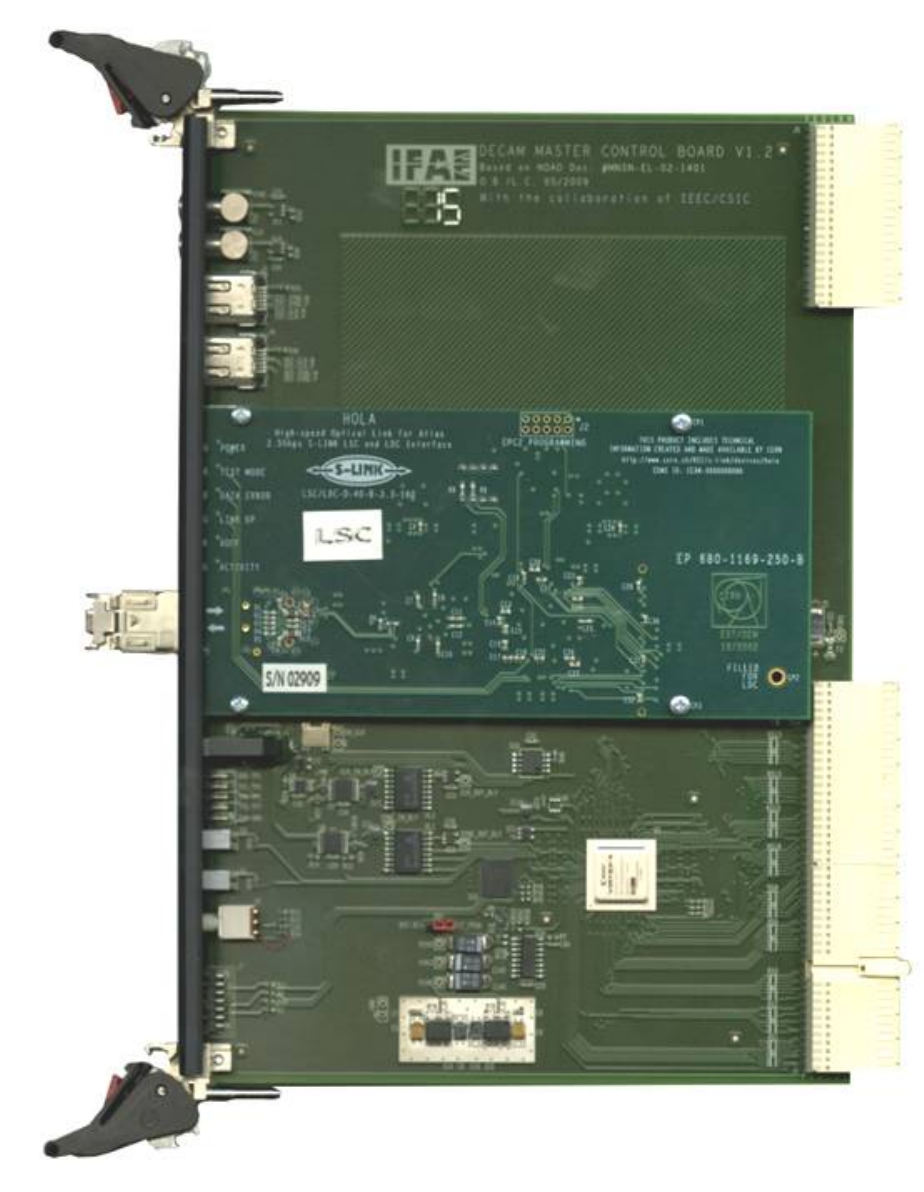}
\caption{The DECam Master Control Board main module.}
\label{fig:mcb} 
\end{center}
\end{figure}

The MCB acts as the bus master for the backplane bus and provides the interface between the peripheral boards in the Front-End Electronics (FEE) crate and the PAN computer, which is connected via a bi-directional gigabit fiber optic link.  Normally, software running on the PAN computer sends a command to the MCB, which performs a read or write operation on the backplane and optionally returns the requested data to the PAN.  Repetitive operations such as CCD exposures benefit from using a programmable onboard sequencer on the MCB as noted below.  During normal data taking, there are many repetitive operations taking place on the backplane bus; for instance, the MCB writes to toggle CCD clock lines. While these repetitive operations can be controlled directly from software on the PAN, the preferred method is to offload these tasks to a programmable sequencer in the MCB Field Programmable Gate Array (FPGA).  The MCB sequencer programs are written in a type of macro assembly language, compiled on a PC, and downloaded into memories in the MCB FPGA.  The sequencer assembly language features user variables and conditional branching as well as arithmetic functions.  Loop count registers in the FPGA are readable and writable from the PAN as well as the sequencer code.  Changing the values in the loop count registers thus modifies operation of the sequencer without the need to recompile and download, and this is particularly useful when changing the size (or region of interest) of a CCD.  The assembly language, assembler compiler and hardware interpreter are proprietary and were developed for the NOAO MONSOON system.  Figure~\ref{fig:mcb} shows a photograph of a Master Control Board.

Communication between the PAN computer and the MCB takes place over a fiber-optic link 
called SLINK~\footnote{For details on the SLINK modules see \url{http://hsi.web.cern.ch/HSI/s-link/}}.  
Originally developed at CERN for readout of the ATLAS particle detector systems, this fiber optic link format is proprietary and uses a custom PCI card (FILAR) in the PAN computer and a custom mezzanine card (HOLA) on the MCB.  At the most basic functional level, the SLINK system can be best described as a pair of 32-bit wide first-in first-out memory buffers (FIFOs).  The PAN computer writes data and commands into one FIFO, which is read out and processed by the FPGA on the MCB.  In the other direction the MCB fills a FIFO and the data are transmitted to the PAN computer and placed directly into system memory by the FILAR card.  

Since the DECam system is comprised of multiple readout crates, it is critical to provide a mechanism in hardware to synchronize operations across them.  A single MCB board is designated as the master and sends a copy of its 40MHz clock and a synchronization signal to a daisy chain of slave MCBs.  Adjustable digital delay lines and phase-shifting clock buffers are controlled by registers on the MCB boards.  These delay parameters must be adjusted once in order to synchronize the system to nanosecond precision.

\subsubsection{Clock Board and Transition Modules}
\begin{figure}[h]
\begin{center}
\includegraphics[width=3in]{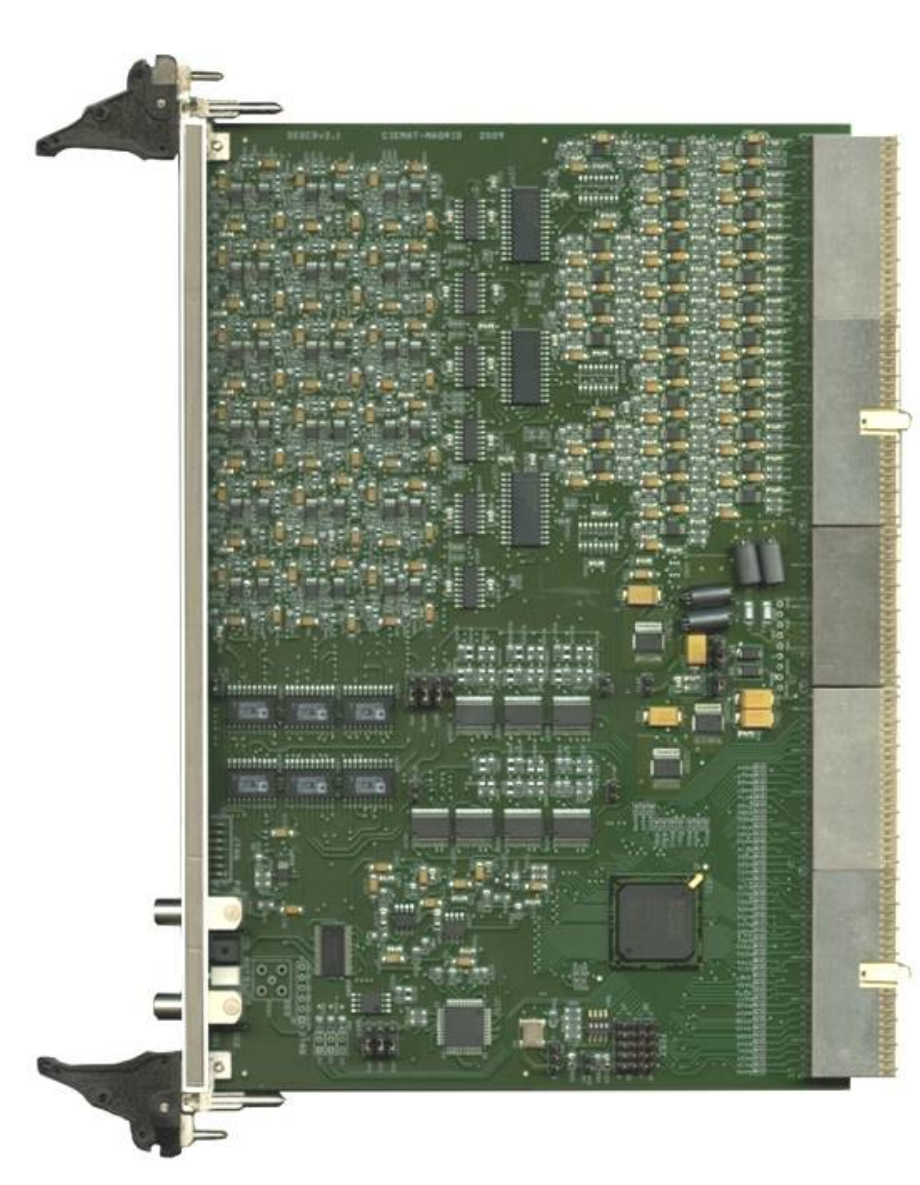}
\includegraphics[width=3in]{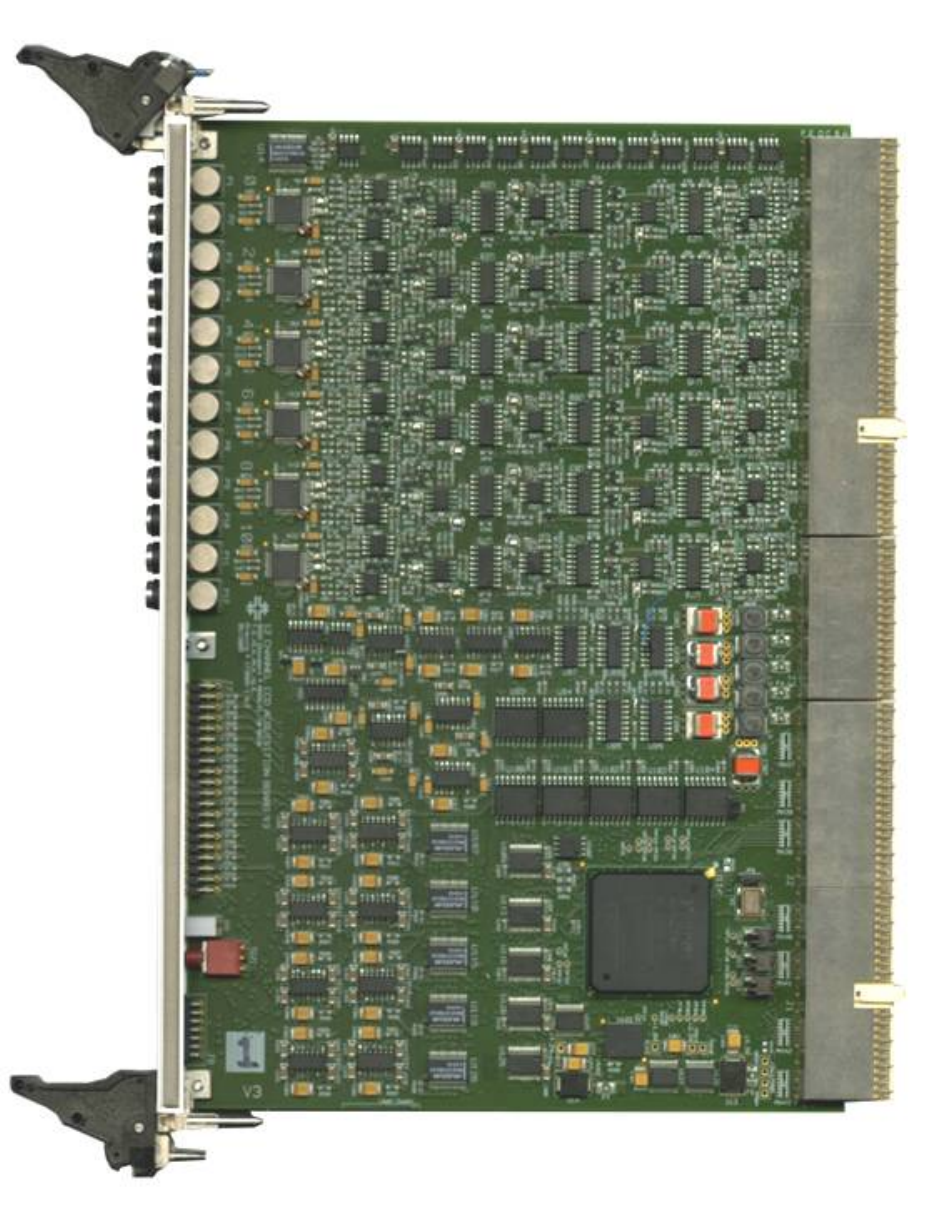}
\caption{The DECam Clock Board main module (left) and 12-channel Acquisition Board (right) main module.}
\label{fig:twocards} 
\end{center}
\end{figure}

Each DECam Clock Board can provide all clock levels needed by nine CCDs, although they can be programmed in groups of three. In practice, we found that the DECam CCDs operate optimally with the same clock levels. The Clock Board Transition Module (CBT) plugs into the rear of the crate behind the Clock Board main module.  The CBT provides filtered analog power for the Clock Board through non-bussed backplane pins.  The module also contains low pass filters, or waveshaping components, for each of the 135 clock outputs as well as the clock signal output connectors for the cables to the VIB. Figure~\ref{fig:twocards} shows a photograph of the main Clock Board.

\subsubsection{Acquisition Boards and Transition Modules}
The primary function of the 12-channel Acquisition module is to digitize the analog video signals from the CCDs and send those data over the backplane to the Master Control Board.    Secondary functions include generating and reading back CCD bias voltages, monitoring temperatures, and storing calibration data.  The acquisition board contains 60 independent digital to analog converter (DAC) channels that are buffered and connected to a dedicated telemetry providing independent monitoring.  Figure~\ref{fig:twocards} also shows a photograph of the Acquisition Board main module.

\begin{figure}[h]
\begin{center}
\includegraphics[scale=0.6]{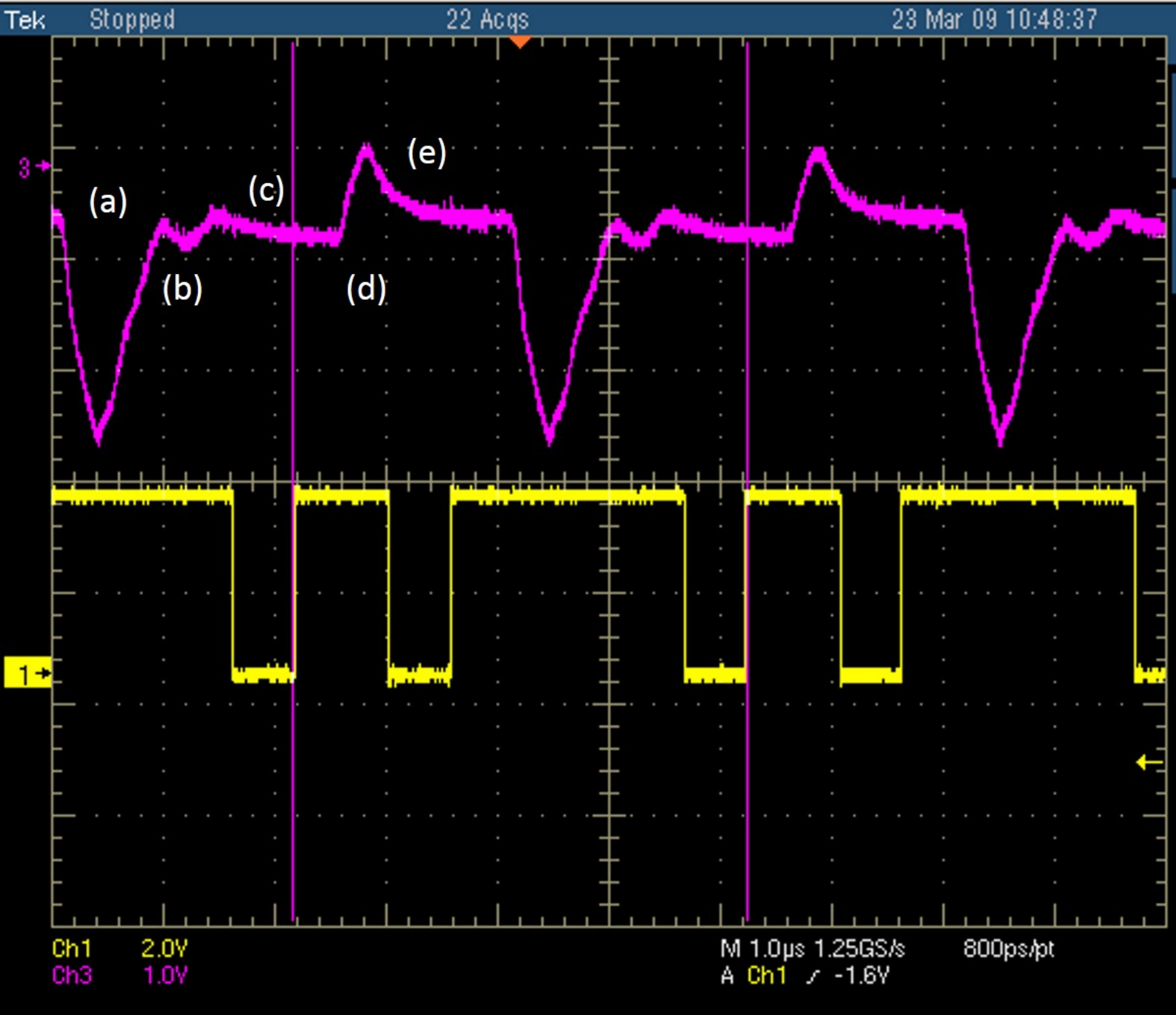}
\caption{Two complete pixel read cycles. The upper trace (2V per division) shows the video input signal. The lower trace (1V per division) shows the integration gate generated by the acquisition card. First (a) on the video trace is the $\sim 4$V feedthrough of the CCD output amplifier reset signal onto the video signal.  The wiggle at (b) is feedthrough of the horizontal clocks during the shift that readies the next pixel's charge to be moved to the integration node.  The ``before" part of the CDS measurement (c) occurs during the first integration window. Next the summing well moves the charge onto the integration node resulting in feedthrough (d). Then after the second integration window the signal is digitized (e). The gap between the timing marks is $4.08\mu{\rm s}$.}
\label{fig:video} 
\end{center}
\end{figure}

Video signals from the CCDs are sent through an analog front end that performs correlated double sampling (CDS) of the video signal.  The relatively complex analog circuitry used in the front end requires several analog switches that are controlled via digital signals from CDS registers in the FPGA. Figure~\ref{fig:video} shows the video signal and the digitizing gates for two read cycles.  Pixel values are digitized with a fast 18-bit analog to digital converter (ADC) and stored in registers on the acquisition board.   Only one set of pixel values may be stored on the acquisition board at a time.  The MCB can read these registers directly using conventional backplane read cycles; however, this is a relatively slow process.  Instead, a complex sequencer on the MCB controls simple sequencers on the clock and acquisition boards to reduce the amount of backplane traffic needed to read out groups of CCDs. The 18-bit digitized data are sent to the MCB, truncated to 16-bits, and ultimately the PAN computers (see Section~\ref{subsec:decamonline}), where the data are recorded to disk. 

The acquisition board sequencer is a finite state machine that when enabled, controls various CDS switches and ADC control signals.  Each vector or state in this sequencer then has a delay parameter ranging from 25 ns to 6.4 $\mu$s.  A total of 64 16-bit vectors or states are stored in a memory accessible from the backplane and may be read or written at any time.  The acquisition board supports a pipeline data transfer mode, where the pixel values are quickly written to the MCB data FIFO at maximum speed without intervention immediately after ADC conversion completes.  Pixel data bus arbitration amongst multiple acquisition cards is controlled by a priority scheme.  Block transfers of pixel data are also supported in burst mode which is initiated by the MCB writing to a control register on each acquisition board.  Redirection registers on the board specify which pixels are sent and in which order for each acquisition cycle.

After the CCD pixels have been digitized, any remaining image persistence is eliminated using an erase/clear sequence after the CCDs have been readout. The substrate voltage is lowered from 40V down to 0V.  Then the vertical clocks are all raised to their maximum of 8V and kept high for 0.5 seconds. Next the substrate voltage is raised back up to 40V and finally the CCD is cleared by transferring any remaining charge off the active pixels.  This erase mechanism fully eliminates image persistence even from fully-saturated CCDs. 

The 12-channel transition board (rear module) is responsible for providing clean analog power to the acquisition board as well as filtering bias voltage outputs to the CCDs and receiving and buffering the video signal from the CCDs.  Cable connections on the rear of the transition board connect to the VIB on the imager vessel.

\subsection{CCD Heater Crate}
A fourth crate, the Heater Controller Crate, is used to maintain the temperature of the CCDs in the Dewar. It drives 12 25-$\Omega$ resistive heaters, each mounted on a separate cooling braid within the Dewar. Each heater's output voltage (maximum 20 V) is controlled by a single-ended input signal supplied by a National Instruments card and controlled by the Slow Controls Computer using a PID loop. The heaters have thermostatic protection so that the current will shut down automatically if they exceed their rated temperature. The construction of the heater crate is similar to that of a readout crate but employs a different set of low-voltage power supplies. An essential goal of the heater crate was to prevent heaters from introducing any additional noise into CCD readings. 

\subsection{AC Power and Crate Grounding Scheme}
The entire Prime Focus Cage assembly is electrically isolated from the rest of the telescope and building's grounding configuration. The electrical isolation at the mechanical attachment points (the location where the support fins attach to the cage's rib beams) is accomplished by using G-10 fiberglass insulating washers and plates. All data communication for the DES readout and slow controls is done via optical links.  Mechanical connection of the liquid nitrogen transport pipes leaves them isolated from the cage. 

A single shielded power cable supplies the 3-phase AC power to the cage's power distribution chassis and provides both a safety ground connection (through the safety ground conductor) and a low impedance, high frequency ground connection (through the cable's shield braid). At the service end of this power cable, the grounding connection is made at an AC power distribution panel mounted on the top of the Cage, above the camera Dewar. This is the only connection between the building and/or telescope's grounded metal and the cage assembly's metal, thus ensuring that no large ground loop can be formed. 

\subsection{Readout Performance} \label{subsec:ro17}
In all respects the CCDs and electronics meet or exceed the requirements shown in Table~\ref{tab-ccdspecs}.  In particular, all 62 imaging CCDs and the 8 focus CCDs are digitized in 17 seconds with 6 to 9 electrons RMS readout noise, much better than the specification shown there.  Including the erase/clear cycle, the full readout takes 20 seconds, usually less than the settling time of the telescope when it is slewed to a new position.

\clearpage
\section{Filter Changer, Shutter, and Active Optics System}\label{s6:Mechanical}

This section describes the moving mechanical systems of DECam:  the filter changer, the shutter and the active optics system (hexapod).  Both the shutter and filter changer designs were derived from designs developed for PanStarrs, but scaled up in size to match the DECam requirements.  In DECam the shutter is bolted to the filter changer and this assembly is installed through the large slot in the barrel between lenses C3 and C4.  Housings fit over the protruding ends of the assembly providing light and air-tight seals.

\subsection{Shutter}
DECam required a lightweight (mass $< 35$ kg) shutter with a 600 mm diameter circular aperture. It was required to be essentially light-tight when closed. The precision measurements required by DES placed stringent demands on the exposure times. The shutter exposure time uniformity was required to be better than 10 ms (i.e. the actual exposure time anywhere on the focal plane should not be more than 10 ms different from anywhere else), with a repeatability of $< 5$ ms.  The exposure  time accuracy was required to be $< 50 $ ms and measured with accuracy $< 10$ ms. It is expected to have a mean time between failure of more than 1,250,000 cycles.

\begin{figure}[h]
\begin{center}
\includegraphics[scale=0.7]{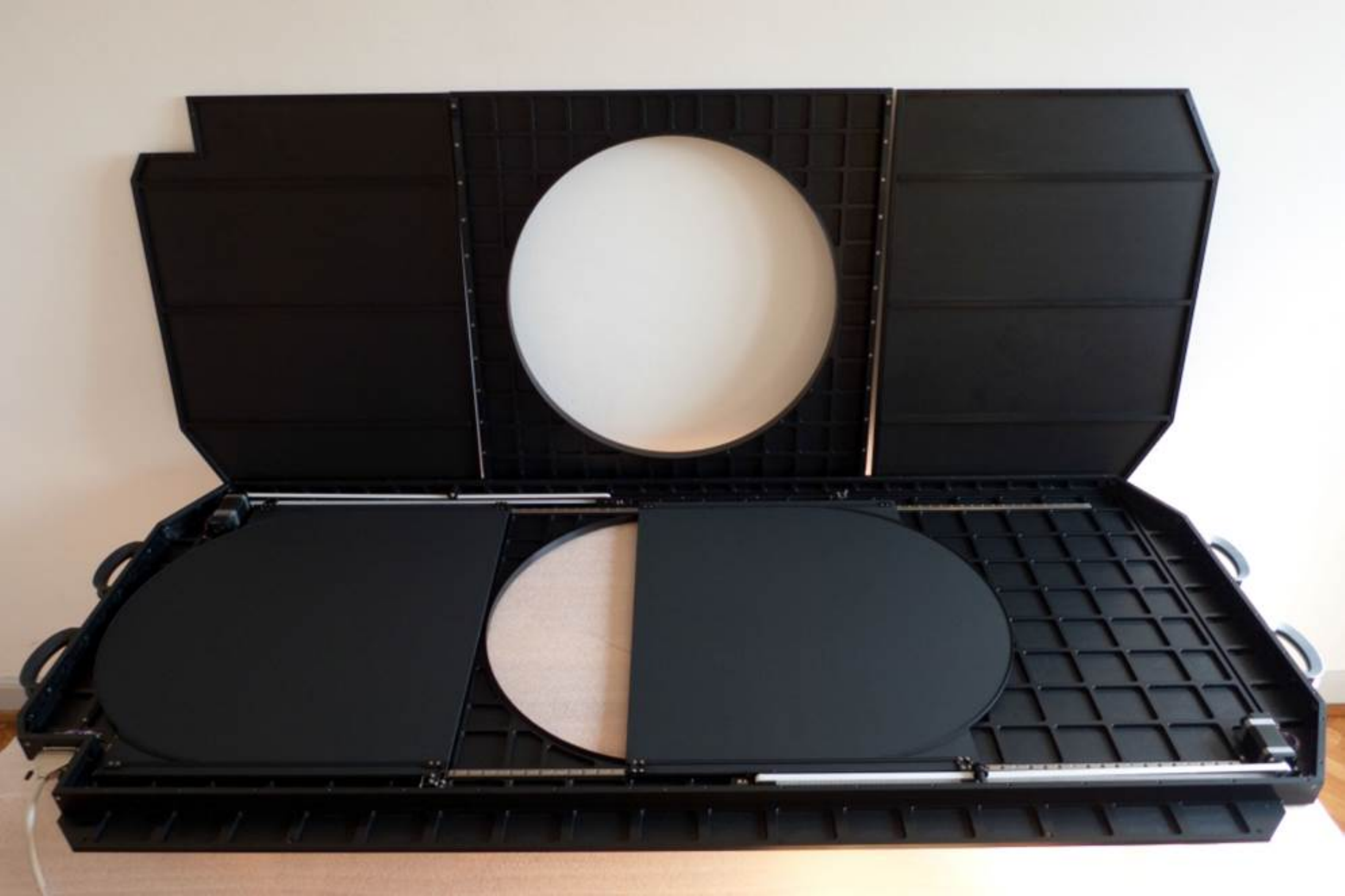}
\caption{Photograph of the DECam shutter with the top cover unbolted and tilted back. }
\label{fig:shutterphoto} 
\end{center}
\end{figure}

The DECam shutter is a slit-type shutter with a 600 mm diameter circular aperture, designed and fabricated by the group led by Klaus Reif at Bonn University and the Horer List Observatory~\footnote{Eventually that group became Bonn Shutters UG (http://www.bonn-shutter.de/).}.   Prior to construction of the DECam 600mm shutter, the largest shutter built by Bonn was the 480 mm $\times$ 480 mm aperture PS1 shutter for the Pan-STARRS telescope.    The shutter has two lightweight blades made from a sandwich of carbon fiber and foam.  Before an exposure, one blade fills the aperture and the other blade is stored to one or the other side of the aperture. At the start of the exposure the first blade moves out of the aperture in the direction away from the stored blade. At the end of the exposure the stored blade moves into the aperture. Thus each part of the focal plane is exposed the same amount of time.   The shutter does not have a preferred direction of movement; for consecutive exposures the shutter blades move first from left to right and then from right to left.   Fig.~\ref{fig:shutterphoto} shows a picture of the DECam shutter with the cover open. 
 
The DECam shutter weighs 35 kg. A single aluminum plate with a 600 mm aperture in the center provides the mounting base. A thin aluminum top with the same aperture provides the cover. The two blades move on a pair of linear bearings.  Stepper motors drive the blades by means of toothed belts. The shutter is controlled by four microcontrollers: one for each shutter blade stepper motor, one for host communication and one for input signal filtering. The firmware on the motor microcontrollers controls the blade movement and is identical for both motors. The firmware on the communication microcontroller provides control through a RS232 line. The firmware on the signal filtering microcontroller prevents signal bouncing and limits the shortest exposure pulse to about $300\mu$s.

 The DECam shutter  is an impact free, low acceleration (i.e. low power) device. Instead of driving the shutter blades at high speed/acceleration the $<1$ ms timing accuracy is achieved by a simple yet very precise motion control of both blades: The generation of every single stepper motor micro-step (16424 for the 600mm aperture) follows a precise time table which is derived from a given velocity profile.  Incremental encoders are mounted on the motors shafts. Comparison of the number of commanded motor steps with the counted encoder increments provides the primary check of proper shutter operation. This check is done at all times during each blade movement. Both blades are driven with identical time tables (i.e. velocity profiles), a prerequisite for uniform exposures. With the preset velocity profiles the full blade motion takes $\sim1.1$ seconds.

\subsection{Filter Changer}
\label{ss:filterchanger}
\begin{figure}[h]
\begin{center}
\includegraphics[scale=0.7]{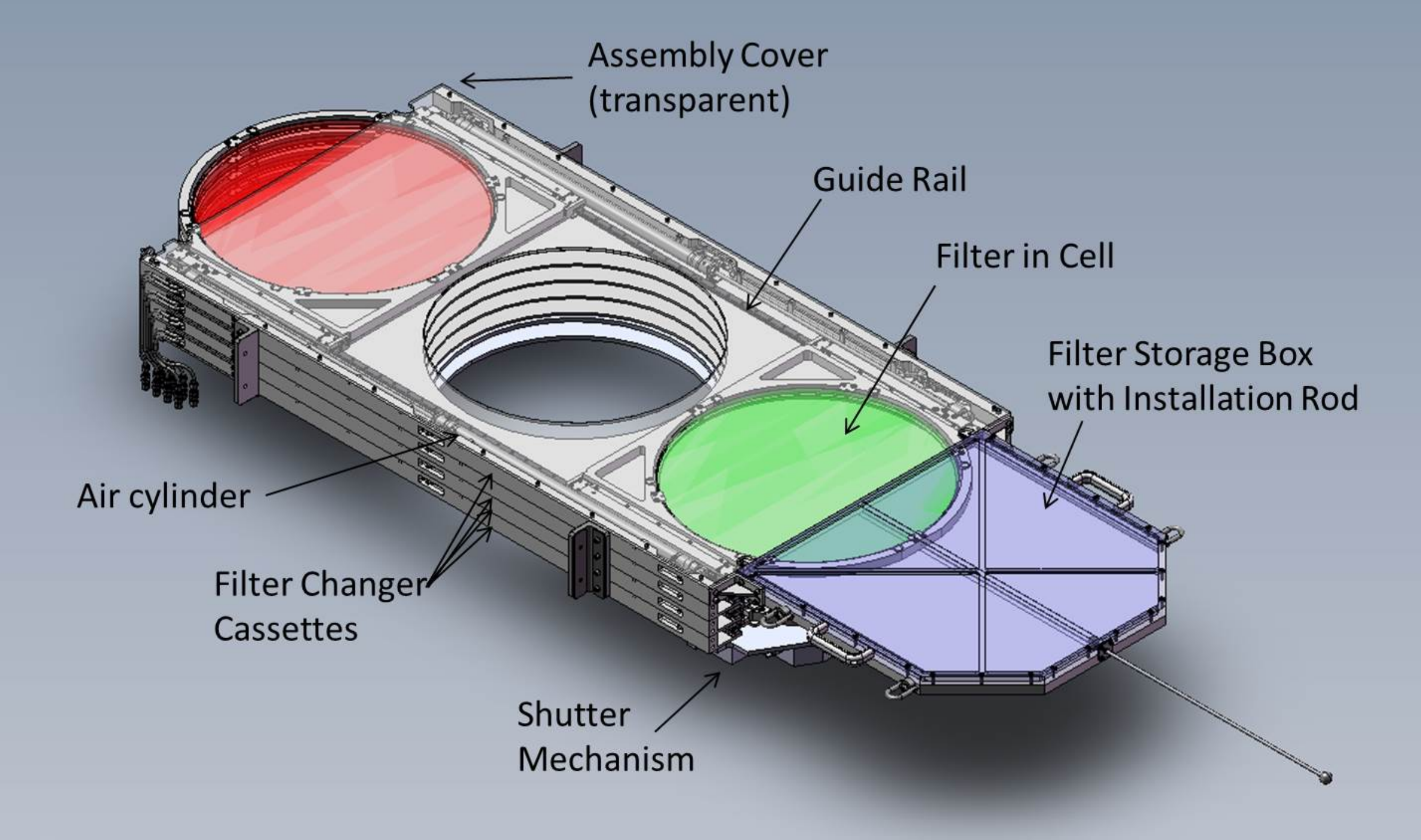}
\caption{The DECam filter-changer with shutter attached (underneath). The overall dimensions are 2000mm width, 900mm height, 232mm thick (including the shutter) and the mass  is $325$ kilograms. Each of the four layers carries two filter plus cell units in two carriages. The carriages slide on guide rails and are each pushed-in or pulled-out by compressed air. The figure calls-out one of each of the two guide rails and compressed air cylinders as well as a Filter Storage Box with insertion/removal rod. }
\label{fig:fc-cartoon-text} 
\end{center}
\end{figure}

The DECam Filter Changer Mechanism (FCM) was designed and fabricated by the DES group at the University of Michigan~\citep{SPIEtarle2010}.    The design was derived from the design of the PanSTARRS FCM\footnote{For details on the PanSTARRS FCM see \url{http://www.amostech.com/TechnicalPapers/2006/Pan-STARRS/Ryan.pdf}}.  The DECam FCM provides positions for eight filters.  At this time there are 7 filters and the 8th slot is occupied by an aluminum filter/cell dummy coated by anti-reflective black paint, known as the ``block" filter.  The block  filter is typically inserted when exposures are not being taken.  Each filter is 13 mm in thickness, 620 mm in diameter, and has a mass of $9.95$ kilograms.  They are housed in four stacked cassette mechanism sub-assemblies. Each cassette houses two filters, and uses compressed-air cylinders to deploy or stow the filters. The compressed-air supply comes from off-telescope at 100 PSI. The air cylinders have integral air cushions at the end of travel to absorb energy of motion and integral needle valves for safety and speed control. Control valves are accessible on the sides of the FCM stack when the FCM is not mounted in the barrel.  Filter position information is provided by reed switches that inform the operators whether each filter is stowed, deployed, or in an intermediate position. Figure~\ref{fig:fc-cartoon-text} shows a schematic of the FCM. It shows the four cassettes, filters in the ``out" position, and the position of one of the filter storage boxes.

Each layer in the filter exchange mechanism consists of an open aluminum channel base plate, which carries two THK linear bearing rails for guiding filter motion, and two Bimba air cylinders to provide individual filter actuation (for insertion of the filters into the active position in the center of the FCM).  The THK rails extend the full length of the FCM, such that two carriage assemblies can share a single set of linear guides.  Each FCM carriage is powered by its own Bimba air cylinder (one mounted on each side of the FCM).  The air cylinders include limit switches for signaling the state of the cylinder (extended or retracted), and Bimba flow control valves, which allow the actuation speed to be adjusted manually.  The combination of air-powered actuator and electrical control valve results in zero heat dissipation during use.  Power is only consumed momentarily, to toggle the air valve, and hence the air cylinder, between the inserted and retracted positions. 

%

\begin{figure}[h]
\begin{center}
\includegraphics[scale=0.5]{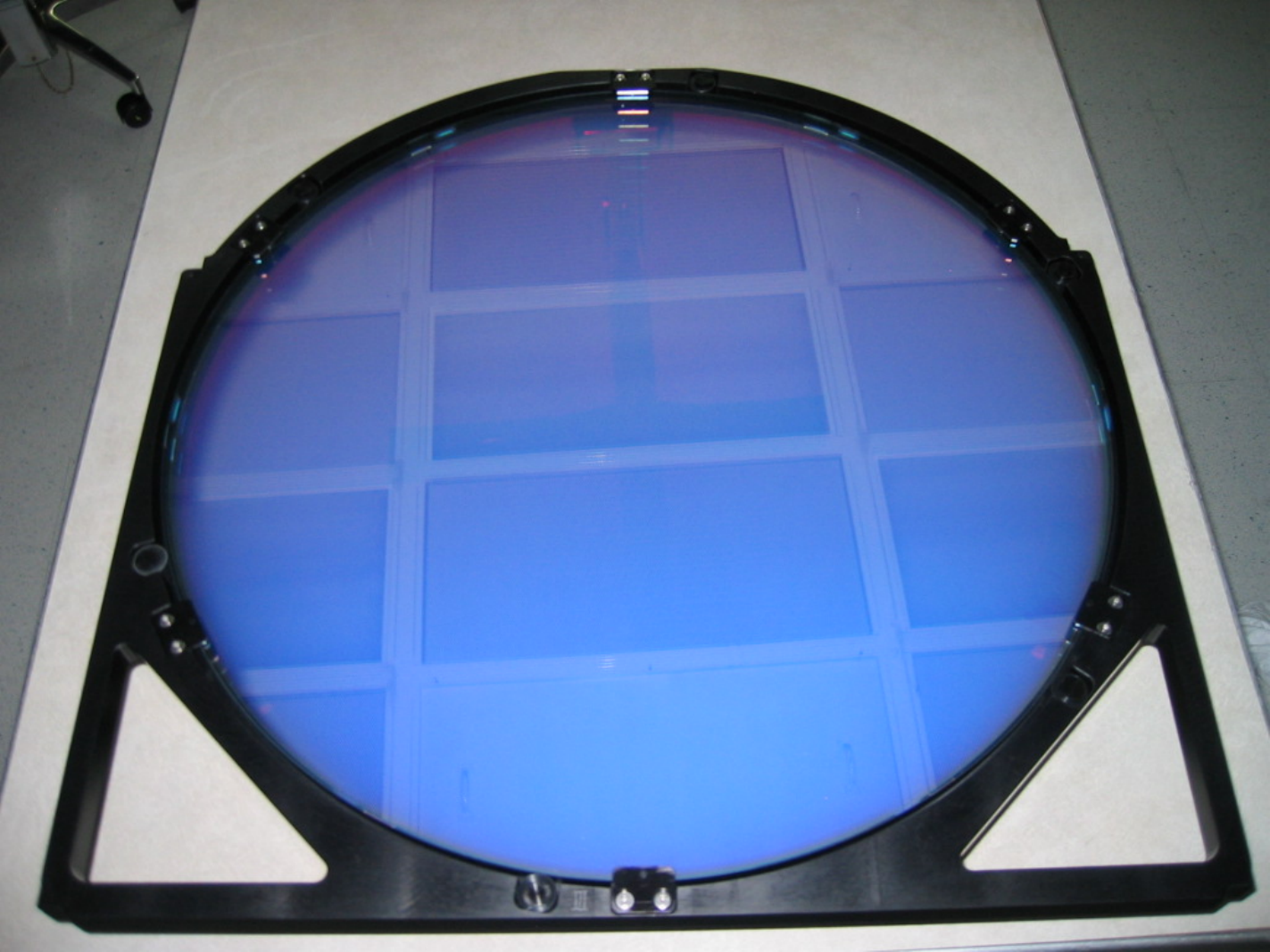}
\caption{The DES i-band filter and frame assembly, showing six radial spacer and cushion disk constraints, as well as frame guides and mounting screw holes. The rectangular "tiles" on the filter are actually reflections of the ceiling of the clean room.}
\label{fig:i-band-filter} 
\end{center}
\end{figure}

The filters are carried in 7075-T7351 aluminum cells.  Six ultra-high molecular weight polyethylene (UHMW-PE) radial spacers evenly distributed around the circumference of the filter define the radial filter boundary.  The frame and radial spacer materials and sizing were set to achieve an athermal design that canceled the effects of thermal expansion on the assembly and minimized loads transmitted to the filters. In addition to the radial spacers, twelve UHMW-PE cushion disks were used to define the position of each filter in the axial direction. Each defining cushion disk has an opposing preload disk, to keep the filter centered and fixed in the frame regardless of gravity orientation.  The radial spacers and cushion disks were fabricated from plastic to avoid glass-metal interfaces.  The filter-to-frame installation was performed in the CTIO cleanroom by DES personnel. Fig.~\ref{fig:i-band-filter} shows a photograph of the i-band filter in its assembly as well as the components that hold it in place.  While inside the filter-changer the filter cells are bolted to their respective filter carriages.

During the times that the filters are not in the filter changer, they are stored in heavy-duty Filter Storage Boxes. These are black-anodized aluminum cases with a slot that can accommodate the filter and its cell. These boxes can be bolted to the filter changer so that a rod can be poked through a hole in the box and screwed into the end of the cell. Then the filter cell can be pushed into or pulled out of the filter changer. That box and push-pull rod are also shown in Fig.~\ref{fig:fc-cartoon-text}.  


\begin{figure}[h]
\begin{center}
\includegraphics[scale=0.5]{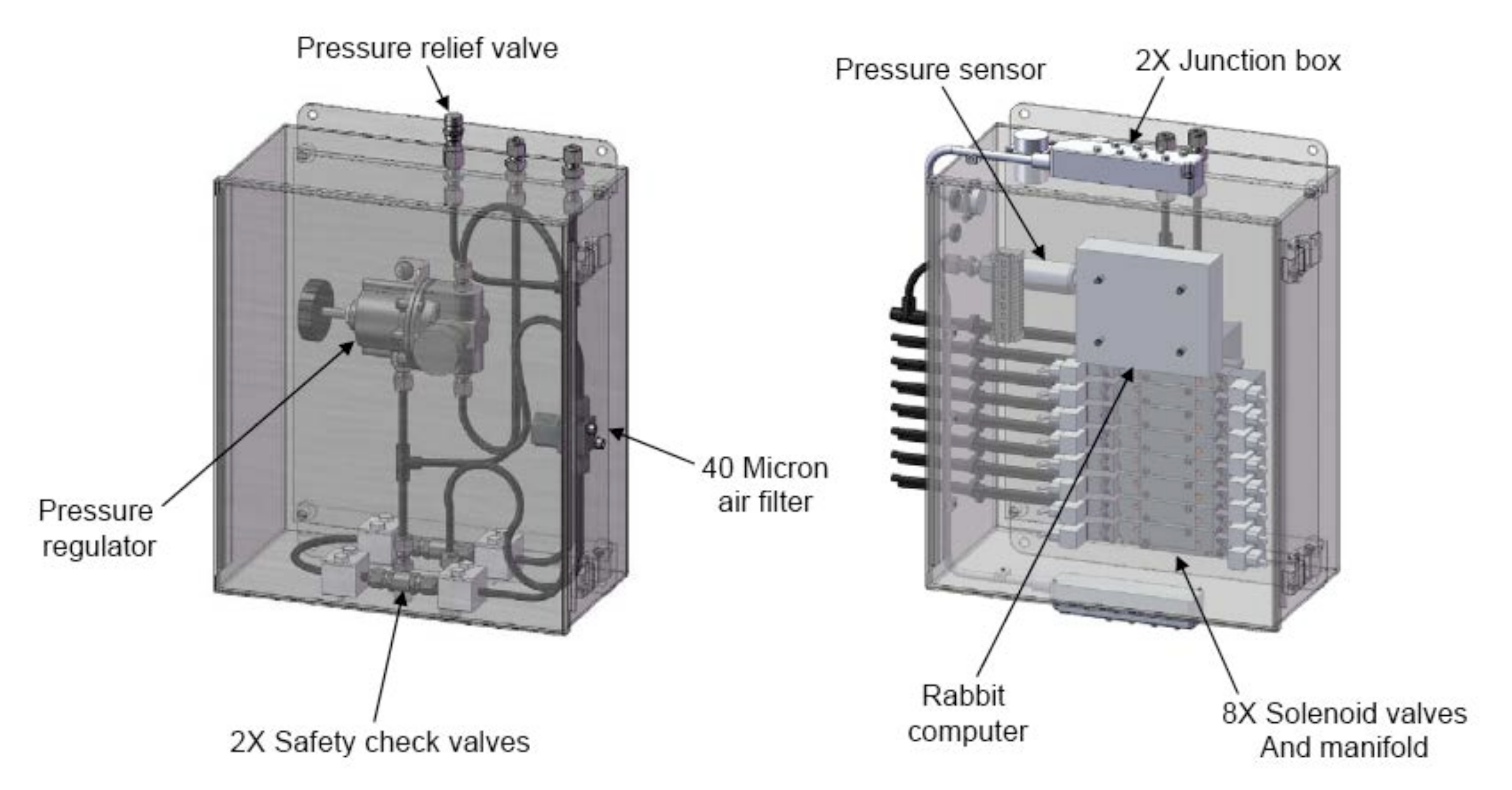}
\caption{Pneumatic control boxes for the filter changer. Note that the enclosures are shown slightly transparently so that the internal components can be seen. A  one-liter 110 p.s.i. gas volume (not shown) was added to the interior box on the left as a retrofit. This provides a buffer volume that is important when the filters are moved frequently. The control box on the right contains the rabbit computer as well as solenoid valves. }
\label{fig:fc-boxes} 
\end{center}
\end{figure}

The positions of the filters are controlled by electronics housed in boxes mounted on the barrel body in the vicinity of the filter changer. One control box contains both the required solenoid valves and a Rabbit BL2100 computer board (see Fig.~\ref{fig:fc-boxes}). Communication with the board is achieved using Ethernet TCP/IP. The power requirements for the computer are 24 Vdc with an average power of 6W and a peak power of 8.5W. It takes 4 to 6 seconds to insert/remove a filter, depending on the orientation of the filter changer.    Firmware prevents filter collisions. The other control box contains a small, pressure-regulated 100 PSI gas storage tank that provides a local buffer. Telemetry recorded by the electronics provides a monitor of the valve air supply line pressure, the filter enclosure humidity and temperature.  

\subsubsection{Filter Changer Fabrication and Testing}
The primary machined components of the FCM were fabricated at Leonard Machine Tool Systems in Warren, MI. The majority of machined components were built with stress-relieved 7075-T7351 aluminum. This aluminum alloy was used due to its relatively high yield strength. All aluminum parts were anodized black after machining to reduce reflection off of metallic surfaces inside of the telescope barrel.  The primary challenge with the fabrication of the FCM was the large footprint (i.e. 1.64 m $\times$ 0.87 m) of the relatively thin ($\sim$40 mm) base plate required for each FCM layer. Asymmetric machining of this plate to form the channel structure that houses the remainder of the assembly was found to cause a significant out-of-plane distortion or bow of the plate upwards of 1 -- 3 mm. Thus, a fabrication process was developed~\citep{SPIEtarle2010}  that minimized these distortion effects and achieved the required flatness of 0.25 mm. 

The FCM was subjected to testing for a full 10\% of its expected lifetime cycles spanning a full range of orientations and operating temperatures.  Each of the four filter changer cassettes was motion-tested to verify proper stow and repeatable deploy operation of both filters.  An aluminum mass model was mounted in each filter frame being tested, since actual filters were not available at the time.  All FCM carriages easily met the position repeatability requirement of 0.5mm in the four different orientations tested over the temperature range of $-10^{\circ}$C to $20^{\circ}$C.  

The FCM and shutter were shipped to CTIO in June 2011 and operationally tested in the Coud\'e room prior to installation in August 2012.  Shortly after installation it was discovered that the Bimba Cylinders in the FCM had small red LEDs mounted near their ends that indicated when the cylinder was closed. They had been covered up with some black plastic by the vendor, but when we operated the filter-changer in the very dark dome we were able to see them glowing dimly if we looked at the hardware from the side at a particular angle. Though there was no evidence any of this light could make it to the focal plane, the filter-changer was removed and a small drill bit was applied to each LED, ending their ability to glow.

\subsection{Eliminating Scattered Light from the Filter Changer and Shutter Assemblies}
\begin{figure}[h]
\begin{center}
\includegraphics[scale=0.5]{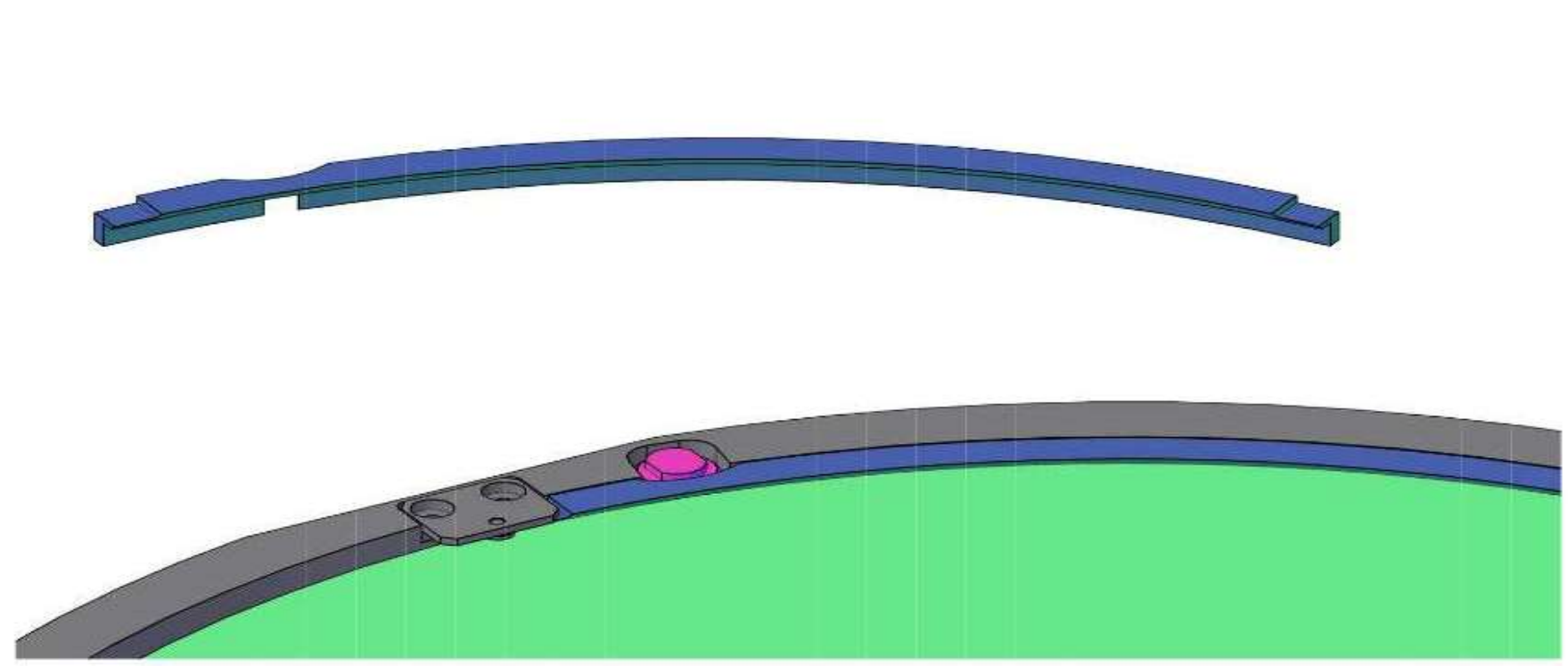}
\caption{Each filter cell was retrofitted with 6 L-shaped arc sections that were loosely trapped into the gap between the filter and the filter cell.}
\label{fig:filterbaffles} 
\end{center}
\end{figure}

\begin{figure}[h]
\begin{center}
\includegraphics[scale=0.8]{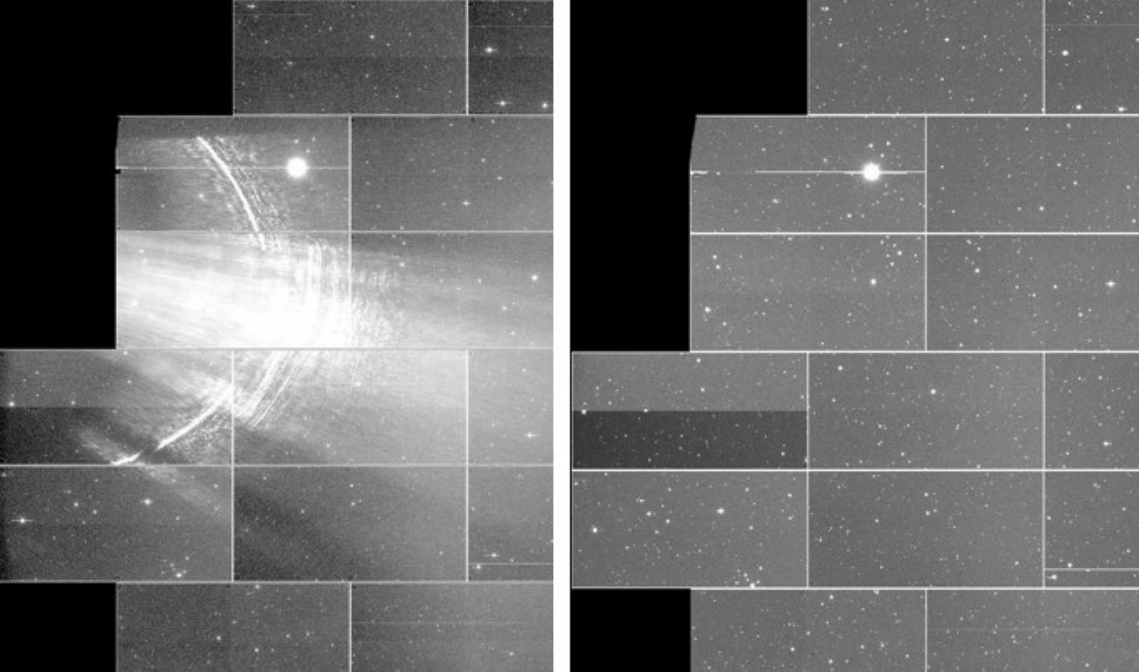}
\caption{These images show a portion of the focal plane from exposures taken with the bright star Canopus located about 1.5 degrees from the edge of the focal plane. On the left is an exposure from before the anti-reflective paint was applied. On the right is the same part of the focal plane after the fix. No trace of the reflections remains. The exposures were 90 s long, taken using the z-band filter. }
\label{fig:canopus} 
\end{center}
\end{figure}

During commissioning we carried out a systematic study of stray-light sources. Arcs and various-shaped smears of stray light were noticeable when there was a very bright star about $1.5 \deg$ off-axis.  It was found that stray light could scatter off the very small uncoated areas at the extreme edges of the filter cells, and then bounce twice off the inside surfaces of the filters into the images, even though these uncoated areas were $> 9$ mm out of the clear aperture.  Thin, L-shaped plastic baffles were added to each filter cell between the edge of the filter and the cell. Fig.~\ref{fig:filterbaffles} shows a cartoon of that retrofit, which  was done in April 2013.  Other sources of scattered light were eliminated by painting the cylindrical-shaped interior edges of the apertures of the filter-changer and shutter with anti-reflective paint (Aeroglaze~\textregistered \  Z306).  That change was performed by CTIO staff in mid-March 2014. All of these sources of scattered light have been eliminated.  Figure~\ref{fig:canopus} shows an example of the change in the amount of the scattered light from the star Canopus, located about 1.5 degrees off of the edge of the focal plane. 

\subsection{Hexapod and Active Optics Control}
\label{ss:hexapod}
\begin{figure}[h]
\begin{center}
\includegraphics[scale=0.5841]{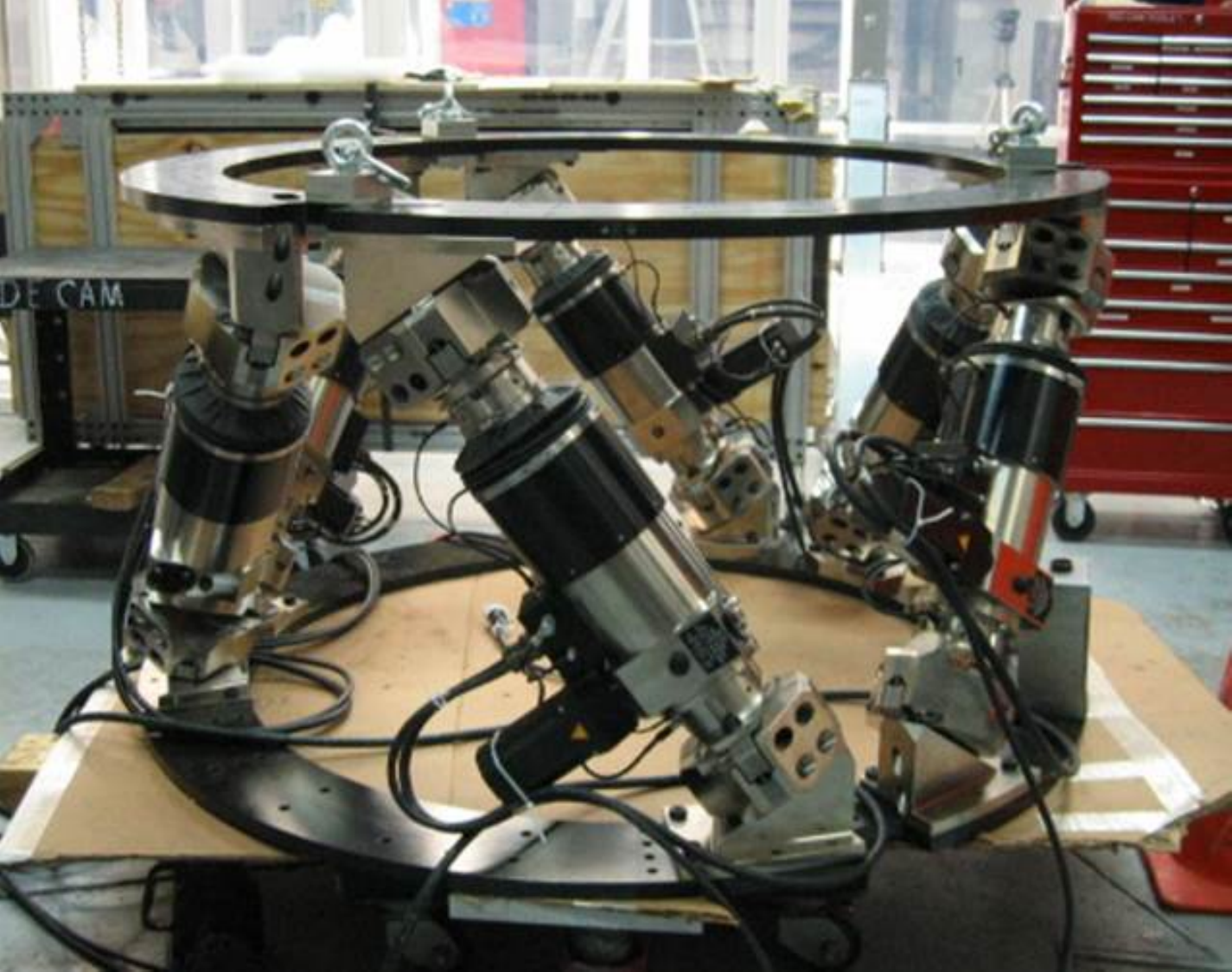}
\includegraphics[scale=0.4062]{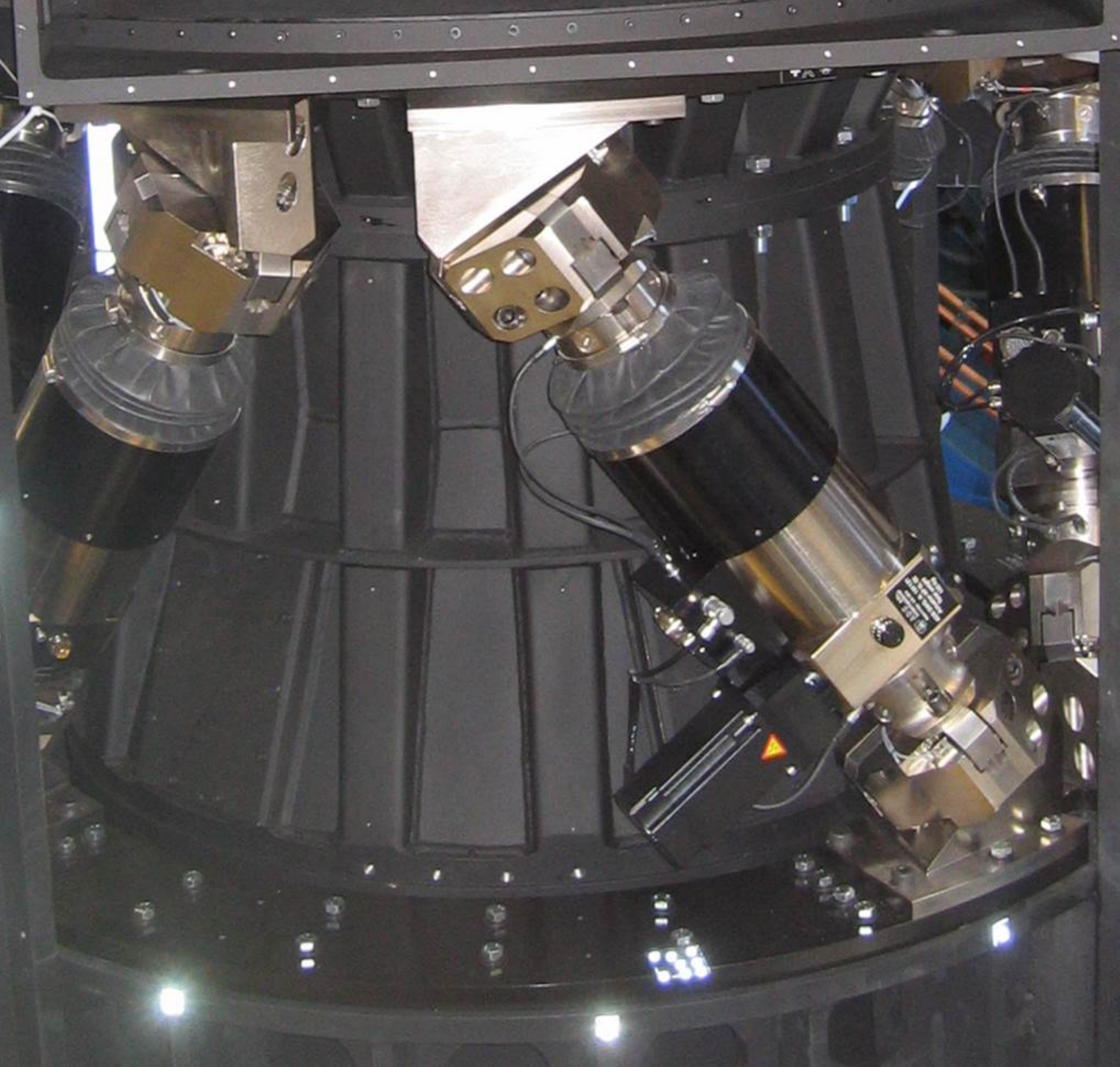}
\caption{The assembled hexapod at CTIO. The picture on the left shows the six legs with the mobile flange on top and the fixed flange, which bolts to the cage, on the bottom. The picture on the right  shows the hexapod, in the same orientation, with the fixed flange bolted to the cage (bottom) and the mobile flange to the body of the barrel. The flex joints are at both ends of each hexapod leg.}
\label{fig:hexapod} 
\end{center}
\end{figure}

To obtain excellent image quality, the focus, lateral alignment, and tilt between the primary mirror and the DECam corrector can be adjusted between exposures using the DECam hexapod.   Although the Blanco telescope has an equatorial mount with a classic Serrurier truss designed to compensate for the relative sag between the prime focus cage and the support of the primary mirror, the use of the hexapod allows for additional corrections. In operations this was shown to be a very effective arrangement. The size and direction of the corrections are determined through analysis~\citep{SPIEroodman2010,SPIEroodman2012} of out-of-focus ``donut"  images on four pairs of 2k $\times$ 2k CCDs located $1500 \mu $m above and below the focal plane.  This subsection describes the parameters of the hexapod.  Fig.~\ref{fig:hexapod} shows the hexapod at the factory prior to shipping to the Fermilab in 2011 as well as its configuration in the assembled camera.

The specifications for the hexapod motions along the optical axis were that the range must be at least $\pm 21$ mm, with an accuracy of $\pm 7.5 \mu$m within that stroke. The lateral drift, and tilt and twist angle accuracies within that stroke must be less than $\pm 25 \mu$m  and $\pm 3 \mu$m each. The minimum incremental motion was required to be $1 \mu$m or less. The speed is fast enough that the normal range of hexapod motions between images does not delay the start of the next exposure. That amounts to $150 \mu$m in less than 7.5 seconds.  There are specifications for the speeds of lateral motions, and for tip and tilts, as well.

The DECam hexapod and its controls were designed and built by ADS International in Valmadrera, Italy.  It was the biggest payload system that they had made up to that time, and combined with the required accuracy, it was considered a rather special design.   The assembled hexapod includes the 6  legs (actuators), whose length is adjustable, and two large flanges: the fixed flange is the base of the hexapod and is bolted to the prime focus cage; the mobile flange is attached to the DECam corrector barrel, as shown in Figure~\ref{fig:cage}.  Joints that are flexible in both cross-sectional dimensions are used at the attachment points between the legs and the flanges to allow 6--axis motion of the mobile flange (on the camera) with respect to the fixed flange (on the cage). That is, when the length of one of the legs of the hexapod is changed with respect to the others, all of the joints must be able to bend to accommodate the new orientation of the flanges with respect to each other.  Note that the CCD vessel is aligned and bolted to the corrector and the combination camera/barrel system moves as a single unit. 

The hexapod system provides both the focus adjustments and the capability to keep the corrector laterally aligned to the primary mirror.   The large load (3500kg) and tight positioning requirements presented a challenge to traditional hexapod designs.  ADS International developed a new hexapod for DES.  The design uses large, modular flex joints to attach the hexapod actuators to the mobile and fixed flanges.   These flex joints provide excellent repeatability ($<2\mu $m) with no backlash.  The maximum range of the hexapod is determined by limiting the flexure angle to less than 1.5 degrees.  

The actuator design is based on a roller screw and is not reversible and thus it assures the actuator will keep its position even when switched off without the need of a brake. The selected components assure virtually backlash free positioning. Actuator structural parts are made of stainless steel AISI 304 and aluminum where possible and compatible with thermo-elastic effects.  An absolute rotary encoder with 2048 counts per rotation is mounted on the roller screw axis. The encoder has an optical resolution of $0.976 \mu$m/cts, and can be interpolated 4096 times, giving the actuator a resolution up to 0.24 nm. Two electrical limit switches are used to define the actuator safe stroke, with mechanical safe stops placed beyond them. 

The hexapod control software limits the motion to within the range of safe bends of the flexures.  The focus stroke limit (z-direction) is $\pm 21$ mm. Tips and tilts are limited to $\le 500 \arcsec$ each.  Motions in the direction perpendicular to the focus direction (x/y) are limited to 11.3 mm for the focus range $| \rm{z}| \le 5$ mm.  The x/y motion limits decrease linearly with increasing $| \rm{z}|$ to $0$ mm at $| \rm{z}| =21$ mm. No rotations of the hexapod are allowed. These limitations are within the safe region of hexapod motions. This range of motion has been shown through operation on the telescope to be more than sufficient.  

The hexapod receives commands to move from the camera control software  (SISPI) between the images.  All of the speed-of-motion specifications were met. For instance, where the requirement for the speed of a z-stroke was 7.5 seconds, a $150 \mu$m movement is accomplished in about one second. 

The focus and alignment algorithms are described in Section~\ref{s7:sispi}, as part of the instrument control and data acquisition software (SISPI). 


\clearpage
\section{Instrument Control and Data Acquisition}\label{s7:sispi}

The DECam mountain-top instrument control and data acquisition system is named Survey Image System Process Integration or SISPI. SISPI is implemented as a distributed multi-processor system with a software architecture based on the Client-Server and Publish-Subscribe design patterns. The underlying message passing protocol is  based on PYRO\footnote{For details on PYRO see \url{http://www.pythonhosted.org/Pyro}}, a powerful distributed object technology system written entirely in Python. A distributed shared variable system was added to support exchange of telemetry data and other information between different components of the system. A detailed technical description of SISPI can be found in these references ~\citep{SPIEhonscheid2010, SPIEeiting2010, SPIEhonscheid2012, 
SPIEhonscheid2014}.

\begin{figure}[h]
\begin{center}
\includegraphics[scale=0.5]{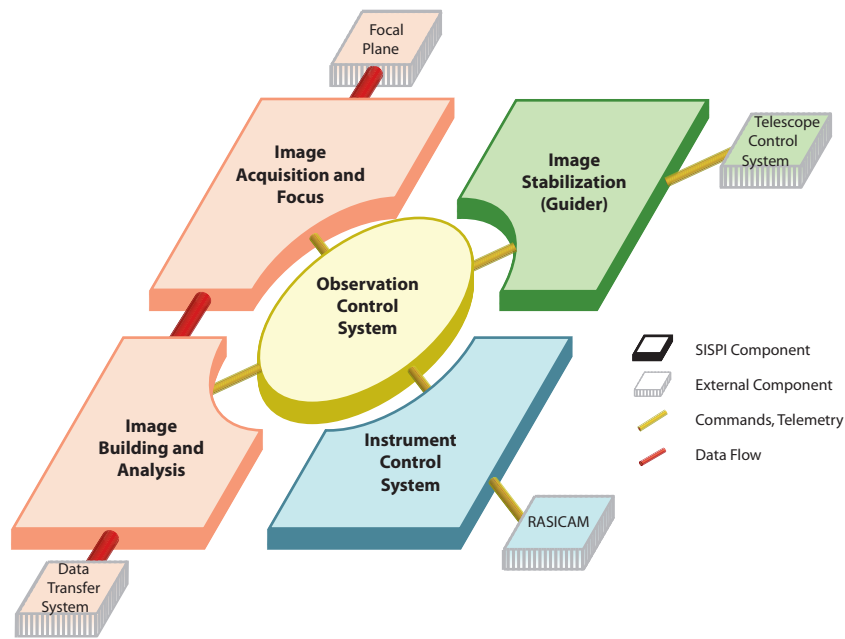}
\caption{Block diagram of the DECam readout and control system (SISPI).} 
\label{fig:sispi-overview}
\end{center}
\end{figure}
 
A schematic overview of the system is shown in Fig. \ref{fig:sispi-overview}. Image data flows from the focal plane through the image acquisition and image-building systems to the data transfer system where the images are buffered and sent off the mountain. The Observation Control System (OCS), which interacts with the observer via a console application, orchestrates the exposure sequence. DES uses an automated survey strategy tool that selects the best next exposure based on survey history and current observing conditions (see section \ref{sec:sispi-imagepipeline}). The OCS is assisted by the instrument control system that monitors individual hardware components. The image stabilization system includes the guider and the focus and alignment system. The image acquisition system is pipelined to maximize throughput allowing the OCS to start processing the next exposure as soon as all pixels are digitized. The performance requirements for the DECam read-out system are set by the size of the focal plane, the read-out time and the typical DES exposure time of 90 seconds. With 62 science CCDs or 520 Mpixels and 16 bits per pixel the size of a DECam exposure is approximately 1 GByte. At a rate of 250 kpix/s and since there are two amplifiers per CCD, the readout/digitization  takes about 17 seconds. Including the time needed for shutter operations (1 s) and to erase and clear the CCDs (3 s), disk I/O (2 s) as well as system overhead (2 s), the minimum time between exposures is 25 seconds. During this time the telescope slews to a new position and the hexapod position is adjusted for best alignment. These activities proceed in parallel, with the slowest setting the actual time between consecutive exposures. Important SISPI components not shown in Fig. \ref{fig:sispi-overview} include the central instrument database and the web-based user interfaces for the observer.

\subsection{System Architecture and Implementation}
\label{sec:sispi-infrastructure}

SISPI is based almost entirely on open source software. For historical reasons parts of the front-end readout system were developed in LabVIEW. Python 2.7 was selected as the main programming language with only a few compute-intensive applications written in C(++). The central code repository is svn-based and
 eUPS\footnote{For details on eUPS see \url{http://dev.lsstcorp.org/trac/wiki/Eups}} was selected for installation and version management. With no open source data distribution services available at the time, the SISPI middleware software was developed around PYRO. PostgreSQL, SQLalchemy and psycopg2 are used for the SISPI database. The DECam/SISPI graphical user interfaces are web-based using HTML5, Javascript and the SproutCore framework~\footnotemark. Computer system management, hardware maintenance and network security services are provided by the host observatory. 
\footnotetext{For details on these software packages see \url{http://www.postgresql.org}, \url{http://pythonhosted.org/psycopg2}, \url{http://www.sqlalchemy.org}, and \url{http://sproutcore.com}}

\begin{figure}[h]
\begin{center}
\includegraphics[scale=0.8]{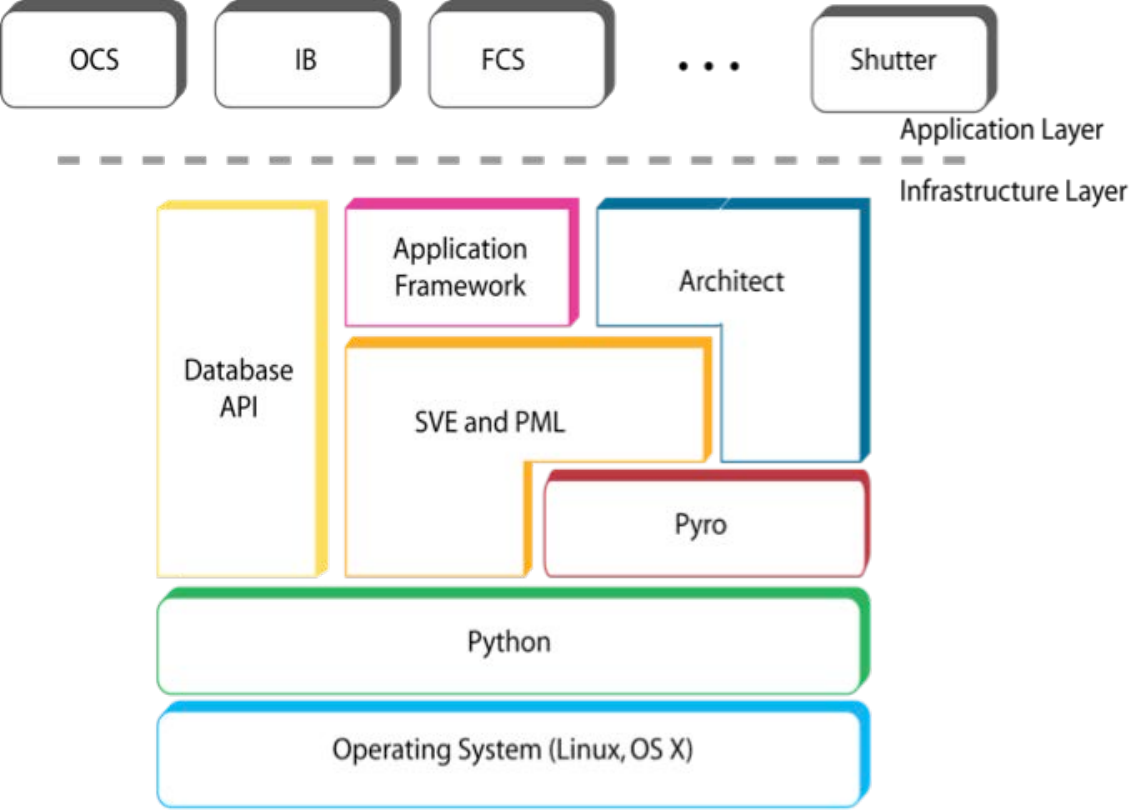}
\caption{Application and infrastrucure layers of the SISPI software stack.} 
\label{fig:sispi-stack}
\end{center}
\end{figure}
 
The SISPI software stack is shown schematically in Fig. \ref{fig:sispi-stack}. We distinguish between the infrastructure layer that includes all system services and the application layer that provides the software to operate the camera. Supported operating systems are Linux (including Raspbian on Raspberry PIs and Debian on Beagle Bone Black controllers) and Mac OS X.
 
A key feature of the SISPI infrastructure software is its support for different hardware configurations. SISPI processes can locate and establish communication links with other processes irrespective of the underlying hardware architecture. In test systems, for example, it is common for applications to share one or two computers whereas on the mountain everything is spread out over many computers to maximize performance. Being able to reconfigure SISPI without requiring changes to the code proved to be very important not only for development but also during commissioning and operations of the instrument. The designs of the SISPI communication and configuration systems both reflect this approach. 

Due to SISPI’s distributed architecture inter-process communication takes a central role in the design of the infrastructure software. The Python Remote Objects (PYRO) software package provides an object-oriented form of remote procedure calls similar to Java's Remote Method Invocation. It allows modules to interact just like standard Python objects even when spread over different computers on the network. PYRO handles the network communication transparently. A name server supports dynamic object location, rendering (network) configuration files obsolete. We distinguish between Command messages and Telemetry data. Commands are used to request information from a remote application or to activate a remote action. The Command or Message Passing system is implemented using a Client-Server design pattern with a thin software layer (PML) on top of PYRO. PML introduces the concepts of Component and Device to provide a uniform naming scheme for SISPI applications. The telemetry system is based on the publish-subscribe design pattern. Again built upon the core functionality provided by PYRO, SISPI uses a concept called Shared Variables. Consisting of a client stub library and a central server (Shared Variable Engine, SVE) this system allows user applications to publish information such as temperature readings or readout status to a virtual data space. Other applications can subscribe to information placed in this virtual data space and will receive updates whenever a publisher submits a new value. The shared variable system supports asynchronous callbacks, guaranteed delivery, multiple publishers of the same shared variable and group subscriptions.

Initialization and configuration of a complex distributed system such as SISPI is a multi-step process. SISPI employs a subversion-based code management and distribution system to support concurrent code development. eUPS organizes SISPI applications and system software into products. Simple command line tools support the selection of specific versions with automatic configuration of environment variables such as PYTHONPATH. The eUPS concept of current versions allows the code manager to define a standard configuration for regular use while software developers can select a specific version of a product to debug and test new features. With the software installed on the computer system, it is the job of the Architect, the SISPI configuration system, to load and start the SISPI processes in the correct order and with the correct arguments. SISPI configurations are described by initialization files using standard Windows .ini file notation. The collection of processes started by the Architect is called a SISPI Instance. Multiple instances can be run on the same set of computers as long as there are no hardware conflicts. An important feature is that individual processes can be stopped and restarted without having to shut down the whole system. For situations that require a full shutdown, the entire SISPI instance can be stopped with a single command. A simple, fast, and reliable restart procedure is critical for efficient observing since a DAQ expert is not always immediately available. The full DECam online system can be restarted in 3-4 minutes. 

The SISPI Application Framework is a Python class that serves as a base class for all SISPI applications. It gives the same basic structure to all applications and brings together all of the SISPI services such as shared variables, remote commands, alarms, logging and configuration management. Additional functionality provided by the Application Framework includes a heartbeat that can be used to monitor the overall state of the system and a standardized management interface with process control functions and remote access to an application's status and configuration. The framework also provides interfaces to the alarm and constants database. 

The SISPI/DECam facility database is based on PostgreSQL, a freeware database that is widely used in the Linux community. We developed a set of routines and libraries using the Python psycopg2 and sqlalchemy modules that hide SQL and other database details from the application programmer. Database access is fully integrated with the SISPI Application Framework. The facility database is located on Cerro Tololo with a mirror database deployed at Fermilab. Given the value of the telemetry archive we use a second computer and the hot standby feature provided by PostgreSQL to implement a redundant database system. This architecture was put to a test when an unusual set of disk failures wiped out the RAID array of the main database server. Once the problem was diagnosed we were able to switch to the standby server in short order and could continue the observing night.

\subsection{Web-based User Interfaces and Remote Access}
DECam observers interact with SISPI using a set of web-based graphical user interfaces running on a multi-screen workstation. Having the user interface based on a web browser provides many desirable features such as platform independence, remote access, a large number of third party tools, no installation required, a certain level of security and vendor-provided maintenance. At the time of development there were, however, some concerns regarding usability and performance in a real time environment. Near real-time performance is needed to provide quick access to specific information and to achieve the responsiveness expected from a modern system. Recent advances in browser technology have addressed this issue. Thanks to faster rendering and JavaScript execution as well as new standards such as HTML5 and the websocket API, that allow for advanced functionality such as bidirectional socket communication and 2D drawing contexts, web browsers are now on a near equal footing with desktop GUI toolkits.

The SISPI GUI architecture follows the Model-View-Controller (MVC) pattern first developed for Smalltalk but now in common use for large applications. The MVC pattern is based on the realization that all applications are, essentially, interfaces that manipulate data. Standard web browsers are used to render the views. For the implementation of the controller component we developed a special SISPI application called the GUIServer. 

The GUIServer application acts as the bridge between the mostly Python, real-time world of SISPI and the JavaScript, HTML based world of the browser. The GUIServer provides several services: getting the GUI code to the client, i.e. serving HTML and JavaScript, sending data from the browser to SISPI applications, and delivering data to the browser such as exposure information, shared variables, user info, etc. The GUIServer is based on the Twisted 
framework \footnote{For details on the Twisted framework see \url{https://twistedmatrix.com/trac}}. The client side (web browser) code is based on HTML/CSS and JavaScript and standard web browsers are used to render the views. We used SproutCore, a modern JavaScript application framework, to develop the SISPI user interfaces. SproutCore is an HTML5 application framework inspired by the Cocoa GUI framework used by MacOS X. It implements many standard, desktop-like user-interface objects and provides a powerful framework for web-based applications. It also supports so-called thick clients. In thick client applications the JavaScript code is loaded during initialization and only data updates are sent in real-time. Websockets, defined as part of the recent HTML5 standard and now supported by all major browsers, are used in SISPI to provide the most responsive user experience. Conceptually similar to traditional TCP/IP sockets, they allow SISPI to push updates and state changes directly to the GUI and so remove the need for the browser to poll for updates. Using this technology the GUIServer is able to deliver SISPI shared variable updates as soon as they become available. All of these technology choices mean that the SISPI GUIs are very responsive even when used remotely over long distance connections.

\begin{figure}[h]
\begin{center}
\includegraphics[scale=1.1]{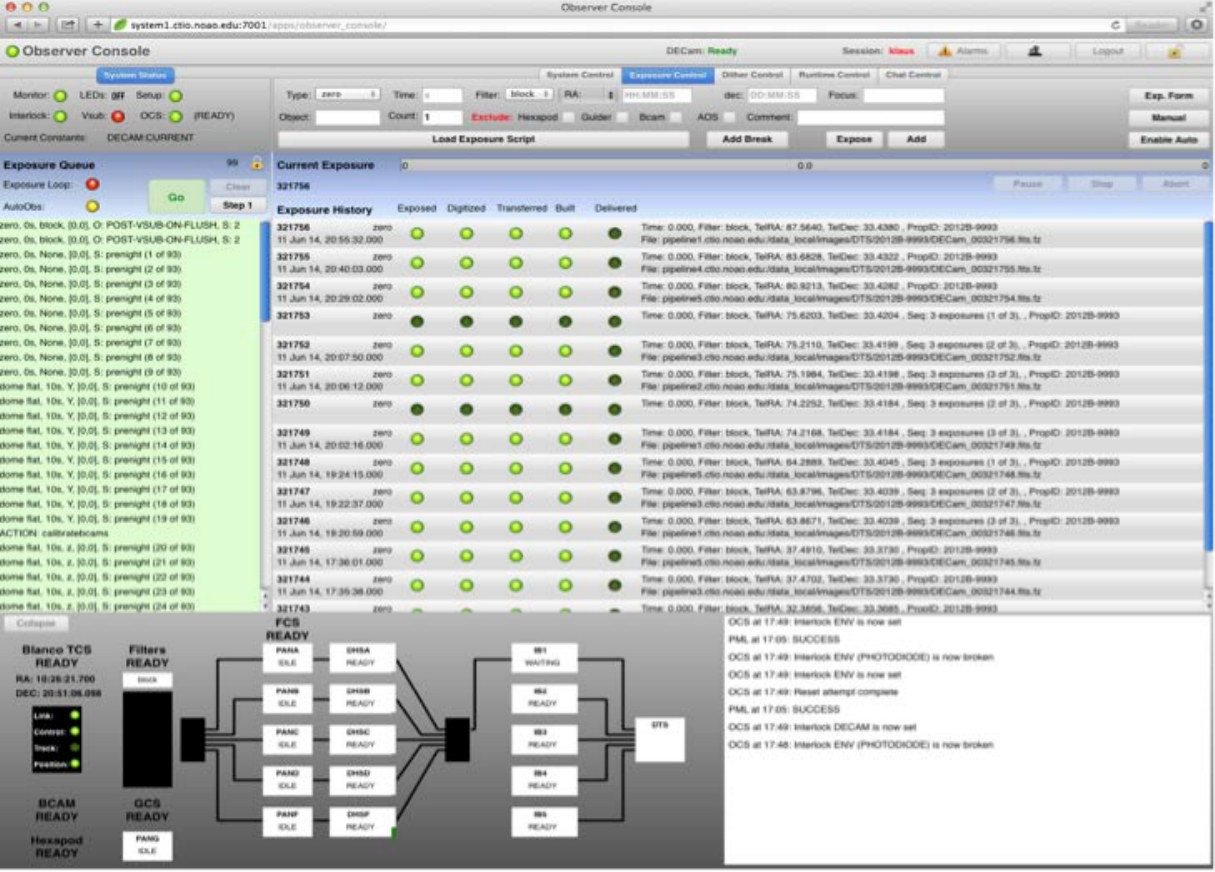}
\caption{The Observer Console User Interface. Across the top few rows from left to right are the system status and a set of controls that allow one to construct exposures to be put on the exposure queue or to select a preconstructed exposure script. In the middle on the left, the field with the green background lists the upcoming exposures. The Exposure History, to the right, shows the status of the previous many exposures. Across the bottom is a display that tracks the construction of the image that is currently being read out. While there are several other pages, observers usually follow this one during  data-taking.} 
\label{fig:sispi-console}
\end{center}
\end{figure}

Figure \ref{fig:sispi-console} shows a screen shot of the most important DECam user interface, the Observer Console. From here the observer configures SISPI, controls the exposure queue and can follow the progress of each exposure as it progresses through the system. Other GUIs available to DECam observers are used to display alarm and exposure histories, to monitor the status of the instrument, to view images and to control the Architect, the SISPI configuration system.

\subsubsection{Remote Access, Authentication and the Master Console}
Due to the web-based architecture of the SISPI user interfaces, remote access is readily available. During commissioning and science verification but also during normal operations this has been an absolutely critical feature. It is uncommon for experts to be on the mountain when a problem with SISPI or the camera occurs. If the situations cannot be addressed by the observer or the telescope operator, an expert is contacted by email, phone or in case of SISPI problems most frequently via Skype. The expert at the remote location can use a web browser to view the same information presented to the observers, fix the problem or suggest a course of action. While remote access is desirable for many reasons it also introduces a large number of security issues that must be considered. To prevent unauthorized access and malicious attacks, SISPI leverages the existing VPN authentication systems used at CTIO by not accepting any commands from outside the CTIO subnet. Of course, this subnet-based discrimination is only a defense against remote malicious users. To protect SISPI from button punchy users a second level of authentication separates the regular user from true experts and system administrators. 

\begin{figure}[h]
\begin{center}
\includegraphics[scale=0.8]{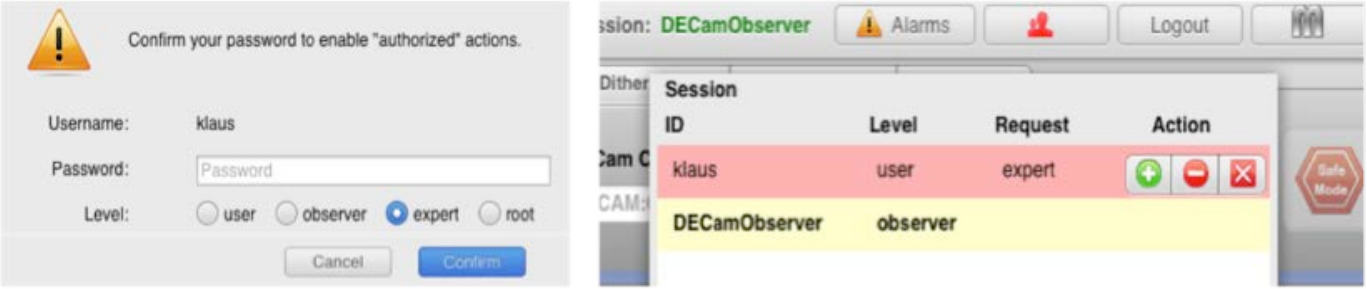}
\caption{The left panel is presented to a remote user requesting privileged access. The right panel shows the session control options at the master console. A request has to be approved before privileged access is granted.} 
\label{fig:sispi-remoteaccess}
\end{center}
\end{figure}
 
This level of security is implemented at the interface level. It is designed to prevent non-expert users from accessing critical configuration settings. When starting a SISPI GUI, users are presented with an authentication screen where they enter a username and password. This is checked against a user database maintained by SISPI. The default authentication level is user. A privileged user can raise the level to observer, expert or root. With increasing level additional functionality becomes available. This mechanism is blocked for offsite users. To allow for remote expert access under controlled circumstances an additional authentication layer and the concept of a master console were added. The master console is typically the DECam Observer Console in the Blanco control room and explicit approval from the observer sitting at this workstation is required before the authentication level of a remote user is changed. This ensures that the observers are aware of any experts accessing the system. The master console concept also includes the ability to terminate remote sessions at any time (Fig.~\ref{fig:sispi-remoteaccess}).

\subsection{The DECam Online System} \label{subsec:decamonline}
Built upon the software infrastructure described in the previous section, the DECam online system consists of the image pipeline, the guider, the focus and alignment system, the interface to the Blanco telescope control system and real-time data quality assessment.
\subsubsection{Image Pipeline}
\label{sec:sispi-imagepipeline}
Built on top of the infrastructure software, the Observation Control System or OCS is the central component of SISPI coordinating all aspects of camera operation and the observation sequence. Connected to the OCS is the tactical observation package (ObsTac) used by the DES collaboration to determine an optimized sequence of pointings for the telescope based on a number of inputs. A typical DES algorithm would take as inputs:  1) current date and time to compute moon position and twilight status, 2) survey status, in the form of fields previously observed, and 3) the current observing conditions as recorded in the telemetry database. The algorithm will then locate candidate tiles from the survey status information, and compute air-masses for the candidates. Given this information it computes a metric for each candidate and then chooses the best tile to observe. The OCS receives this information and proceeds to take these images unless the sequence is overwritten by the observer. Besides survey mode (ObsTac) and manual mode, the OCS also supports simple scripting of a given series of exposures and provides standard features such as general purpose dither patterns. 
During observing, SISPI reads entries from an observing queue and performs the corresponding exposures. Entries on the queue contain all the data needed for the observation, including the coordinates of the field, the exposure time, and the filter. The queue may be edited at any time. Whenever SISPI completes an exposure, it pops the next observation request from the queue and (re-)starts the exposure sequence.

At the end of an exposure, data are read from the CCDs and digitized using Monsoon-based electronics described in Section~\ref{s5:Electronics}.  Image data are transferred from the front-end crates to pixel acquisition nodes (PAN) via an optical data link. Our readout software (panview) is based on the ArcVIEW package~\citep{SPIEarcview2002}. Recall (Section~\ref{subsec:ro17}) that it takes 17 seconds to readout the front-end electronics and about 3 seconds to de-scramble the pixel data and to save the image to disk in FITS format.  The focal plane is configured such that the data from all 8 focus and alignment CCDs are read by the same PAN. The out-of-focus donut images are analyzed in real time to compute adjustments to the hexapod support system in order to maintain the best possible image quality. In order to complete the hexapod motion before the beginning of the next exposure, any adjustment commands must be issued within 20 seconds after the shutter closes. Since it takes about 8.5 s to readout the $2\rm{k}\times 2 \rm{k}$ alignment CCDs the focus algorithm has about 8 seconds to complete. A separate PAN computer is allocated to read the guide CCDs and to execute the guide algorithm. To provide guide signals to the telescope, the guide CCDs are operated independently of the image CCDs by using a separate Monsoon system for their readout. In order to achieve the required update frequency of the guider correction signals of about 1 Hz the guide CCDs use window or region-of-interest (ROI) readout mode which has been successfully tested with the imager. The guider algorithm itself was tested in a dedicated observing run using a DECam $2\rm{k}\times 2 \rm{k}$ CCD mounted to the CTIO 1m telescope.

Using high speed Gigabit Ethernet links each PAN sends its data to the Image Builder (IB) system which is implemented as a processor farm. The data transfer protocol is optimized for performance and instead of standard TCP/IP sockets we are using netcat~\footnote{For details on netcat see \url{http://netcat.sourceforge.net}} to achieve transfer times of less than 10 seconds for the entire image (1 GB). Throughput is further enhanced by using a memory based file system for temporary files. The images are saved to disk in compressed FITS format.  Each image builder process, written in Python, receives data from every PAN and when an image is complete, it assembles the exposure data in a multi-extension FITS format using PyFITS~\footnote{For details on PyFITS see \url{http://www.stsci.edu/institute/software\_hardware/pyfits}}. While processing the image the IB completes the primary FITS header by inserting information such as telescope position, airmass, filter, etc. published by other applications. Before the data are written to disk each image is passed through a quality assurance process that analyzes the image data and determines standard quantities such as sky noise and seeing. Assuming a multi-core CPU is used for the IB processor farm, allowing data I/O to proceed in parallel, the available processing time ${\rm t_{Proc}}$ is simply ${\rm t_{Proc} = t_{Exp} \times N}$ where ${\rm t_{Exp}}$ is the exposure time and N is the number of IB nodes. We have estimated that 4-5 nodes will be sufficient for DES but the system can easily be expanded. The IB nodes are controlled by the Image Builder Supervisor (IBS) which is implemented as part of the OCS. The IBS maintains a queue of available IB nodes which is used by the OCS to instruct the PANs where to send the next image data. 

Images are stored on RAID disks local to each image builder. An Image Directory Service provides a catalog of available images and provides users and other applications, for example Quick Reduce (discussed below), with access tools. Once again using netcat or similar tools for the data transfer we obtain significant performance gains compared to an nfs based system. The NOAO Data Transport System (DTS) \citep{SPIEfitzpatrick2010} is used to transfer data from the Cerro Tololo mountain-top to the CTIO campus in La Serena via a 155 Mbits/s microwave link. From there DTS transfers the image data to the NOAO archive in Tucson and the DES Data Management system at NCSA at the University of Illinois.

\subsubsection{Guider}
The DECam guider system is designed to support closed-loop telescope tracking to stabilize the exposures to an accuracy of a few tenths of an arcsecond. The guider uses information from four $2\rm{k}\times 2 \rm{k}$ CCDs located at the edge of the DECam focal plane. During a science exposure the guide CCDs are operated in region-of-interest mode and are read out continuously with a typical exposure time of 600 ms. These postage-stamp images of the guide stars are rapidly analyzed by the guider software, which looks for shifts in the position of the stars detected. Every second a position error signal is sent to the telescope control system to close the tracking loop.

The guide algorithm is the core of the guider software. It determines the change in guide star position and delivers the error signal to the Telescope Control System (TCS). Following this objective the guide algorithm handles 3 main tasks:
Acquisition - Find and set the best reference star for guiding (once), Tracking - Follow the reference star with the best signal to noise, and Compute offset - Determine and deliver the offset to the TCS. 

For each DECam exposure, the guider executes the acquisition task once during the initialization stage. For the rest of the exposure the guider operates in a loop continuously running the tracking and the compute offset tasks. Acquisition is carried out during the initialization phase. The goal is to select the stars best suited for guiding from the stars available in full frame images from the 4 guide sensors. The SExtractor~\citep{sextractor} software is used to identify stars. It has proven to be a fast, precise and reliable tool for this task. On average, SExtractor provides more accurate stellar positions for stars that have a higher signal-to-noise (S/N). Before selecting the highest signal-to-noise stars, the guide algorithm applies a filter to remove the following from the list of detected objects: stars with at least one pixel saturated, double stars,
 stars close to or truncated by a chip boundary, stars with neighbors that are bright and close enough to significantly bias its flux, galaxies, cosmic rays, and detector defects.

In addition, the algorithm searches for isolated stars within a user-defined range, to avoid possible confusion with the guide star during the tracking stage. Once a guide star is selected for each of the available guide CCDs its position is taken as the zero-point location in the detector for this exposure. 

After the acquisition phase when guide stars have been set, the system is reconfigured to readout only a Region of Interest (ROI) around each guide star. This dramatically reduces the readout time of the guide CCDs to few milliseconds. The ROI size is a configuration parameter. The default area is $50\times50$ pixels.
A signal is sent to the online system to start to exposure and to readout the 4 guide sensors in a continous loop. Each set of postage stamp images is analyzed by the guider software the correction signal is sent to the TCS. Tracking corrections are sent at a rate of about 1 Hz. The system is designed to minimize jitter between updates. During the tracking phase, the guider attempts to find the reference guide star selected during the acquisition phase. For this we use PyGuide, a simpler image analysis tool that provides fast and accurate centroids at a constant rate\footnote{For details on PyGuide see \url{http://www.astro.washington.edu/users/rowen/PyGuide}}. The critical task during this step is to correctly match the sources found in the postage stamp images to the initial set of guide stars. For the guider to achieve the required performance, it is very important to avoid losing or mismatching the reference guide stars. For this purpose the guider algorithm employs a user adjustable search range surrounding the zero point position set during the acquisition phase. The search range is large enough to have the true guide star inside (larger than the maximum tracking error) and small enough to cover just few stars from the several available in the ROI image. Ideally only the reference guide star will be within the search range as non-isolated stars are discarded at the Acquisition stage.

From the remaining stars, the algorithm chooses the star with most similar flux to the reference guide star flux. The guide star flux is updated after every iteration to safeguard against changing sky conditions during the exposure. In general, only the reference star is located in the search range but in case we find more than one star, this method is very efficient in identifying the guide star.

With the guide stars detected in the postage stamp images, computing the offset from a single CCD is as simple as measuring the distance from the current position of the guide star found and the previous zero point. It is slightly more complicated when multiple guide stars have to be considered.
A simple way to combine the information from the multiple guide stars would be to average the offsets. However, since it is known that the position accuracy is higher in high S/N stars, the offset is averaged by weighting the offset with the corresponding guide star's S/N.  By using the weighted mean, the more reliable stars have more importance in the final offset. In addition, the guider algorithm uses information about the agreement/discordance among the tracking error information given by the different CCDs. This means that if three CCDs provide a very similar signal whereas the fourth suggests a different correction, the algorithm discards the outlier and does not include this information when calculating the correction offset. This 
\begin{figure}[h]
\begin{center}
\includegraphics[width=0.8\linewidth]{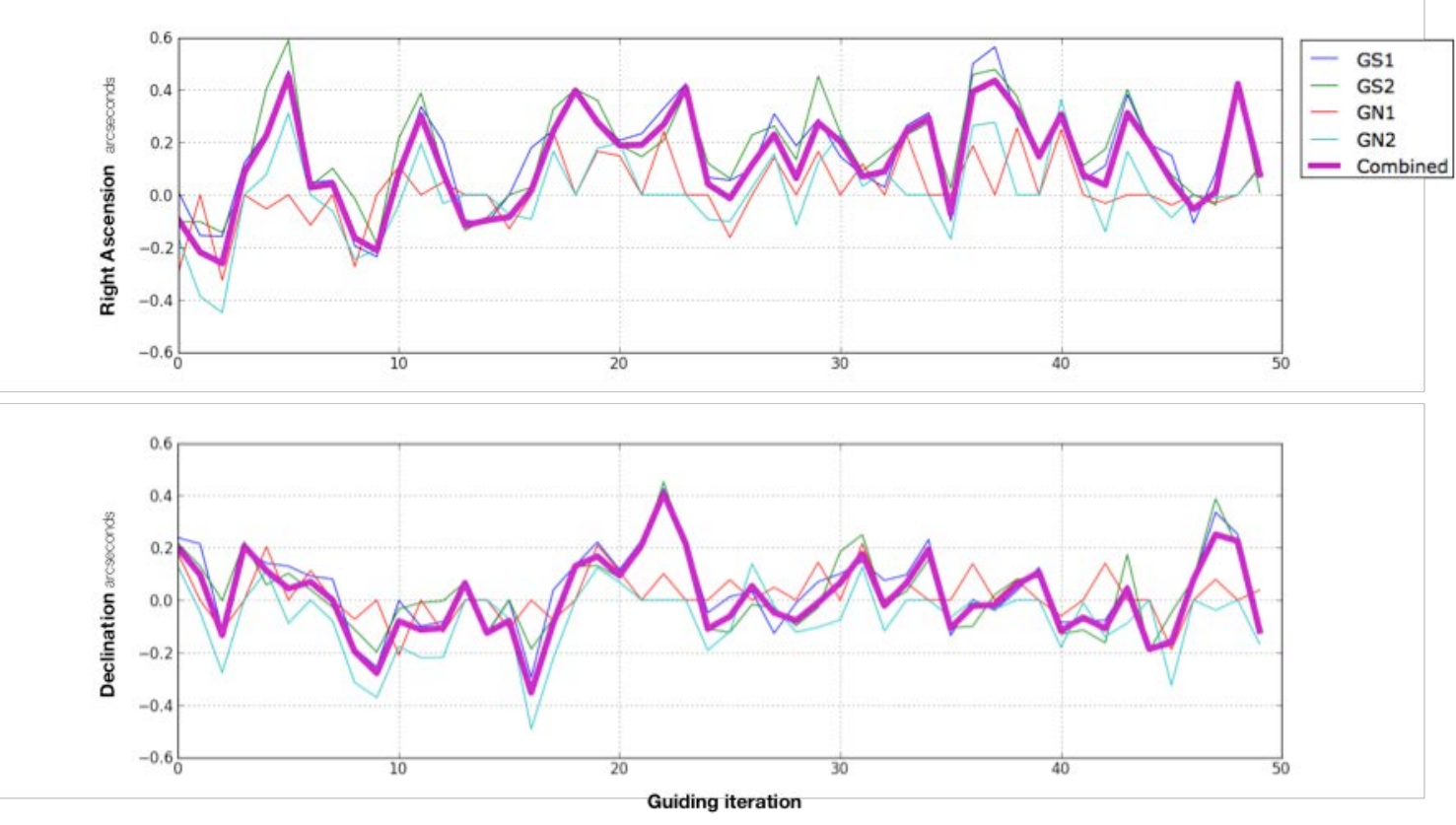}
\caption{Time history chart for the guider corrections in RA and dec for a typical DECam exposure. For each coordinate the plot shows the combined offset and the contributions from the four guide CCDs.}
\label{fig:sispi-guidestars}
\end{center}
\end{figure}
selection is accomplished by measuring the Euclidean distance between the offsets from all guide stars. If the discordance exceeds a threshold, the highest discordant guide star is discarded. The threshold is adjusted for every iteration depending on the correction signal differences found among the available guide CCDs. In other words, more discordant signal offsets will have a more tolerant threshold than very accurate signals. Having an adaptive threshold is important as different atmospheric conditions may lead to very different discard thresholds.
During the main guiding loop, the algorithm removes a guide CCD from further processing if it has to be discarded for more than 5 consecutive iterations. It will also be disabled if the guide star cannot be found for the same number of consecutive times. The number of failed iterations can be adjusted by the observer.
\begin{figure}[h]
\begin{center}
\includegraphics[width=0.8\linewidth]{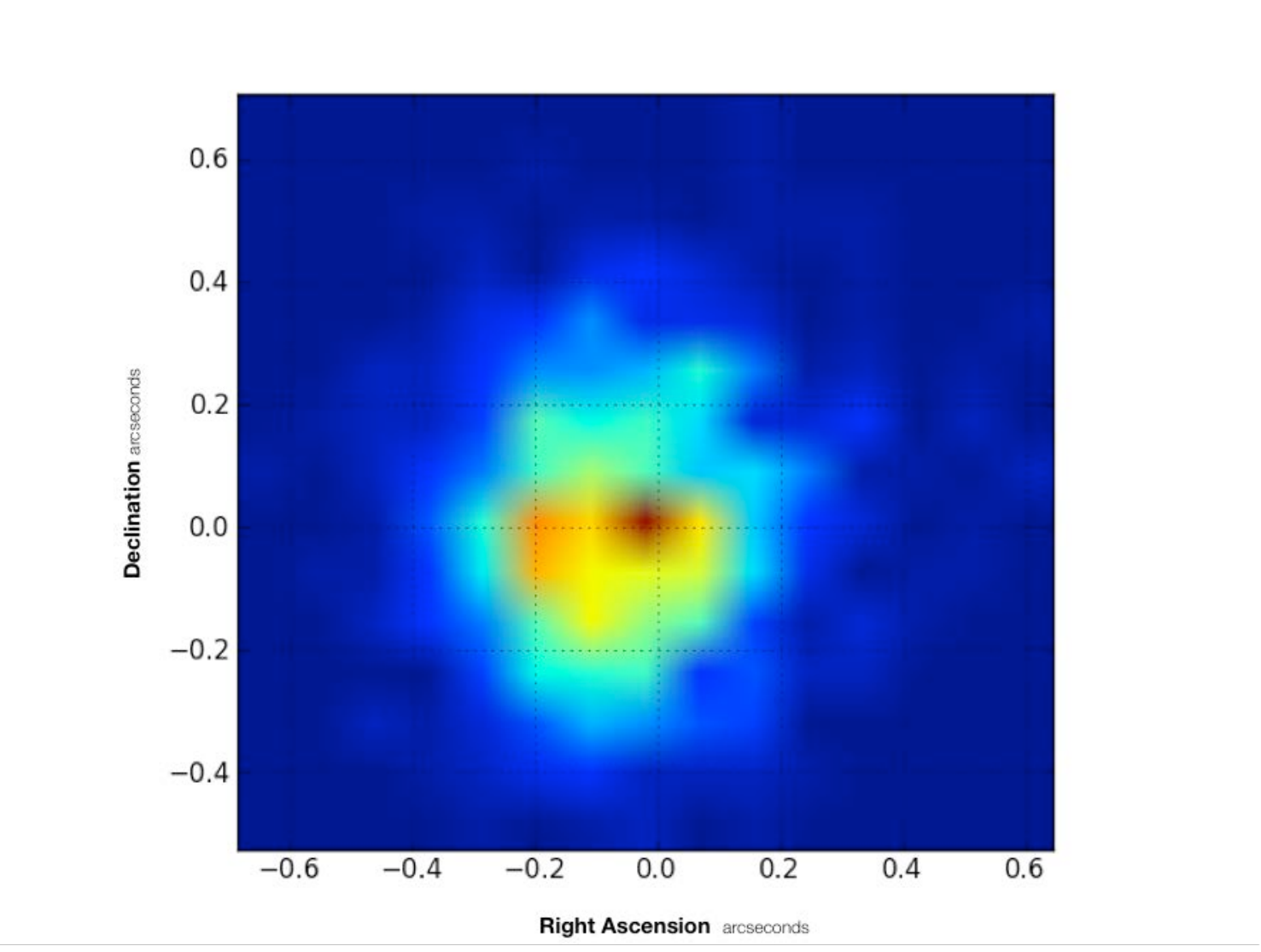}
\caption{Distribution of of correction offsets in arcseconds in declination and right ascension for a typical DECam exposure.} 
\label{fig:sispi-corrections}
\end{center}
\end{figure}

The combined correction offsets as well as the results from individual guider CCDs for a typical DECam exposure are shown in Fig. \ref{fig:sispi-guidestars}.
This example shows how the multiple guide signals follow a random pattern around the center (driven by the minor telescope tracking errors) with standard variation of about 0.2\arcsec, similar to the pixel size. In this particular example, guide CCD GN1 shows a worse performance due to a lower S/N guide star. The integrated view of the correction signals (Fig. \ref{fig:sispi-corrections}) shows the RA/Dec corrections peak at the target coordinates which is another indication of the excellent tracking performance of the Blanco telescope.

\begin{figure}[h]
\begin{center}
\includegraphics[width=1.0\linewidth]{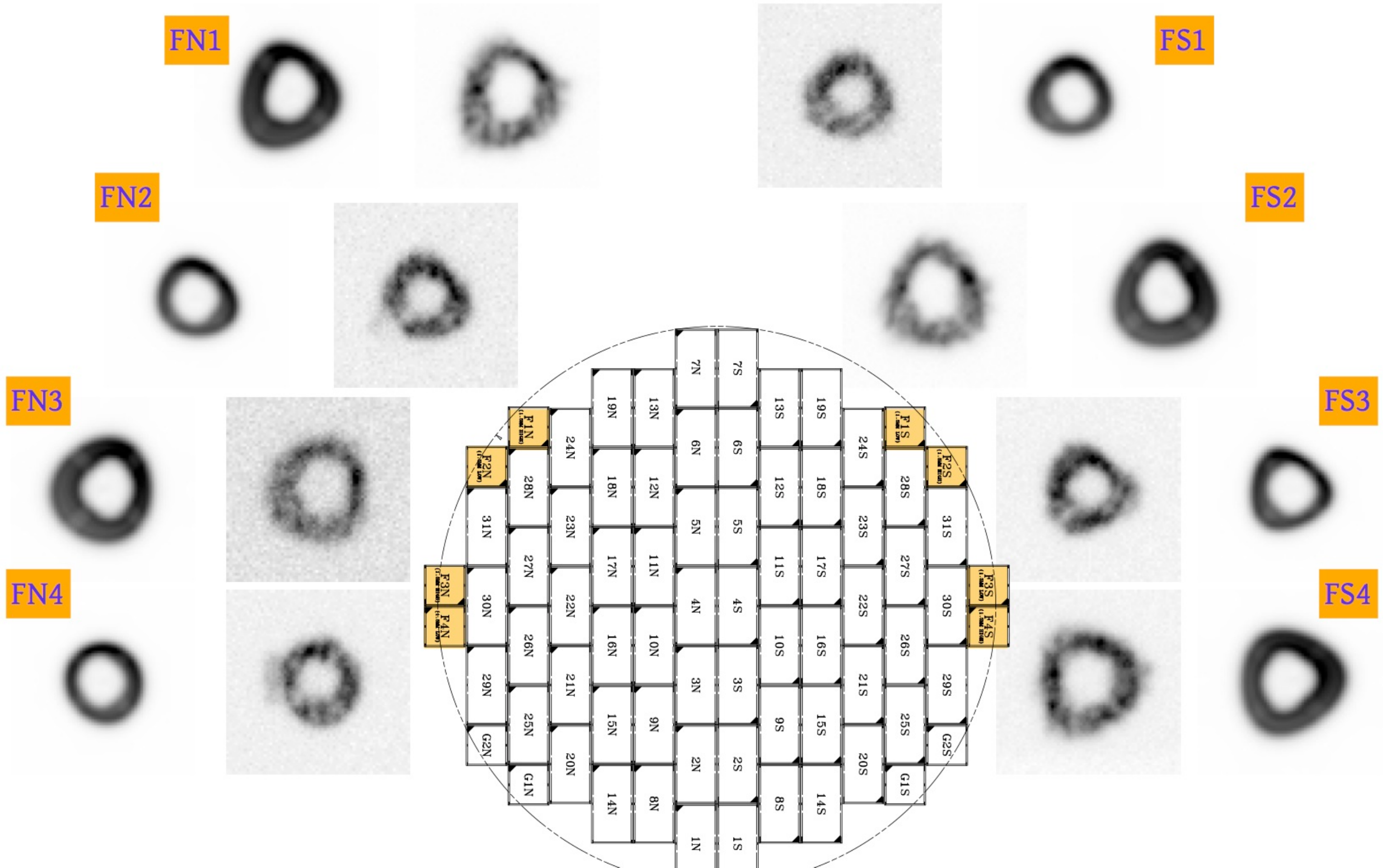}
\caption{The DECam focal plane, with the 8 wavefront sensors shown (shaded yellow) and an example out-of-focus star from each (interior), along with the corresponding fitted model (exterior).  These donuts come from an image taken during DECam commissioning; the camera was roughly $200 \mu m$ out of focus at the time. The four $2{\rm k} \times 2 {\rm k}$ CCDs used for guiding are located near the bottom-right and bottom-left of the focal plane.  } 
\label{fig:sispi-focalplane}
\end{center}
\end{figure}

\subsubsection{Active Optics System} \label{subsec:focus}
The DECam active optics system (AOS) uses four pairs of intra- and extra-focal CCDs as wavefront sensors to correct the camera's focus and alignment in closed loop.  Out-of-focus stars in these sensors, termed donuts, appear as aberrated images of the telescope pupil. The donuts are analyzed in terms of pupil-plane Zernike polynomials, corresponding to the primary aberrations, and these measurements of the aberrations of the Blanco plus DECam optical system can be transformed into a measurement of the defocus and misalignment of DECam.  Typical donuts along with images of their Zernike fitted model are shown in Fig.~\ref{fig:sispi-focalplane}. A description of the studies performed to develop the AOS may be found in \citet{SPIEroodman2010, SPIEroodman2012}, and a more complete description of the AOS, the donut fitting algorithm and the system performance is given in \citet{SPIEroodman2014}.

\begin{figure}[h]
\begin{center}
\includegraphics[width=0.5\linewidth]{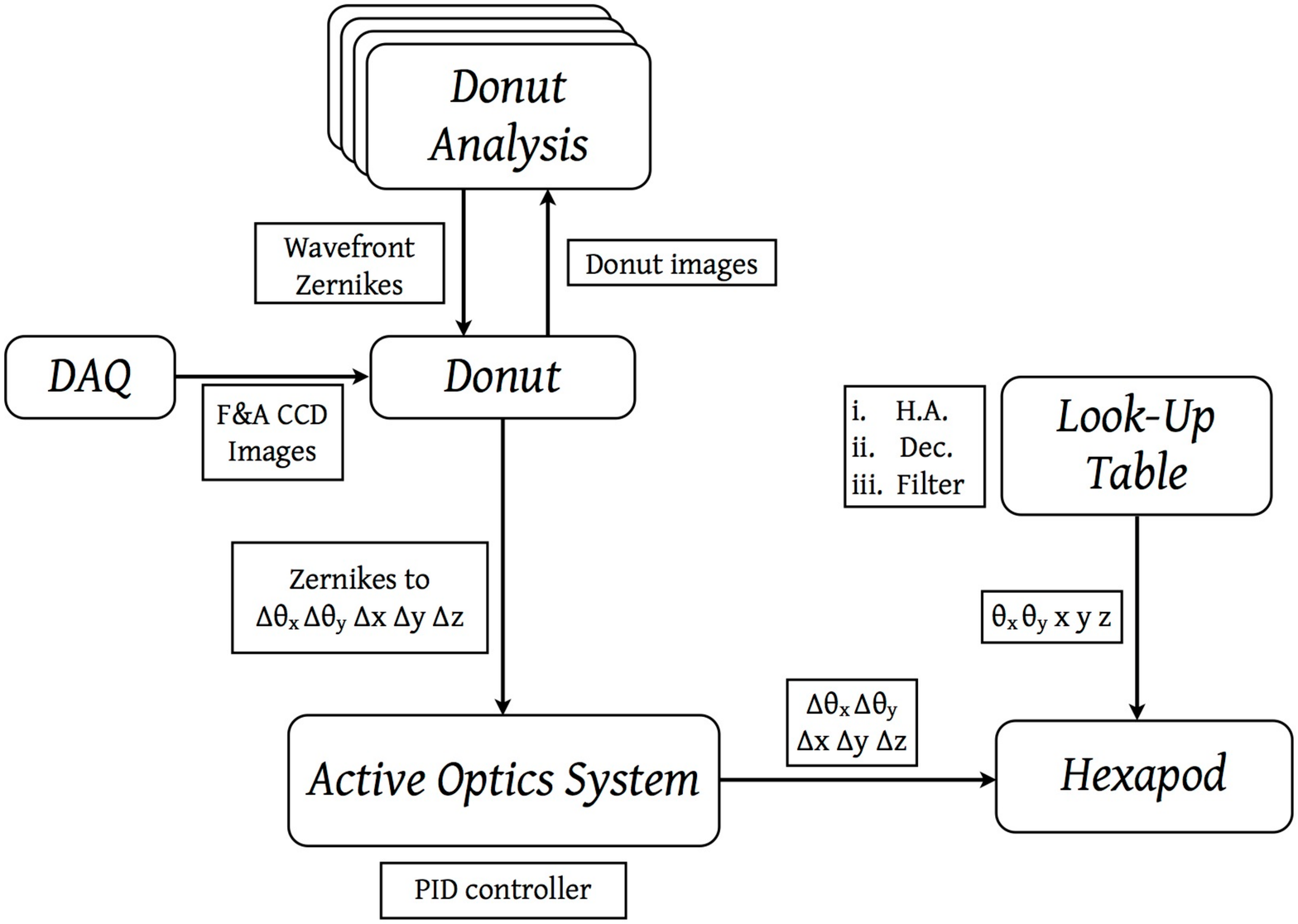}
\caption{Block Diagram of the DECam Active Optics System.} 
\label{fig:sispi-aosblock}
\end{center}
\end{figure}

 The AOS is implemented as part of the SISPI system and consists of a number of software modules which implement the following steps: 1) receive the eight $2\rm{k} \times 2\rm{k}$ wavefront CCD images, 2) locate usable donuts by finding contiguous pixels which have local flux maxima and satisfy total flux, size and shape requirements, 3) distribute small cut-out images of the donuts to processing farm nodes, 4) run the wavefront retrieval algorithm on this ensemble of donuts, 5) collect the fitted Zernike coefficients and distill them into DECam focus and hexapod alignment displacements, 6) pass these values into a PID control loop for each of the five degrees of freedom (focus, x and y decenter, tip and tilt) and hence 7) produce hexapod adjustments to be applied prior to the next image.  All steps are performed in a 7 second time window between the time when the wavefront CCDs are read-out and the start of the next image.  Since the DECam science CCDs are $2\rm{k} \times 4\rm{k}$ and so take double the time to read-out, the AOS operates with no extra deadtime. A block diagram of the AOS is shown in Fig.~\ref{fig:sispi-aosblock}.  On average 30 donuts are analyzed per image, with at least 10 donuts available 97\% of the time. 

\begin{figure}[h]
\begin{center}
\includegraphics[width=0.4\linewidth]{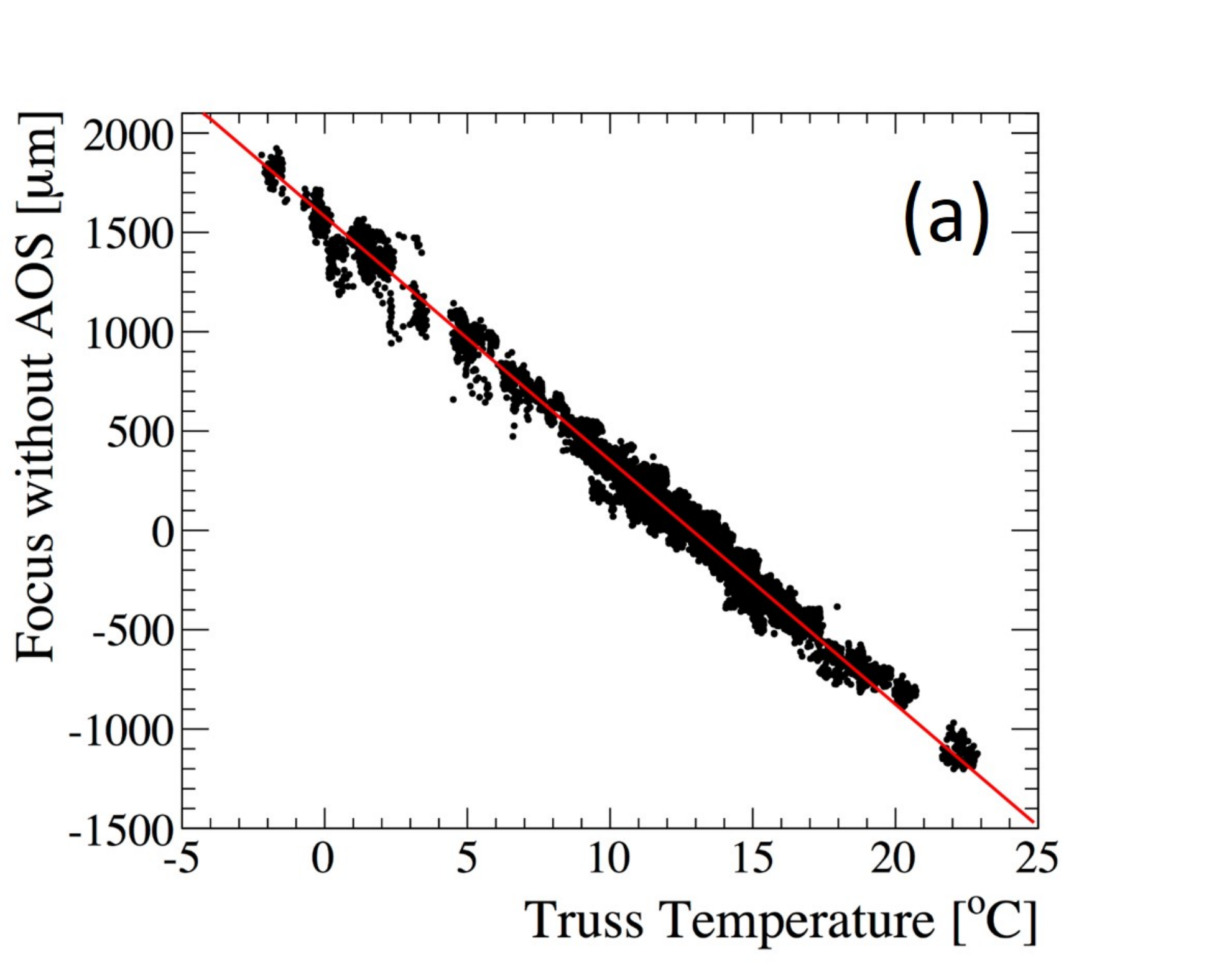}
\includegraphics[width=0.4\linewidth]{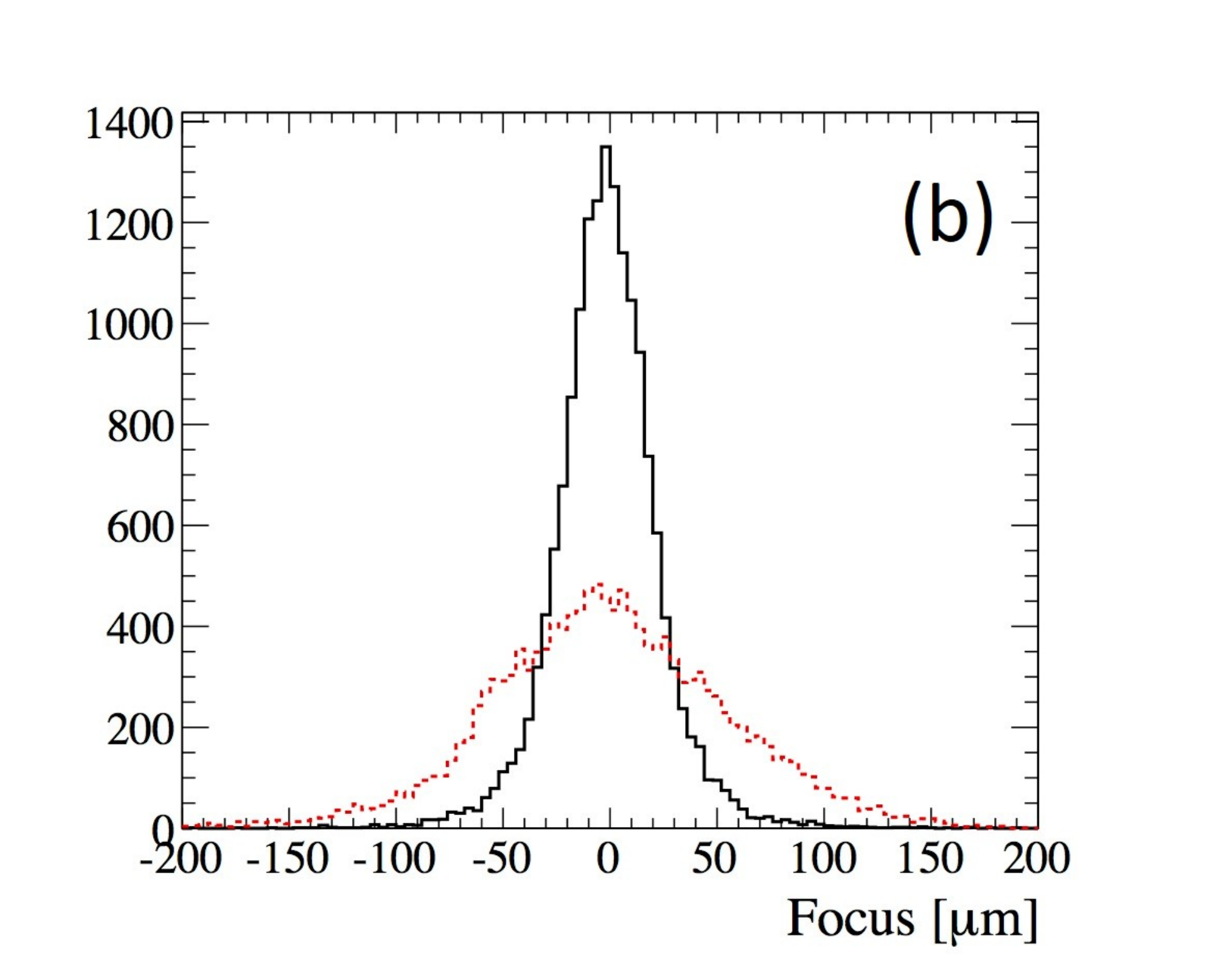}
\end{center}
\caption[] 
 { \label{fig:sispi-focus}
 a) Distance from nominal focus, with only the focus LUT, vs. truss temperature.  The slope of these points indicate a change in focus of $120 \mu m/^{o}C$. b) Distance from nominal focus for the AOS (black, solid line) and for a linear temperature correction (red, dotted line).  Images from the DES Year 1 wide-field survey are used in these plots.  The focus RMS for the AOS is $24 \mu m$, compared to $60 \mu m$ for a linear temperature correction. }
\end{figure} 

\begin{figure}[h]
\begin{center}
\includegraphics[width=0.4\linewidth]{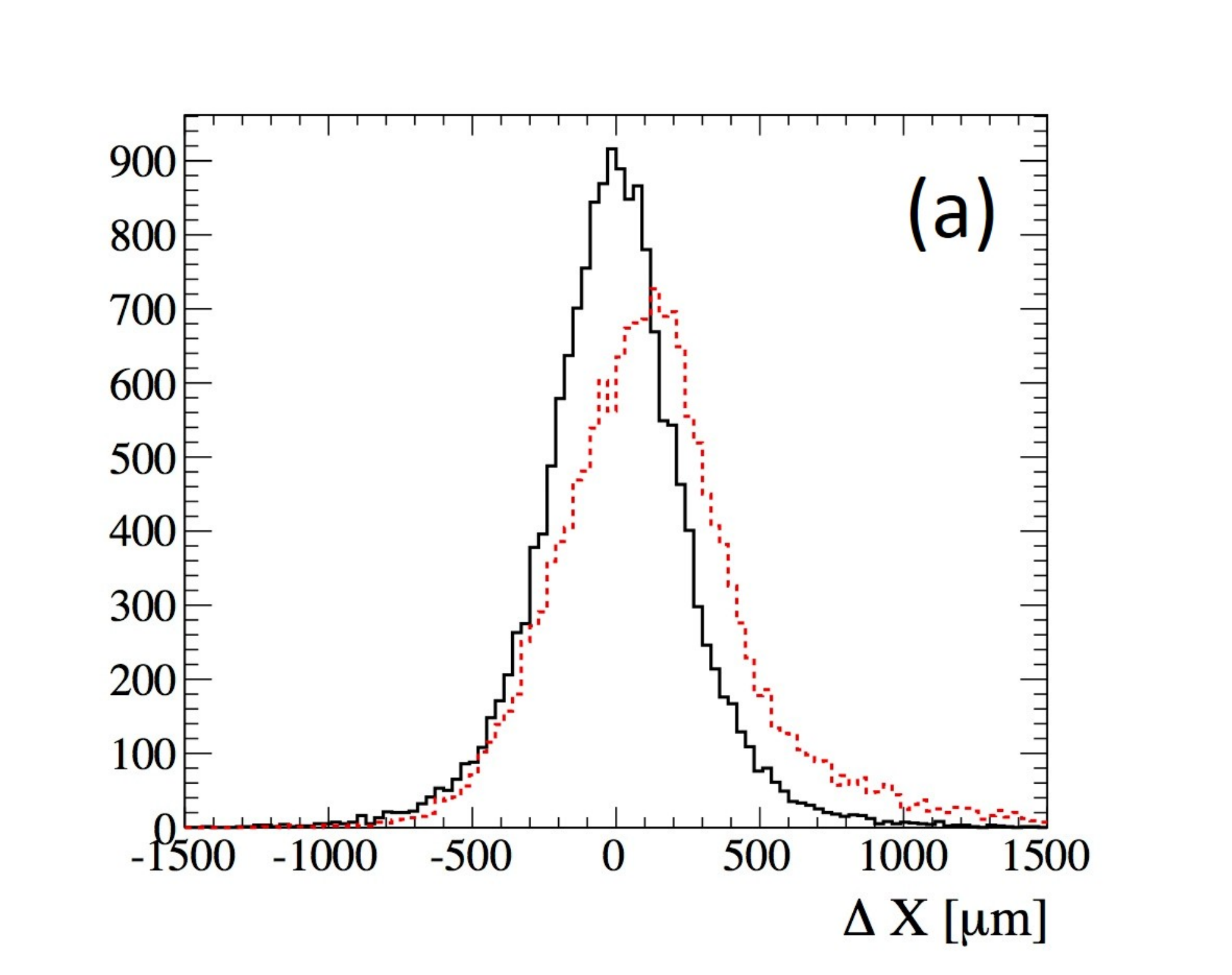}
\includegraphics[width=0.4\linewidth]{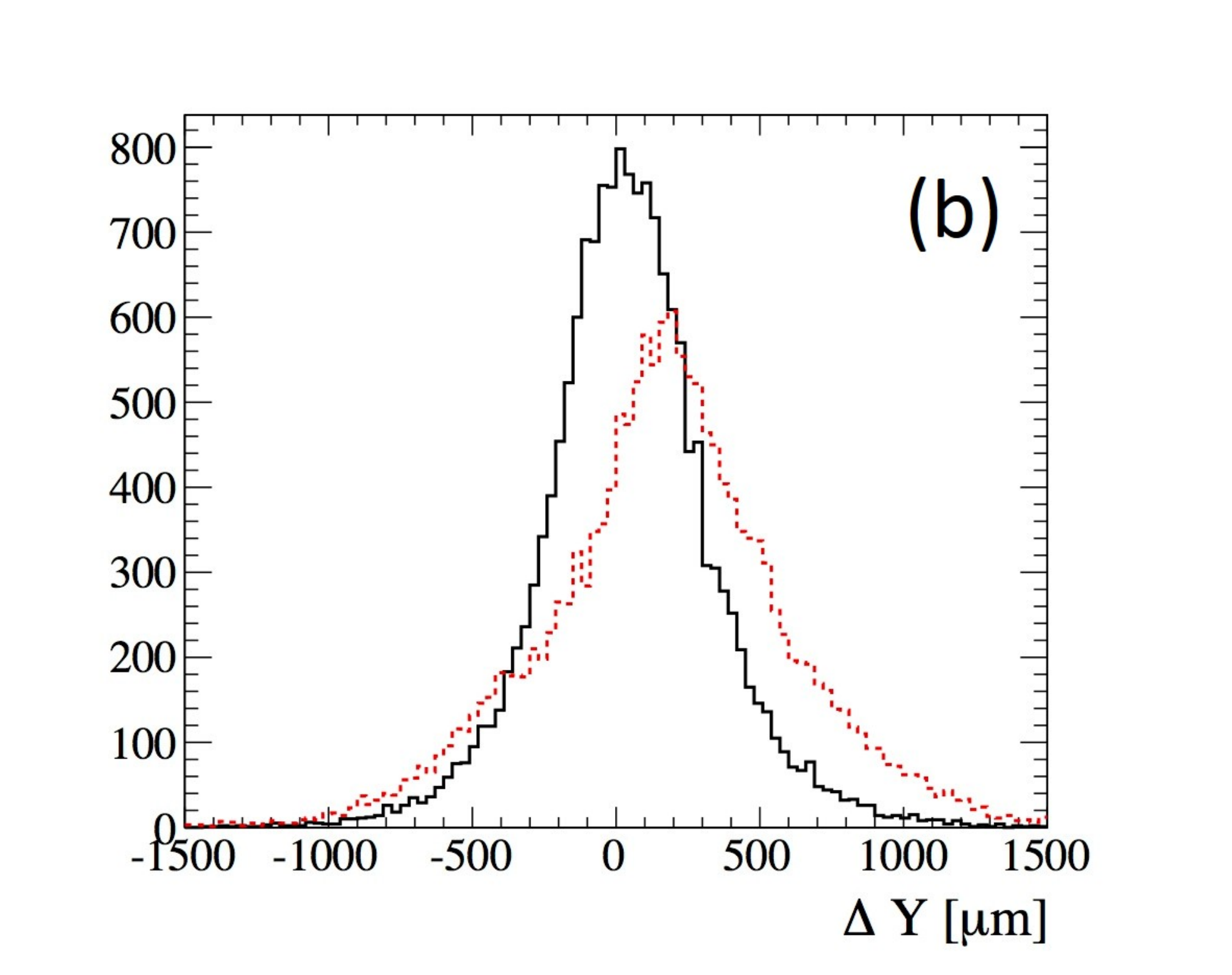}
\includegraphics[width=0.4\linewidth]{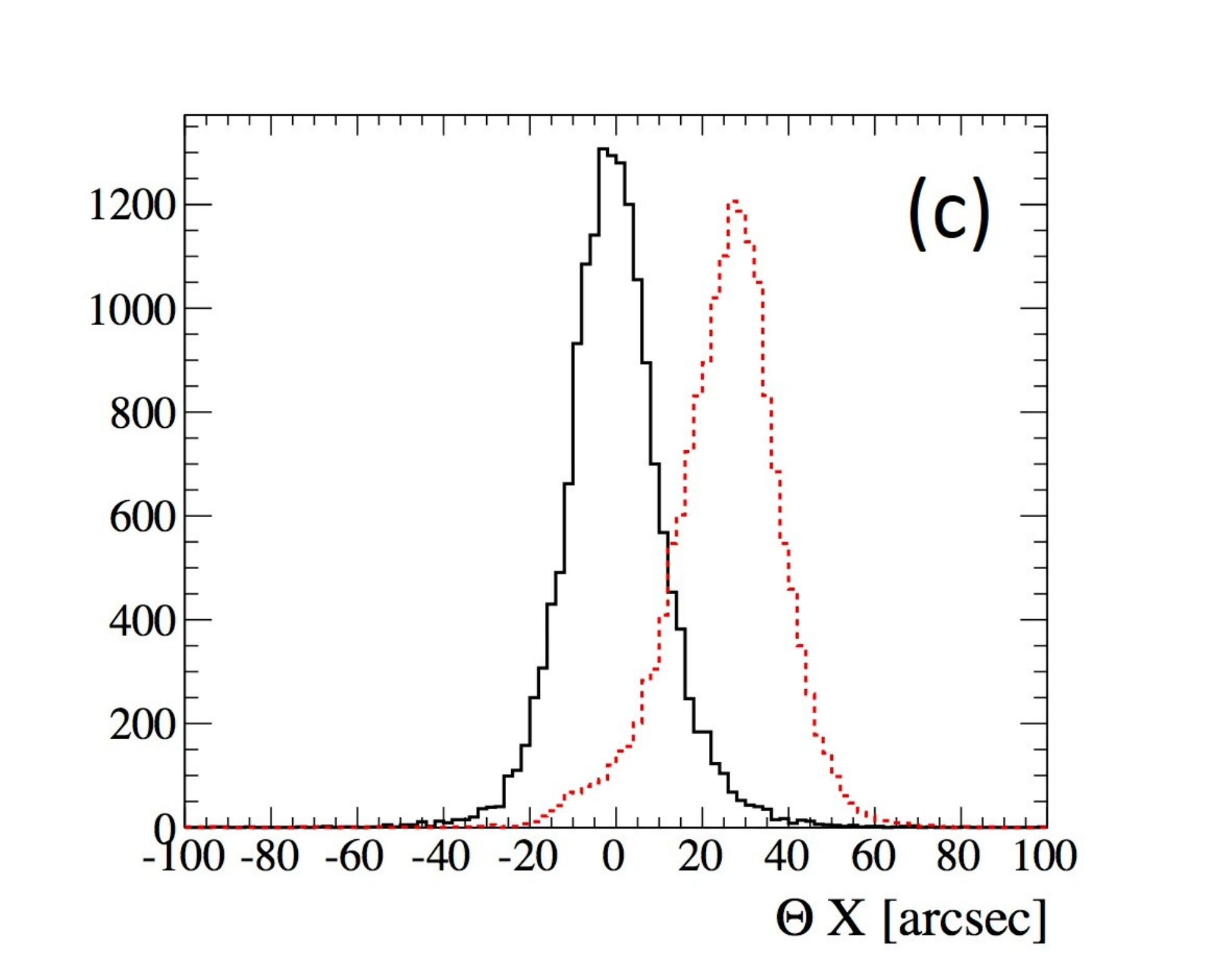}
\includegraphics[width=0.4\linewidth]{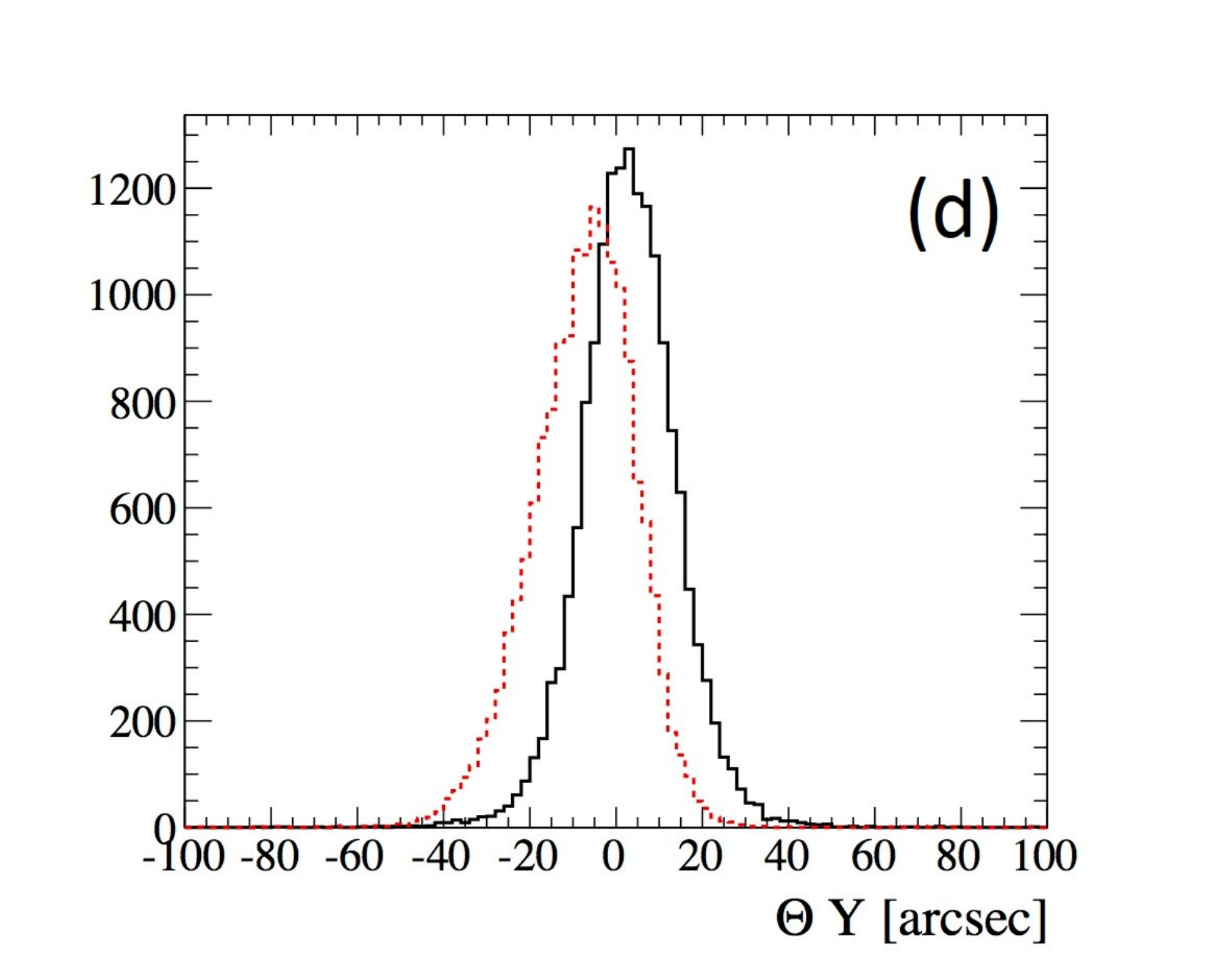}
\end{center}
\caption[] 
 { \label{fig:sispi-alignment}
 The hexapod alignment degrees of freedom, showing the distance from nominal decenter or tilt for the AOS (black, solid line) compared to the LUT only (red, dotted line) for a) $\Delta X$, b) $\Delta Y$, c) $\theta X$, d) $\theta Y$.}
\end{figure}

The DECam hexapod position is commanded from two sources: a fixed look-up-table (LUT) and the AOS.  The LUT contains static focus and alignment settings indexed by telescope orientation, hour angle and declination, and in the case of focus only also by filter.  The LUT values were determined from a suite of engineering images, collected with the entire focal plane out of focus, and covering a grid of points in altitude and azimuth.   In the time between images, the LUT hexapod settings are applied as appropriate for the following image and then the AOS adjustments are also applied, adding to the LUT values.  We find that deviations from the LUT values do accumulate over time, and so the AOS adjustments are continually updated throughout each night of observing.  Changes in focus are mostly due to temperature variations; however, the closed loop operation of the AOS yields a significant improvement in the focus for DECam, as shown in Fig.~\ref{fig:sispi-focus}.  Likewise the control of alignment in the AOS greatly improves upon the collimation from the LUT alone, as shown in Fig.~\ref{fig:sispi-alignment}.

The DECam AOS operates without any human intervention and observers need not take focus scans. The AOS achieved closed-loop operation for control of focus during the DECam commissioning period, and for control of alignment during DES science verification.  Closed loop operation was made the default condition for all observers at the end of DES SV, and it has remained in stable problem-free operation since that time.

\subsubsection{Interface to the Blanco Telescope Control System}
The telescope control system of the Blanco 4m telescope was upgraded as part of the CTIO facilities improvement program. In order to meet the strict requirements imposed by the Dark Energy Survey on tracking precision, slewing speed and execution efficiency, both the servo controller hardware and the control software have been replaced~\citep{SPIEwarner2012} with modern systems. The connection between SISPI and the Blanco 4-m telescope control system (TCS) is provided by the TCS Interface application. It basically functions as a protocol translator between the message passing architectures used by the two systems. Through this interface SISPI commands the TCS to move the telescope to a new position.  In addition, the TCS Interface gives SISPI access to the Blanco dome environmental system including the CTIO MASS/DIMM, an external weather station, and RASICAM, the radiometric all-sky infrared camera developed to support the Dark Energy Survey (see Section~\ref{subsec:rasicam}).

\subsubsection{Telemetry, Image Health and Quick Reduce} \label{subsubsec:telemetry}
For the DECam/SISPI telemetry system we adopted an approach more commonly known from high energy physics experiments. The basic idea is very simple: every bit of information recorded at the time an image was taken can potentially be useful to understand systematic uncertainties during offline analysis. All non-image data that are relevant to the operation of DECam are stored in the SISPI facility database. This includes a number of different types of data such as camera calibration and configuration information, alarm and error messages as well as environmental information from the instrument and telescope control systems. 

\begin{figure}[h]
\begin{center}
\includegraphics[scale=0.8]{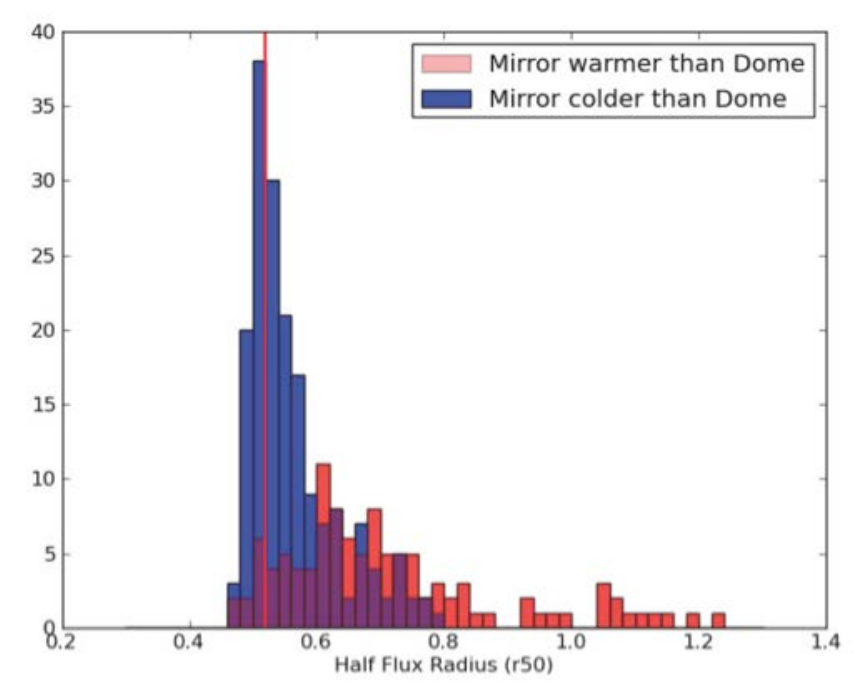}
\caption{Data Mining example: Average half flux radius ($\approx 0.52 $ times the Seeing FWHM) for DECam exposures taken early during commissioning for different environmental conditions as explained in the text.} 
\label{fig:sispi-telemetry}
\end{center}
\end{figure}

The archived information has proven to be an immensely valuable tool to diagnose problems with the camera, the image quality, or telescope performance. An example from the DECam commissioning period is shown in Fig. \ref{fig:sispi-telemetry}. Plotted is the half flux radius, a measure of image quality related to seeing, for exposures taken under different environmental conditions. While for exposures taken with the primary mirror temperature below the dome temperature the half flux radius peaks around 0.5$\arcsec$, corresponding to a seeing of $\approx$0.9 $\arcsec$ FWHM, the image quality is much worse when the mirror is warmer than the dome environment. This particular issue was resolved by modifying the daytime mirror cooling algorithms and other measures that reduced dome seeing effects. The archived telemetry information is not only used for offline viewing and data mining but is presented online as well, mostly in the form of time charts. This provides the observer an easy way to check instrument performance and to detect trends that potentially could indicate problems. 

Continuous monitoring of both the instrument and the image quality is required to control systematic uncertainties to achieve the science goals of the Dark Energy Survey and to allow continuous, error-free operation of DECam. Several SISPI components are designed to implement these quality assurance procedures. ImageHealth, a first check of the image quality of every exposure, runs on every exposure as part ofthe image builder process. The image health algorithms determine mean and noise values for each amplifier in both the overscan and data regions. Simple object detection is used to find star like objects on each CCD for a PSF analysis. Both PSF size and ellipticity are measured. The average PSF size for each image is a measure for the current seeing condition and together with the average sky background. It is reported to the ObsTac tool that selects the targets for next exposures. 

\begin{figure}[h]
\begin{center}
\includegraphics[scale=0.4]{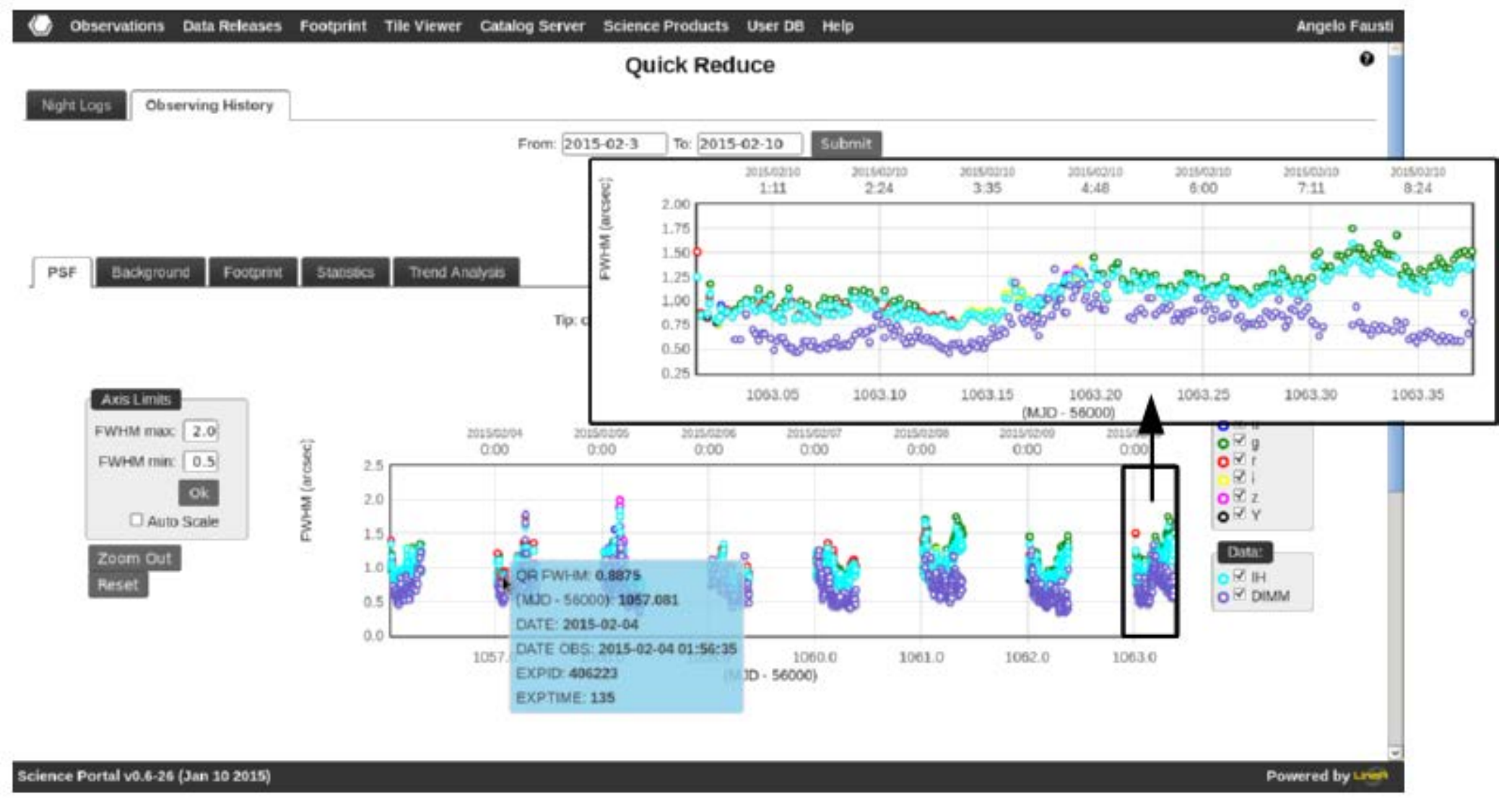}
\caption{Quick Reduce results exported to the DES Science Portal at Fermilab. The plot shows the variation of the FWHM (arcsec) in the selected period of time. The points in the inset correspond to measurements on each exposure. The QR measurements are compared with the Image Health (cyan) and the DIMM seeing (violet) for reference.} 
\label{fig:sispi-qr-history}
\end{center}
\end{figure}

Quick Reduce (QR) is a web based system developed by DES-Brazil to monitor the quality of the DECam exposures in real time. For every image a user configurable set of CCDs will be processed with a simplified version of the Data Management algorithms. The QR pipeline includes overscan, bias, and flat field corrections and produces a source catalog using the SExtractor algorithm. Several quality checks are performed on a sample of high S/N and good quality flag objects. QR presents a complete log showing the raw and reduced images, the variation of the PSF along the focal plane, the PSF distortion, number counts and sky brightness tests for each exposure. The median, mean and RMS values for each CCD are stored in a database and presented by interactive plots as new exposures are observed. Alarms can be configured with predefined limits for each test and sent to the SISPI alarm system. QR results are sent to the DES Science Portal at Fermilab on a daily basis allowing collaborators to easily evaluate the quality of the data over different time periods as shown in Fig. \ref{fig:sispi-qr-history}. QR is available for both DES survey operations, community observing and for remote observers with authorized access.

Additional quality assurance tools available to the observer include a real time display that automatically shows a down-sampled image of every exposure and the observer workstation. SISPI will copy every image to the observer workstation where it can be accessed by standard tools such as ds9 or iraf for an in depth interactive analysis. The observer can choose to process an image with his private, customized algorithms. These operations are completely decoupled from the SISPI image pipeline and do not affect the overall throughput.

\subsection{Performance}
During the first 1.5 years of operation SISPI performed very well. Our design choices to use Python, open source middleware, web-based GUIs using HTML5, websockets and JavaScript,  and a distributed architecture proved to be up to the task. The observing experience provided by SISPI has been well received by observers from both the DES collaboration and the community program. The system was ready in time for DECam commissioning and we have achieved our performance goals. System reliability and availability improved steadily. The number of support calls declined rapidly; only 1.5\% \ of the total observing time (13 hours out of 888 hours) available during the first year~\citep{htdY1} of the Dark Energy Survey were lost due to SISPI problems.  For comparison, 90 hours were lost due to bad weather (10\%) and 18 hours were lost because of problems with the telescope and infrastructure support (2\%). Losses of observing time due to issues with DECam systems excluding SISPI added up to another 13 hours (1.5\%). These are all excellent reliability numbers for a new instrument in its first year of operation. The second DES observing season has been completed, with negligible downtime due to either DECam or SISPI.  

\clearpage
\section{Calibration Systems}\label{s8:Calibration}

%
%

The Dark Energy Survey requires several systems devoted to calibrations separate from DECam itself.  We supplied a new flat-field screen located inside of the Blanco dome and new light-emitting diode projectors for each filter band-pass, a separate powerful spectrophotometric flat-field projector that allows us to measure the relative throughput in narrow wavelength bands, and three separate systems that determine whether or not a given night is photometric.  This section describes the hardware for those systems. 

\subsection{Flat Field System}
The DECam calibration system, DECal, produces broadband flat fields that can be acquired daily to correct the pixel-to-pixel variations across the DECam detectors. The DECal system includes a new highly reflective Lambertian dome flat field screen that is illuminated by an array of high-power light-emitting diodes (LEDs). One LED has been selected to illuminate each filter bandpass used in the DES survey with a wavelength centered on the filter bandpass when possible.  In addition, a ``white'' LED provides illumination across multiple filters in the blue part of the spectrum.
More information about the daily dome flat field system can be found in \citep{jenetal}.

\subsection{Spectrophotometric Flat Field System}
DECal also includes a spectrophotometric calibration system that is used to measure the relative throughput of the complete telescope + instrument system as a function of wavelength and position on the focal plane. The spectrophotometric calibration is accomplished by imaging nearly monochromatic light (1 - 10nm bandwidth) incident on the flat field screen with DECam while at the same time monitoring the amount of light on the screen with NIST calibrated photodiodes placed around the top ring of the telescope in four separate stations. The signal received by the photodiodes is proportional to the intensity of light illuminating the flat field screen. The DECam images are compared with the photodiode output to determine the relative sensitivity of the entire telescope + instrument optical system as a function of wavelength and focal plane position. Note that the system provides a relative (not absolute) throughput measurement. The full spectrophotometric characterization (a scan versus wavelength for one or more filters) is performed  several times per year. As the fluxes of the nearly monochromatic reflected light are low and it is impossible to make the inside of the dome completely dark during the day time, these measurements are performed  during cloudy nights. The data are used to spectrophotometrically calibrate the DES survey data, and to monitor the DECam instrument's sensitivity as a function of time. More information about the DECal spectrophotometric calibration system can be found in \citep{SPIErheault2010, SPIErheault2012}.

\begin{figure}[h]
\begin{center}
\includegraphics[scale=0.5]{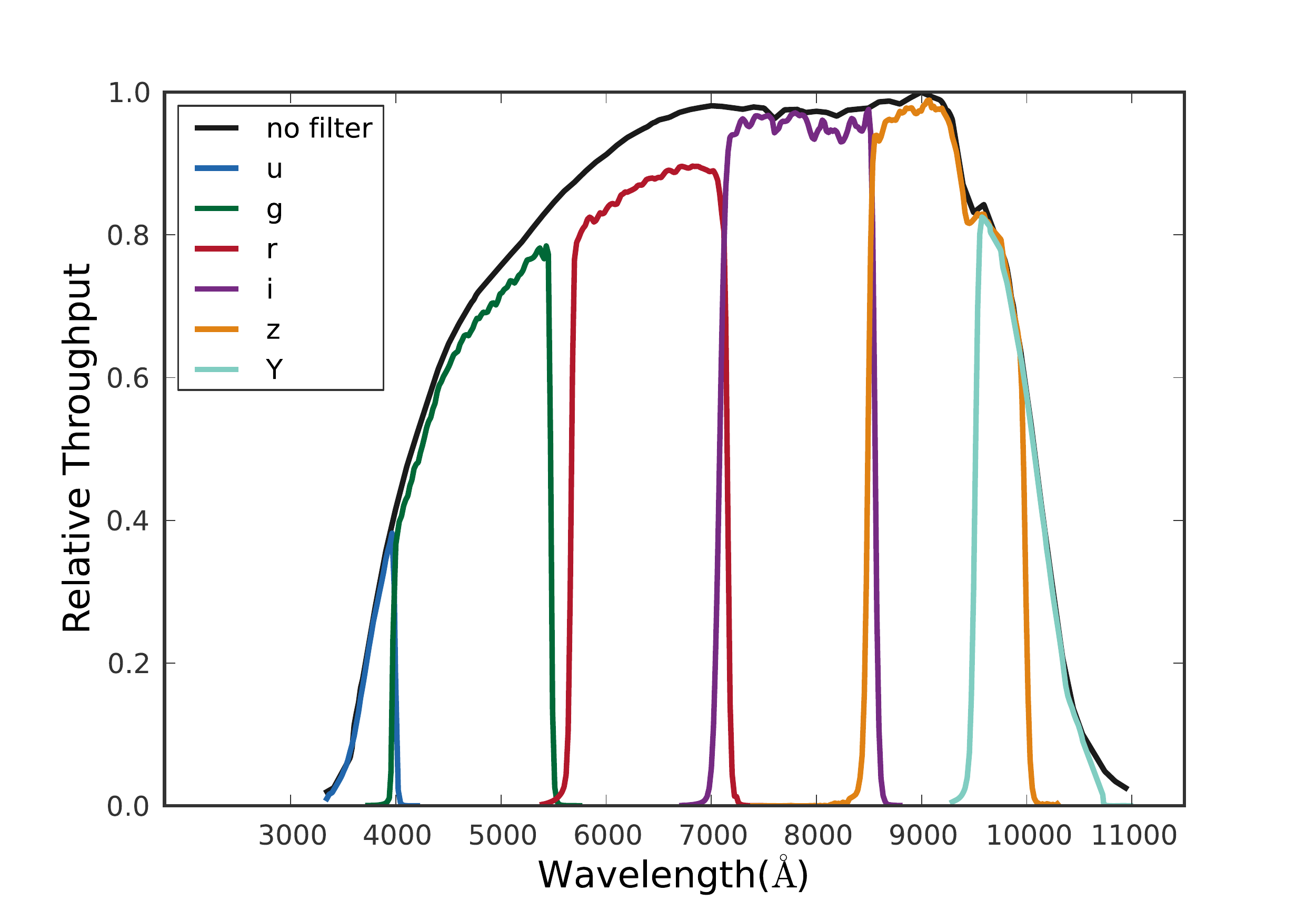}
\caption{Throughput as a function of wavelengths of the DECam optical train, including the various filters DES uses. These throughputs are calculated relative to the use of no filter at 9000 Angstroms. }
\label{fig:des_ugrizY} 
\end{center}
\end{figure}

The DECal system has been used to perform multiple spectrophotometric calibration scans of the DECam system since its commissioning in late 2012. Figure~\ref{fig:des_ugrizY} shows the results of DECal scans obtained during September-November 2013.  As the DES survey progresses, the dataset will be examined to monitor potential changes in the overall throughput by comparing these baseline measurements with future scans.

\subsection{RASICAM} \label{subsec:rasicam}
Meeting the goal of 1\% photometric precision requires careful monitoring of the observing conditions to determine whether or not a given night is photometric. 

The DECam project has supplied the Radiometric All-Sky Infrared CAMera (RASICAM), an all-sky infrared camera system~\citep{SPIElewis2010,SPIEreil2014}. RASICAM monitors the sky using the wavelength range $9.7 < \lambda (\mu {\rm m}) < 12.5$. In this wavelength range, relatively warm clouds are easily distinguished from cold clear skies. RASICAM can distinguish a sky with thin, high-altitude cirrus clouds from truly photometric conditions. In humid conditions a strong air mass signal is also easily seen.

RASICAM consists of a FLIR A325 thermal IR $320 \times 240$ pixel microbolometer camera embedded in a two-mirror optical system. The mirror system provides $2\pi$ sky coverage from the horizon to 76 degrees elevation with 80\% reflectivity. An automated composite enclosure protects the optics from weather during inclement conditions and from daylight. Each night, the camera is opened and closed by the CTIO telescope operators via a web interface. Every 90 seconds each night RASICAM collects an integrated image of sky conditions. The online, realtime image processing algorithm calibrates, background-subtracts and normalizes the thermal image. The analysis software makes an evaluation as to whether the sky is photometric at the time of the image.  Both sky variability and IR power are provided. This includes a local measure of the sky in the direction the Blanco telescope is pointed as well as the global full sky measures. The resulting image and analysis are shared with observers on the mountaintop via web interface and local graphical interface. The summary variables are also sent to the telescope control system (TCS) and included in the FITS headers of DECam images.

\begin{figure}
\begin{center}
\begin{tabular}{|c|c|c|}
\hline
\includegraphics[width=0.2\textwidth]{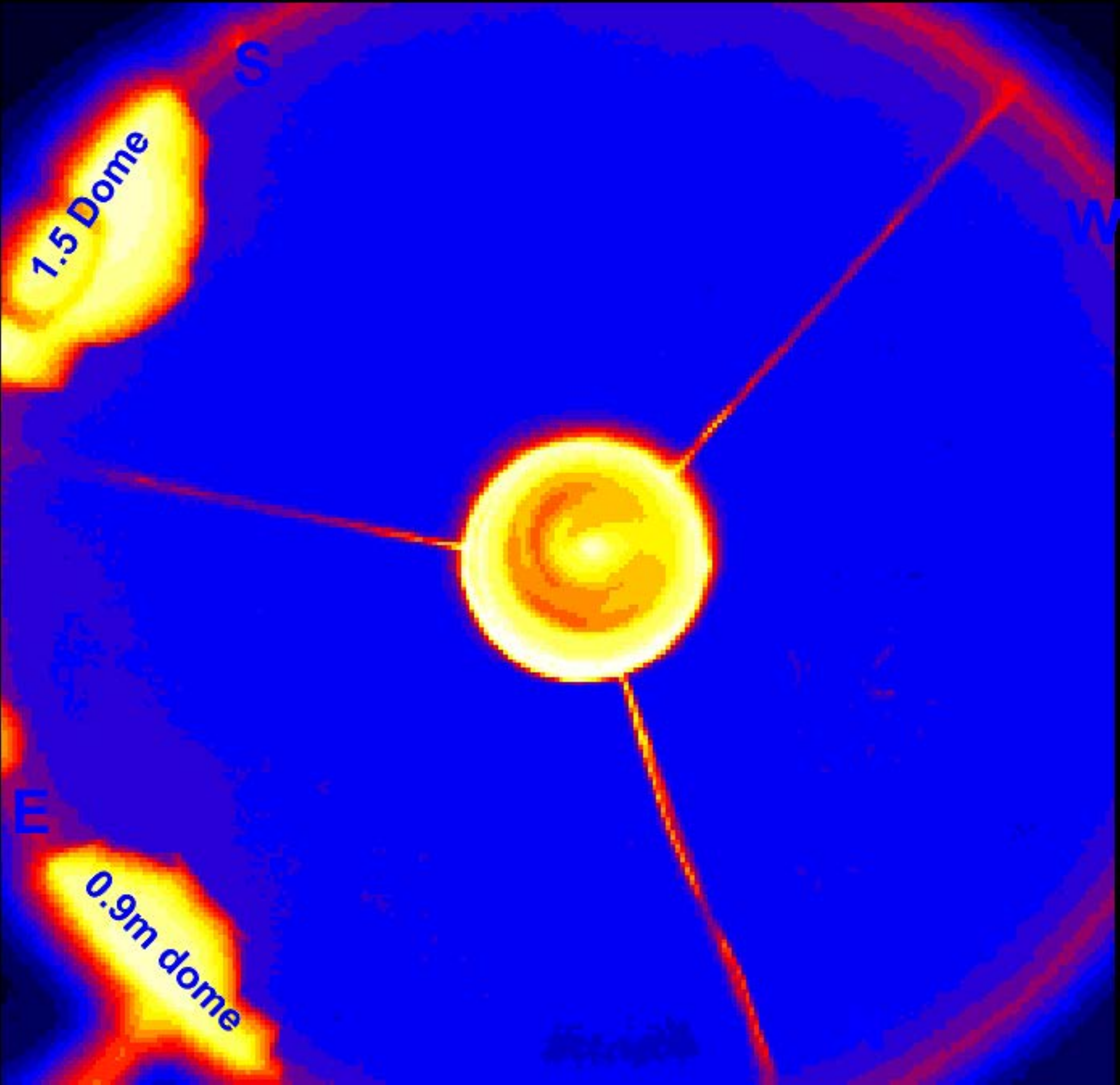} &
\includegraphics[width=0.2\textwidth]{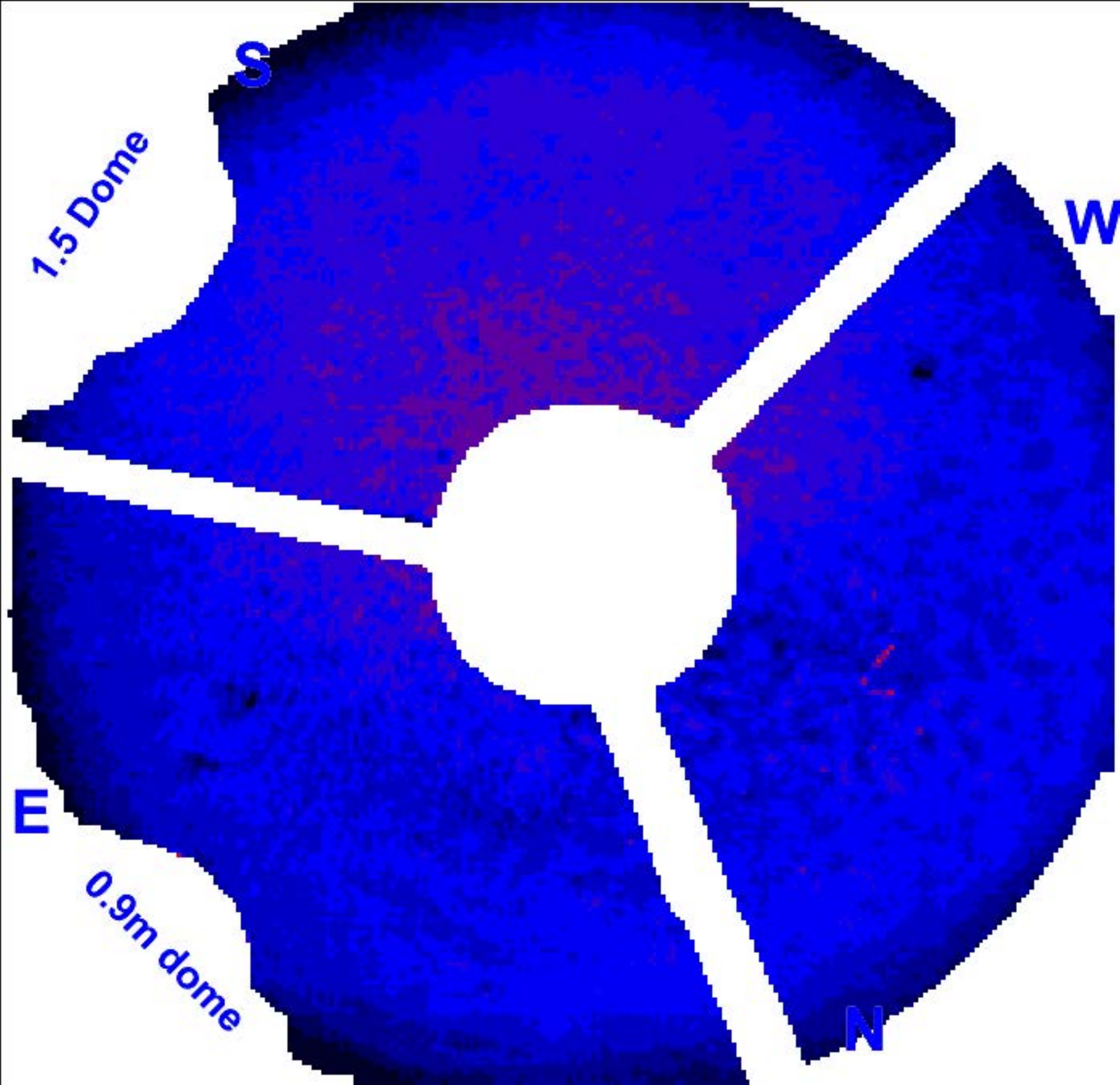} &  \includegraphics[width=0.2\textwidth]{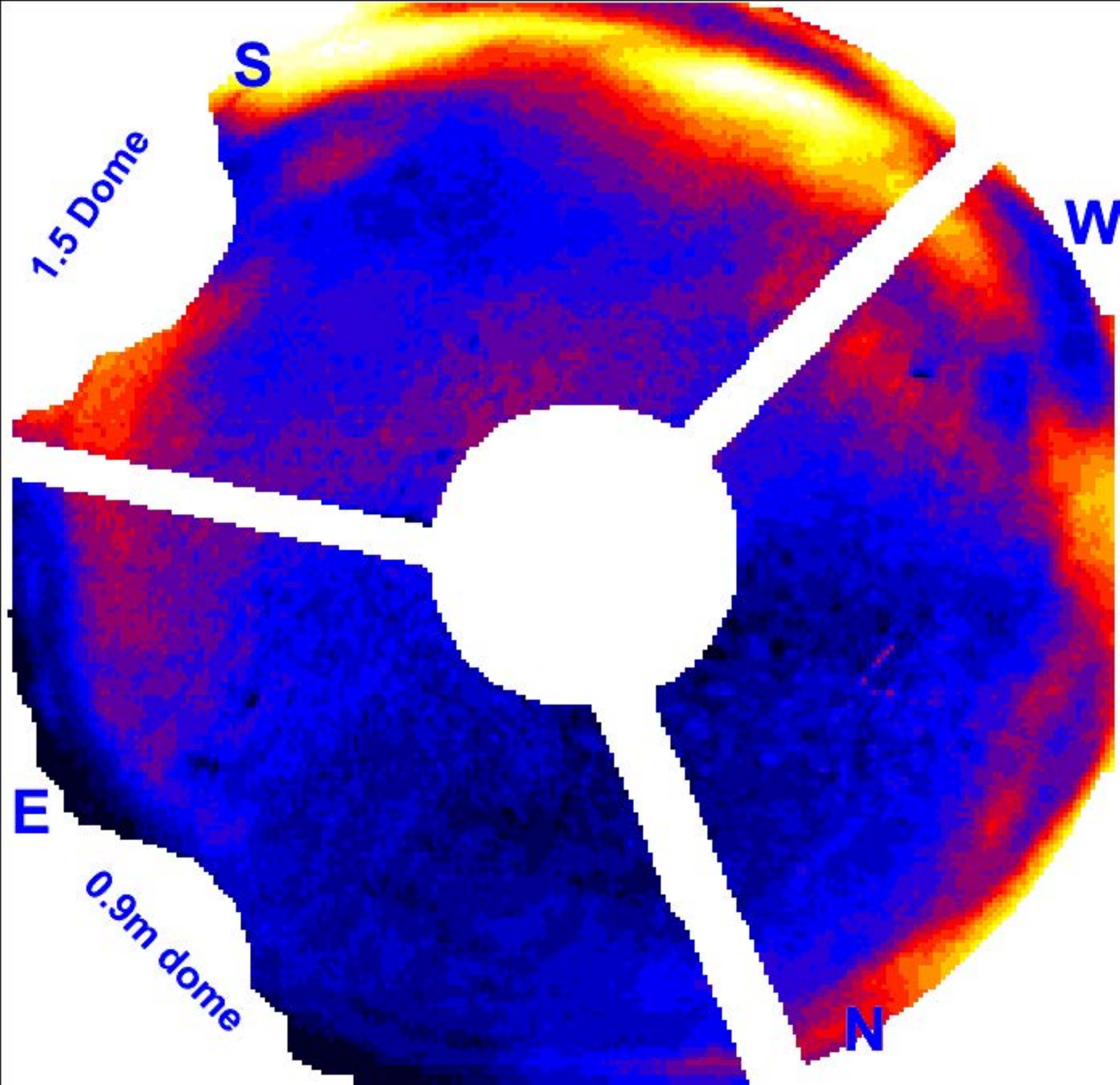} \\
\hline
a. Raw Image of & b. Clear Sky        & c. Partly Cloudy \\
Clear Sky             &                             & Sky \\
\hline \hline
\includegraphics[width=0.2\textwidth]{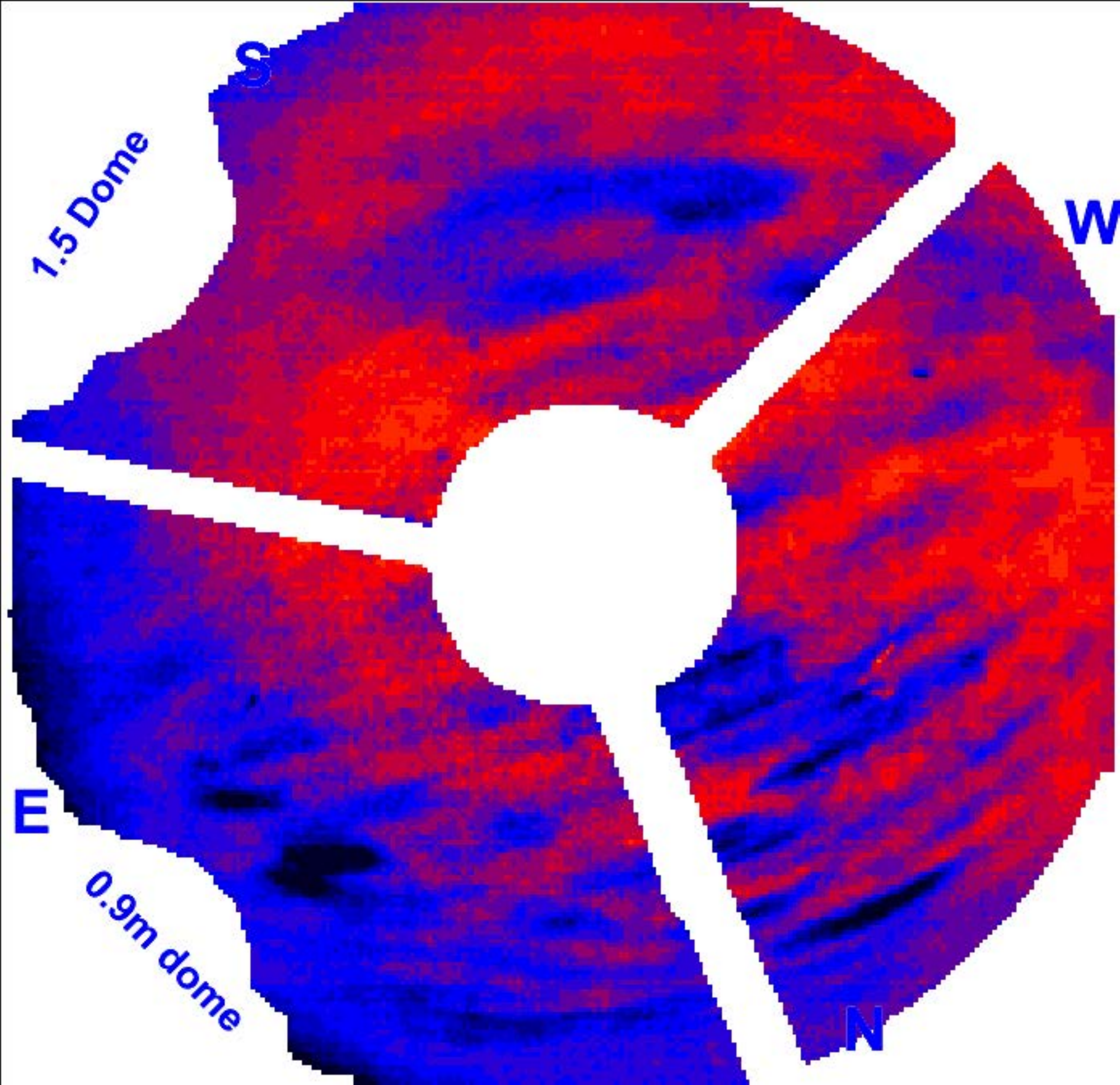} & \includegraphics[width=0.2\textwidth]{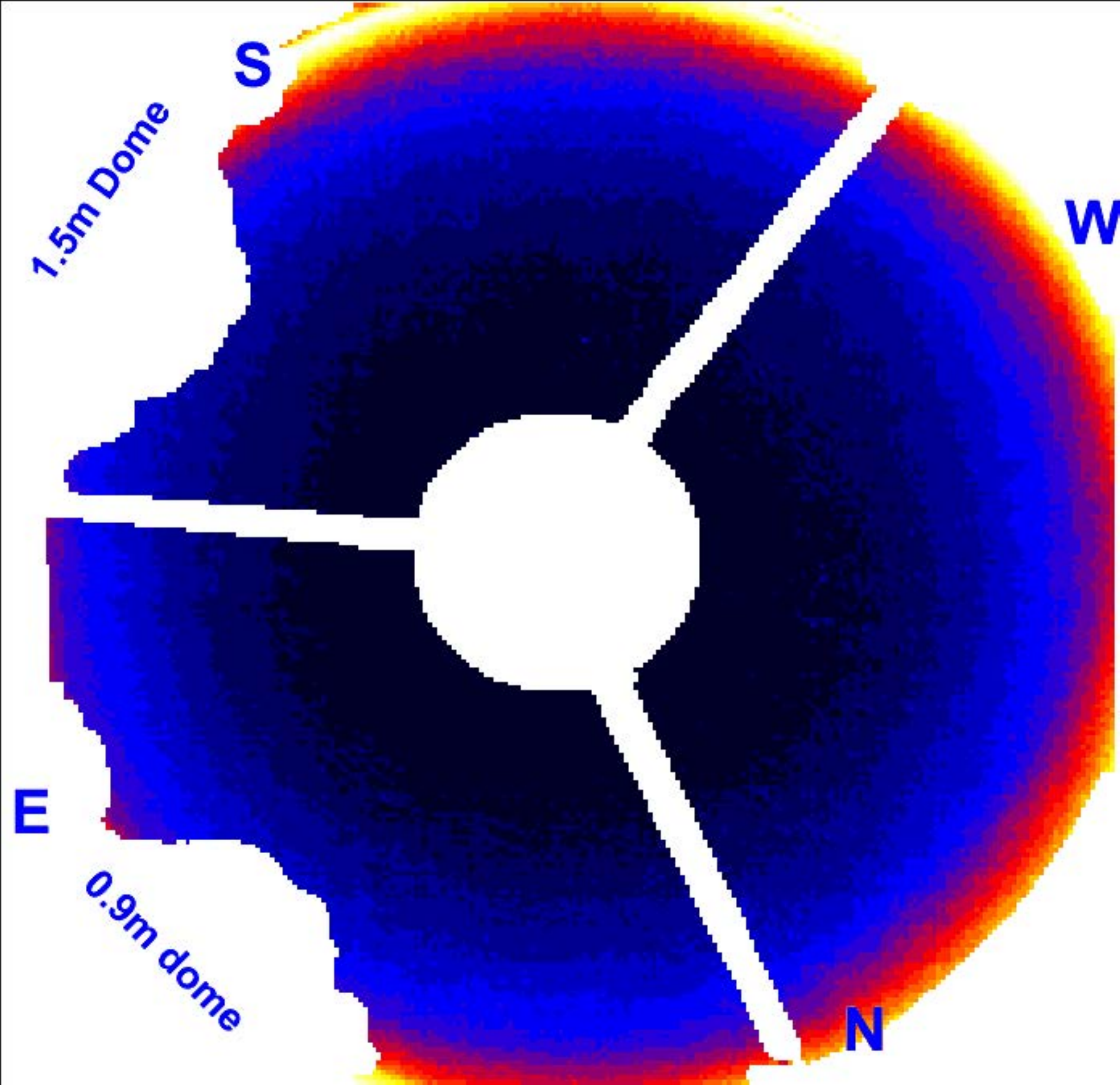} & \includegraphics[width=0.2\textwidth]{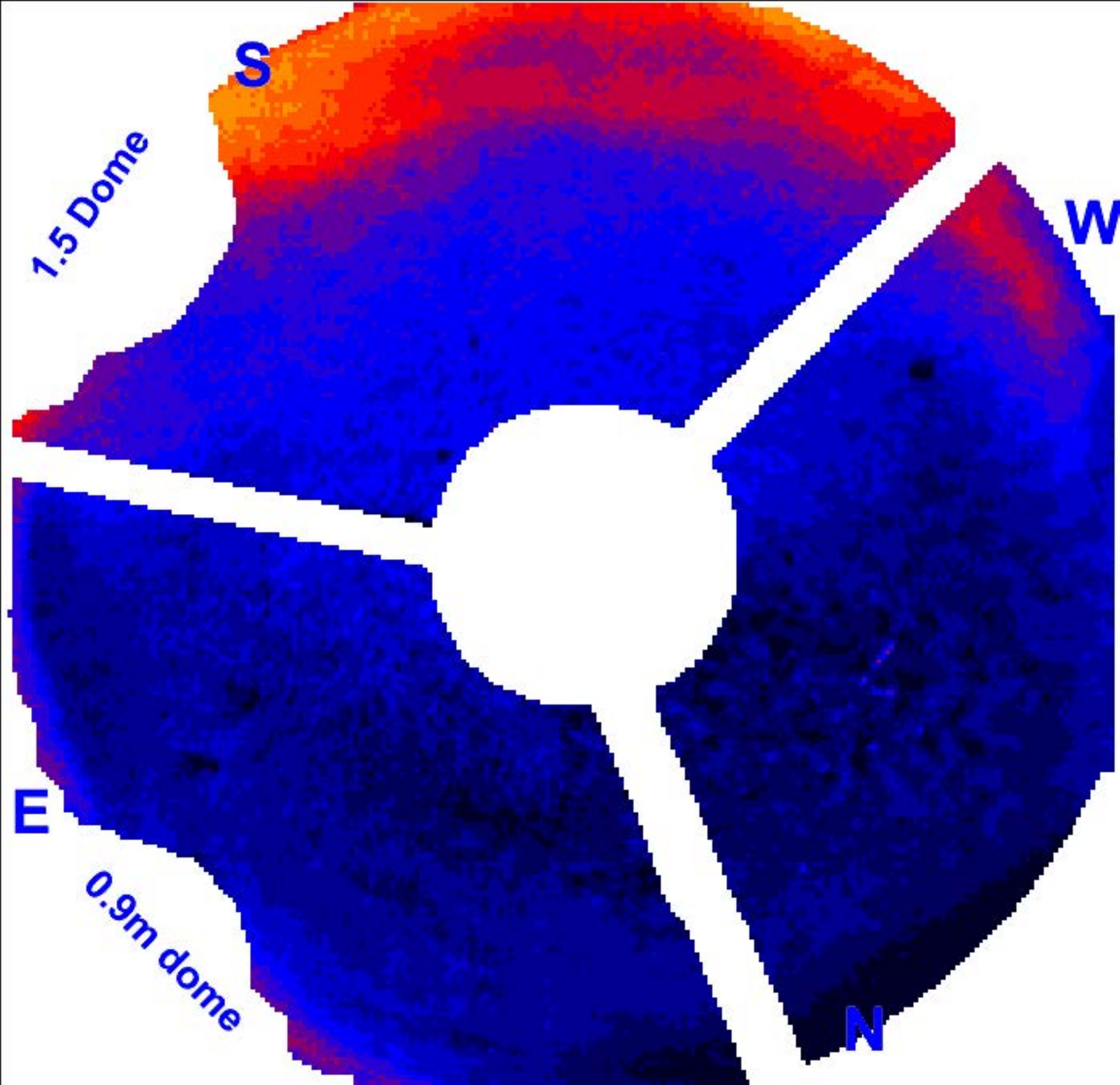} \\
\hline
d. Cloudy Sky & e. Humid Sky & f. Clouds to \\
                       &                      & South \\
\hline
\hline
\includegraphics[width=0.2\textwidth]{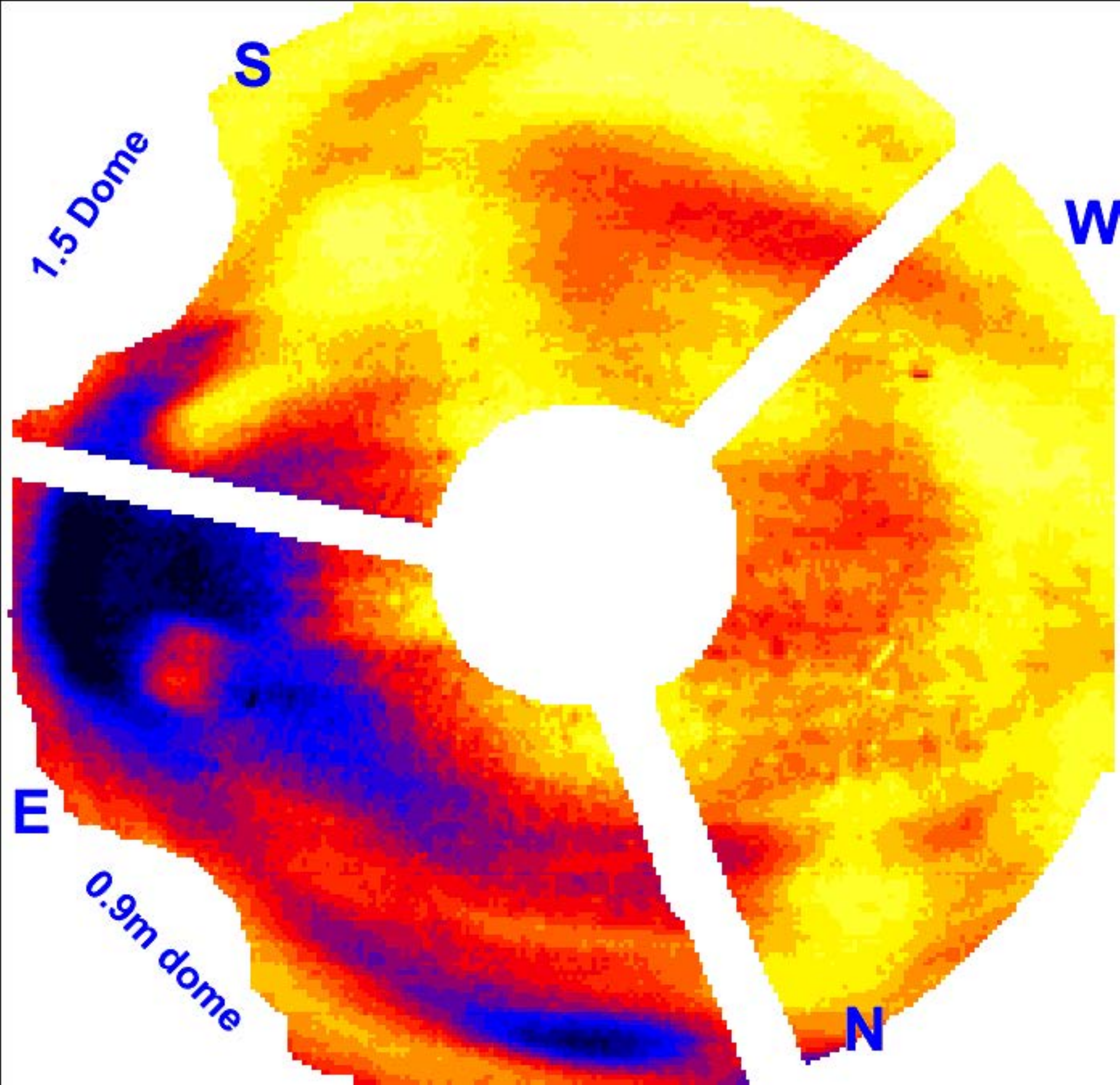} & \includegraphics[width=0.2\textwidth]{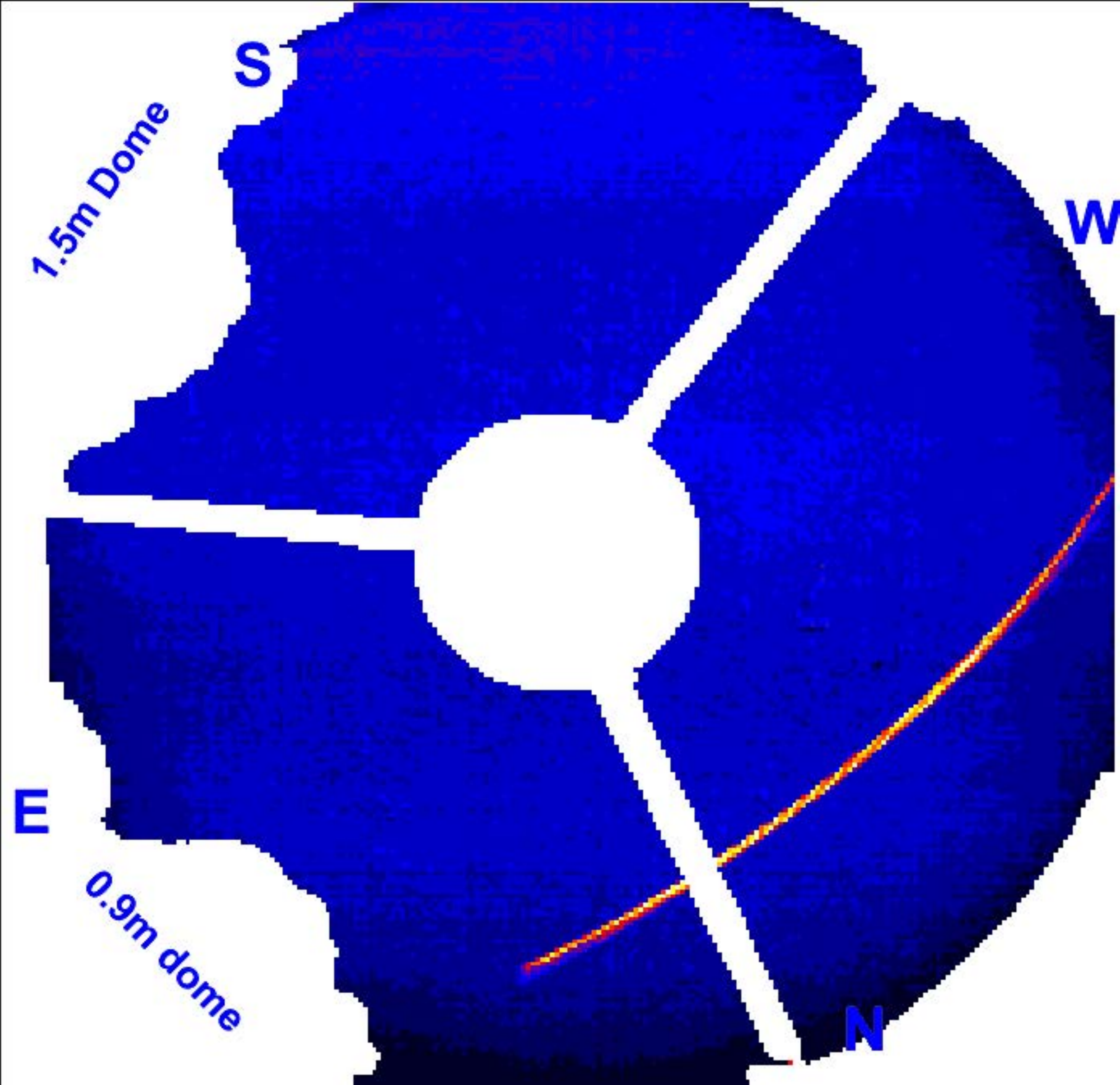} & \includegraphics[width=0.2\textwidth]{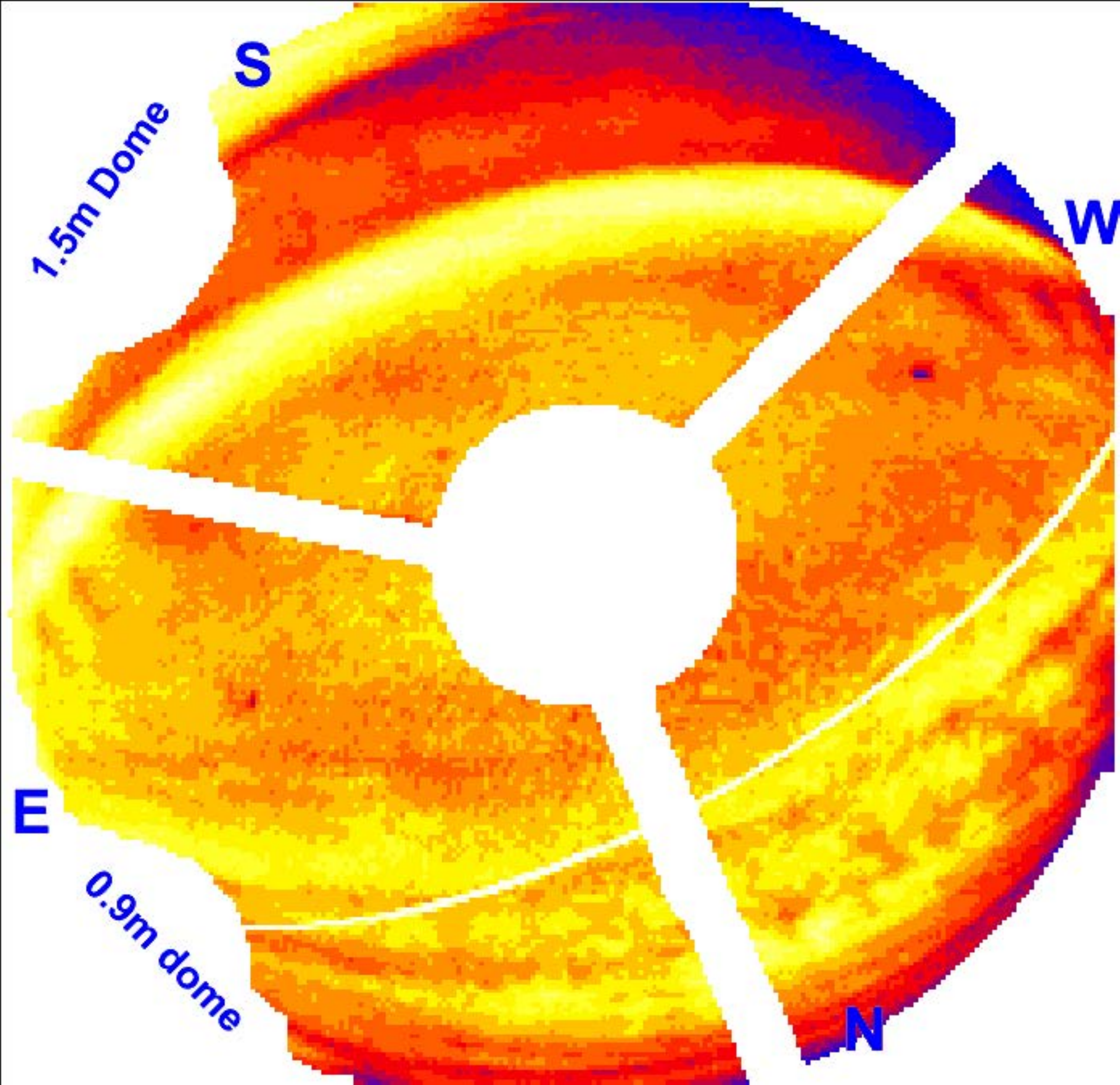} \\
\hline
g. Overcast Sky & h. Clear Night & i. Cloudy Night \\
& max(I(x,y)(t)) & max(I(x,y)(t)) \\
& Note moon track & E/W winds \\
\hline
\end{tabular}
\end{center}
\caption{Infra-red images from RASICAM under various sky conditions.  The image (a) in the upper left is taken under clear sky condtions. One can identify thermal sources such as the domes on the mountaintop. The secondary mirror at the center and its three-armed support are identifiable.  This background is subtracted from the other images that are shown.  The obvious sources in the background exposure are masked from the other examples.   Thus, the example (b) of a clear sky shows very little IR signal.  Example (c) shows a partly cloudy sky; yellow indicates more water vapor than does red. Example (e) shows a humid sky; there is more water vapor seen as one looks through more atmosphere in the directions  towards the horizon. Example (h) and (i) shows the maximum of the intensity in each pixel over the course of the night. }
\label{fig:rasicam-9-color} 
\end{figure}

Example images under various sky conditions are shown in Fig.~\ref{fig:rasicam-9-color}. The same figure also shows two summary images where the maximum intensity seen in each pixel over an entire night are shown. The images and results of the analysis are stored online for further use. A seasonal reprocessing of the RASICAM data is also provided to the DES calibration group to assist in DES data processing.  Additionally, each morning a movie of the prior night's sky images with summary plots is uploaded to a public server  to allow maximum community access to the data~\footnote{Rasicam Server \url{http://www.youtube.com/user/rasicam2}}.


\subsection{GPSMon}
In order to provide a cross-check of the amount of precipitable water vapor (PWV) in the atmosphere, DES has supplied a high-precision dual-band Global Positioning System (GPS). PWV causes a shift in arrival time of the GPS signal via the increased index of refraction~\citep{blakeshaw}.  A Trimble NetR9 GNSS receiver and GPS antenna mounted on the exterior of the 1.5-meter telescope building on CTIO has sufficient sensitivity to provide a PWV measurement.  The data are collected 24 hours per day and sent to a centralized Suominet\footnote{For details on Suominet see \url{http://www.suominet.ucar.edu/}} processing center. PWV results are made public within a few days.

\subsection{aTmCam}
The Atmospheric Transmission Monitoring Camera (aTmCam) is a new calibration system designed to enable the improved photometric calibration of data acquired by the DECam. aTmCam is housed in a new dome on the CTIO summit and consists of a Paramount telescope mount and four small telescopes (commercial photographic camera lenses) and CCD detectors each with a different narrow-band filter that monitor the brightness of suitable standard stars throughout the night.  Each narrowband filter is selected to monitor a different wavelength region of the atmospheric transmission, including regions dominated by the precipitable water vapor and aerosol optical depth. The colors of the stars are measured by this multi narrow-band imager system simultaneously. The measured colors, a model of the observed star, and the measured throughput of the system can be used to derive the atmospheric transmission of the site on sub-minute time scales.  A full description of the aTmCam system can be found in Refs.~ \citep{SPIEtingli2012,SPIEtingli2014}.

\clearpage
\section{Assembly and Installation}\label{s9:Installation}

The assembly of the camera required coordination between all of the members of the camera and telescope project teams.  While the final installation at CTIO took place in 2011 and 2012, the planning and preparations began well before that.  The work was challenging and was done with a great deal of review and care.  This section presents a summary.  

\subsection{Early Preparations at CTIO}
As part of the preparations for DECam, CTIO undertook improvements~\citep{abbott-inst} to the infrastructure and the telescope facility.  The major elements of that are described.

The telescope's primary mirror has 24 radial supports equally-spaced around the edge and glued into place. These supports provide a counterweight that holds the primary mirror in the center of the mirror cell even as the telescope is moved in declination.  A longstanding problem was that the supports broke away from the mirror at a rate of one or two a year.  When that happened the mirror tended to have unpredictable lateral movements. These supports were redesigned so that they would not peel away from the mirror. They were all replaced in 2009. The mirror has has had repeatable movements since that time, now compensated-for by motions of the hexapod.  The primary mirror was re-aluminized in March 2009.

The Blanco computer room and control rooms on the ground floor were both relocated, enlarged, and modernized. The new rooms have raised floors and active temperature control. The new rooms were commissioned in April 2011.  A 15' by 20' enclosed clean room was installed in the Coud\'e Room on the main floor of the telescope dome.  A new telescope control system (TCS) was installed~\citep{SPIEwarner2012} in 2011 so as to remove obsolete components. An area in the ground floor garage space was converted into a dust-free space suitable for final assembly of the cage and corrector and the integration with the hexapod. This included the addition of a 15-ton overhead crane.  The roof of the old control room on the upper level of the dome was reinforced so that it could support the camera's LN2 tank.

A new 40T glycol chiller was installed outside of the dome and the glycol circulation system was upgraded. A new air compressor and dessicator was installed for those systems that require dry compressed air.

\subsection{Preassembly and Testing at Fermilab and University College London}
DECam was first assembled and tested at Fermilab on a full-scale reproduction of the telescope top-rings and fins that we called the ``Telescope Simulator"~\citep{SPIEdiehl2010,MSS-TIPP}. The Telescope Simulator had pitch and roll capability allowing us to position the camera in any orientation that it might meet on the Blanco.  A copy of CTIO's secondary mirror handling ``Northwest" platform was manufactured and converted into the Imager Handling System for DECam. A new f/8 handling system and platform was designed at Argonne National Laboratory, necessary because the Blanco's inner top-end ring would no longer be able to pivot 180 degrees on its axle (thus facing the f/8 mirror towards the primary mirror) after DECam was installed.

The initial use for the telescope simulator was to develop the procedures for using the f/8 handling system to install and remove a dummy f/8 mirror and the counterweight that is normally mounted on the bottom of the of Prime Focus Cage when the f/8 is not in place there.  The new f/8 platform was the first component of the DECam project that was sent to CTIO. It was installed in the dome during January 2011.  Next the simulator was used to develop the cage installation procedure and to gain experience with the mechanically over-constraining fins that connect the cage to the inner top ring of the telescope. 

Meanwhile the lenses were installed in their cells and carefully aligned using a rotary table at University College London.  The lens plus cell assemblies were then inserted into their respective barrel sections, the cells having been previously aligned in the barrel sections using a coordinate measuring machine (CMM) at Fermilab. These assemblies were also checked using the rotary table at UCL. They were pinned so that the positioning could be reproduced. The barrel sections were separated  from each other and then sent to CTIO. The rotary table was sent with them so that the optical alignments could be verified before the barrel sections were reassembled. 

Assembly and testing continued at the Telescope Simulator. A full camera assembly was made, using an identical copy of the barrel  with dummy weights in place of the optical elements. The hexapod was used to unite the ersatz barrel with the cage. The final imager Dewar held engineering grade CCDs in place of the final, science grade CCDs.  A flat-field generator and star pattern projector~\citep{SPIEjhao2010} were used as the light sources. The emphasis was on integration and  tests of the readout electronics, the filter-changer (without using the final filters, which were still being manufactured), shutter, hexapod, the LN2 cooling system controls, mechanical specifications such as the stiffness of the barrel and the flatness and alignment of the focal plane, and tests of the software control systems (SISPI). These tests were completed by May 2011. After that all of the camera components were sent to Chile apart from the imager Dewar, which was retained at Fermilab for testing while we loaded it with the science grade CCDs.  The imager Dewar was the last major component to be shipped~\citep{SPIEderylo2012}, arriving in December 2011. 

\subsection{Final Assembly at CTIO}
CTIO led~\citep{abbott-inst,freddy-inst} the camera installation with intense and active participation by the DECam design and construction team. SISPI was installed on the new DECam computers in April 2011. From June to August 2011 the camera infrastructure was installed into the Blanco Dome, including the LN2 system that cools the CCDs to operating temperature, and the chillers for the readout electronics.  The camera components arrived in  December and were staged for installation. The filter changer, shutter, and hexapod were individually retested using  the SISPI controls.  

In January 2012 the optical corrector, which had arrived at CTIO, was assembled in the dust-free tent on the ground floor. The alignment of the components was tested using the rotary table.

\begin{figure}[h]
\begin{center}
\includegraphics[scale=0.5]{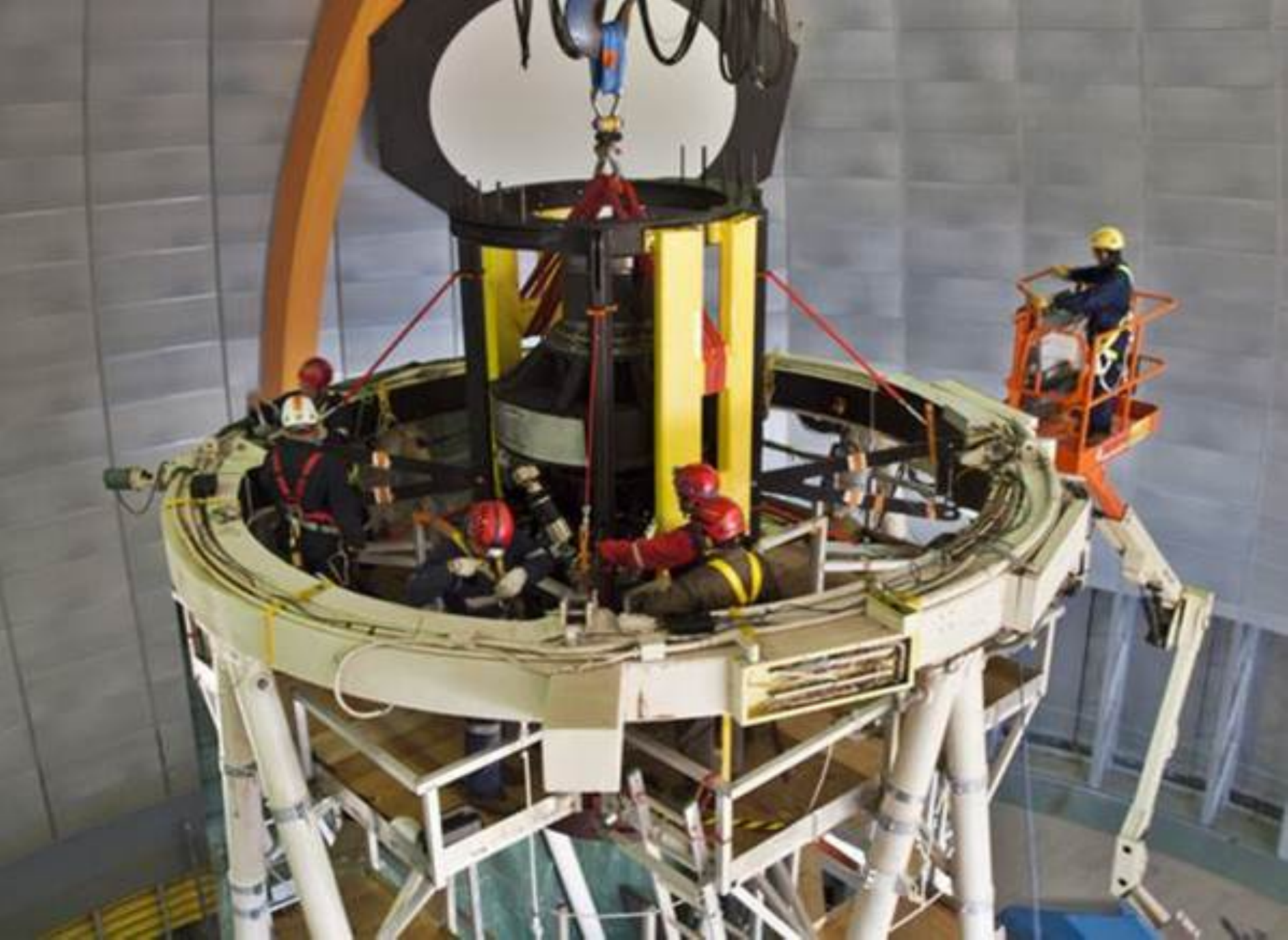}
\caption{The DECam cage and barrel assembly being installed at the top of the Blanco in May 2012. The cage is picked up from and stiffened by two (yellow-colored) strong-back frames clearly visible in the photograph. The technicians are standing on platforms designed and installed for working on the cage installation. }
\label{fig:CageInsertion} 
\end{center}
\end{figure}

In early March 2012 the telescope was moved to zenith and locked into place. Work platforms were installed in the upper part of the telescope that enabled access to the fin connections to the telescope rings.  The primary mirror was removed and set aside, with protection, on the main floor of the dome. Then the old Prime Focus cage was lowered to the floor and removed. The new prime focus cage was assembled using the copy of the barrel mentioned above and a dummy imager. That assembly was used to test the installation and fitting procedure and then removed. The cage attachment legs (fins) were reused from the original cage support.  In May the DECam Cage, hexapod, and barrel (including the optics), but using the dummy imager instead of the real thing, were raised up the center of the telescope and attached. Figure~\ref{fig:CageInsertion} shows the new cage being inserted into the telescope upper structure. Engineers and technicians from NOAO in Tucson made up part of the team for the cage installation work. The primary mirror (with its cell) was returned and then the cage was optically centered from the Cassegrain hole in the center of the primary. Because DECam is more than 2500 kgs heavier than the previous camera and cage assembly, and has an asymmetric 
 moment arm due to the LN2 cable wraps on the SW side, maintaining telescope balance around the RA/DEC axis required additional weight to be added to the Cassegrain Cage. A total counterweight of 5600 kg was needed, requiring a new Cassegrain Cage~\citep{SPIEschurter2014} with a reinforced design. Lead (Pb) and steel plates were used as the counterweight material. The new Cassegrain cage was installed in June.  The telescope remained pointing at zenith throughout this work. 

\begin{figure}[h]
\begin{center}
\includegraphics[scale=0.3]{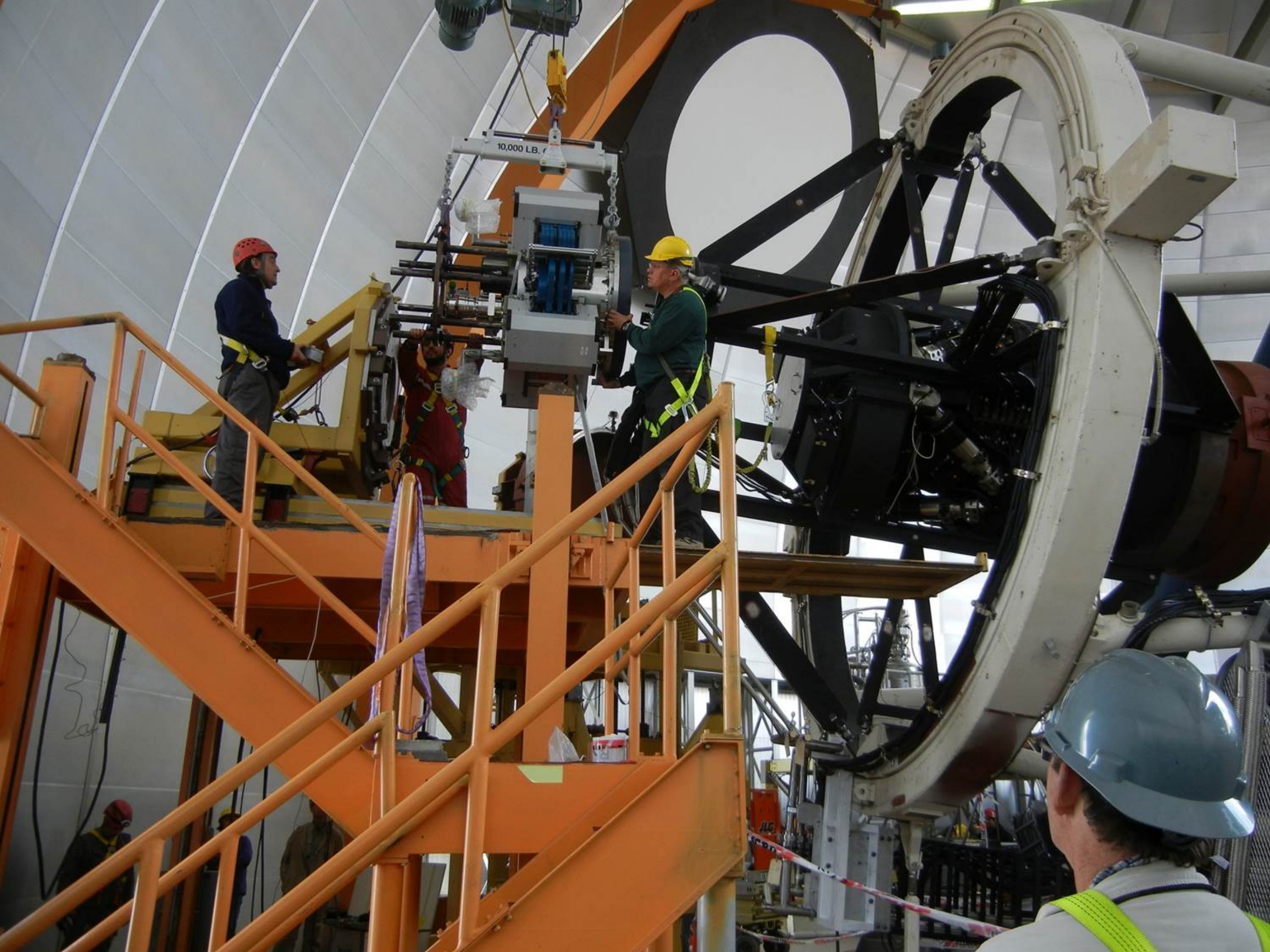}
\includegraphics[scale=0.3]{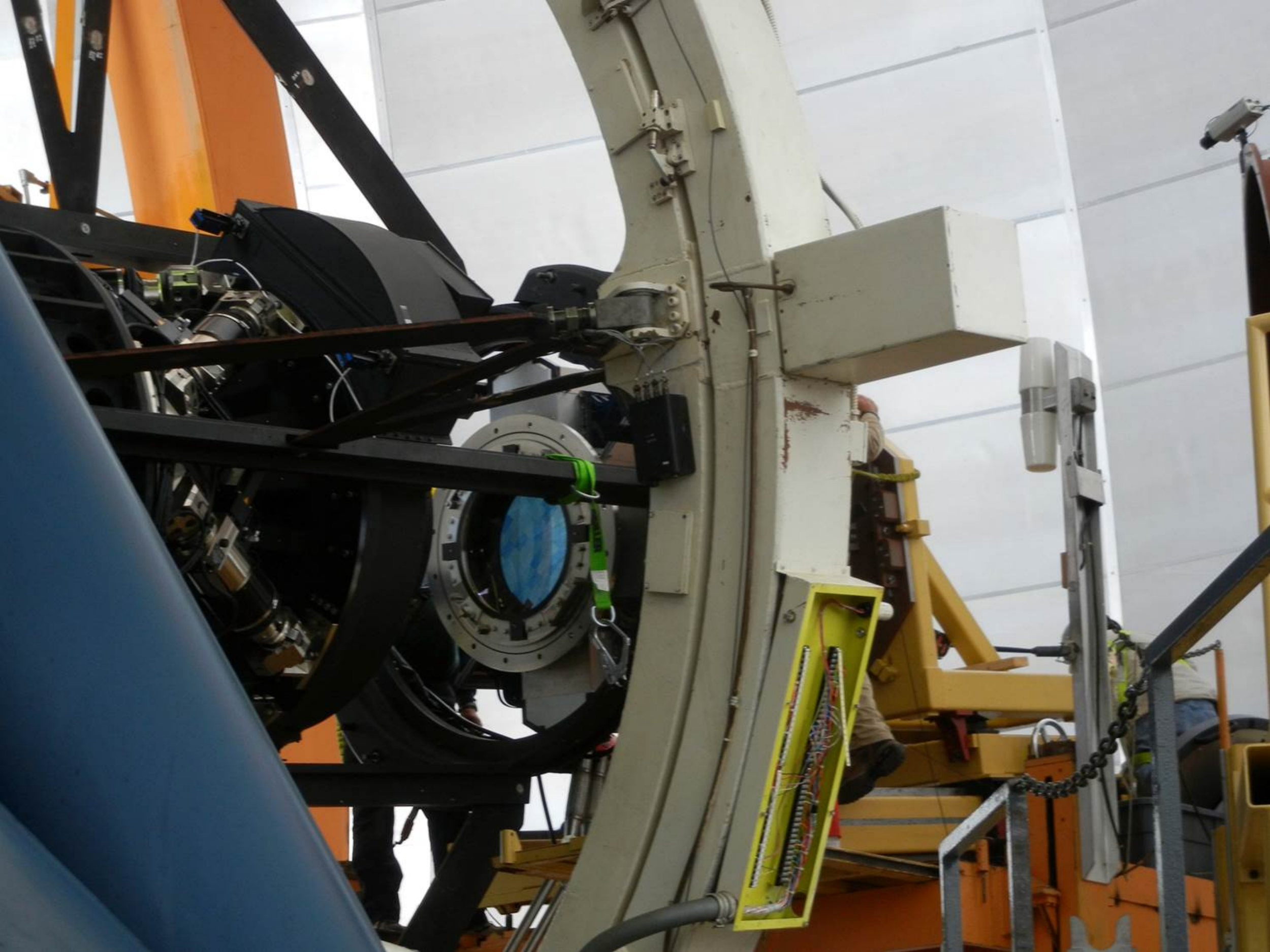}
\caption{The imager was picked up using the dome crane and fastened to the Imager Mounting System (IMS) on top of the Northwest Platform (orange-colored steel structure). Rods screwed onto the back of the Imager were bolted to holes on the IMS (left). The imager was then (right) drawn up to and attached to the barrel. Then the mounting rods and frame were detached from the imager.  }
\label{fig:mountcamera} 
\end{center}
\end{figure}

\begin{figure}
\begin{center}
\includegraphics[height=3.3in, width=3.3in]{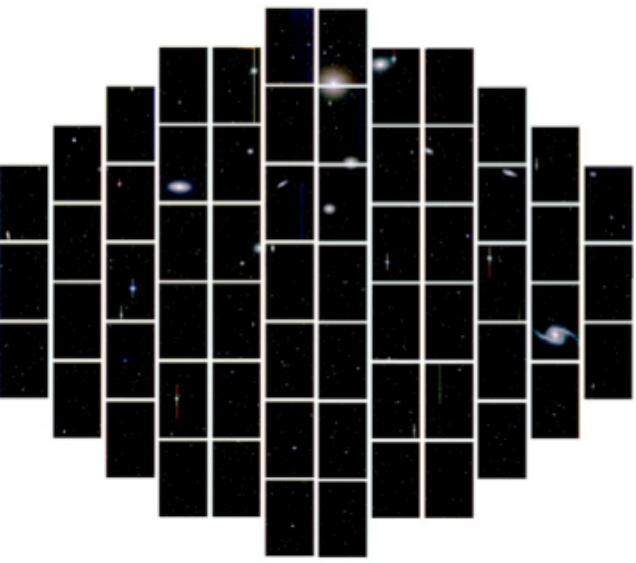}
\includegraphics[height=2.5in]{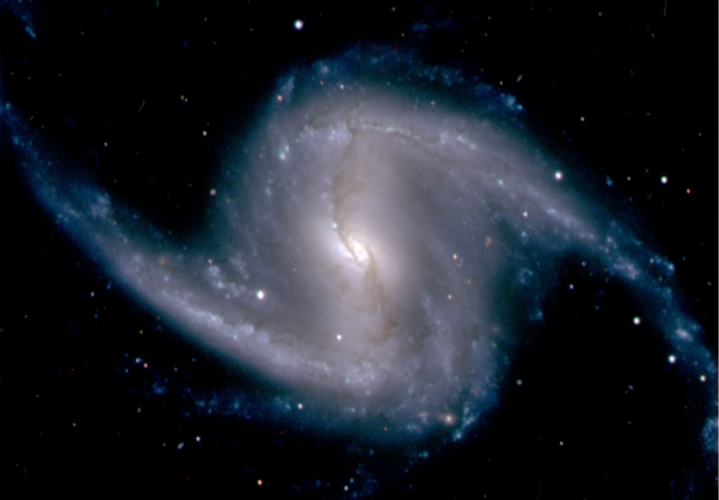}
\caption{ First light night images of the Fornax galaxy cluster (upper)  and NGC1365 (lower) from DECam on the Blanco Telescope.}
\label{fig:firstlight1and2}
\end{center}
\end{figure}

After the new cage was installed the telescope was moved to the Northwest Platform. Over the next couple of months the telescope was dressed with the vacuum-jacketed liquid nitrogen lines, water-glycol, vacuum, compressed-air and other utilities. The imager and readout electronics, which had been under continuous test in the Coud\'e Room, were attached to the barrel on August 30, 2012. See Fig.~\ref{fig:mountcamera}. Next the electronic and electrical connections were installed on the back of the camera. ``First Flight" of the camera, when the telescope was actively driven around in the dome in RA and DEC, occurred within a few days. The LED cones, 
 photodiodes and fibre projector units of the calibration system  (DECal - see Section~\ref{s8:Calibration}) were installed on the telescope top ring during early September 2012.  The filter changer and shutter  were installed last.  Official ``First Light" was on the night of September 12 (UT), 2012.  We achieved 1.2\arcsec~seeing across the focal plane within 45 minutes of the start of observations with the new camera. Among the objects observed were the globular cluster 47 Tucanae (NGC 104), the Fornax Galaxy Cluster, and the Small Magellanic Cloud.  Fig.~\ref{fig:firstlight1and2} shows a color-coadd image of the Fornax galaxy cluster.

\clearpage
\section{Summary}\label{s10:Summary}

The Dark Energy Camera (DECam) is a new 2.2-degree diameter field-of-view imager  located at the Prime Focus of the Victor M. Blanco 4m telescope on Cerro Tololo at 2207 meters elevation in the Andes Mountains near La Serena, Chile.  The camera was constructed by the Dark Energy Survey Collaboration, and meets or exceeds the stringent requirements designed for the wide field and supernova surveys for which the collaboration is using it.  That the Blanco is an equatorial design with massive mount and a strong Serrurier truss tube structure, in common with all large telescopes built in the period 1950-1980, allows installation of a much heavier top-end 
 structure than was originally planned. DECam is one of a new generation of very large imagers on moderate to large sized telescopes (PanStarrs~\citep{panstarrs}, HyperSuprimeCam~\citep{hsc}, eventually LSST~\citep{LSSToverview, LSST32giga}) and until LSST the only one in the southern hemisphere. Currently, it has the largest corrector optics, filters, and shutter aperture of any astronomical instrument. 

DECam was designed and built to be an NOAO facility instrument, in order to carry out the Dark Energy  Survey (DES) and community programs.  As of this date, DES has completed the second of its five scheduled half-year seasons.  The camera has performed well and been very reliable~\citep{htdY1} with little downtime. During the nights of the second DES observing season (mid-August 2014 to mid-February 2015), camera systems were operational during 99.7\% of the 930 observing hours available.  The Blanco telescope and associated systems had a similar reliability over that period.  These achievements are the result of considerable efforts to provide reliable  equipment and readily deployable spare parts, as well as  efficient and effective operation. 

This paper provides a technical description of the camera including the  engineering, construction, installation,  and the current status. It includes descriptions of  the five lens optical corrector, the lens cells, and the barrel structures that house them. The DES and DECam filters, the filter changer, the shutter with a 60 cm aperture, are described.  The DECam imager Dewar and the infrastructure that maintains vacuum and cools the CCDs are detailed. The  570 Mpixel  CCD focal plane comprises  62 2k$\times$4k for imaging, and 12 2k$\times$2k 250-micron thick fully-depleted CCDs for guiding and focus.  The CCDs have  $15 \mu {\rm m} \times 15 \mu {\rm m}$ pixels with a plate scale of 0.263\arcsec per pixel. A hexapod system provides state-of-the art focus and alignment capability. The camera is read out in 20 seconds with 6-9 electrons readout noise.   The instrument controls and data acquisition systems have been described, including the user interface, the image pipeline, the interface to the telescope, and the hardware and algorithms used by the guider and active optics focus and alignment system. The calibration and auxillary systems provided with DECam are described. 

It is anticipated that the DECam will be a productive instrument on the forefront  of astrophysical research for years to come, serving not only  the DES Collaboration,  but the whole astronomical community.

\clearpage

\section{Acknowedgements}
The installation of DECam on the Blanco Telescope, and associated preparatory work, took of order 50 person-years of effort.   We would like to thank the CTIO Telescope Operations Group, led sequentially by Oscar Saa, Gale Brehmer and Esteban Parkes, the CTIO Mechanical Engineering Group led by Andres Montane, the KPNO-based  team who assisted in the installation, led by Will Goble, the Telescope Control System project team led by German Schumacher, and the Computer Infrastructure Support Group led by Ronald Lambert.  The NOAO safety officer, Chuck Gessner, together with CTIO safety officers and external specialist consultants, ensured that telescope modifications and DECam installation proceeded safely

Funding for the DES Projects has been provided by the U.S. Department of Energy, the U.S. National Science Foundation, the Ministry of Science and Education of Spain, 
the Science and Technology Facilities Council of the United Kingdom, the Higher Education Funding Council for England, the National Center for Supercomputing 
Applications at the University of Illinois at Urbana-Champaign, the Kavli Institute of Cosmological Physics at the University of Chicago, 
the Center for Cosmology and Astro-Particle Physics at the Ohio State University,
the Mitchell Institute for Fundamental Physics and Astronomy at Texas A\&M University,
Financiadora de Estudos e Projetos, 
Funda{\c c}{\~a}o Carlos Chagas Filho de Amparo {\`a} Pesquisa do Estado do Rio de Janeiro, Conselho Nacional de Desenvolvimento Cient{\'i}fico e Tecnol{\'o}gico and 
the Minist{\'e}rio da Ci{\^e}ncia e Tecnologia, the Deutsche Forschungsgemeinschaft and the Collaborating Institutions in the Dark Energy Survey. The DES participants from Spanish institutions are partially supported by MINECO under grants AYA2012-39559, ESP2013-48274, FPA2013-47986, and Centro de Excelencia Severo Ochoa SEV-2012-0234, some of which include ERDF funds from the European Union.

The Collaborating Institutions are Argonne National Laboratory, the University of California at Santa Cruz, the University of Cambridge, Centro de Investigaciones Energeticas, 
Medioambientales y Tecnologicas-Madrid, the University of Chicago, University College London, the DES-Brazil Consortium, the Eidgen{\"o}ssische Technische Hochschule (ETH) Z{\"u}rich, 
Fermi National Accelerator Laboratory, the University of Edinburgh, the University of Illinois at Urbana-Champaign, the Institut de Ciencies de l'Espai (IEEC/CSIC), 
the Institut de Fisica d'Altes Energies, Lawrence Berkeley National Laboratory, the Ludwig-Maximilians Universit{\"a}t and the associated Excellence Cluster Universe, 
the University of Michigan, the National Optical Astronomy Observatory, the University of Nottingham, The Ohio State University, the University of Pennsylvania, the University of Portsmouth, 
SLAC National Accelerator Laboratory, Stanford University, the University of Sussex, and Texas A\&M University.

\end{document}